%% file: main_arxiv.tex
\documentclass{article}
\usepackage{spconf,amsmath,graphicx}
\usepackage{lipsum}

\include{misc/user_colors}

\usepackage{graphicx}
\usepackage{bm}
\usepackage{booktabs}

\usepackage{rotating}
\usepackage{multirow}
\usepackage{subcaption}

\usepackage{amssymb}

\usepackage[pagebackref,breaklinks,colorlinks]{hyperref}
\usepackage[linesnumbered,ruled,lined,boxed, vlined]{algorithm2e}
\usepackage{algorithmicx} 
\usepackage{algpseudocode}

\usepackage[capitalize]{cleveref}
\crefname{section}{Sec.}{Secs.}
\Crefname{section}{Section}{Sections}
\Crefname{table}{Table}{Tables}
\crefname{table}{Tab.}{Tabs.}
\usepackage{amsthm}
\newtheorem{theorem}{Theorem}
\newtheorem{lemma}[theorem]{Lemma}

\def\wplus{$\mathcal{W}^+$ }
\def\wstar{$\mathcal{W}^{\star}$ }
\def\wplussp{$\mathcal{W}^+$}

\def\wstarc{$\mathcal{W}^{\star}_c$ }
\def\wstarcsp{$\mathcal{W}^{\star}_c$}

\title{Video coding using learned latent GAN compression}

\name{Mustafa Shukor, Bharath Bhushan Damodaran*\thanks{Corresponding author: Bharath.Damodaran@interdigital.com}, Xu Yao, Pierre Hellier}
\address{InterDigital, Inc., France}

\usepackage[utf8]{inputenc}

\usepackage{fancyhdr}
\usepackage{kantlipsum}

\begin{document}
\setcounter{page}{1} 

\maketitle

\begin{abstract}
 We propose in this paper a new paradigm for facial video compression. We leverage the generative capacity of GANs such as StyleGAN to represent and compress a video, including intra and inter compression. Each frame is inverted in the latent space of StyleGAN, from which the optimal compression is learned. To do so, a diffeomorphic latent representation is learned using a normalizing flows model, where an entropy model can be optimized for image coding. In addition, we propose a new perceptual loss that is more efficient than other counterparts. Finally, an entropy model for video inter coding with residual is also learned in the previously constructed latent representation. Our method (SGANC) is simple, faster to train, and achieves better results for image and video coding compared to state-of-the-art codecs such as VTM, AV1, and recent deep learning techniques. In particular, it drastically minimizes perceptual distortion at low bit rates.
\end{abstract}

\section{Introduction}

With the explosion of videoconferencing, the efficient transmission of facial video is a key industrial problem. In addition, building digital worlds raises the question of transmitting facial representations, where perceptual distortion matters more than exact fidelity.  Image compression can be formulated as an optimization problem with the objective of finding a codec which reduces the bit-stream size for a given distortion level between the reconstructed image at the receiver side and the original one. The distortion mainly occurs due to the quantization in the codec, because the entropy coding method \cite{arithm_code, entropy_information} requires the discrete data to create the bit-stream. The compression quality of the codec depends on the modeling of the data distribution close to the real data distribution since the expected optimal code length is lower bounded by the entropy \cite{shanon}. Existing compression methods suffer from various artefacts. Especially at very low bits-per-pixel (BPP), blocks and blur degrade the image quality, leading to a poorly photorealistic image.

Motivated by the appealing properties of the StyleGAN architecture for high quality image generation, we propose a new compression method for facial images and videos. The intuition is that the GAN latent representation associated to any face image is somehow disentangled, and a perceptual compression method should be easier to train using the latent, both for intra and inter coding. In addition, at extremely low BPP, a compressed latent code should always lead to a photorealistic image, hence leading to a compression technique that is perceptually more pleasant. To the best of our knowledge, no work has studied how the latent space of StyleGAN can be used for efficient perceptual compression.

However, we acknowledge that the method relies on the hypothesis that a GAN inversion technique can approximate accurately any input facial image. While this hypothesis is currently a limitation, we note that the performances of GAN inversion and generation methods have significantly improved recently and is still a hot topic \cite{Hu2022,Dinh2022}. We take the leap of faith that the hypothesis will become valid in the near future. In addition, specialized GANs could be fine-tuned for a limited set of faces, leading to a valid GAN inversion in dedicated applications. For such applications of personalized videoconferencing, our system is valid if the fine-tuned GAN is known in advance at the receiver side. Finally, for the scenario of digital copies of our physical world (\textit{e.g.}, metaverse), transmitting facial videos requires perceptual distortions to be minimal, even at the cost of some loss of fidelity to the original face.

For real natural images, a StyleGAN encoder (GANs inversion) \cite{encodinginstyle, wei2021simplebase} have been proposed to project any image onto the latent space of StyleGAN. Our objective is to compress this latent representations. %
{Although, training GANs encoders and generators are computationally heavy, which is not desirable for compression}.

\begin{figure*}[htbp]
     \centering
     \includegraphics[width=0.8\linewidth]{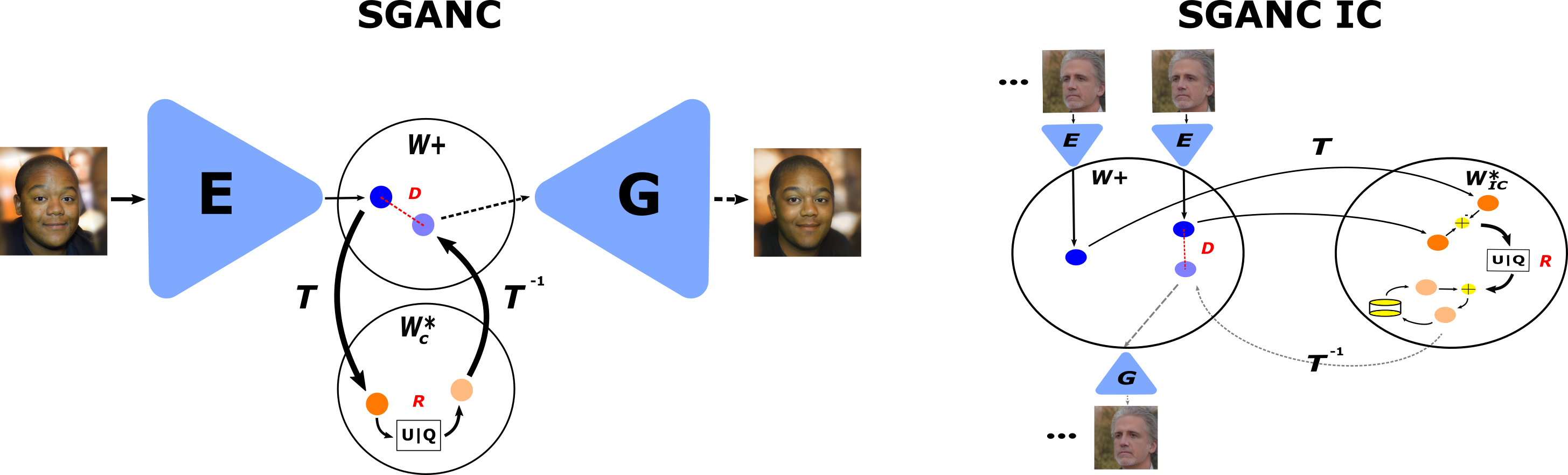}
    \caption{Illustration of the proposed approach. $E$ and $G$ are the pretrained and fixed StyleGAN2 encoder \cite{encodinginstyle} and generator \cite{styleganimporveing} respectively. Only the mapping $T$ and the entropy model are trained using the joint rate (R) and distortion (D) loss. \textbf{On the left} we show SGANC for image compression; On the encoding side, the images are projected in the latent space (\wplussp) then mapped to \wstarc where the quantization/compression is done. On the decoding side, the obtained bit-stream is decompressed, then mapped to \wplus before generating the reconstructed image. \textbf{On the right} we show SGANC for video compression using inter coding (SGANC IC); we quantize the consecutive latent code difference and reconstruct the current latent code from this difference and the last reconstructed latent code.}
   \hfill
    \label{fig:main}
\end{figure*} 
In order to overcome the above challenges, we propose to learn a new proxy latent space representation using a diffeomorphic mapping function. This new latent representation is precisely learned to optimize the compression efficiency. The approach is illustrated in Figure \ref{fig:main}. The advantages of such space are three-fold; (a) it makes the approach very efficient as we can use off the shelf pretrained StyleGAN encoder and generator without the need to retrain them for each quality level. (b) it allows to learn a representation optimal for compression, associated with an efficient entropy model. (c) it allows to obtain high quality photorealistic images thanks to StyleGAN. Finally, we extend this approach to videos and propose a new method to optimize for the inter-coding with residuals in the new latent representation, as illustrated in Figure \ref{fig:main_inter}.

{While we focus in this paper on facial video compression, we think the method we propose here could be extended to different types of scenes. Recently, \cite{sauer2022stylegan} has proposed to extend StyleGAN to ImageNet classes. As a result, learning scene-specific compression paves the way towards semantic perceptual compression methods.}

Our contributions can be summarized as follows: 
\begin{itemize}
    \item We propose a new paradigm for facial video compression, leveraging the generative power of StyleGAN for artifact-free and high quality image compression.
    \item Without any GAN retraining, we propose to learn a proxy latent representation that effectively fits the entropy model, while using off-the-shelf pretrained GAN encoder/decoder models. We propose to learn a model for intra and inter video coding.
    \item We propose a new perceptual distortion loss that is more efficient to compute and leverages the multiscale and semantic representation in the latent space of StyleGAN.
    \item We show high quality and lower perceptually-distorted reconstructed images for low BPP compared to traditional and deep learning based methods for image compression. We show better qualitative and quantitative results compared to the most recent state-of-the-art methods  for video compression.
\end{itemize}
The rest of the paper is organised as follows: section \ref{sec:related works} presents related works on image and video compression. Our proposed method is detailed in section \ref{sec:method} and section \ref{sec:experiments} presents the experimental results.

\section{Related Work}
\label{sec:related works}

\subsection{Image Compression:} Traditional image codecs such as JPEG \cite{JPEG}, JPEG2000 \cite{JPEG2000} have been carefully human-engineered to extract and compress image features.

Recently, deep learning based codecs \cite{balle2017end_iclr, balle2017end, ganextreme, mentzer2020highgancomp} have gained significant attraction in the community due to their superior performance.
These approaches learn nonlinear transformations of the input data and an entropy model jointly with the objective of maximizing compression ratio while maintaining high reconstruction quality. 
{To this end,} several methods use variational autoenconder (VAE) type architecture \cite{balle2017end, minnen2018joint, cheng2020learned} and achieve impressive performance at higher BPP, however they are sub-optimal at low BPP.
Some papers have targeted the entropy model:  \cite{balle2018variational} propose to use a hyper prior as a source of side information to capture the spatial dependencies. \cite{minnen2018joint} follow a similar approach and augment the hierarchical model with an autoregressive one. \cite{cheng2020learned} build the previous approach with a discretized Gaussian mixture entropy model with attention modules. Few papers have improved the model structure and the architecture \cite{li2018learning, cheng2018deep, cheng2019deep} and some used RNNs \cite{toderici2017full, johnston2018improved}.

Others have leveraged the power of generative adversarial networks (GAN) \cite{gan} and used adversarial losses, especially for low bitrates \cite{rippel2017real, ganextreme}. These approaches produce high quality and lower perceptually-distorted image reconstructions \cite{generativecompression, mentzer2020highgancomp}, even for high bitrates \cite{tschannen2018deep}. These methods are computationally  heavy as they require adversarial training for each quality level. We argue that this is not practical especially for data compression, where each quality level requires to train a new model from scratch.  

The choice of distortion loss is crucial for better reconstruction, traditionally PSNR or MS-SSIM \cite{msssim_met} are used. These metrics capture the pixel wise distortion and poorly capture the perceptual distortion. Several works have tried to remedy these limitations. Motivated by the success of perceptual losses such as LPIPS \cite{zhang2018unreasonable} or VGG16 \cite{johnson2016perceptual} for various applications \cite{superresolution, dosovitskiy2016generating, styletransferpercept}, some papers \cite{generativecompression, chen2020perceptually} propose to include a perceptual distortion in addition to the pixel wise ones. These perceptual losses are based on computationally expensive networks such as VGG16 or learned perceptual metrics such as LPIPS. Moreover, the backbone networks are pretrained for an unrelated and discriminative task (\emph{e.g.} Image classification pretraining on ImageNet).

\subsection{Video Compression:} Deep Video Compression systems also minimize the rate-distortion loss. In addition to the spatial redundancy (SR), temporal redundancy (TR) is reduced by incorporating motion estimation modules.
Traditional methods \cite{h264, hevc} rely on handcrafted and block based modules. Recently, deep learning based video compression systems proposed to replace the traditional modules by learned ones \cite{lu2019dvc}. Motion estimation \cite{dosovitskiy2015flownet} and compensation are often used to address TR. Several improvements have been made to reduce TR; \cite{lin2020m} leverage multiple frames to improve the motion compensation, and \cite{hu2020improving} use multi resolution flow maps to effectively compress locally and globally. 
{However, training motion estimation modules make these compression systems less efficient during training}.
Some approaches perform TR reduction in latent space \cite{feng2020learned, interframe, hu2021fvc}. Others propose to perform frame interpolation \cite{interframe}, or an interpolation in the latent space of GANs \cite{generativecompression}. Recently, \cite{Liu_2021_CVPR} have proposed to train a variational autoencoder with an adversarial constraint. Temporal encoding is tackled using an LSTM network in latent space.

\subsection{Videoconferencing:}
This is becoming ubiquitous nowadays due to the pandemic. Many papers have targeted facial video synthesis \cite{wang2021one, Zhou_2021_CVPR, Liuvideoappl, zakharov2019few}, or compression artifacts removal \cite{Zhang2020DAVDNetDA}. However, there are few papers targeting specifically facial video compression. \cite{wang2021one} was the first to propose a compression framework based on extracted keypoints and learned appearance model for video synthesis. Similarly, in \cite{lowbandcompress, oquab2021low} a compression framework based on keypoints and adversarial training was proposed. 

{Our approach is significantly simpler and more efficient to train as we do not incorporate pose, expression nor keypoints extraction, which is desirable to support a wide range of quality levels, as for deep learning based compression systems, for each quality level a new model should be trained from scratch}.

\begin{figure*}[t]
     \centering
     \includegraphics[width=\linewidth]{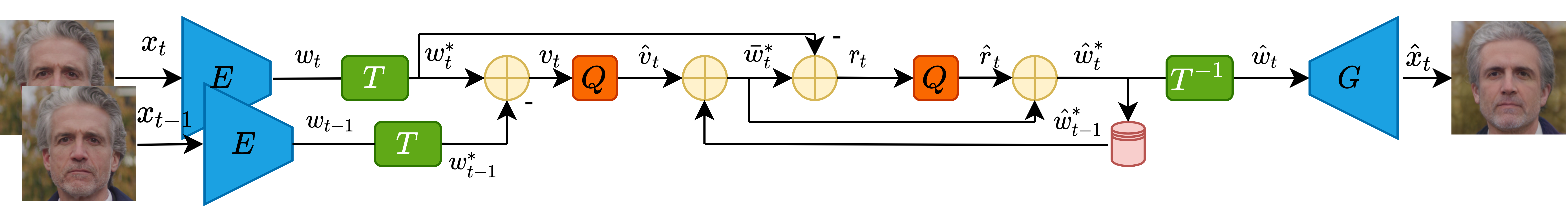}
    \caption{An illustration of our proposed approach for Video compression using Inter Coding with residual during test (Section \ref{sec:sganc_ic_method}). $Q$ corresponds to quantization, compression encoding and decoding. Here $g=1$.}
    \hfill
    \label{fig:main_inter}
\end{figure*}

\section{Learning Proxy Representation for Compression}
\label{sec:method}
In this section, we describe our proposed method to learn the latent space dedicated for compression, illustrated in Figures \ref{fig:main} and \ref{fig:main_inter}. The method can be summarized as follows: an input facial video frame is first projected in the latent space of StyleGAN \wplus. Our method consists in learning a transformation $T$ from the original latent space  \wplus to a new proxy representation denoted as \wstarc. The transformation $T$ is defined as a diffeomorphic mapping (\emph{i.e.}, normalizing flows model \cite{papamakarios2019normalizing}) and is learnt so that the intra image compression and the inter video compression are optimal in \wstarc. It is noted that, learning a proxy representation in \wstarc for compression is also motivated by the fact that the rate distortion loss can not be optimized in the original space \wplus without training the encoder/generator. 

We argue that the benefit of such approach can be explained by three factors. First, it leverages the generative power of StyleGAN2, known to outperform other GANs and generative models. Second, this approach separates, to some extent, the image reconstruction problem and learning the compression bottleneck. We optimize solely for the latter in a learned latent representation, thus leading to very efficient compression. Third, the new learned representation is optimal for compression but preserves the appealing properties of the original StyleGAN latent space to generate high-quality photorealistic images, thanks to the homeomorphism of the normalizing flow model.

Below, we first briefly describe some background materials (section \ref{sec:methodbackground}), and then present our methods for images or intra compression (section \ref{sec:methodintra}) and for videos or inter compression (section \ref{sec:methodinter}).

\subsection{Background}
\label{sec:methodbackground}
\subsubsection{StyleGAN Generator} \quad StyleGAN \cite{stylegan} is the state of the art unconditional GAN for high quality realistic image generation. It consists of a mapping function that takes a vector sampled from a normal distribution ($\bm{z} \sim \mathcal{N}(0, \bm{I})$) and maps it to an intermediate latent space $\mathcal{W}$ using a fully connected network ($M$) before feeding it to multiple stages (\emph{i.e.}, \wplus space) of the generator ($G$) to generate an image ($\bm{x}$) from the distribution of the real images ($\bm{x} \sim p\_\bm{x}$):
\begin{equation}
    \bm{x} = G(\bm{w}) \quad \bm{w} = M(\bm{z})  \quad \bm{z} \sim \mathcal{N}(0, \bm{I})
\end{equation}
It is shown that the latent space of GAN is semantically rich and enjoys several properties such as semantic interpolation \cite{radford2016unsupervised}. In addition, the latent vector encoded in \wplus space of StyleGAN captures a hierarchical representation of the projected image. In our case, we use StyleGAN2 \cite{styleganimporveing} which is an improved version.

\subsubsection{StyleGAN Encoder} \quad The StyleGAN encoder \cite{encodinginstyle} is a deterministic function denoted as $E$. Its role is to project a real image into the latent space of StyleGAN (\emph{e.g.}, $\mathcal{W}$ or \wplussp), in such a way that the reconstructed image by the StyleGAN generator is minimally distorted ($\hat{\bm{x}} = G(E(\bm{x})) \approx \bm{x}$). In our case, the image is projected in \wplus with dimension of $(18\times512)$ where each dimension controls a different convolution layer of the StyleGAN generator. Currently the encoding based GAN inversion approaches are not ideal, which explains the difference between the projected and the original image.

\subsection{Image Compression (SGANC)}
\label{sec:methodintra}
\label{sec:image_compression}

We assume that we have a pretrained and fixed StyleGAN2 generator $G$ that considers a latent code $\bm{w} \in \mathcal{W}^{+}$ and generates a high resolution image of size $1024 \times 1024$, and an encoder $E$ (pretrained and fixed) that embeds any given image $\bm{x}$ to the latent code $\bm{w}$ in \wplus such that $G(E(\bm{x})) \simeq \bm{x}$. Our objective is to learn a new latent space $\mathcal{W}^{\star}_c$ optimal for image compression.
The $\mathcal{W}^{\star}_c$ is obtained using a bijective transformation $T: \mathcal{W}^{+} \rightarrow \mathcal{W}^{\star}_c$, which is parametrized as a normalizing flows model (our work only requires the bijectivity, as such, no maximum-likelihood term is included in the training objective). $T$ maps a latent code $\bm{w} \in \mathcal{W}^{+}$ into $\bm{w}^{\star}_c \in \mathcal{W}^{\star}_c$ such that the latent vectors $\bm{w}^{\star}_c$ has the minimum entropy and the sufficient information to generate the original image with minimal distortion from the inverted latent code $T^{-1}(\bm{w}^{\star}_c)$ using $G$. 
The entropy model is trained on the latent codes $\bm{w}^{\star}_c \in \mathcal{W}^{\star}_c$ by minimizing the following rate loss after the quantization:

\begin{equation}
\label{eq:rate-loss}
    \mathcal{R}=-\mathbb{E}\sum_i^D \log_2 p_i(Q(T(\bm{w}) ))=-\mathbb{E}\sum_i^D \log_2 p_i(T(\bm{w}) + \bm{\epsilon} )
\end{equation}

For \eqref{eq:rate-loss} to be differentiable, following \cite{balle2017end_iclr}, we relax the hard quantization $Q$ by adding uniform noise $\bm{\epsilon}$ to the latent vectors $T(\bm{w})$. $p_i$ is the $i^{th}$ dimension of the probability density function in \wstarcsp, $D$ is the latent vector dimension, $\bm{w} = E(\bm{x})$ where $\bm{x}$ is the input image and $\bm{\epsilon}$ is sampled from a uniform distribution $\mathcal{U}_{[-0.5, 0.5]}$. The entropy model $\bm{p}$ is modeled as the fully factorized entropy model and it is also parameterized by a neural network as in \cite{balle2018variational}. 
\label{sec:sganc_ic_method}

The transformed/quantized latent codes should also have sufficient information to reconstruct the image, and this is achieved by minimizing a distortion loss. In general the distortion loss is computed in the image space, although, we propose a new efficient distortion loss directly in the latent space:
\begin{equation}\label{eq:distort-loss}
    \mathcal{D}= d\left(\bm{w}, T^{-1}\left(T(\bm{w}) + \bm{\epsilon} \right) \right),
\end{equation}
where $d$ is any distortion loss, in this paper we use the mean squared error (MSE) loss. The \eqref{eq:distort-loss} can be seen as a perceptual distortion loss, and it is motivated by the fact that the latent space of GANs is semantically rich. We argue that this is true especially for StyleGAN, where the latent code is extracted from several layers of the StyleGAN encoder which allows to capture multiscale and semantic features/representations. Compared to existing perceptual losses, our loss does not requires to generate the images during training or to compute heavy losses such as VGG16 or LPIPS in the image space.
The total loss used to learn our proposed latent space $\mathcal{W}^{\star}_c$ is a trade-off between the rate \eqref{eq:rate-loss} and distortion \eqref{eq:distort-loss} as shown below
\begin{equation}
    \mathcal{L}= \mathcal{R} + \lambda \mathcal{D}
\end{equation}
Where $\lambda$ is the trade-off parameter, and the transformation $T$ and the entropy model $\bm{p}$ are jointly optimized. Once the optimization is completed, to create the bit-stream the latent codes in $\mathcal{W}^{\star}_c$ are quantized with $Q$ using rounding operator. 
\begin{figure*}[t!]
     \centering
     \begin{subfigure}[b]{0.24\linewidth}
         \centering
         \caption{ {PSNR}}
         \includegraphics[width=\linewidth, height=4cm]{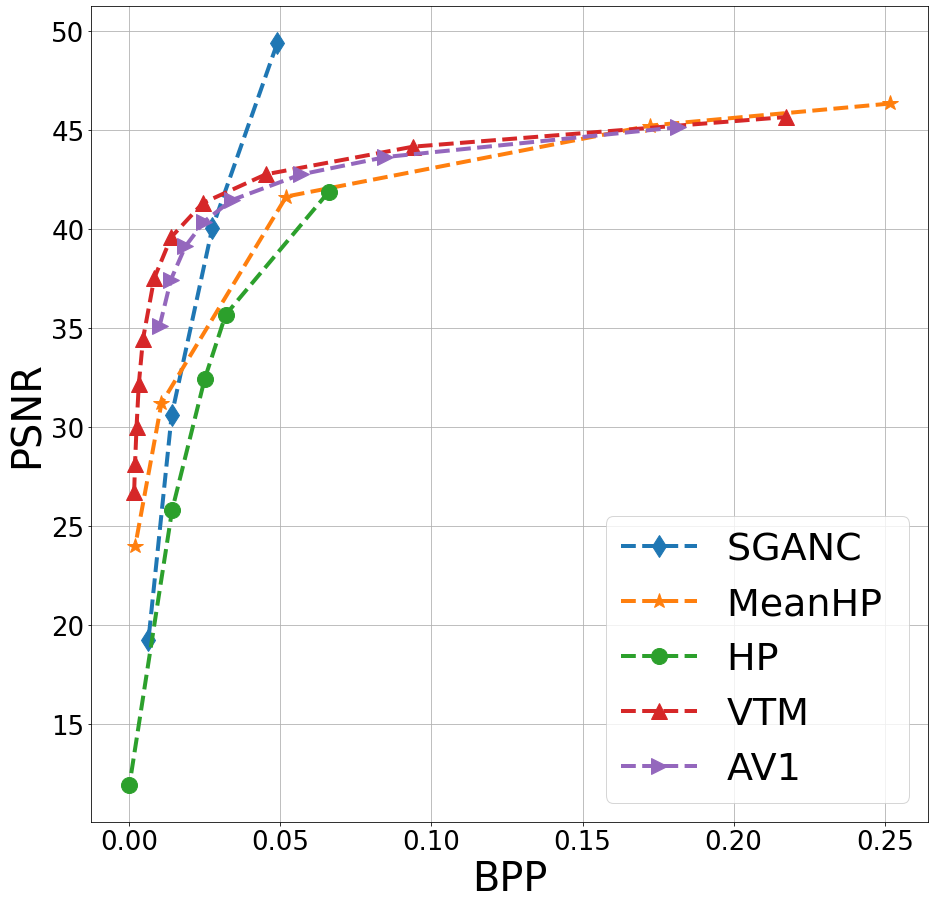}
     \end{subfigure}
     \hfill
     \begin{subfigure}[b]{0.24\linewidth}
         \centering
         \caption{{MS-SSIM}}
         \includegraphics[width=\linewidth, height=4cm]{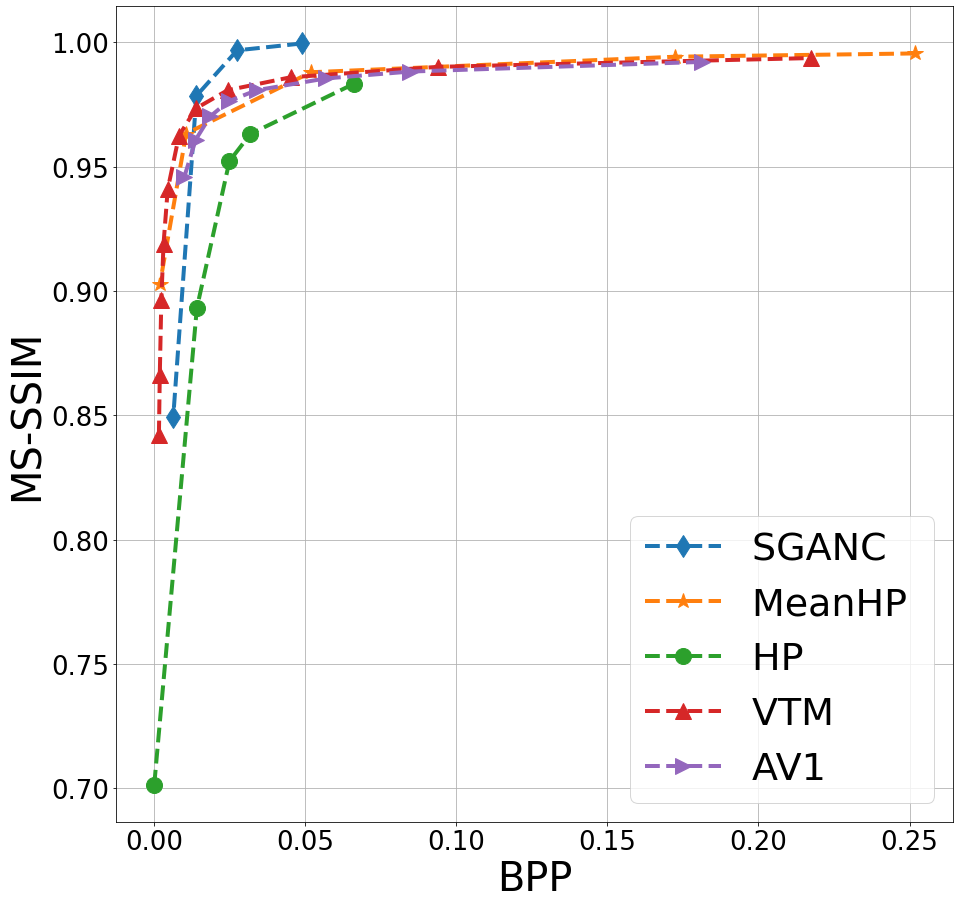}
     \end{subfigure}
     \hfill
     \begin{subfigure}[b]{0.24\linewidth}
         \centering
         \caption{ {LPIPS \cite{zhang2018unreasonable}}}
         \includegraphics[width=\linewidth, height=4cm]{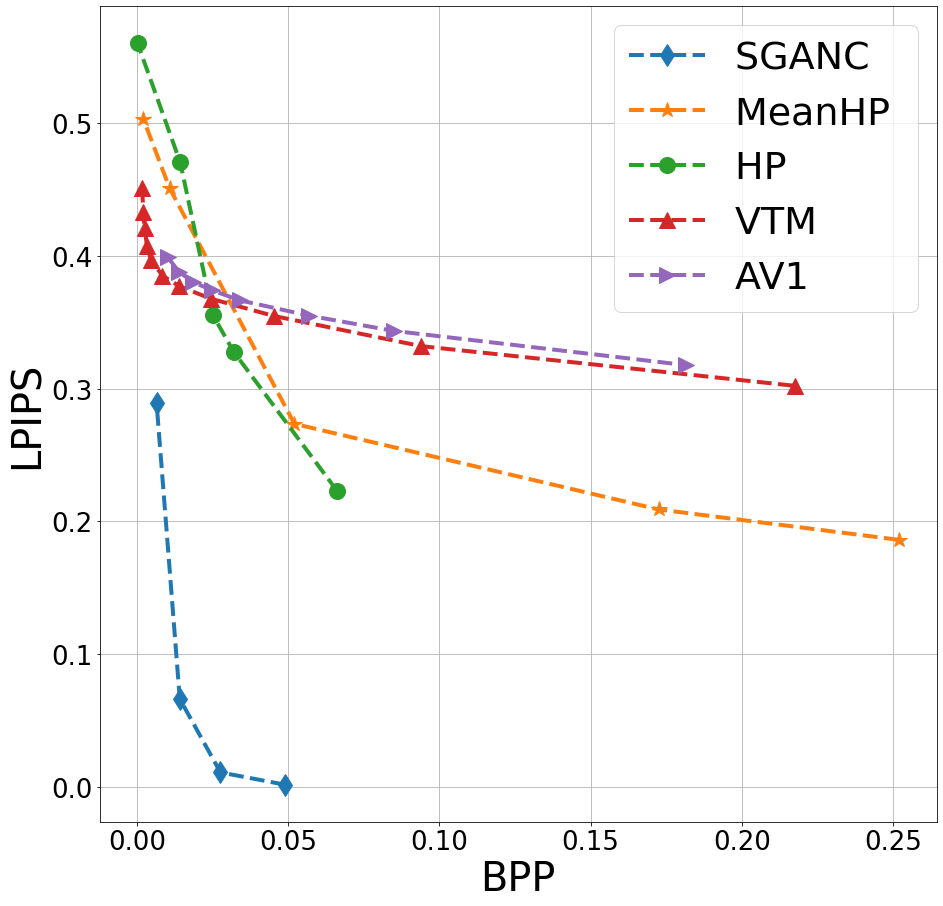}
     \end{subfigure}
     \hfill
         \begin{subfigure}[b]{0.24\linewidth}
         \centering
         \caption{ {PIM \cite{pim}}}
         \includegraphics[width=\linewidth, height=4cm]{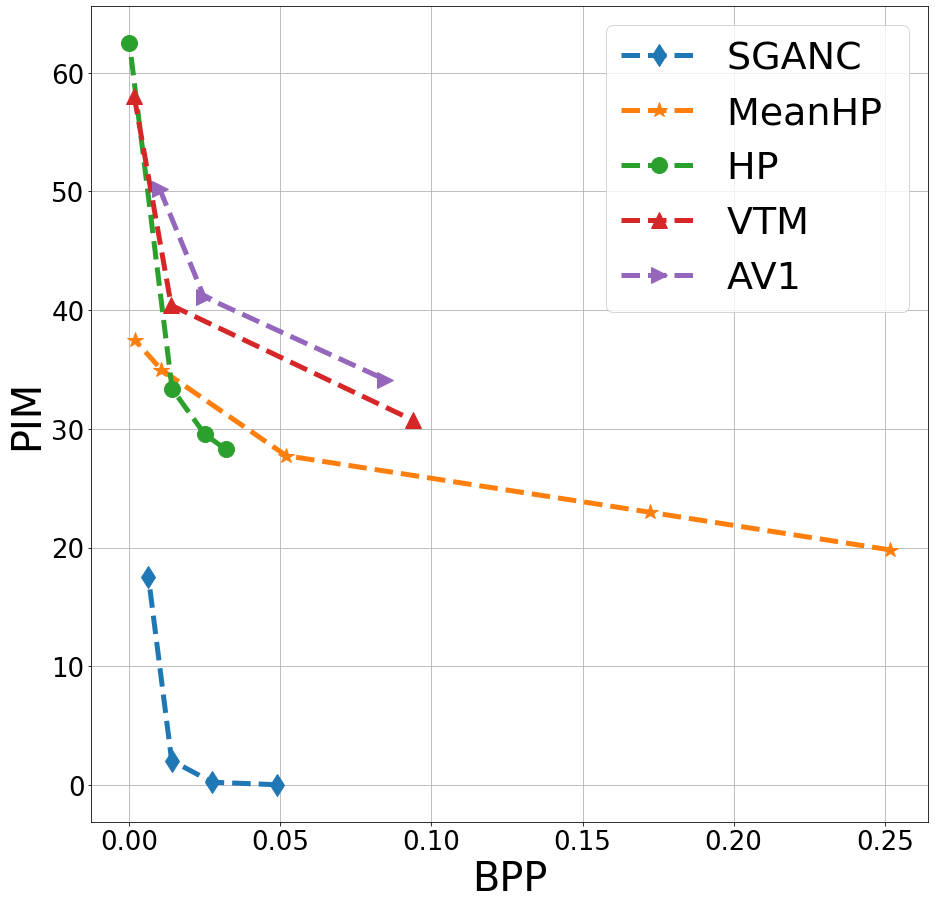}
     \end{subfigure}

        \caption{Rate distortion curves on MEAD intra dataset: for medium and large BPP, our method (SGANC in blue) is better in terms of LPIPS and MS-SSIM than VTM (Red), AV1 (Purple), MeanHP (Orange) and HP (Green). For high BPP, our method is better in terms of PSNR. The perceptual metrics LPIPS and PIM clearly show that our method outperforms existing methods perceptually.}
        \label{fig:results_MEAD_intra}
\end{figure*}

\subsection{Video Compression (SGANC IC)}
\label{sec:methodinter}
Here, we propose our approach for video compression, denoted as SGANC IC, using learned inter coding with residuals. Video compression methods and standards rely on motion estimation and compensation modules to leverage the temporal dependencies between frames. As our approach is formulated in the latent space, our inter coding schema is based on the successive latent differences, since these differences reflect the temporal changes.

Specifically, inter coding with residuals is performed in the latent space (Figure \ref{fig:main_inter}). Given a sequence of frames $\{\bm{x}_1, \bm{x}_2, ..., \bm{x}_{t-1}, \bm{x}_{t}, ... \}$, a pretrained encoder $E$ is first used to obtain the latent representation. Then, similarly to intra coding, a transformation $T$ is learned to map these frames to $\mathcal{W}^{\star}_{IC}$, leading to a sequence of latent codes $\{\bm{w}^{*}_1, \bm{w}^{*}_2, ..., \bm{w}^{*}_{t-1}, \bm{w}^{*}_{t}, ... \}$. The mapping $T$ is learnt such that the sequence of latent codes in $\mathcal{W}^{\star}_{IC}$ is optimal for inter-coding by taking the temporal dependencies into account. Our approach can be summarized as follows (the complete description of the algorithm can be retrieved in the supplementary material).
\begin{itemize}
\setlength\itemsep{0em}
    \item The first latent code is coded using the method described in section \ref{sec:methodintra}: $\hat{\bm{w}}_{0}^{*}$ (using the same entropy model or preferably another one trained for image compression).  The following steps are repeated until the end of the video; 
    \item The difference between two consecutive latent codes are computed and quantized; $\hat{\bm{v}}_t = Q(\bm{w}^{*}_t - \bm{w}^{*}_{t-1})$.
    \item From the previous reconstructed code, an estimate of the latent code at frame $t$ is obtained; $\bar{\bm{w}}_t = \hat{\bm{w}}_{t-1}^{*} + \hat{\bm{v}}_t$.
    \item The residual between the estimated and the actual latent code is computed and quantized as $\hat{\bm{r}}_t = Q(\bm{w}^{*}_t - \bar{\bm{w}}_t^{*})$ (for all the frames or each gap (\emph{i.e.}, $g$ frames).
    \item The quantized difference $\hat{\bm{v}}_t$ and the residual $\hat{\bm{r}}_t$ are compressed using an entropy coding and sent to the receiver. 
    \item On the receiver side, the current latent code is reconstructed from the residual and the estimated latent code (or only from the estimated latent code each $g$ frames): $\hat{\bm{w}}_t^{*} = \bar{\bm{w}}_t^{*} + \hat{\bm{r}}_t$.
    \item The latent codes in $\mathcal{W}^{\star}_{IC}$ are remapped to \wplus to generate the images using the pretrained StyleGAN2 $G$. 
\end{itemize}
To learn the new latent space $\mathcal{W}^{\star}_{IC}$ for inter-coding, the transformation $T$ and the entropy model ($\bm{p}$) are learned to optimize the rate-distortion loss:
\begin{align}\label{eq:IC-rate-dist}
    \mathcal{L}_{IC} &= \lambda d\left(\bm{w}_t, T^{-1}\left(\hat{\bm{w}^{*}}_t \right) \right) -\Bigg(\mathbb{E}\left[\sum_i^D \log_2 p_i\left(\hat{\bm{v}}_t \right)\right] \nonumber\\
    &+ \mathbb{E}\left[\sum_i^D \log_2 q_i\left(\hat{\bm{r}}_t \right)\right] \Bigg) 
\end{align}

As similar to section \ref{sec:image_compression}, we replace the quantization $Q$ by adding uniform noise, and the entropy model is modeled as in \cite{balle2018variational}. While \eqref{eq:IC-rate-dist} has two entropy models, we  show below, that it is sufficient to learn only one entropy model on the differences $\hat{\bm{v}}_t$, as $\hat{\bm{r}}_t$ admits the explicit probability distribution known as Irwin-Hall distribution (the proof of the following Lemma \ref{lemma} can be found in the supp. material). 
\begin{lemma}\label{lemma}
Let $Q(x) = x+\epsilon$ be the continuous relaxation of the quantization, where $\epsilon$ follows the uniform distribution $\mathcal{U}_{[-0.5, 0.5]}$ . Let $\bm{w}^{*}_0 \in \mathcal{R}^n$,$\hat{\bm{w}}^{*}_0 = Q(\bm{w}^{*}_0)$, and $\bar{\bm{w}}_t^{*} = \hat{\bm{w}}_{t-1}^{*} + Q(\bm{w}^{*}_t - \bm{w}^{*}_{t-1}) $
 , $t = 1, \ldots, n$.   If $\hat{\bm{r}}_t = Q(\bm{w}^{*}_t - \bar{\bm{w}}_t^{*})$ is the quantized residual defined for every $t$ such that $t \equiv 0 \pmod{g}$, then $\hat{\bm{r}}_t$ follows the Irwin-Hall distribution with the parameter $3+(g-1)$ and the support being shifted by $-0.5*(3+(g-1))$.
\end{lemma}
This leads to optimizing our latent space $\mathcal{W}^{\star}_{IC}$ for the rate-distortion loss only with one entropy model for $\hat{\bm{v}}_t$ by discarding the last term in \eqref{eq:IC-rate-dist}. During test, the explicit distribution of the residuals can be used for entropy coding.

\begin{figure*}[t]
    \centering
    \begin{subfigure}[b]{0.2\linewidth}
    \centering
      \includegraphics[width=\linewidth]{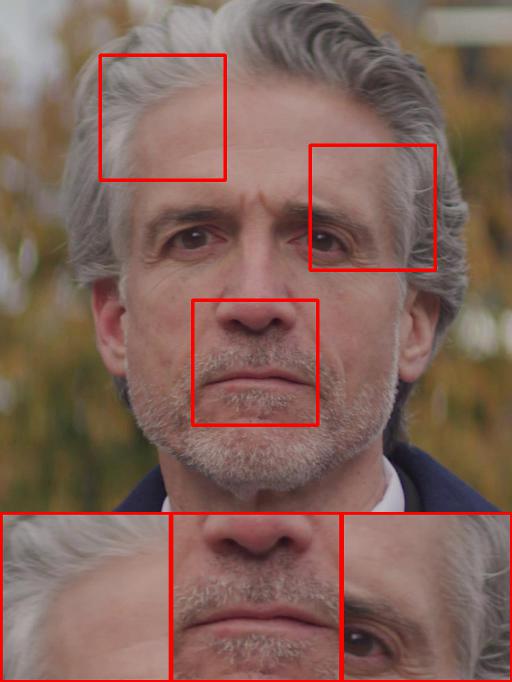} 
      \caption{{Original}\\
      \centering }
    \end{subfigure}
\begin{subfigure}[b]{0.2\linewidth}
\centering
\includegraphics[width=\linewidth]{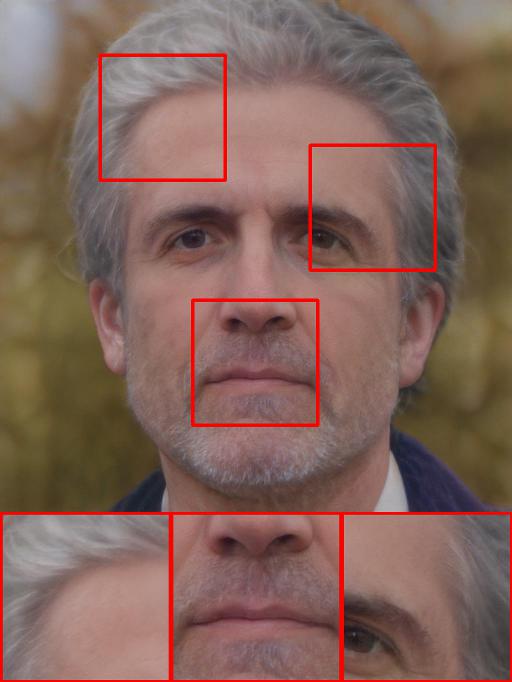}
\caption{{Projected}}
\end{subfigure} 
\begin{subfigure}[b]{0.2\linewidth}
\centering
\includegraphics[width=\linewidth]{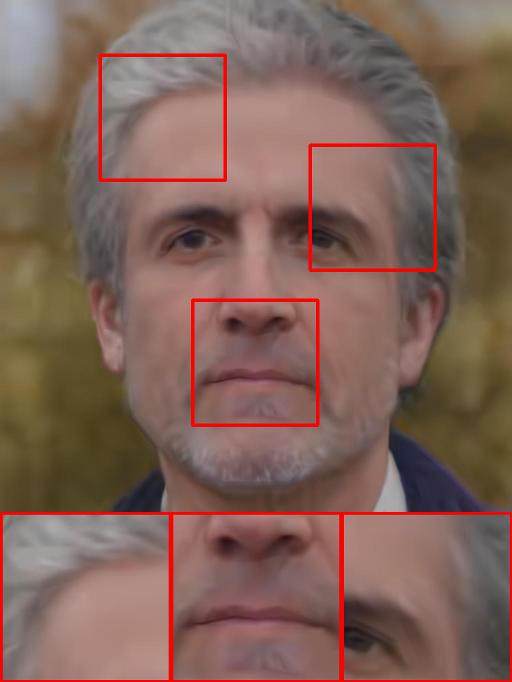} 
\caption{{VTM}, \tiny{BPP:0.0155} \\
\centering \tiny{PSNR$\uparrow$:37.4, MS-SSIM$\uparrow$:0.951, PIM$\downarrow$:21.1}}
\end{subfigure}
\\
\begin{subfigure}[b]{0.2\linewidth}
\centering
\includegraphics[width=\linewidth]{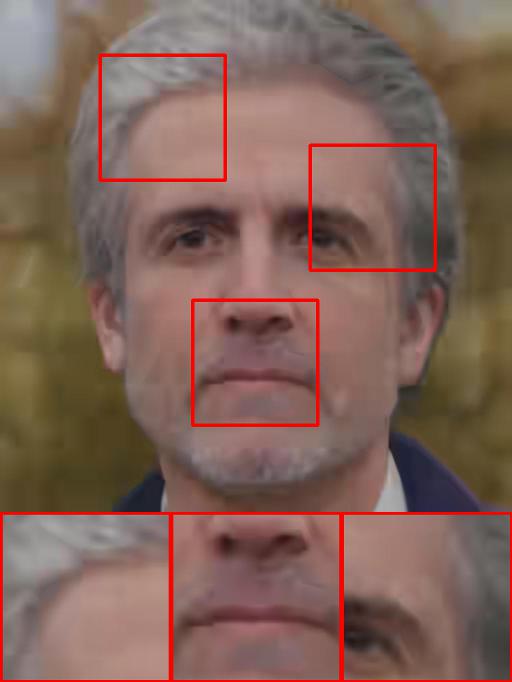} 
\caption{{AV1}, \tiny{BPP 0.0156}
\\
\centering \tiny{PSNR$\uparrow$:32.1, MS-SSIM$\uparrow$:0.934, PIM$\downarrow$:24.2}}
\end{subfigure}
\begin{subfigure}[b]{0.2\linewidth}
\centering
\includegraphics[width=\linewidth]{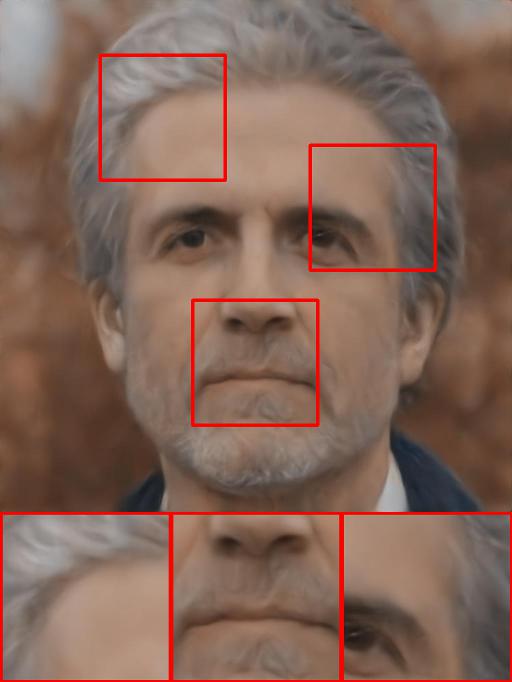} 
\caption{{MeanHP}, \tiny{BPP 0.0119} \\
\centering \tiny{PSNR$\uparrow$:35.9, MS-SSIM$\uparrow$:0.953, PIM$\downarrow$:39.7}}
\end{subfigure}
\begin{subfigure}[b]{0.2\linewidth}
\centering
\includegraphics[width=\linewidth]{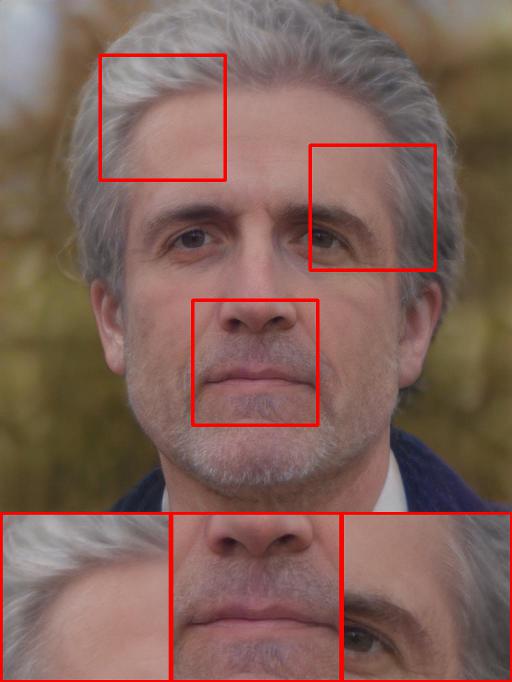} 
\caption{{SGANC (Ours)}, \tiny{BPP 
0.0146} \\
\centering \tiny{PSNR$\uparrow$:29.3, MS-SSIM$\uparrow$:0.970, PIM$\downarrow$:1.77}}
\end{subfigure}
    \caption{Qualitative results for image compression with comparable BPP for all methods (images are better seen zoomed). Other methods introduce blocking artifacts and blurring. Our method (SGANC) leads to high quality reconstruction  and perceptually lower distortion.}
    \label{fig:qual_results_FP006734MD02_1}
\end{figure*}

\subsubsection{Stage-Specific entropy models:} In \cite{stylegan}, it is shown that each stage/layer  of the StyleGAN generator corresponds to a specific scale of details. To leverage this hierarchical structure, we propose stage specific entropy models. Specifically, the first layers, which correspond to coarse resolution (\emph{e.g.}, $4^2-8^2$) affect mainly high level aspects of the image, such the pose and face shape, while the last layers affect the low level aspects such as textures, colors and small micro structures. Here, we propose to leverage this hierarchical structure and weight the distortion $\lambda$ differently for each layer of the generator (Note that, the latent code in \wplus or $\mathcal{W}^{\star}_{IC}$ consists of 18 latent codes of dimension 512 and each one corresponds to one layer in the generator, hence its dimension is $(18, 512)$). For practical reasons, we split the $18$ layers in three stages ($1-8$, $8-13$, $13-18$), and learn one transformation and entropy model for each group. The distortion $\lambda$ is  chosen to be higher for the first layers and decreases subsequently.

\section{Experiments}
\label{sec:experiments}
\subsection{Datasets}
\label{sec:datasets}

\noindent\textbf{Celeba-HQ} \cite{karras2018progressive}, a dataset consists of 30000 high quality face images of $1024 \times 1024$ resolution.

\noindent\textbf{FILMPAC}, a dataset consists of 5 video clips with high resolution and the length varies between 60 and 260 frames. These videos can be found on the filmpac website \cite{filmpac} %
by searching their names (FP006734MD02, FP006940MD02, FP009971HD03, FP010363HD03, FP010263HD03). 

\noindent\textbf{MEAD} \cite{mead}, a high resolution talking face video corpus for many actors with different emotions and poses. The training dataset for inter-coding consists of 2.5 k videos with frontal poses. For evaluation, we created  \textbf{MEAD-inter dataset} consisting of 10 videos (selected from  MEAD) of different actors with frontal pose. We also created \textbf{MEAD-intra dataset} consisting of 200 frames selected from these videos with frontal pose for evaluating image compression.

\subsubsection{Dataset preprocessing:}  All the frames are cropped and aligned using the same prepossessing method as that of FFHQ dataset \cite{stylegan}, on which the StyleGAN is pretrained. As we compare the reconstructed image with the projected one for SGANC, we project all the frames (encode the original images and reconstruct them using StyleGAN2). All frames are of high resolution ($1024\times1024$).

\begin{figure*}[htbp]
     \centering
     \begin{subfigure}[b]{0.24\linewidth}
         \centering
         \caption{{PSNR}}
         \includegraphics[width=\linewidth, height=4cm]{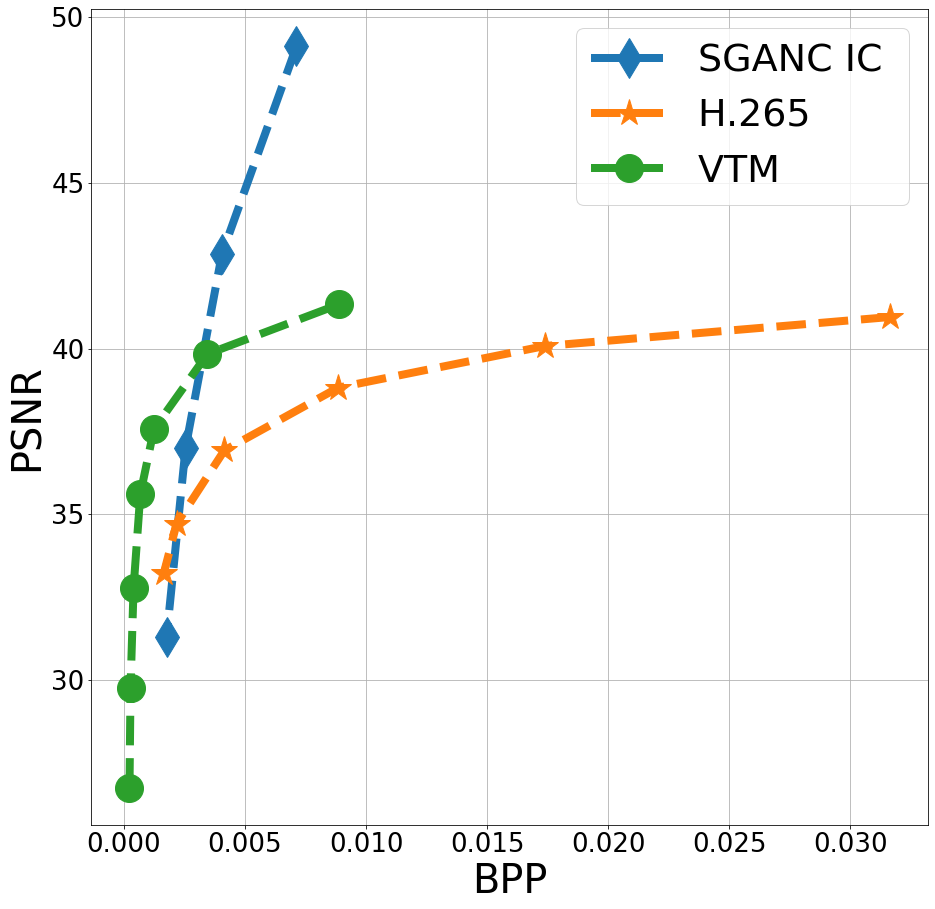}
     \end{subfigure}
     \hfill
     \begin{subfigure}[b]{0.24\linewidth}
         \centering
         \caption{ {MS-SSIM}}
         \includegraphics[width=\linewidth, height=4cm]{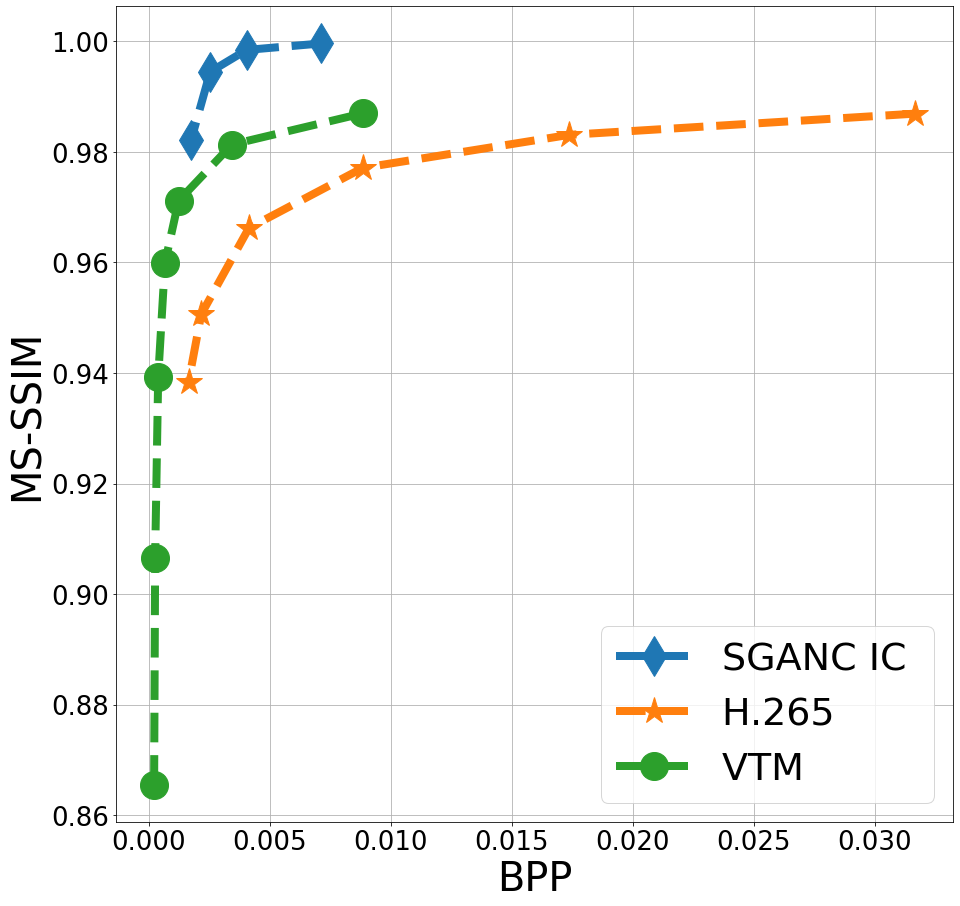}
     \end{subfigure}
     \hfill
     \begin{subfigure}[b]{0.24\linewidth}
         \centering
         \caption{ {LPIPS \cite{zhang2018unreasonable}}}
         \includegraphics[width=\linewidth, height=4cm]{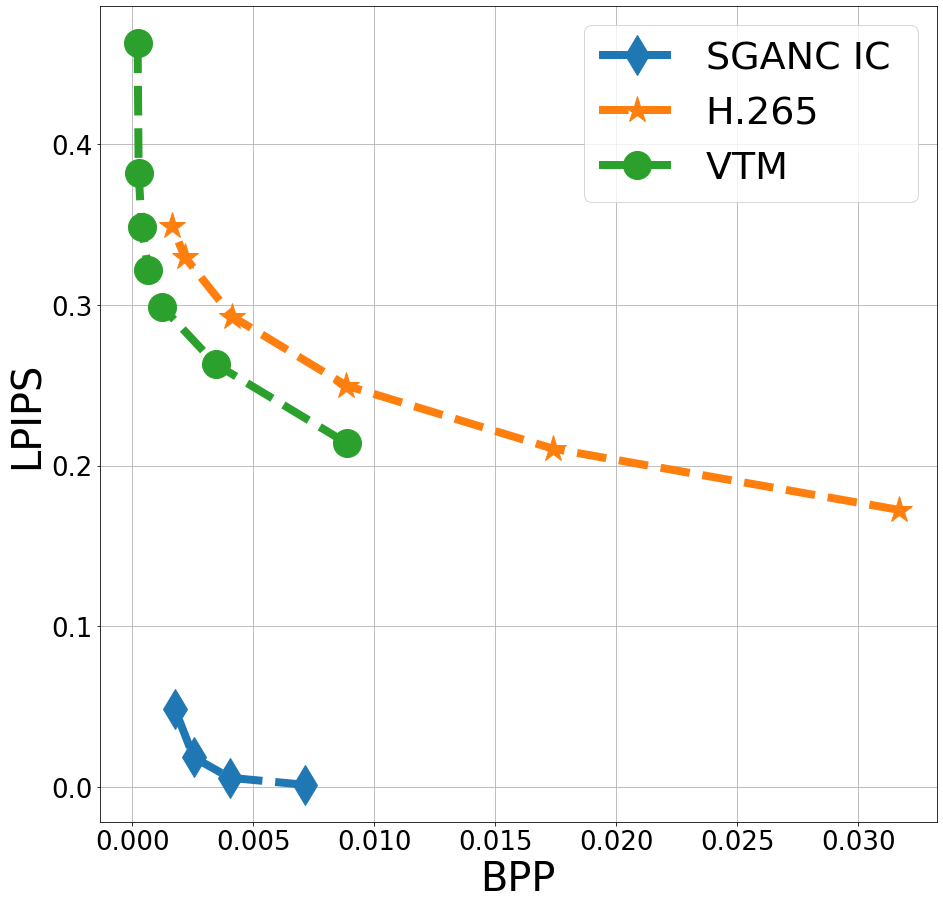}
     \end{subfigure}
     \hfill
     \begin{subfigure}[b]{0.24\linewidth}
         \centering
         \caption{ {PIM \cite{pim}}}
         \includegraphics[width=\linewidth, height=4cm]{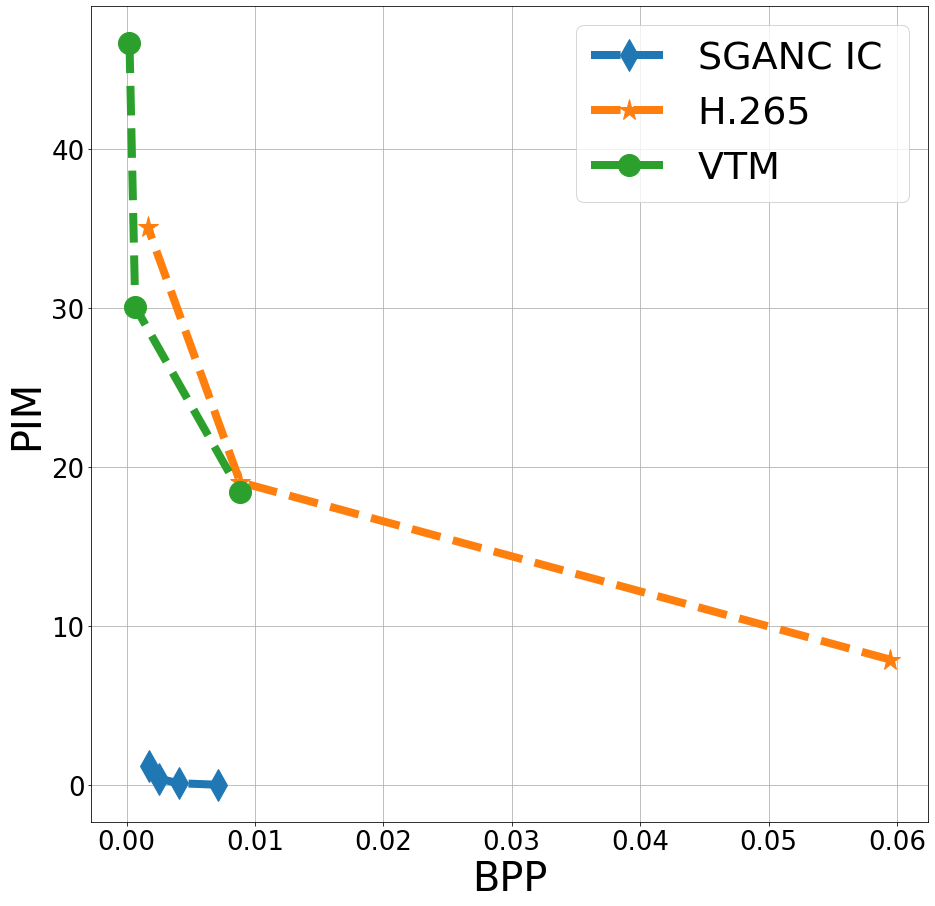}
     \end{subfigure}
        \caption{Rate distortion curves on MEAD inter dataset. Our approach SGAN-IC leads to a significantly lower  perceptual distortion as measured by the perceptual metrics LPIPS and PIM.}
        \label{fig:results_MEAD_inter}
\end{figure*}

\subsection{Implementation details}
\label{sec:implementation_details}
We used StyleGAN2 generator ($G$) \cite{styleganimporveing} pretrained on FFHQ dataset \cite{stylegan}. The images are encoded in \wplus using pSp, a pretrained StyleGAN2 encoder ($E$) \cite{encodinginstyle}. The parameters of the generator and the encoder remain fixed in all the experiments. The latent vector dimension in \wplus and \wstarc is $18\times 512$. The mapping function $T$ is modeled using RealNVP architecture \cite{realnvp} without batch normalization. This NF model consists of $13$ coupling layers, and each coupling layer consists of $3$ fully connected (FC) layers for the translation function and $3$ FC for the scale one with LeakyReLU as hidden activation and Tanh as output one (Total number of trainable parameters=20.5 M). For the entropy model, we have used the fully factorized entropy model \cite{balle2018variational} assuming the dimensions are independent, based on the implementation from the CompressAI library \cite{begaint2020compressai}. For the stage specific entropy model, $\lambda$ is kept constant for the first stage and then decreased linearly to be $1e-2$ smaller for the last layer. For all the experiments, we used Adam optimizer with $\beta_1=0.9$ and $\beta_2=0.999$, learning rate=$1e-4$ and the batch size=$8$. Once the training is completed, we used Range Asymmetric Numeral System coder \cite{duda2013asymmetric} to obtain the bit-stream.

\subsubsection{Training:} We encoded all the images once, and the training is performed solely using the latent codes. Thus, we do not need the generator nor the encoder during training, which makes the approach light and fast to train. For SGANC, We train on Celeba-HQ dataset. For SGANC IC, we trained on 2.5 k videos from the MEAD dataset \cite{mead}, where each batch contains video slices of size = 9 frames. All the frames are preprocessed as in section \ref{sec:datasets}.

\begin{figure*}[htbp]
\setlength\tabcolsep{2pt}%
\centering
\small
\begin{tabular}{p{1cm}ccccc}
\toprule

Model &
VTM\cite{VTM}  & 
H.265\cite{h265} &
DVC\cite{lu2019dvc} &
\textit{Wang et. al}\cite{wang2021one} & 
SGANC IC (Ours) \\
&
\includegraphics[width=0.18\linewidth]{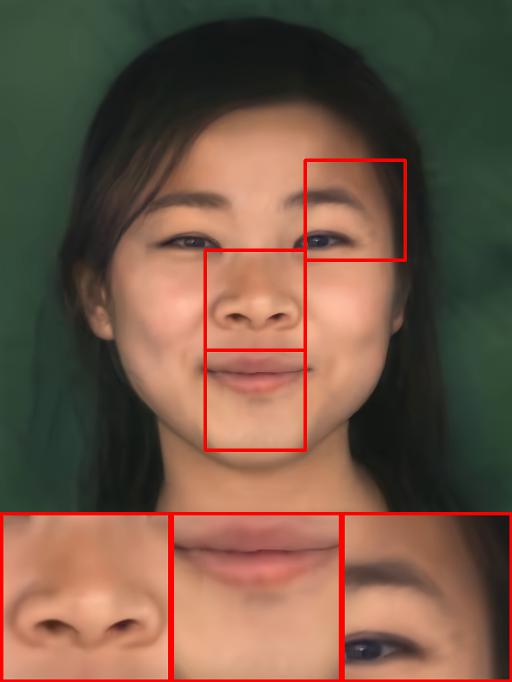}  &
\includegraphics[width=0.18\linewidth]{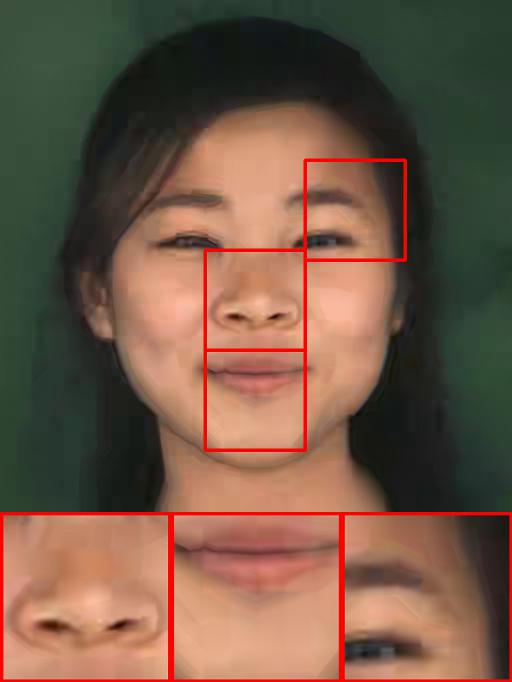} &
\includegraphics[width=0.18\linewidth]{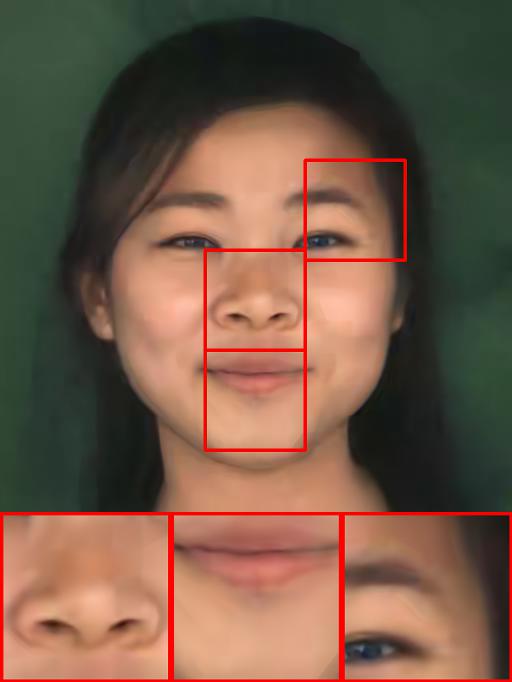}  & 
\includegraphics[width=0.18\linewidth]{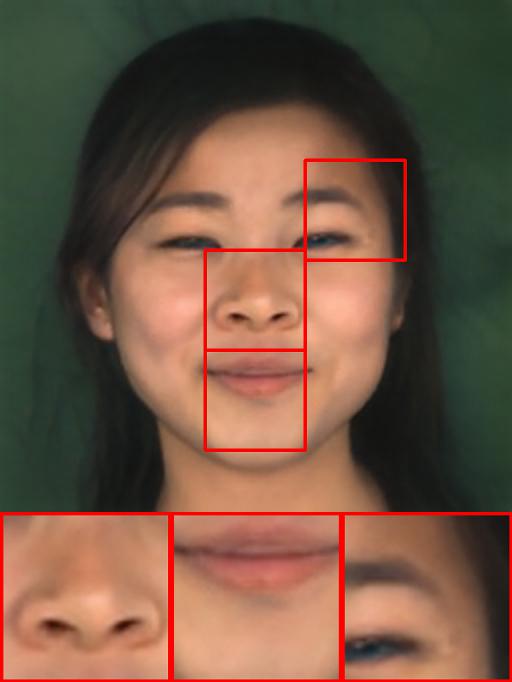}
&
\includegraphics[width=0.18\linewidth]{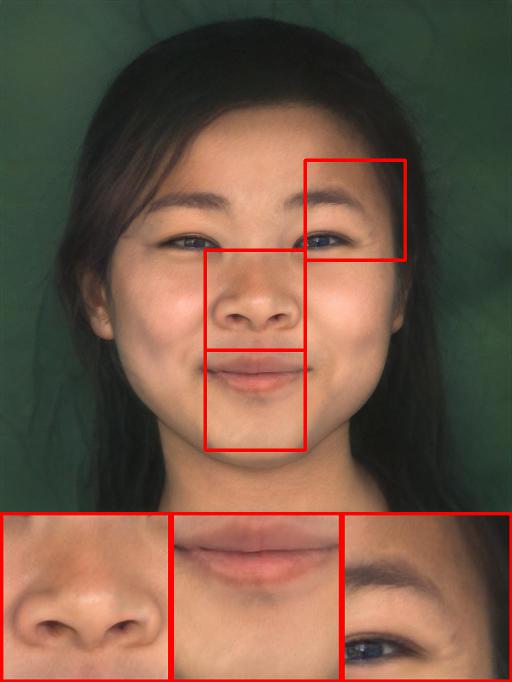} \\ \cline{3-6}
BPP &
0.0018  & 
0.0021 &
0.0175 &
0.0016 & 
0.0017 \\ 
PSNR$\uparrow$ &
\textbf{38.94}   & 
34.87  &
38.36 &
 26.60& 
32.23  \\ 
MSSSIM$\uparrow$ &
0.978   & 
0.953  &
0.972&
 0.899& 
\textbf{0.985}  \\ 
LPIPS$\downarrow$ &

0.2800    & 
0.3267   &
0.3142&
0.3265 & 
\textbf{0.0465}  \\ 
PIM$\downarrow$ &
24.97    & 
29.76    &
24.44&
22.40 & 
\textbf{1.221}   \\ 

\bottomrule
\end{tabular}
\caption{Qualitative results for video compression (images are better seen zoomed). Other methods introduce blocking artifacts and blurring. Our method (SGANC IC) leads to high quality reconstruction and  a significantly lower perceptually distortion, as measured by the metrics LPIPS and PIM. The zoomed versions are displayed in supplementary material.}
\label{fig:qual_results_low_mead_inter_4_main}
\end{figure*}
\begin{figure}
\centering
\begin{subfigure}{0.45\linewidth}
    \includegraphics[width=0.95\linewidth]{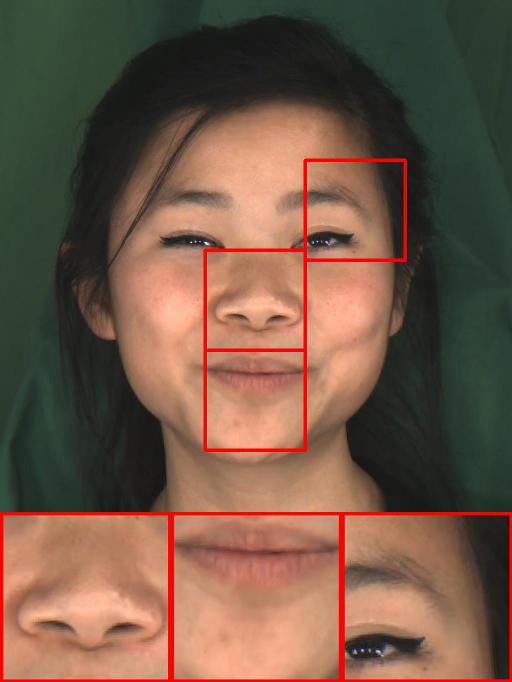}
    \end{subfigure}
    \begin{subfigure}{0.45\linewidth}
    \includegraphics[width=0.95\linewidth]{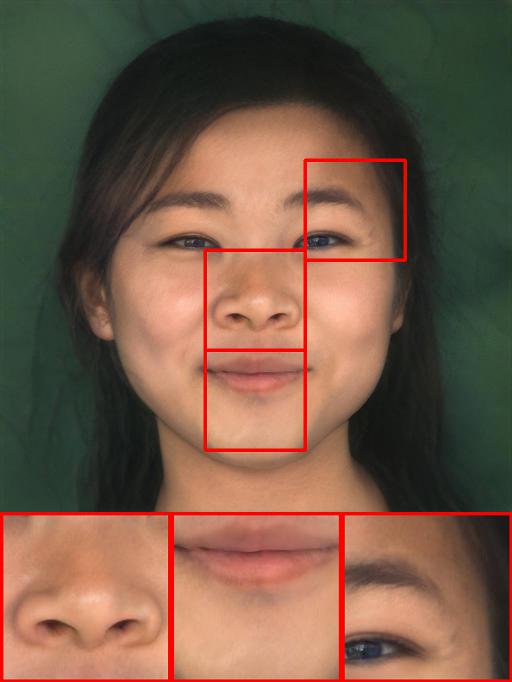}
\end{subfigure}
\caption{Original (left) and projected (right) frame, the reconstructed frame from StyleGAN is not perfect.}
\label{fig:org_proj_frames}
\end{figure}

\subsection{Results}
In this section we present the experimental results of our proposed methods. We used the following metrics to assess the quality of the compression methods: Peak Signal to Noise Ratio (PSNR), Multi Scale Structural Similarity (MS-SSIM) \cite{msssim_met}, Learned Perceptual Image Patch Similarity (LPIPS) \cite{zhang2018unreasonable}, and the Perceptual Information Metric (PIM) with a mixture of 5 Gaussian distributions \cite{pim}.

\begin{figure*}[t]
\setlength\tabcolsep{2pt}%
\centering
\small
\begin{tabular}{ccccc}

Orig &
VTM\cite{VTM}  & 
DVC\cite{lu2019dvc} &
\textit{Wang et. al}\cite{wang2021one} & 
SGANC IC (Ours) \\
\includegraphics[width=0.18\linewidth]{images/inter_qualitative_draw_new/W016/orig/frame0002.jpg}
&
\includegraphics[width=0.18\linewidth]{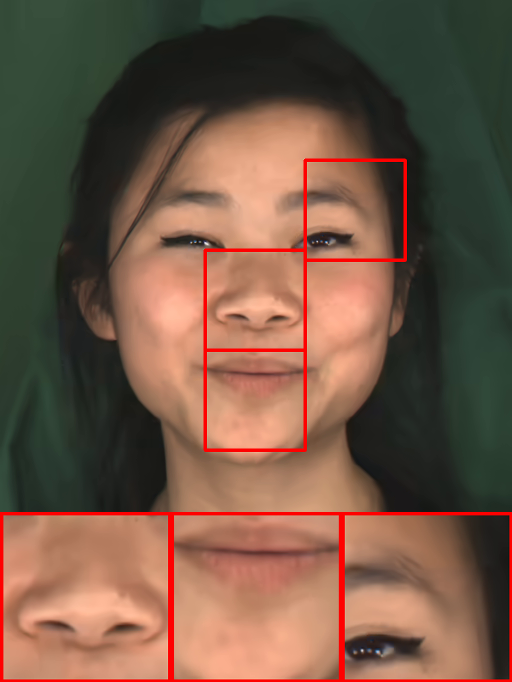}  &
\includegraphics[width=0.18\linewidth]{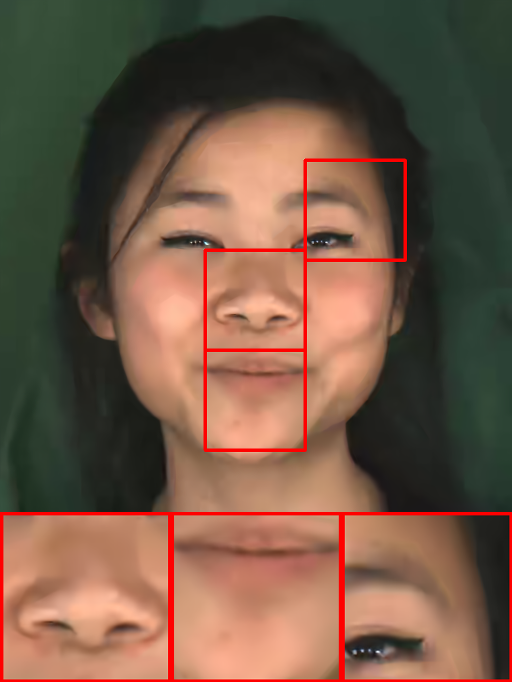}  & 
\includegraphics[width=0.18\linewidth]{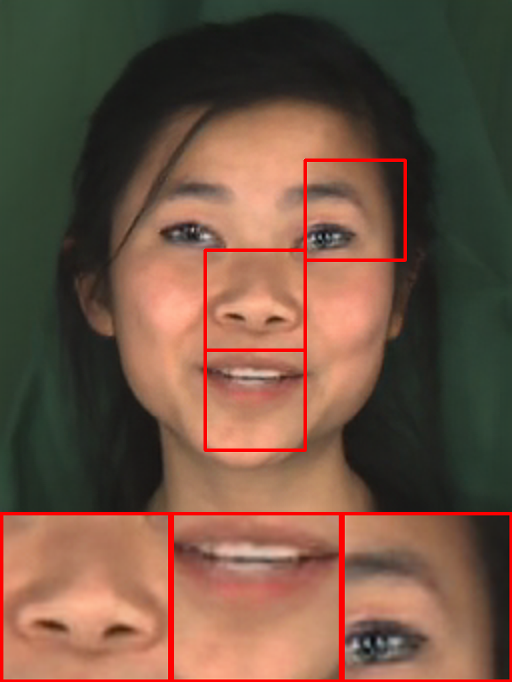}
&
\includegraphics[width=0.18\linewidth]{images/inter_qualitative_draw_new/W016/sganc_ic/l1_6/frame0001.jpg} \\ %
\end{tabular}
\caption{Qualitative comparison with the original frames for video compression (images are better seen zoomed). Other methods introduce blocking artifacts and blurring. Our method (SGANC IC) leads to high quality reconstruction (wrt projected frames Fig. \ref{fig:org_proj_frames}) and  a significantly lower perceptually distortion. The zoomed versions are displayed in supplementary material.}
\label{fig:comp_with_originalframes}
\end{figure*}

LPIPS and PIM are two perceptual metrics; PIM is recently proposed and follows an information-theoretic approach, inspired from the human perceptual system \cite{pim}. The authors of LPIPS, on the other hand, show that it is consistent with the human perception.
We report the size of the compressed images in bits per pixel (BPP), which is common in evaluating image and video compression.
To exclude the distortion coming from the GAN inversion technique, all the frames are projected (\textit{i.e.,} inverted) using the encoder. The projected frames are used for all approaches in the main paper as a baseline to compute the distortion metrics. We would like to stress out that we evaluate in the core paper the reconstruction results with the projected images of the StyleGAN, and not the original frames. This is motivated by the fact that we aim here at measuring the performance of the compression method solely. However, for sake of completeness, results comparing to the original frames are briefly presented in the \ref{subsec:compwrtorgframes} and detailed in the supplementary material.

\subsubsection{Image Compression}\label{sec:res-img-comp}
We compare our method with the most recent state-of-the-art codecs such as VTM \cite{VTM}, AV1 \cite{AV1} and deep compression models such as scale Hyper Prior (HP) \cite{balle2018variational}, factorized entropy model with scale and mean hyperpriors (MeanHP) \cite{minnen2018joint} and the anchor variant of \cite{cheng2020learned}. MeanHP and \cite{cheng2020learned} are trained with the MS-SSIM objective and HP with the MSE (we found that this leads to better results) on the Celeba-HQ dataset. All the methods are evaluated on the independent evaluation dataset: MEAD-Intra and FILMPAC.

\textbf{Quantitative results:} We present the rate distortion curves for different BPP on the \textbf{MEAD-intra} dataset in Figure \ref{fig:results_MEAD_intra}. As can be observed, for the perceptual metrics LPIPS and PIM, our method significantly outperforms the state-of-the-art methods by a large margin. Deep learning based methods are better than traditional codecs for high BPP but they are inferior for lower BPP. This observation highlights the potential of our method to obtain reconstruction with good perceptual quality. For classical metrics, our method is the best at medium and high BPP for MS-SSIM, and has the highest PSNR at high BPP. It is noted that the deep learning based competitors trained with MS-SSIM objective, are still not able to outperform the traditional codecs VTM and AV1. The above observations also holds on the evaluation of the FILMPAC dataset, please refer to the supplementary material.

\textbf{Qualitative results:} Next we discuss the qualitative results, shown in figure \ref{fig:qual_results_FP006734MD02_1}. Perceptually, at a comparable BPP, AV1 introduces blocking artifacts while VTM generates blurry results. The reconstruction of MeanHP is not sharp enough and present distortion in color. Our method delivers perfect reconstruction compared to the projected image and preserves all the details. For example, the details of the eye brows are preserved where as the existing methods failed. Additional results are presented in the supplementary material.

\subsubsection{Video Compression}
We compare our method (SGANC IC) with state-of-the-art, deep learning based and traditional methods:  \textit{Wang.et.al}\cite{wang2021one}, DVC\cite{lu2019dvc}, Versatile Video Coding Test Model (VTM) \cite{VTM}, and H.265 standard \cite{h265}. We used the official implementation for both VTM (Random access with GOP=16) and H.265 (\href{http://ffmpeg.org/}{FFmpeg Library}). The implementation details of all the methods are detailed in the supplementary material. Each method is evaluated on the \textbf{MEAD-inter} dataset. 
\textbf{Comparison with learning based methods:} \quad First we compare our work with \cite{wang2021one} using the implementation \cite{oneshotgit}
at a comparable low bit-rate, the reconstructed frame from the compressed video and average metric over the frames are shown in Figure \ref{fig:qual_results_low_mead_inter_4_main}. It is clear, from the quantitative metrics and visual inspection, that our method significantly outperforms other ones. We encourage the reader to see the compressed videos in the supplementary file, where they clearly show that \cite{wang2021one} suffers from blurred reconstruction, as well as artifacts probably due to unstable landmark detection. 

Next we compare our method with the learning based video compression method DVC
\cite{lu2019dvc} using the pre-trained weights \cite{dvc_git} at low bit-rate ($\lambda=256$) using the I-compressed frames by H.265 ($q=48$) and meanHP using our trained model for intra-compression. It has higher BPP compared to other methods, but still the reconstructions are very blurry and not sharp. The perceptual quality is drastically lower than our method. Despite being simple, our method better restored the edges and texture, and reconstructed frames with high perceptual quality than the learning based ones\cite{wang2021one, lu2019dvc}. 
\textbf{Comparison with traditional SOTA methods:}
Here, we compare our method with traditional approaches. Figure \ref{fig:results_MEAD_inter} presents the quantitative evaluation curves on the \textbf{MEAD-inter} dataset and the metrics are first averaged over the frames in each video, and then averaged across all the videos.  As similar to the observations in section \ref{sec:res-img-comp}, our proposed method for video compression achieved impressive performance and significantly outperformed the state-of-the-art methods especially with respect to perceptual metrics. 
In terms of PSNR, our method is lower than VTM and comparable with H.265 at low bit-rate, but achieved better PSNR at high BPP with both methods. 
Figure \ref{fig:qual_results_low_mead_inter_4_main} compares the reconstruction quality with comparable BPP. Our method is almost artifacts free, photorealistic and perceptually more pleasant, while VTM leads to blurry results and H.265 exhibits blocking artifacts. For more visual results please refer to the supp. material.
Note that, despite the high quality of our method, the reconstruction of StyleGAN (projected vs original) is still not perfect (Fig. \ref{fig:org_proj_frames}). Once this factor is eliminated, the distortion coming from the quantization (projected vs ours) is negligible, which makes our method very promising taking into account of the exponential improvement of GAN generation and inversion.
\subsubsection{Comparison with original frames}\label{subsec:compwrtorgframes}
For the sake of completeness, we provide the visual comparison with original frames in Fig. \ref{fig:comp_with_originalframes}. Despite the high quality and photo-realistic outputs, our method loses the fidelity in the reconstruction. We stress that this behaviour is expected as off-the-shelf StyleGAN inversion is not perfect. Further we computed the FID score between the original frames and reconstructed frames with different methods. Our FID score (109.32) is similar to the learning based methods (DVC 116.09, Wang et.al 110.49). It is also noted that the loss of fidelity is the feature of the learning based compression using GAN loss functions\cite{ganextreme} to generate photo-realistic outputs.
\subsubsection{Computational complexity}
Our method is computationally efficient and it only brings the minor computational burden to the entire workflow with off-the-shelf StyleGAN encoder and decoder. Our evaluation shows that the average computation time to map a point from \wplus to \wstar is $4.7ms$. Regarding the complexity of the styleGAN encoder and generator, there are recent works that propose solutions for real-time StyleGAN encoding and decoding, even on mobile platforms\cite{sergei_belousov, BELOUSOV} and for real-time image editing using StyleGAN\cite{kim_2021}.
The training time of our method is also not longer, since we only retain the parametization (invertible property) of the flow transformation, thus no likelihood maximization and plus the dimension of the latent space is small compared to image space. Our training time is approximately $6$ hours on a single $V100$ GPU.
\section{Conclusion}
\label{sec:discussion}
We have proposed in this paper a new paradigm for facial image and video compression based on GANs. Our framework is efficient to train and leads to perceptually competitive results compared to the most efficient state-of-the-art compression systems. We believe that at low bitrates, our solution leads to a different and more acceptable type of distortion, since the reconstructed image is very sharp and photorealistic. Our approach is not restricted for faces since GANs have been proposed for various natural objects and have been recently extended to ImageNet scenes \cite{sauer2022stylegan}. For a specific category of object, for which a GAN has not been trained already, our approach could be used after training the specific GAN.
The main limitation of our method is the approximation of any input image using StyleGAN. Firstly, We believe that the continuous and impressive improvements of GANs inversion and generation will limit this limitation in the near future. For instance, the third generation of StyleGAN generator was very recently released \cite{karras2021alias}. Secondly, we believe that our method can have an impact in application scenario where perceptual distortion is more important than exact face fidelity, such as extremely low bandwidth videoconferencing, or interaction in digital worlds like the metaverse.

\appendix
\section{Appendix}
\subsection{Outline}

The Appendix is organized as follows:
\begin{itemize}
    \item In section \ref{app:wang_comp}, we show the visual comparison of the reconstructed videos of our approach with the state of art methods, and provide the commands used for traditional codecs.
    \item In section \ref{sec:app_ablation_t}, we discuss the importance of using a bijective transformation $T$.
    \item In section \ref{sec:app_res_lemma_proof}, we prove that the distribution of residuals follows the known Irwing-Hall distribution.
    \item In section \ref{sec:app_sganc_ic} we detail the algorithms for training and testing of the inter coding with residual approach.
    \item In section \ref{sec:app_ablation_img_vid_comp} we detail the ablation study for image and video compression.
    \item In section \ref{sec:app_more_results_img_vid} we provide more results. 
\end{itemize}
\subsection{Visual comparison of our method with the state of the art methods}\label{app:wang_comp}
Here we provide the reconstructed videos of our method and the state of the art ones; Wang \textit{et.al}\cite{wang2021one}, DVC\cite{lu2019dvc}, H265\cite{h265}, and VTM\cite{VTM}. Except for DVC, where BPP is higher, all the reconstructed videos are at comparable BPP. For the state of the art methods, we provide the reconstructed videos both with projected and original frames.

In the supplementary file, the \emph{Videos} folder is organized as follows:
\begin{itemize}
    \item \emph{SOTA method}:
    \begin{itemize}
        \item \emph{original}
        \begin{itemize}
            \item W016\_SoA\_avgbpp\_SGANIC\_avgbpp.mp4
        \end{itemize}
        \item \emph{projected}
        \begin{itemize}
            \item W016\_SoA\_avgbpp\_SGANIC\_avgbpp.mp4
        \end{itemize}
    \end{itemize}
\end{itemize}
Where "SoA" substitutes the name of the method and "avgbpp" substitutes the corresponding BPP. Each video file compares SoA method with our method. We included only one video due to the space limitations.

\subsubsection{Wang et.al \cite{wang2021one}}
 The results of Wang et.al\cite{wang2021one} are obtained using the implementation from \footnote{https://github.com/zhanglonghao1992/One-ShotFree-ViewNeuralTalkingHeadSynthesis}. The reconstructed videos with both original and projected frames are displayed in: \emph{Videos$\rightarrow$Wang}.  
 
 The input frames to Wang et.al in \footnote{https://github.com/zhanglonghao1992/One-ShotFree-ViewNeuralTalkingHeadSynthesis} are of size $256\times 256$, thus the output of our method is also re-scaled to $256\times 256$ for the visualization.

\subsubsection{DVC}
The results of DVC are obtained using the implementation from {https://github.com/GuoLusjtu/DVC}. For the I-frame compression, we used H265 with $q=48$, which corresponds to the lower bpp. 

The pre-trained model from {https://github.com/GuoLusjtu/DVC}, with $\lambda=256$ has the lowest BPP, and we have used this model in our experiments to compare with our method. DVC has higher BPP compared to our method. The reconstructed videos with both original and projected frames are displayed in: \emph{Videos$\rightarrow$DVC}.  

\subsubsection{H265}
The results of H265 are obtained using \textit{ffmpeg} with $GOP=10$ and $q=48$. The videos are displayed in: \emph{Videos$\rightarrow$H265}.
The following is the detailed command to run H.265 (after converting the frames to yuv format using ffmpeg): 

\verb|ffmpeg -s 1024x1024 -pix_fmt yuv420p|\\ 
\verb|-r 25 -i inputvideo.yuv -c:v libx265| \\
\verb|-tune zerolatency -x265-params|\\
\verb|``crf=48:keyint=10:verbose=1"| \\
\verb|outputVideo.mkv|

The $q=48$ is chosen because to have a comparable BPP with our method on the lowest BPP.

\subsubsection{VTM}
The results of the VTM are obtained using the implementation from https://vcgit.hhi.fraunhofer.de/jvet/VVCSoftware\_VTM, with $q=36$. 
The input frames are converted to YUV format before feeding to the VTM encoder. The videos are displayed in: \emph{Videos$\rightarrow$VTM}.

It is clear from the videos that our method is almost artifacts free, photorealistic and perceptually more pleasant compared to the state-of-the-art.

The detailed commande to run VTM  for a video of length $NB$ frames (after converting the frames to yuv format using ffmpeg):

\verb|SEQ=``-i input.yuv --InputBitDepth=8|\\
\verb|--InputChromaFormat=420 --FrameRate=25|\\ 
\verb|--SourceWidth=1024 --SourceHeight=1024|\\
\verb|--FramesToBeEncoded=NB"|

Then:

\verb| $ENC  -c $CFG $SEQ -ip 32 -fs 0| \\
\verb| -q 36 -b bs.bin -o rec.yuv| \\
\verb|--PrintMSSSIM $OPT $>$ enc.log & |

And the following for decoding:
\verb|$DEC  -b bs.bin -o rec.yuv -d 8 |

Where \$ENC and \$DEC are the paths to the encoder and decoder binaries respectively, and \$CFG is the path to the configuration file:\\
\verb|cfg/encoder_randomaccess_vtm_gop16.cfg| from the official repository \footnote{https://vcgit.hhi.fraunhofer.de/jvet/VVCSoftware\_VTM}

The $q=36$ is chosen to have a comparable BPP with our method on the lowest BPP.

\subsection{Parametrization of the transformation $T$}
\label{sec:app_ablation_t}
In this section, we investigate the importance of using a bijective transformation (\emph{i.e.}, Normalizing Flows). To this end, we replace the Real NVP model with an autoencoder (AE) and retrain it with the entropy model on Celeba-HQ with the same implementation details as described in the main paper. The AE consists of $7$ layers with the same dimension (\emph{i.e.}, $512$) for the encoder and $7$ layers with the same dimension for the decoder with ReLU activation functions.
\subsubsection{Results:} 
\begin{figure*}[h]
\setlength\tabcolsep{2pt}%
\centering
\small
\begin{tabular}{p{0.7cm}cc}
\toprule
  &
Original &
Projected \\
 &
\includegraphics[width=0.44\linewidth]{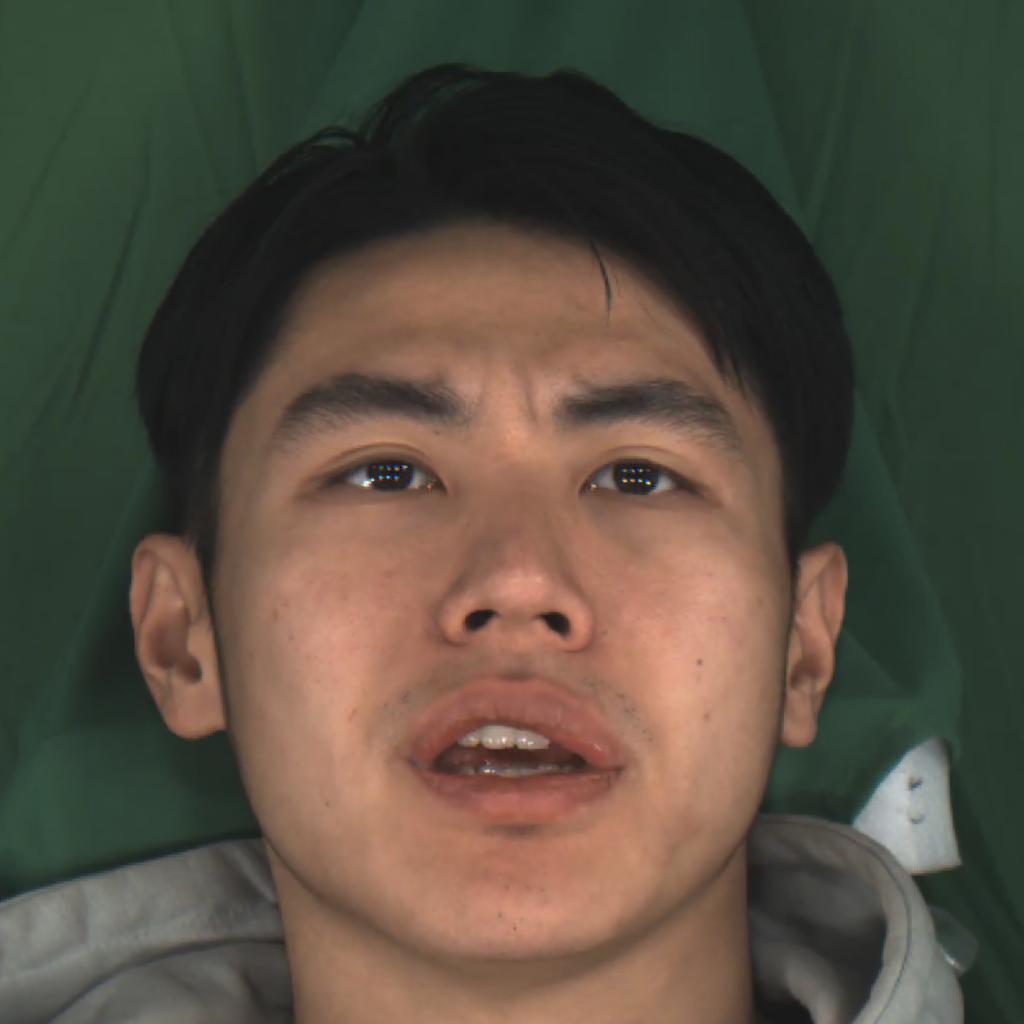}  & 
\includegraphics[width=0.44\linewidth]{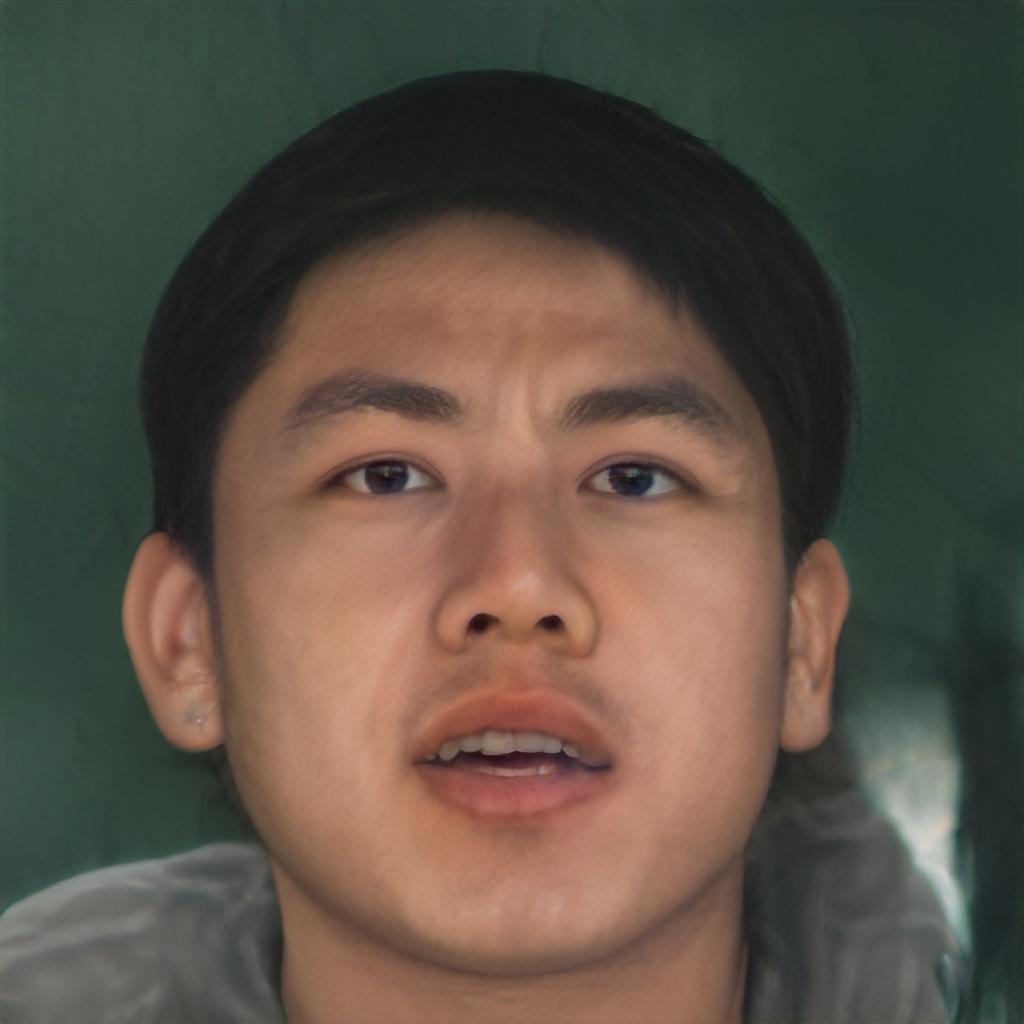}  \\  
  &
 & 
 \\ \midrule
Model &
AEC &
SGANC \\ 
&
\includegraphics[width=0.44\linewidth]{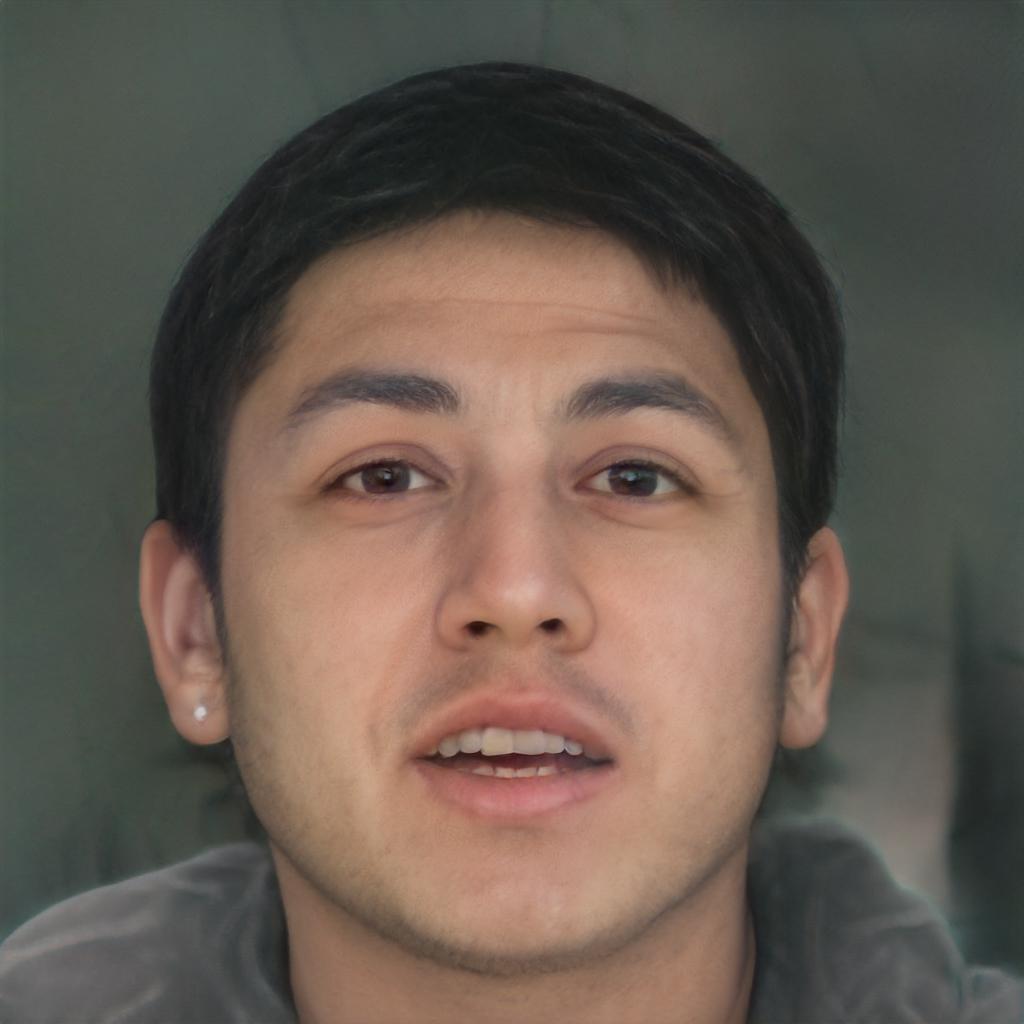} &
\includegraphics[width=0.44\linewidth]{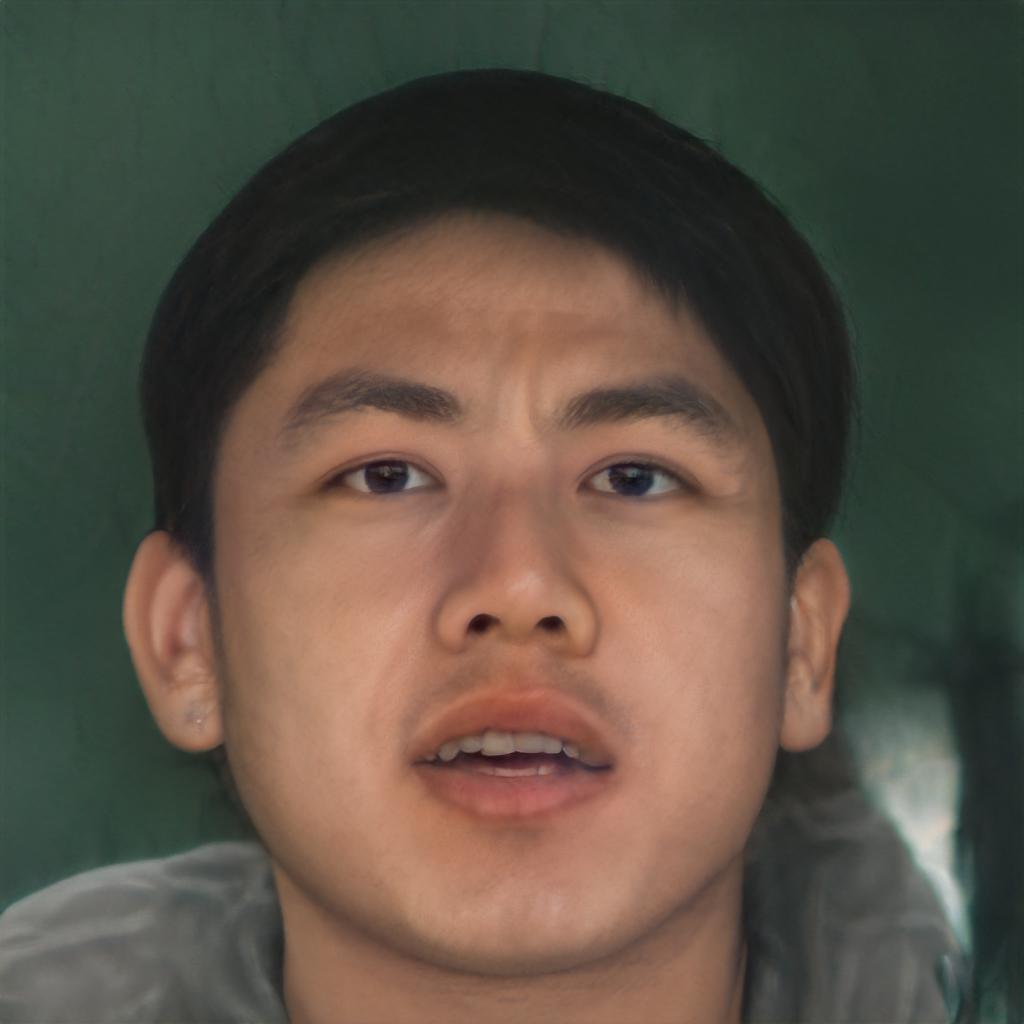} \\
BPP &
0.017 &
0.014 \\
\bottomrule
\end{tabular}
\caption{Qualitative results for image compression: Using an autoencoder (AEC) instead of normalizing flow (NF) leads to high distortion and a change in the facial attributes and the identity of the person. Our method (SGANC) leads to high quality reconstruction  and perceptually lower distortion.}
\label{fig:qual_results_low_mead_inter_4_vs_ae}
\end{figure*}

From Figure \ref{fig:qual_results_low_mead_inter_4_vs_ae}, we can notice that the parametrization of the transformation $T$ as Normalizing Flows (\emph{i.e.}, Real NVP) is key to better results. For instance, some facial attributes are changed when using an AE (such as the age and the skin color), as well as the person's identity.

\begin{algorithm*}[h]
\SetAlgoLined
\KwResult{Transformation ($T$), Entropy model ($p$)}
 Initialization: video dataset encoded as latent codes, where each video= $\{\bm{w}_1, \bm{w}_2, ..., \bm{w}_{t-1}, \bm{w}_{t}, ... \}$, number of frames in each batch ($N$), dataset size ($S$), Encoder (E), Generator (G)\;
 \While{$i < S$}{
 $t, L= 1, 0$\;
 \While{$t < N$}{
 $\bm{w^}{*}_t$, $\bm{w^}{*}_{t-1}$ = $T(\bm{w}_t)$, $T(\bm{w}_{t-1})$ \tcp*{Map the latent codes to \wstarc}
 
 $\hat{\bm{v}}_t = Q(\bm{w^}{*}_t - \bm{w^}{*}_{t-1})$ \tcp*{{Quantize (adding noise) the differences}}
 
 $\bar{\bm{w}}_t^{*} = \hat{\bm{w}}_{t-1}^{*} + \hat{\bm{v}}_t$  \tcp*{Compute an estimate of the latent code}
 $\hat{\bm{w}}_t^{*} = \bar{\bm{w}}_t^{*}$\;
 $L = L + \mathcal{L}$ \tcp*{Compute the loss $\mathcal{L}$ (\eqref{eq:l1_reg})}
 $t = t+1$\;}
 Update the parameters of $T$ and $p$ to minimize $L$\;
 $i = i+1$\;
 }
 \caption{SGANC IC: during training}
 \label{alg:sganc_ic_train}
\end{algorithm*}

\begin{algorithm*}[h]
\SetAlgoLined
\KwResult{Compressed frames sequence: $\{\hat{\bm{x}}_1, \hat{\bm{x}}_2, ..., \hat{\bm{x}}_{t-1}, \hat{\bm{x}}_{t}, ..., \hat{\bm{x}}_{N} \}$}
 Initialization: frames sequence: $\{\bm{x}_1, \bm{x}_2, ..., \bm{x}_{t-1}, \bm{x}_{t}, ..., \bm{x}_{N} \}$, $N$=number of frames, Encoder (E), Generator (G), Transformation $T$, entropy coder (EC) and decoder (ED), quantizer (Q), $g$ for residual coding\;
 $\hat{\bm{w}}^{*}_0$ = $ ED(EC(Q(T(E(\bm{x}_0))))$ \tcp*{Intra Coding of the first frame}
 $t=1$\;
 \While{$t < N$}{
 $\bm{w^}{*}_t$ = $T(E(\bm{x}_t))$ \tcp*{Encode the frames and map them to \wstarc}
 $\bm{w^}{*}_{t-1}$ = $T(E(\bm{x}_{t-1}))$\;
 $\hat{\bm{v}}_t = ED(EC(Q(\bm{w^}{*}_t - \bm{w^}{*}_{t-1})))$ \tcp*{Quantize, Compress and Decompress (Receiver) the differences}
 
 $\bar{\bm{w}}_t^{*} = \hat{\bm{w}}_{t-1}^{*} + \hat{\bm{v}}_t$  \tcp*{Compute an estimate of the latent code}
 
 \eIf{$t \% g == 0$}
  {
   $\hat{\bm{r}}_t = ED(EC(Q(\bm{w^}{*}_t - \bar{\bm{w}}_t^{*})))$ \tcp*{Quantize, Compress and Decompress (Receiver) the residual}
   
 $\hat{\bm{w}}_t^{*} = \bar{\bm{w}}_t^{*} + \hat{\bm{r}}_t$  \tcp*{Reconstruct the latent code}
  }
  {
  $\hat{\bm{w}}_t^{*} = \bar{\bm{w}}_t^{*}$\;
  }

 $\hat{\bm{x}}_t$ = $G(T^{-1}(\hat{\bm{w}}_t^{*}))$  \tcp*{Reconstruct the image}
 $t = t+1$\;
 }
 \caption{SGANC IC: during test}
 \label{alg:sganc_ic_test}
\end{algorithm*}

\subsection{Explicit Distribution of the Residuals: Proof of Lemma 1}
\label{sec:app_res_lemma_proof}

Let $Q(x) = x +\epsilon$ be the continuous relaxation of the quantization, and $\epsilon$ follows the uniform distribution $\mathcal{U}_{[-0.5, 0.5]}$. Let the $t$ be the frame index ranging from $\{0, 1, \ldots, K-1\}$.

Let $\bm{w}^{*}_0 \in \mathcal{R}^n$ and $\hat{\bm{w}}^{*}_0 = Q(\bm{w}^{*}_0)$. Let us define the following
\begin{align}
    \hat{\bm{v}}_t & = Q(\bm{w}^{*}_t - \bm{w}^{*}_{t-1}) = \bm{w}^{*}_t - \bm{w}^{*}_{t-1} + \epsilon, \\
    \bar{\bm{w}}_t & = \hat{\bm{w}}_{t-1}^{*} + \hat{\bm{v}}_t \nonumber,
\end{align}
Similarly, we define 
\begin{align}
    \hat{\bm{r}}_t =\left\{
                \begin{array}{ll}
                  Q(\bm{w}^{*}_t - \bar{\bm{w}}_t^{*}), \quad \forall & t \equiv 0\pmod{g}\\
                  0 & else
                \end{array}
              \right.
\end{align}

Now we prove that $\hat{\bm{r}}_t$ follows the Irwin-Hall or the uniform sum distribution \cite{johnson1995continuous}.
we have:
\begin{align}
\label{eq:proof_main}
\hat{\bm{r}}_t = Q(\bm{w}^{*}_t - \bar{\bm{w}}_t^{*}) &= \bm{w}^{*}_t - \bar{\bm{w}}_t^{*} + \epsilon_1 \nonumber \\
 &= \bm{w}^{*}_t - \left(\hat{\bm{w}}_{t-1}^{*} + \hat{\bm{v}}_t \right) + \epsilon_1 \nonumber \\
 & = \bm{w}^{*}_t - \left(\hat{\bm{w}}_{t-1}^{*} + \bm{w}^{*}_t - \bm{w}^{*}_{t-1} + \epsilon_1  \right) + \epsilon_2 \nonumber \\
 & = \bm{w}^{*}_t - \hat{\bm{w}}_{t-1}^{*} - \bm{w}^{*}_t + \bm{w}^{*}_{t-1} - \epsilon_2   + \epsilon_1 \nonumber \\
  & =  \bm{w}^{*}_{t-1} - \hat{\bm{w}}_{t-1}^{*} - \epsilon_2   + \epsilon_1  
\end{align}
For $t \equiv 0 \pmod{g}$, the reconstructed latent code is computed from the residual, and can be written as:
\begin{align}
\hat{\bm{w}}_{t}^{*} & = \hat{\bm{r}}_{t} + \bar{w}_{t}^* \nonumber\\
& = Q(\bm{w}^{*}_{t} - \bar{\bm{w}}_{t}^{*}) + \bar{\bm{w}}_{t}^* \nonumber\\
& = \bm{w}^{*}_{t} - \bar{\bm{w}}_{t}^{*} + \epsilon + \bar{\bm{w}}_{t}^*  \nonumber\\
& = \bm{w}^{*}_{t} + \epsilon
\end{align}
Otherwise, it is the estimated latent code $\bar{\bm{w}}_{t}^*$, thus $\hat{\bm{w}}_{t}^{*}$ can be written as:
\begin{align}
\label{eq:proof_w_hat}
\hat{\bm{w}}_{t}^{*} = \left\{
              \begin{array}{ll}
                 \bm{w}^{*}_{t} + \epsilon, \quad \forall & t \equiv 0\pmod{g}\\
                  \bar{\bm{w}}_{t}^* & else
                \end{array}
              \right.
\end{align}
For $g=1$, \eqref{eq:proof_main} becomes:
\begin{align}
\hat{\bm{r}}_t & = \bm{w}^{*}_{t-1} - \hat{\bm{w}}_{t-1}^{*} - \epsilon_2   + \epsilon_1 \nonumber \nonumber\\
& = \bm{w}^{*}_{t-1} - \bm{w}^{*}_{t-1} + \epsilon_3 - \epsilon_2   + \epsilon_1 \nonumber \nonumber \\
& = \epsilon_3 - \epsilon_2   + \epsilon_1
\end{align}
For $g>1$, Let us define the following quantity:
\begin{align}
\label{eq:proof_mt}
    \hat{\bm{m}}_t =\left\{
                \begin{array}{ll}
                  \hat{\bm{r}}_t, & \forall t \equiv 0\pmod{g}\\
                  Q(\bm{w}^{*}_t - \bar{\bm{w}}_t^{*}) & else
                \end{array}
              \right.
\end{align}

Using \eqref{eq:proof_w_hat} and \eqref{eq:proof_mt}:
\begin{align}
\hat{\bm{r}}_t & = \bm{w}^{*}_{t-1} - \hat{\bm{w}}_{t-1}^{*} - \epsilon_2   + \epsilon_1 \nonumber \\
& =  \bm{w}^{*}_{t-1} - \bar{\bm{w}}_{t-1}^{*} - \epsilon_2   + \epsilon_1 \nonumber \\
& = \hat{\bm{m}}_{t-1} - \epsilon_2\nonumber \\
& = ... \nonumber \\
& = \hat{\bm{m}}_{t-(g-1)} - \sum_{i=1}^{g-1} \epsilon_i
\end{align}
Similarly to \eqref{eq:proof_main}, we can replace $\hat{\bm{m}}_{t-(g-1)}$:
\begin{align}
\hat{\bm{r}}_t  &= \bm{w}^{*}_{t-g} - \hat{\bm{w}}_{t-g}^{*} + \epsilon_1 - \sum_{i=1}^{g-1} \epsilon_i
\end{align}
As $t-g \equiv 0 \pmod{g} $ (as the residual is used only each $t \equiv 0 \pmod{g}$), from \eqref{eq:proof_w_hat}, we can replace $\hat{\bm{w}}_{t-g}^{*}$ by its relaxed approximation:
\begin{align}
\hat{\bm{r}}_t  & = \bm{w}^{*}_{t-g} - \left(\bm{w}_{t-g}^{*} + \epsilon_2 \right) + \epsilon_1 - \sum_{i=1}^{g-1} \epsilon_i \nonumber\\
& = \epsilon_1 - \epsilon_2 + \epsilon_3 - \sum_{i=1}^{g-1} \epsilon_i
\end{align}
Which is the sum of $3 + (g-1)$ independent random variables following the uniform distribution.

We showed $\hat{\bm{r}}_t$ follows the uniform sum distribution, which is the Irwin-Hall distribution $IH(x;n)$ with parameter $n=3+(g-1)$ and $x \in [-n\times0.5, n\times0.5]$. Since the $\epsilon$ is the uniform distribution $\mathcal{U}_{[-0.5, 0.5]}$, in our case the support of the Irwin-Hall distribution is shifted by $-n\times 0.5$.

\subsection{Inter Coding (SGANC IC)}
\label{sec:app_sganc_ic}
In this section we present the algorithms for video compression using inter coding with residual during training in Alg. \ref{alg:sganc_ic_train} and testing in Alg. \ref{alg:sganc_ic_test}.

\label{sec:app_res_distro}

\subsection{Ablation Study}
\label{sec:app_ablation_img_vid_comp}
In this section we detail the ablation study for image and video compression.
\subsubsection{Image Compression (Distortion loss)}
\label{sec:app_ablation_disto}
In this section we compare different choices of the distortion loss. Specifically, we compare the MSE loss in the image space (Img. D), the MSE loss in the latent space \wplus (Lat. D), the combination of both (Lat.-Img. D), the LPIPS loss in the image space (LPIPS D) and the combination of the MSE and the LPIPS losses in the image space (LPIPS-Img. D). The implementation details are the same as in section 4.2 in the paper, except for the choice of the distortion loss and we use a batch size of 4 when training with LPIPS. We compare on both MEAD intra and FILMPAC dataset.
\subsubsection{Results:} 
\begin{figure*}[t]
     \centering
     \begin{subfigure}[b]{0.32\linewidth}
         \centering
         \caption{\tiny{PSNR}}
         \includegraphics[width=\linewidth, height=5cm]{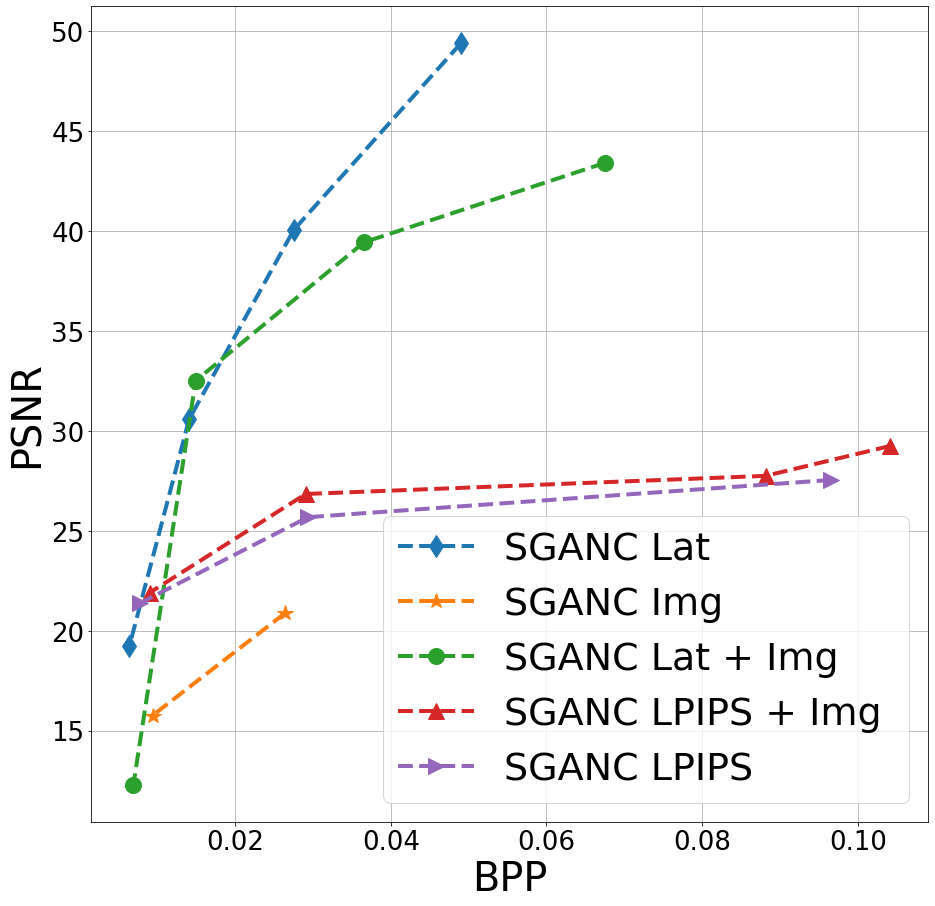}
     \end{subfigure}
     \hfill
     \begin{subfigure}[b]{0.32\linewidth}
         \centering
         \caption{\tiny{MS-SSIM}}
         \includegraphics[width=\linewidth, height=5cm]{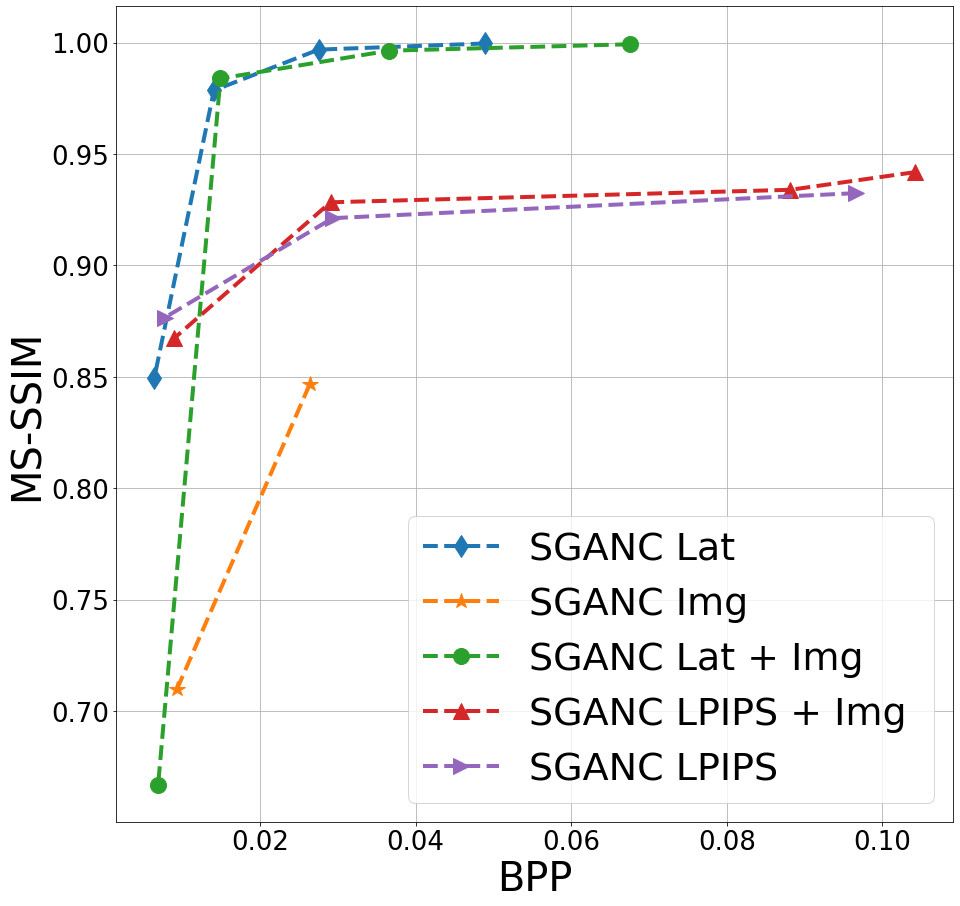}
     \end{subfigure}
     \hfill
     \begin{subfigure}[b]{0.32\linewidth}
         \centering
         \caption{\tiny{LPIPS \cite{zhang2018unreasonable}}}
         \includegraphics[width=\linewidth, height=5cm]{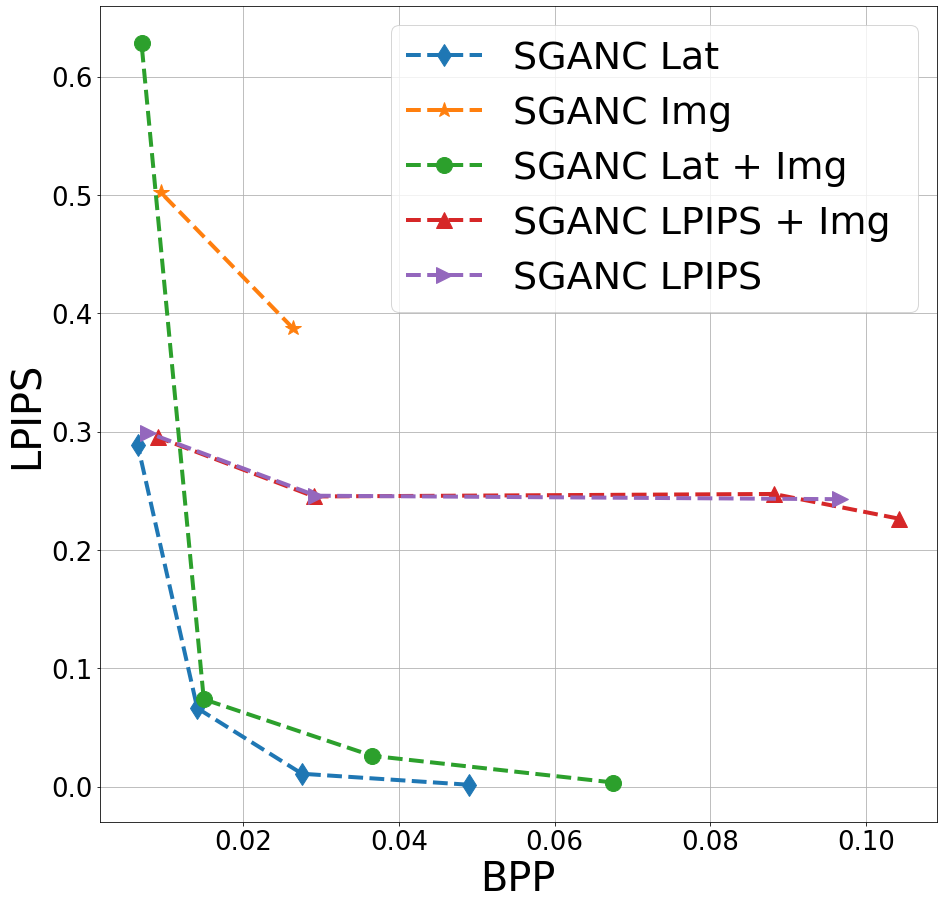}
     \end{subfigure}
        \caption{Ablation study for the choice of the distortion loss on MEAD intra dataset. we compare the MSE loss in the image space (SGANC Img, Orange), the MSE loss in the latent space \wplus (SGANC Lat, Blue), the combination of both (SGANC Lat + Img, Green), the LPIPS loss in the image space (SGANC LPIPS, Purple) and the combination of the MSE and the LPIPS losses as done in other works (SGANC LPIPS + Img, Red). The loss in the latent space (SGANC Lat, ours) outperforms the MSE and LPIPS losses in the image space. Using the MSE loss with our loss does not seem to improve the results (SGANC Lat + Img).}
        \label{fig:distortion_ablation_MEAD_intra}
\end{figure*}

\begin{figure*}[h]
\setlength\tabcolsep{2pt}%
\centering
\begin{tabular}{p{0.5cm}cccccc}
\toprule
\begin{turn}{90}\hspace{0.7cm} Original \end{turn} &
 &
 \includegraphics[width=0.16\linewidth]{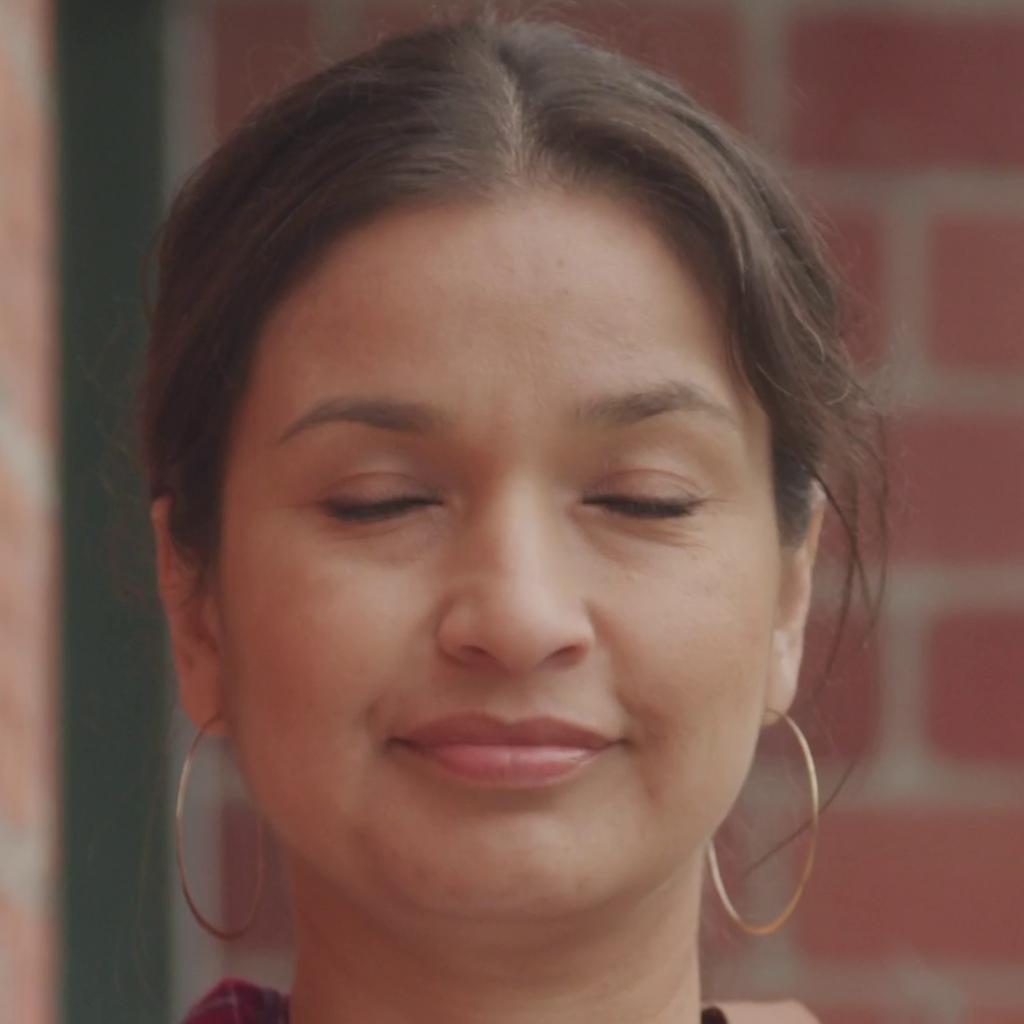}  & 
\includegraphics[width=0.16\linewidth]{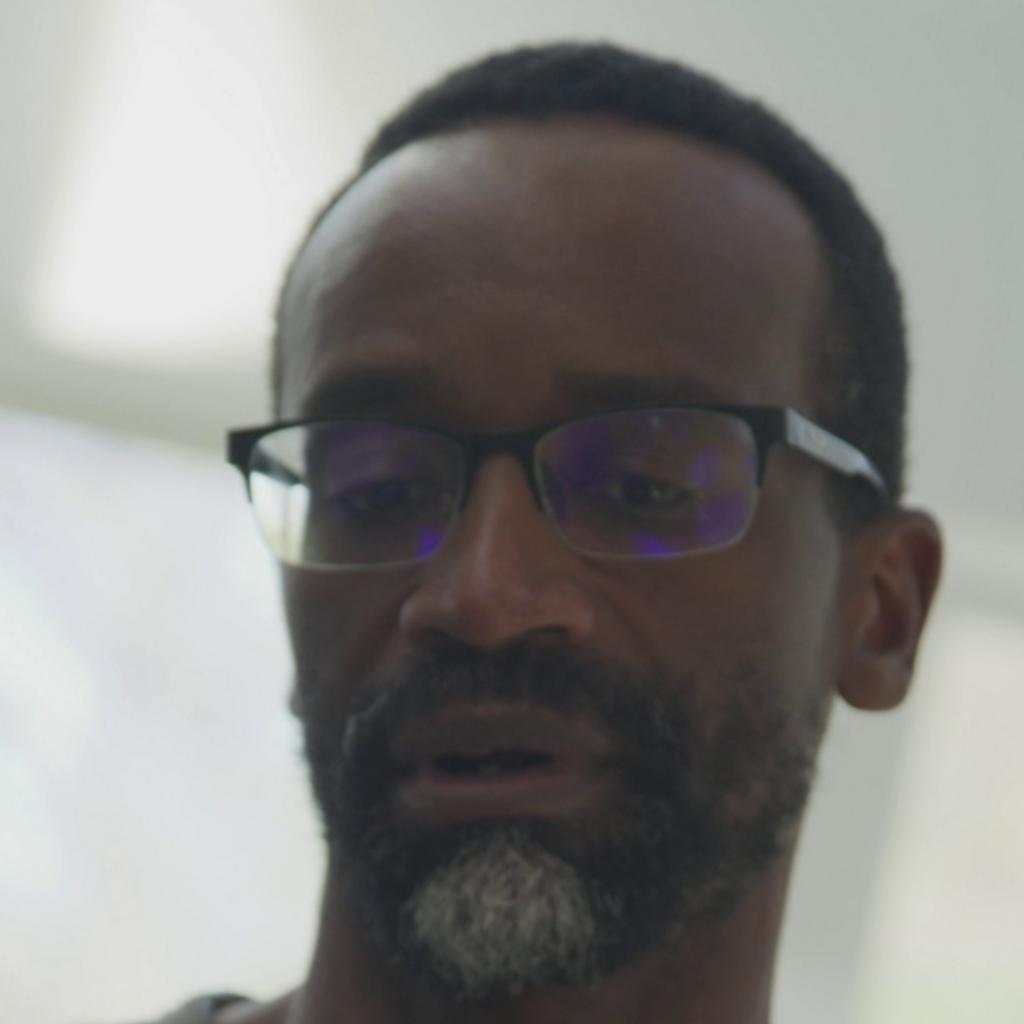}  & 
\includegraphics[width=0.16\linewidth]{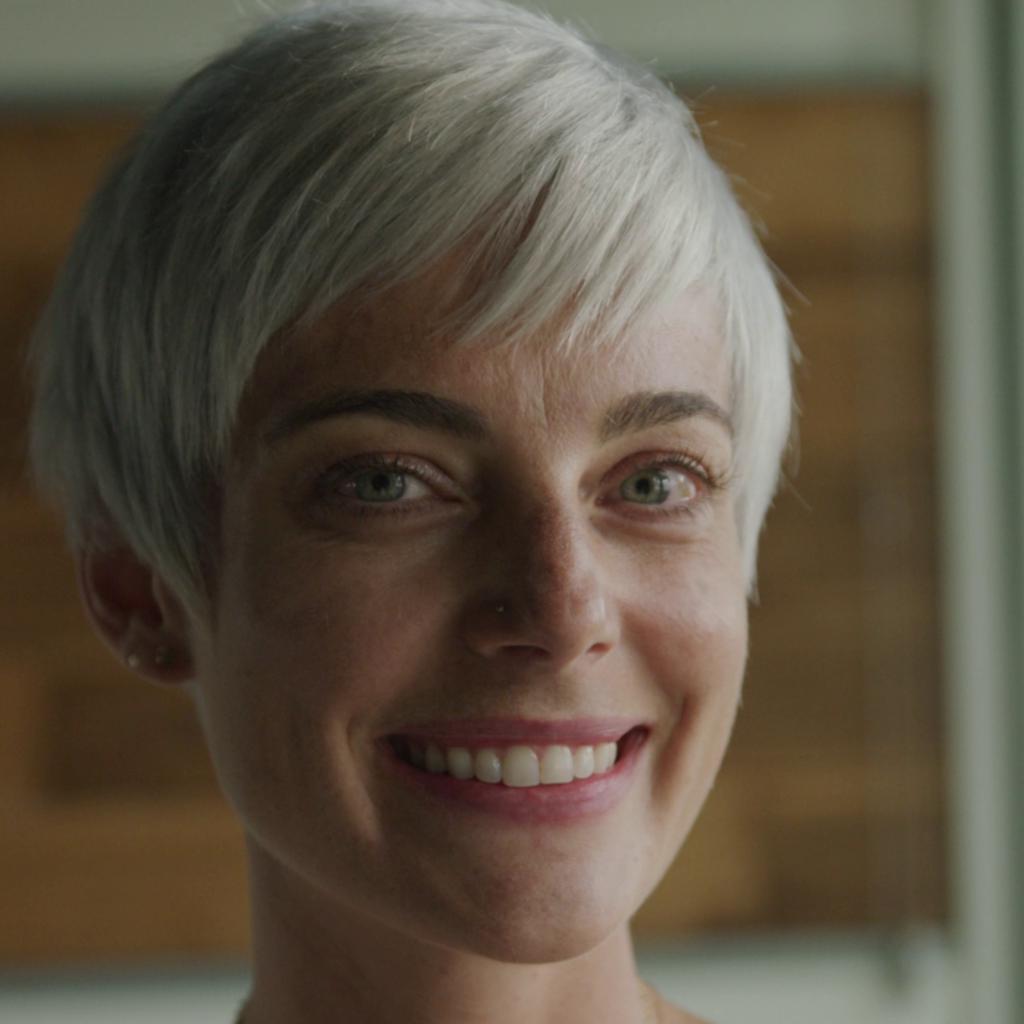} &
\includegraphics[width=0.16\linewidth]{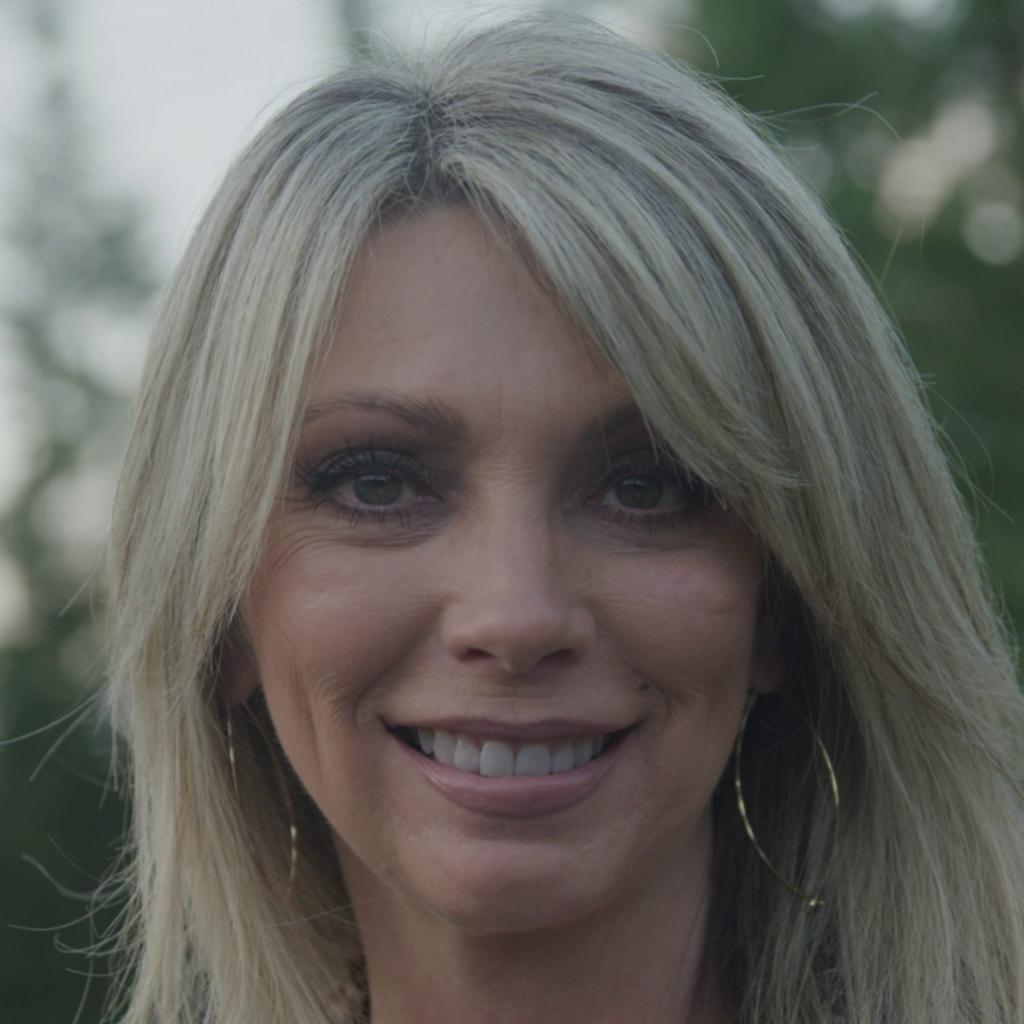} &
\includegraphics[width=0.16\linewidth]{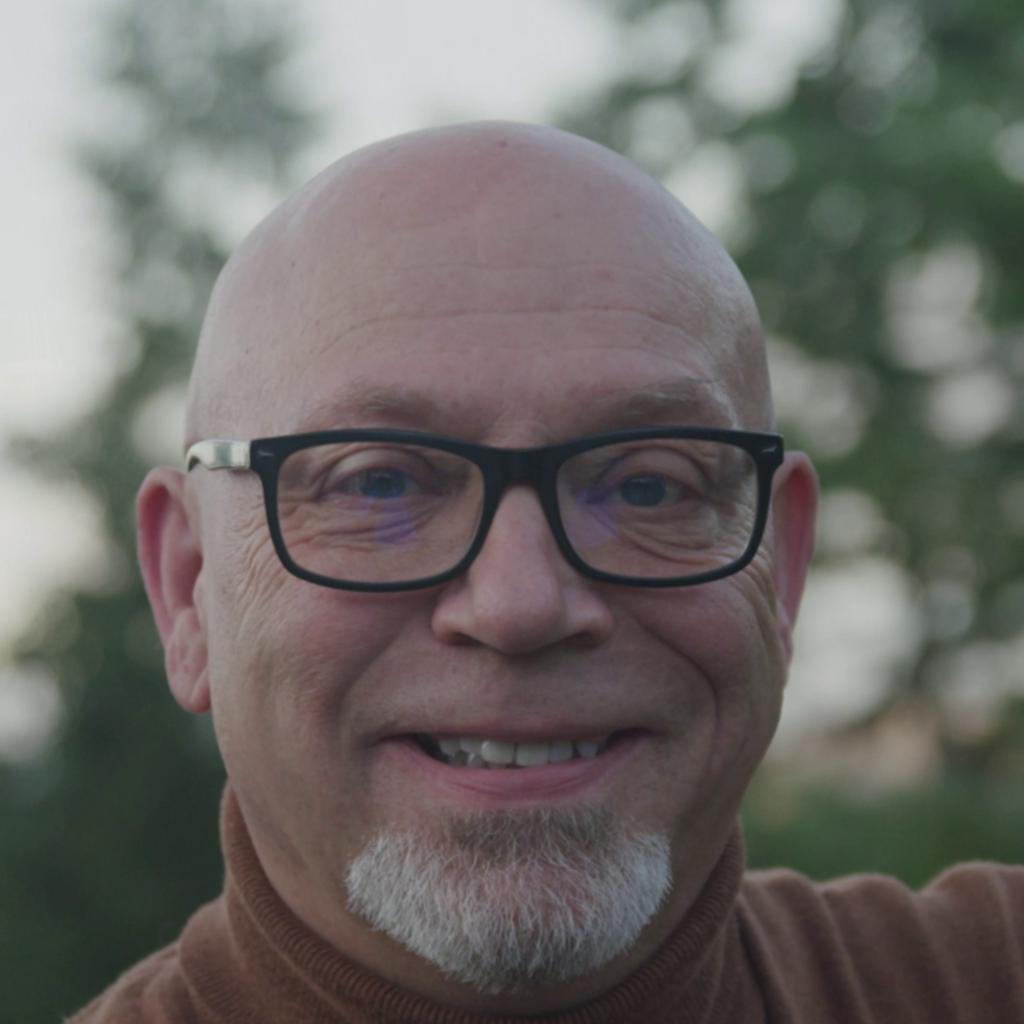} \\ 

 \begin{turn}{90}\hspace{0.7cm} Projected \end{turn} &
 &
 \includegraphics[width=0.16\linewidth]{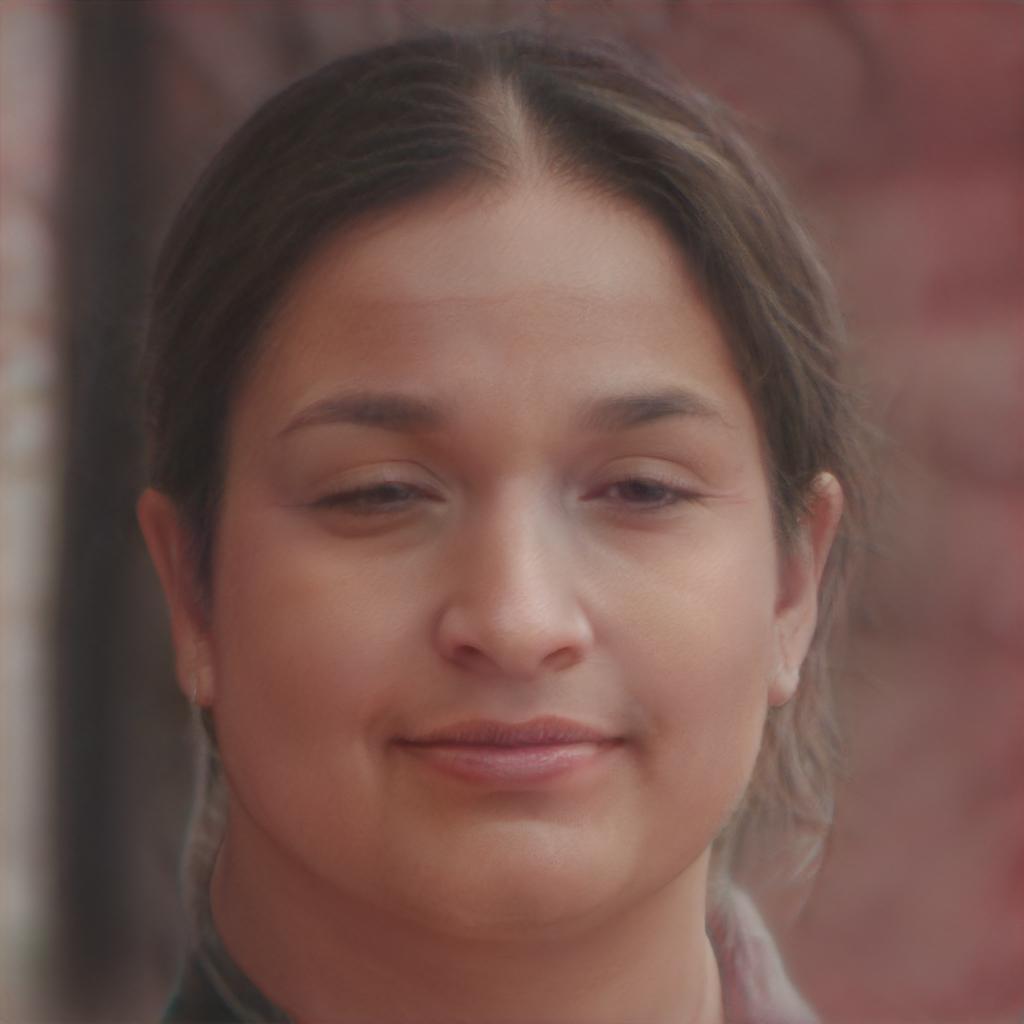}  & 
\includegraphics[width=0.16\linewidth]{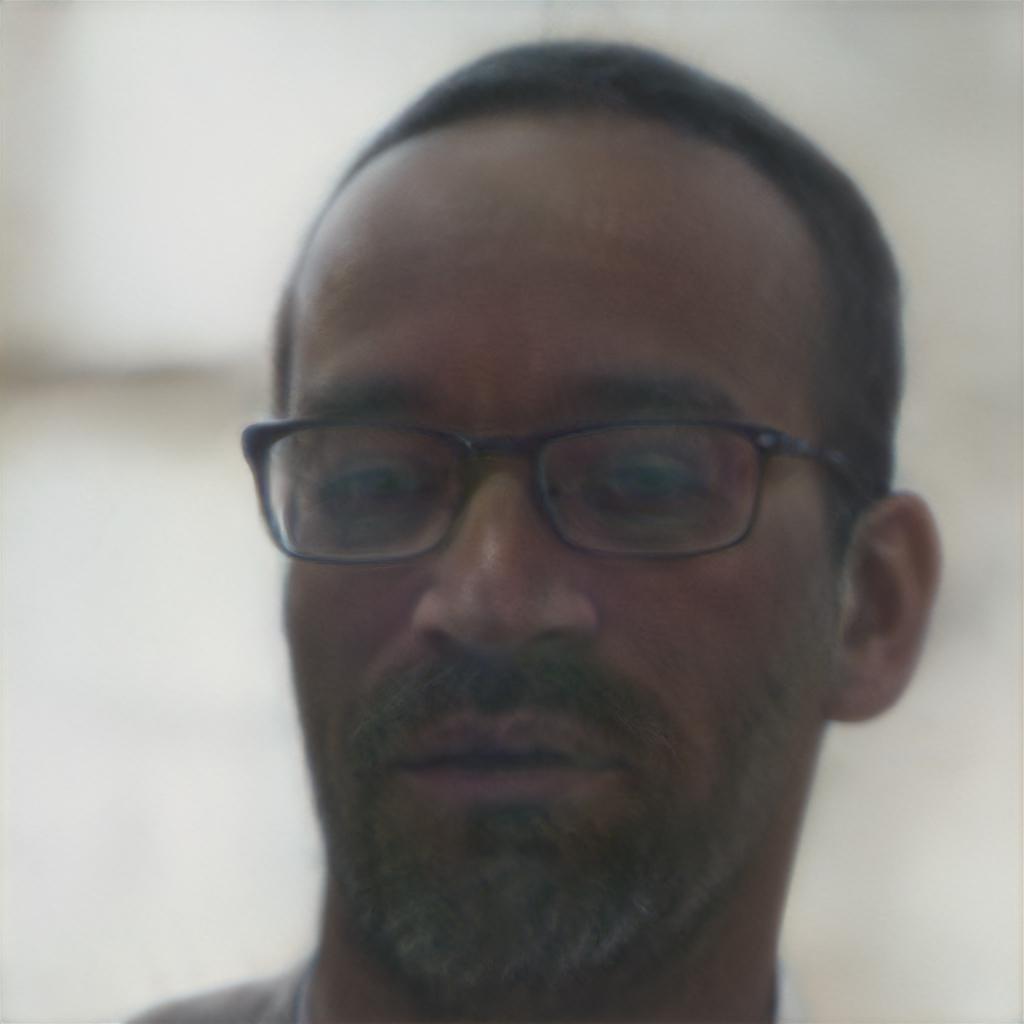}  & 
\includegraphics[width=0.16\linewidth]{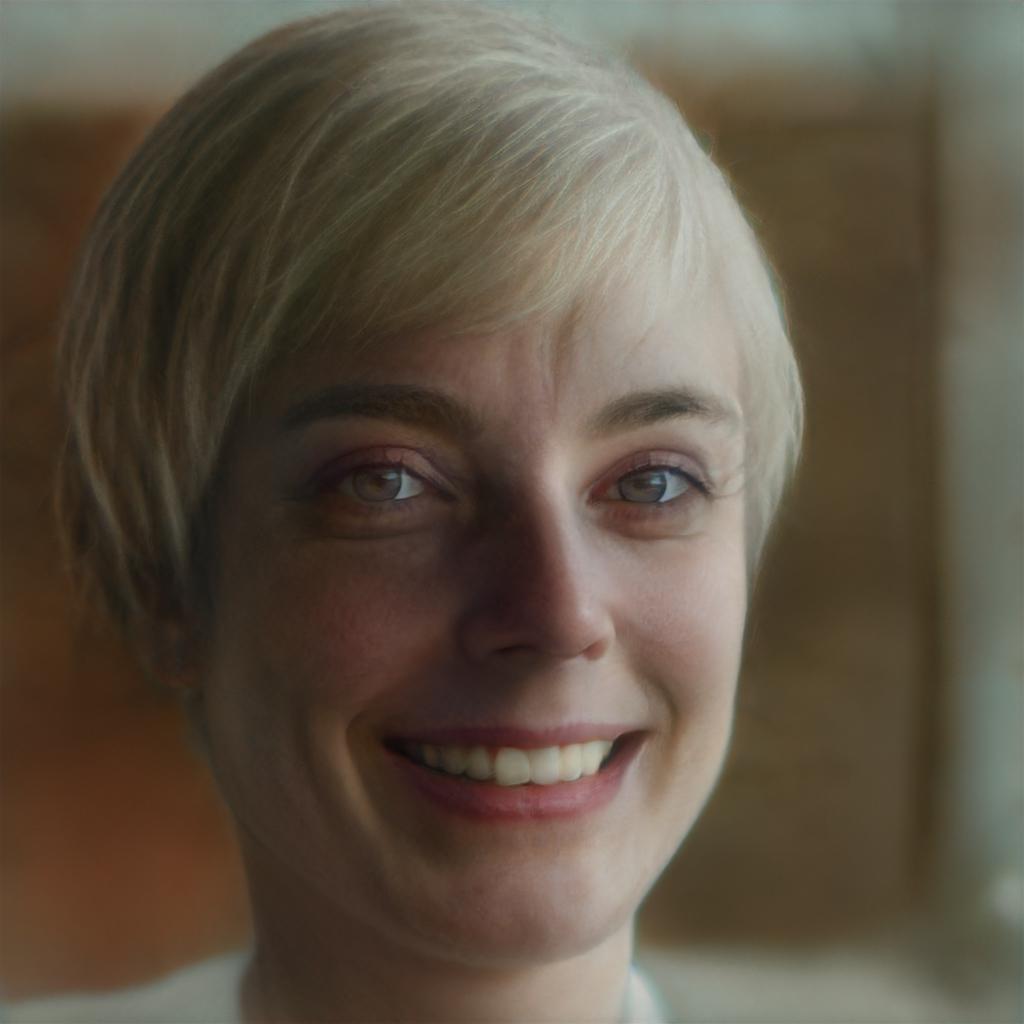} &
\includegraphics[width=0.16\linewidth]{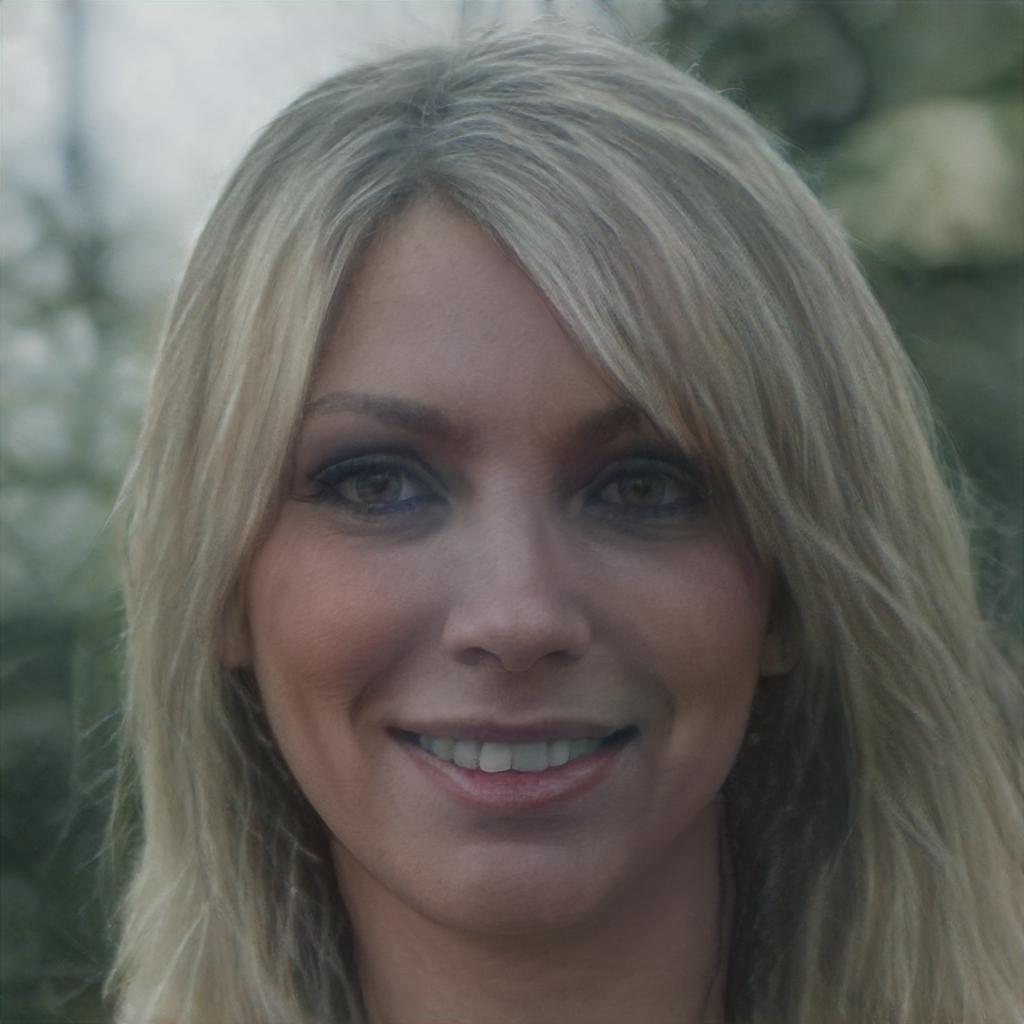} &
\includegraphics[width=0.16\linewidth]{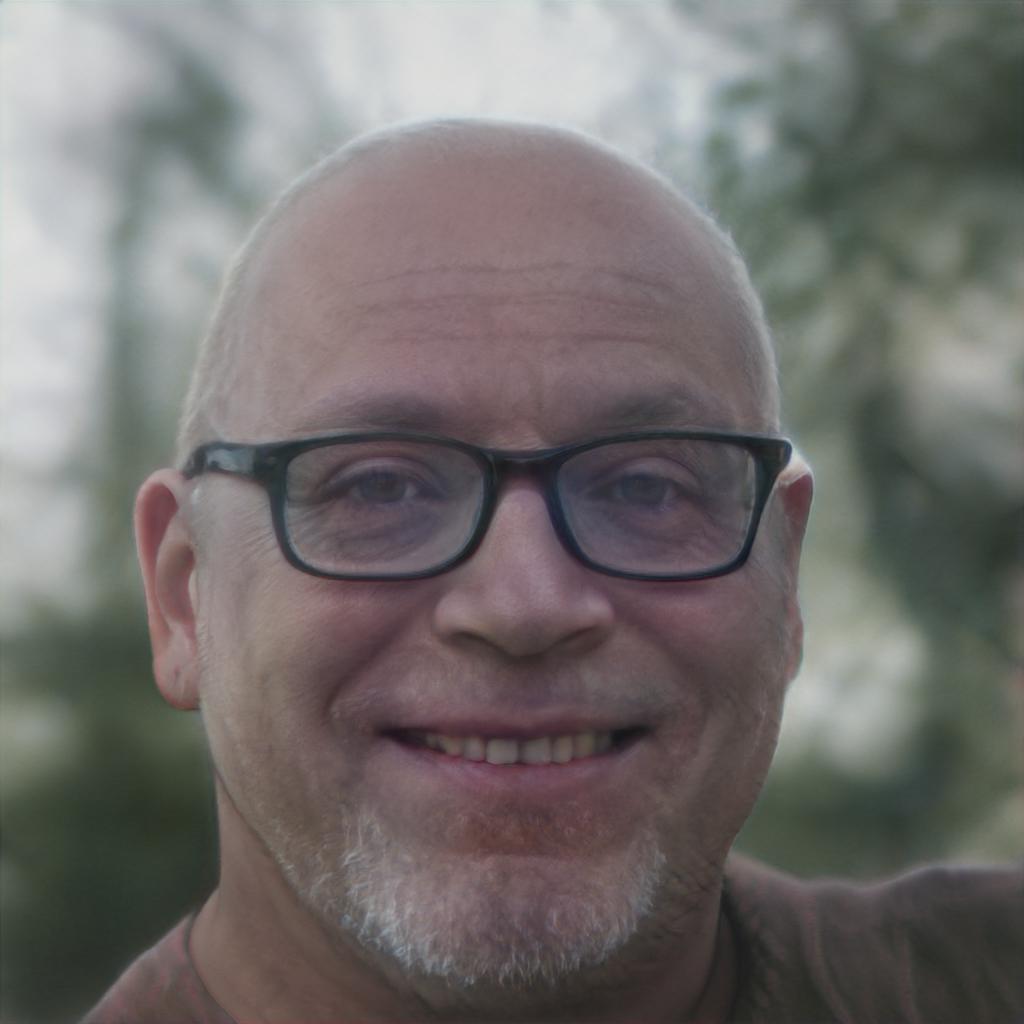} \\

 \begin{turn}{90}\hspace{0.8cm} Lat. D \end{turn} &
\begin{turn}{90}\hspace{0.6cm} BPP=0.016 \end{turn} &
 \includegraphics[width=0.16\linewidth]{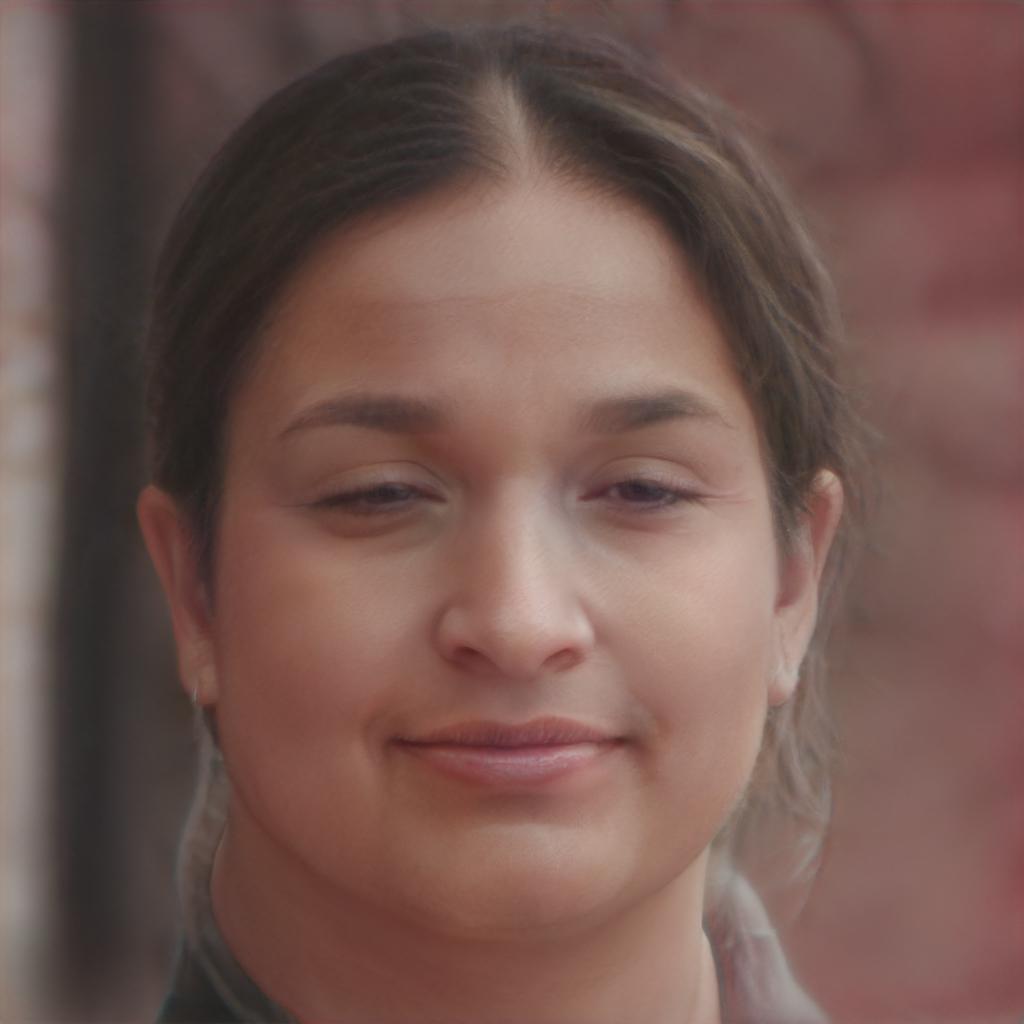}  & 
\includegraphics[width=0.16\linewidth]{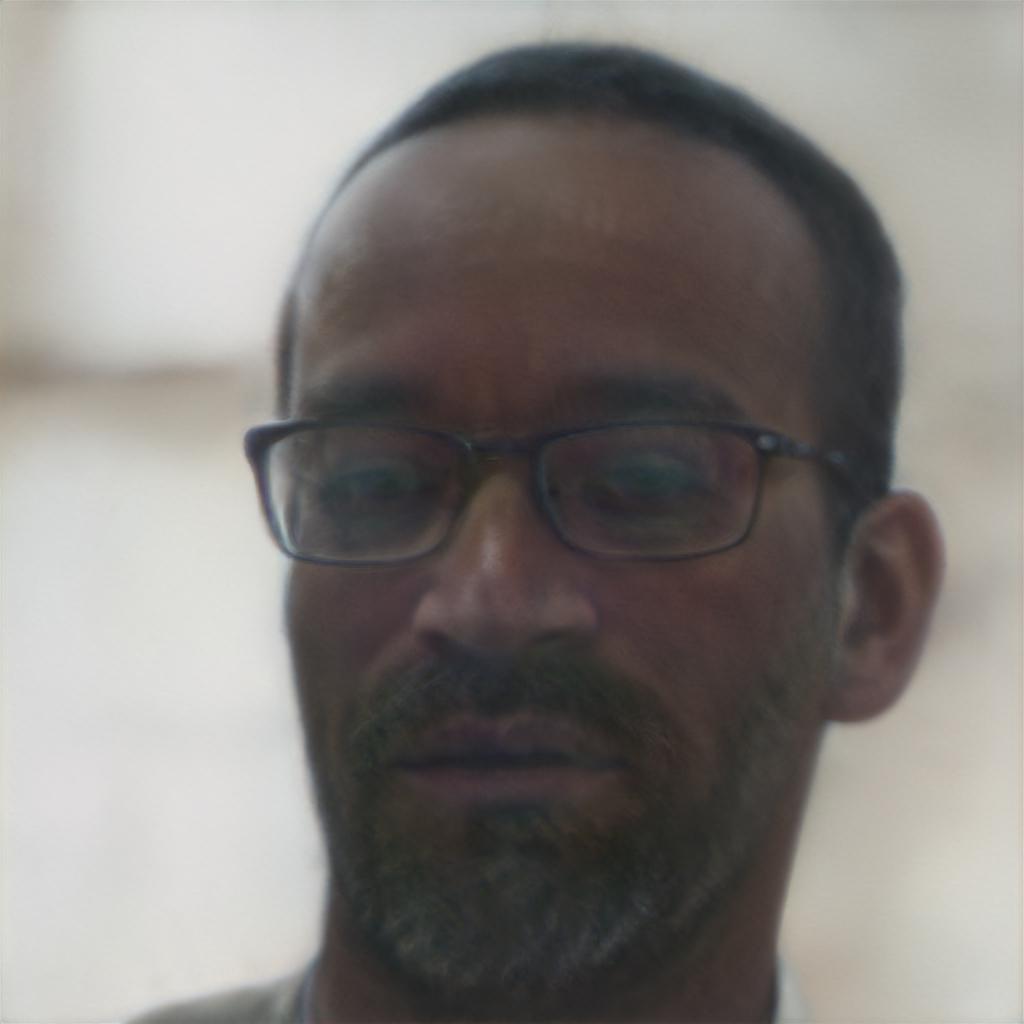}  & 
\includegraphics[width=0.16\linewidth]{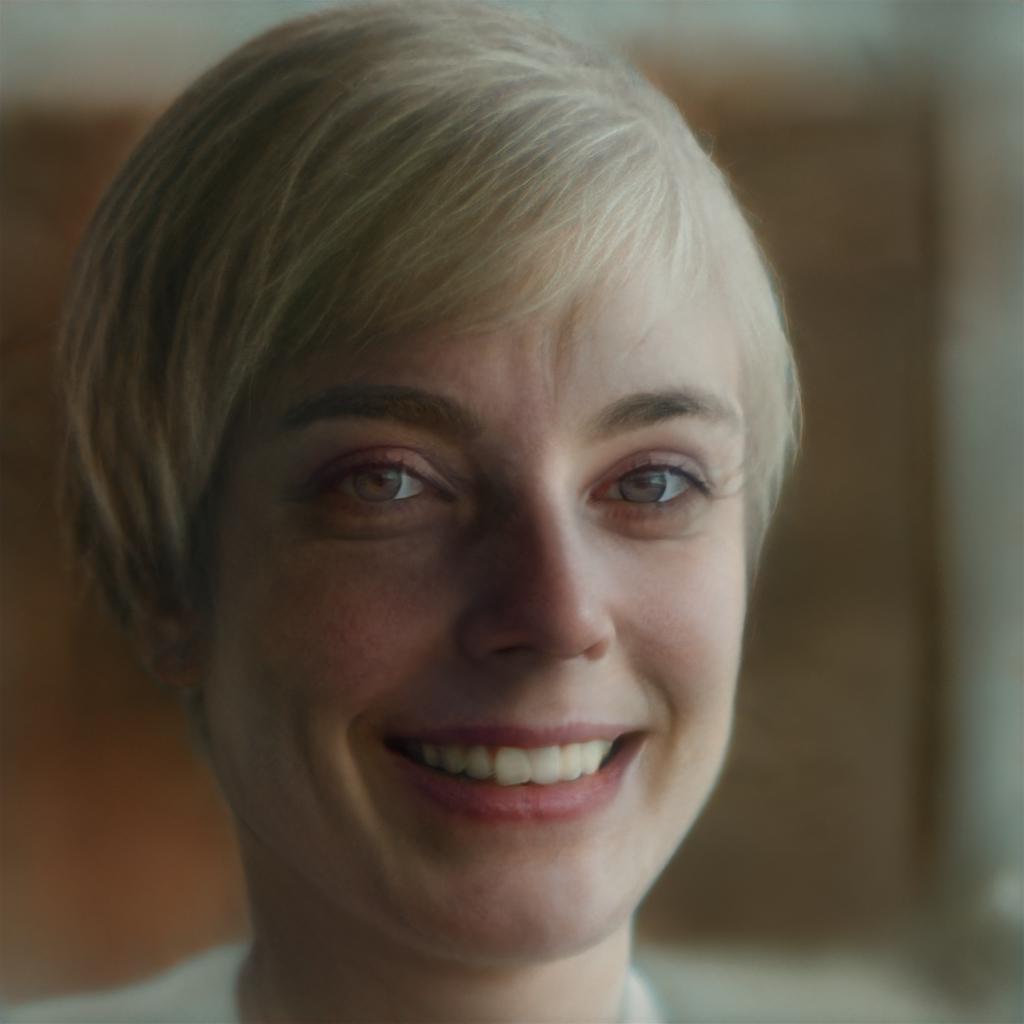} &
\includegraphics[width=0.16\linewidth]{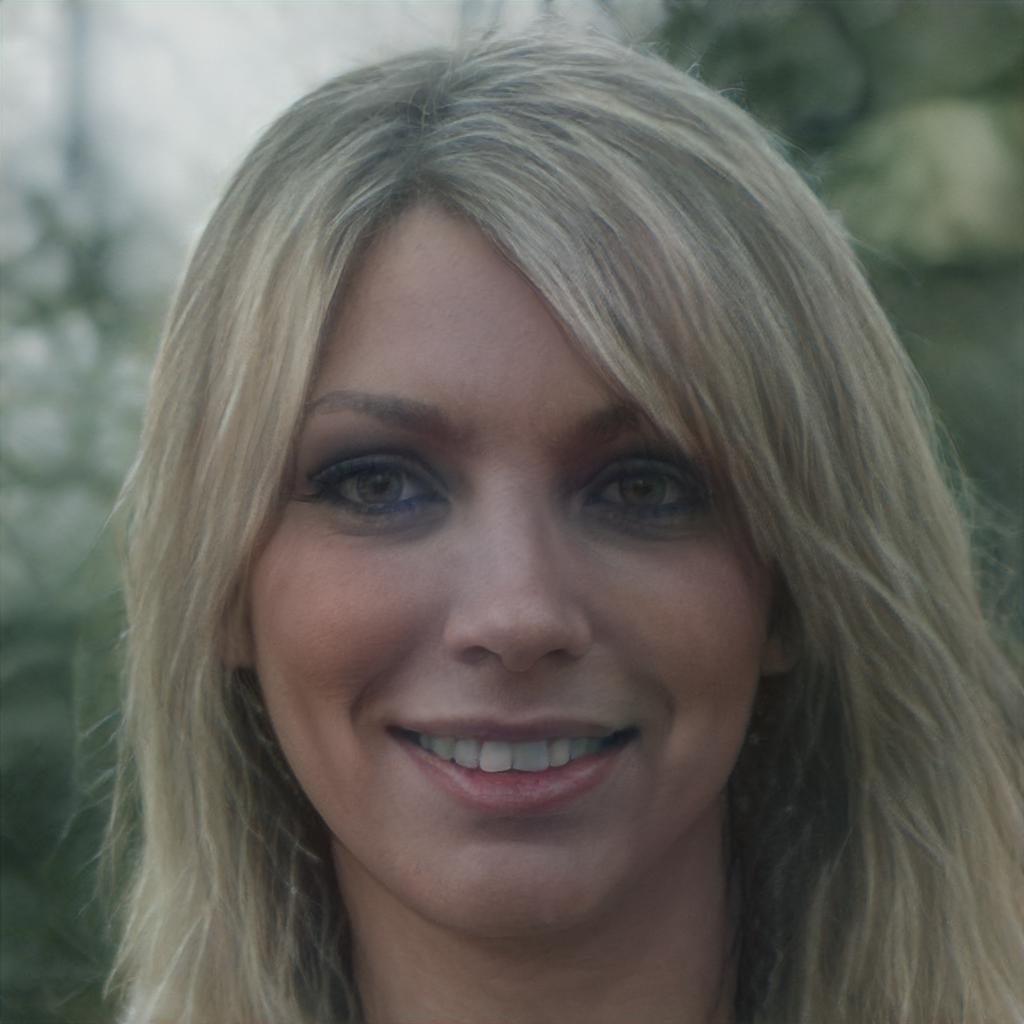} &
\includegraphics[width=0.16\linewidth]{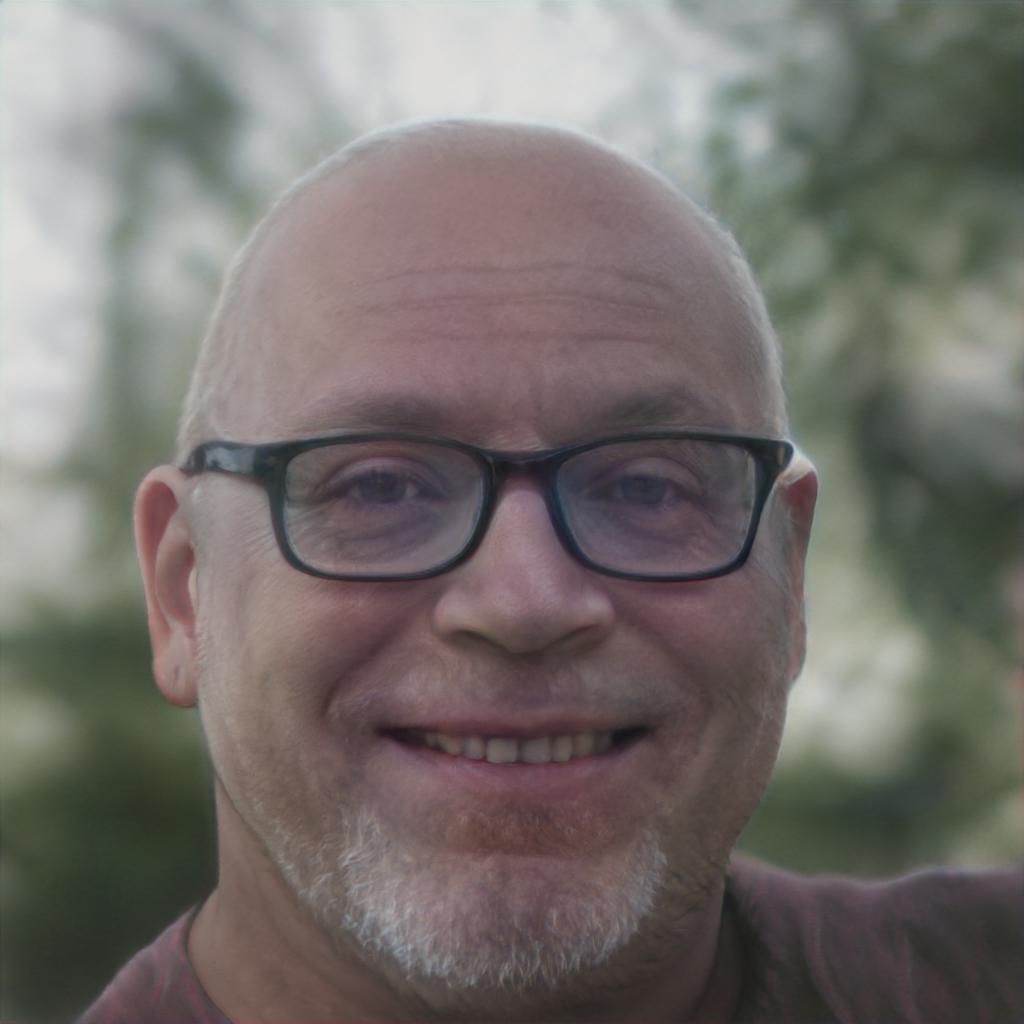} \\

 \begin{turn}{90}\hspace{0.8cm} Img. D \end{turn} &
\begin{turn}{90}\hspace{0.6cm} BPP=0.027 \end{turn} &
 \includegraphics[width=0.16\linewidth]{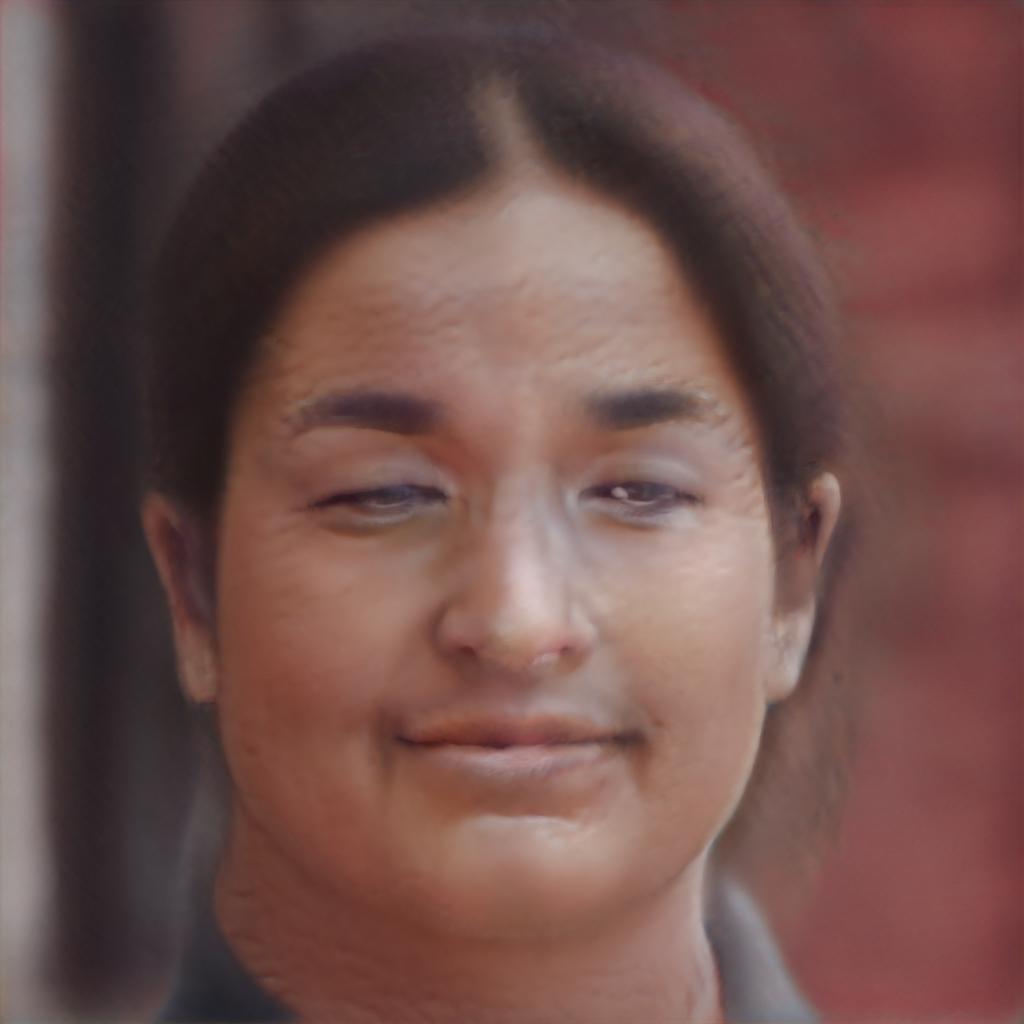}  & 
\includegraphics[width=0.16\linewidth]{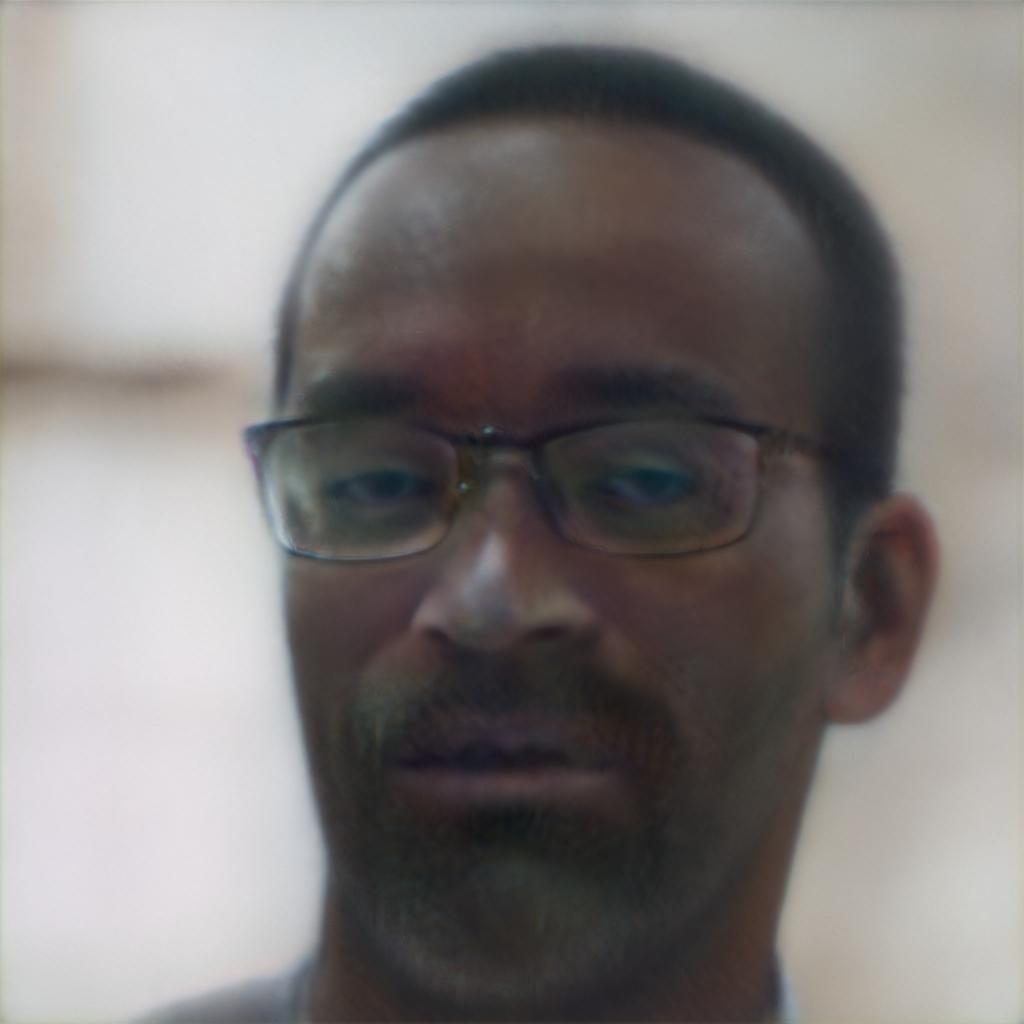}  & 
\includegraphics[width=0.16\linewidth]{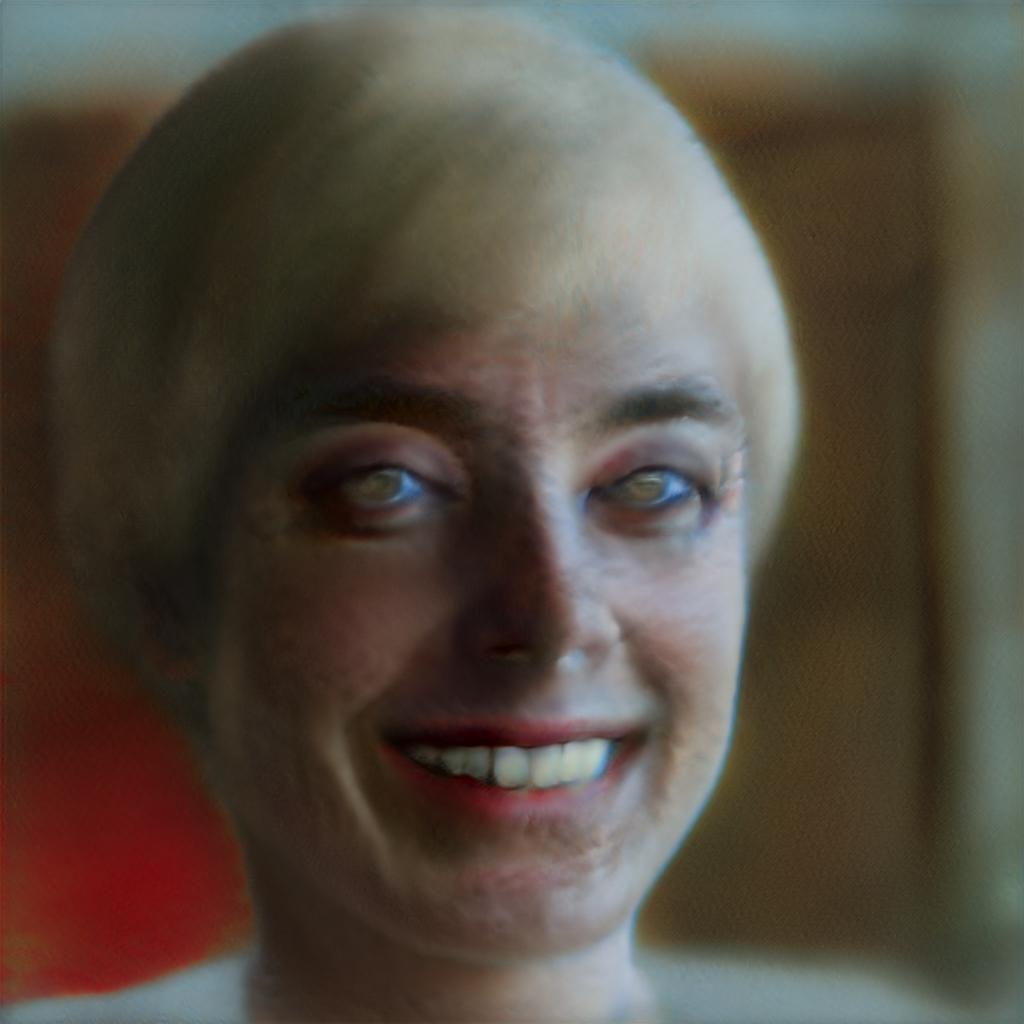} &
\includegraphics[width=0.16\linewidth]{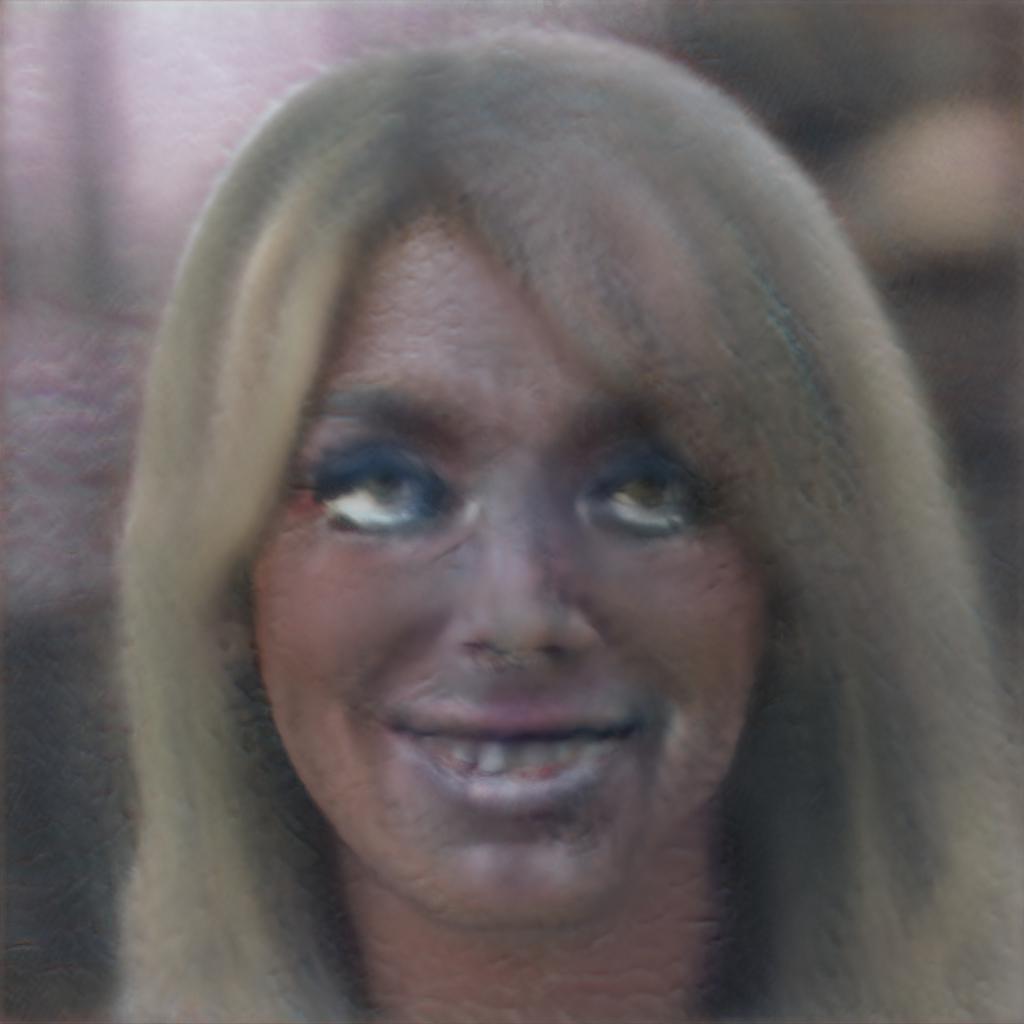} &
\includegraphics[width=0.16\linewidth]{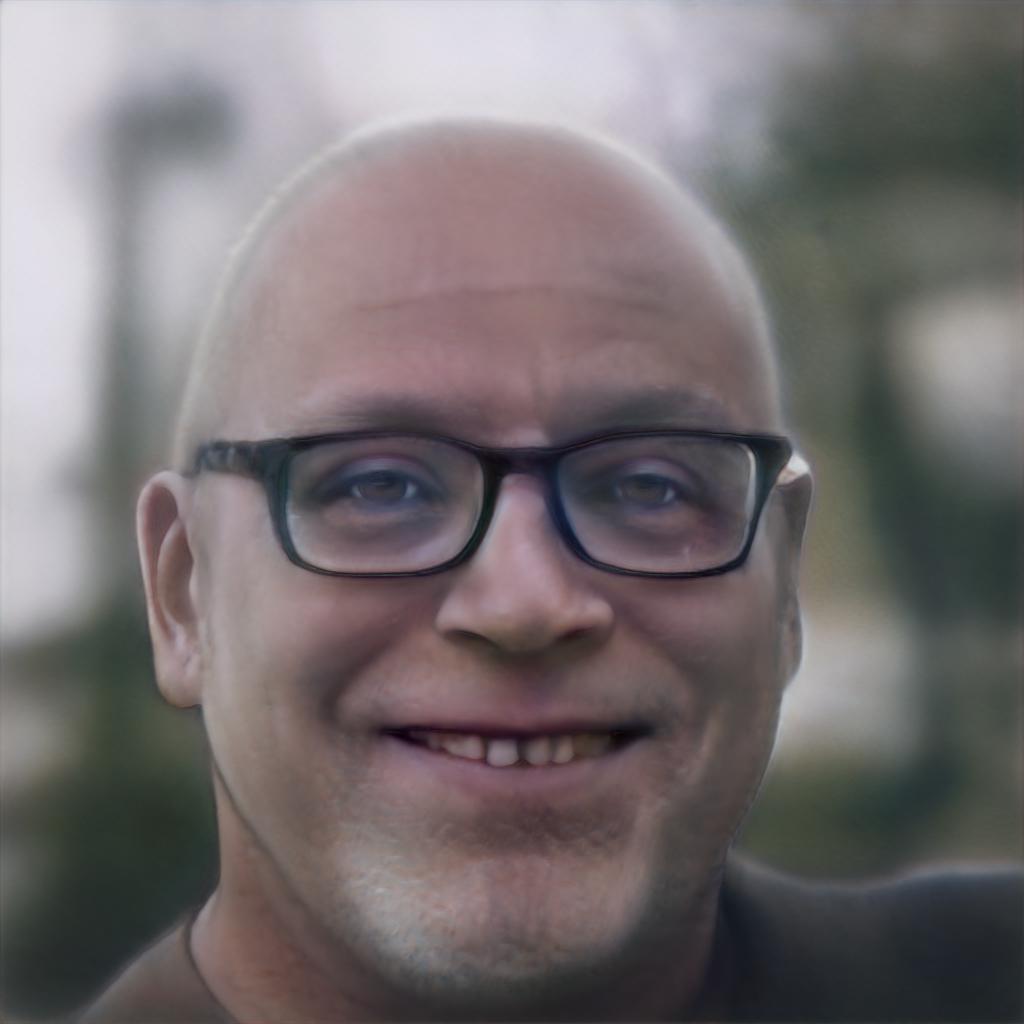} \\

 \begin{turn}{90}\hspace{0.6cm} Lat.-Img. D \end{turn} &
\begin{turn}{90}\hspace{0.6cm} BPP=0.017 \end{turn} &
 \includegraphics[width=0.16\linewidth]{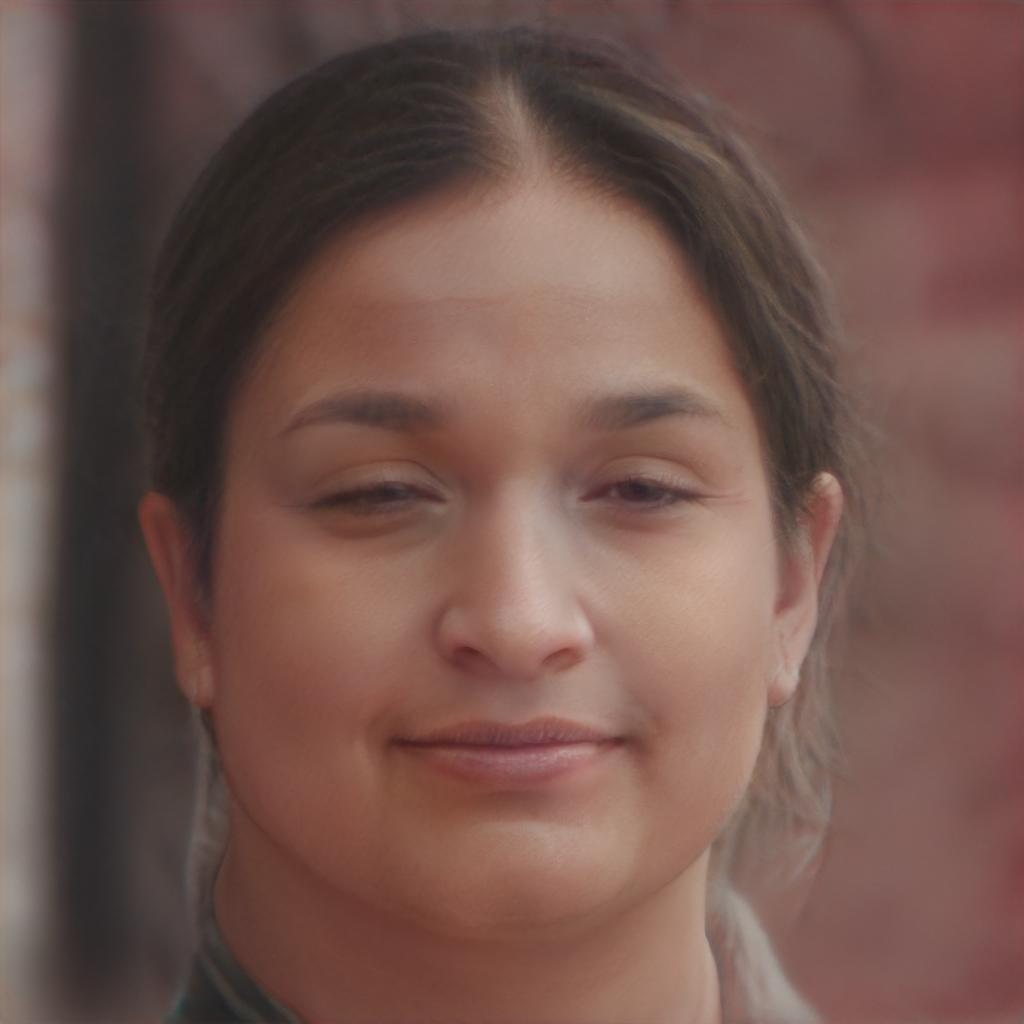}  & 
\includegraphics[width=0.16\linewidth]{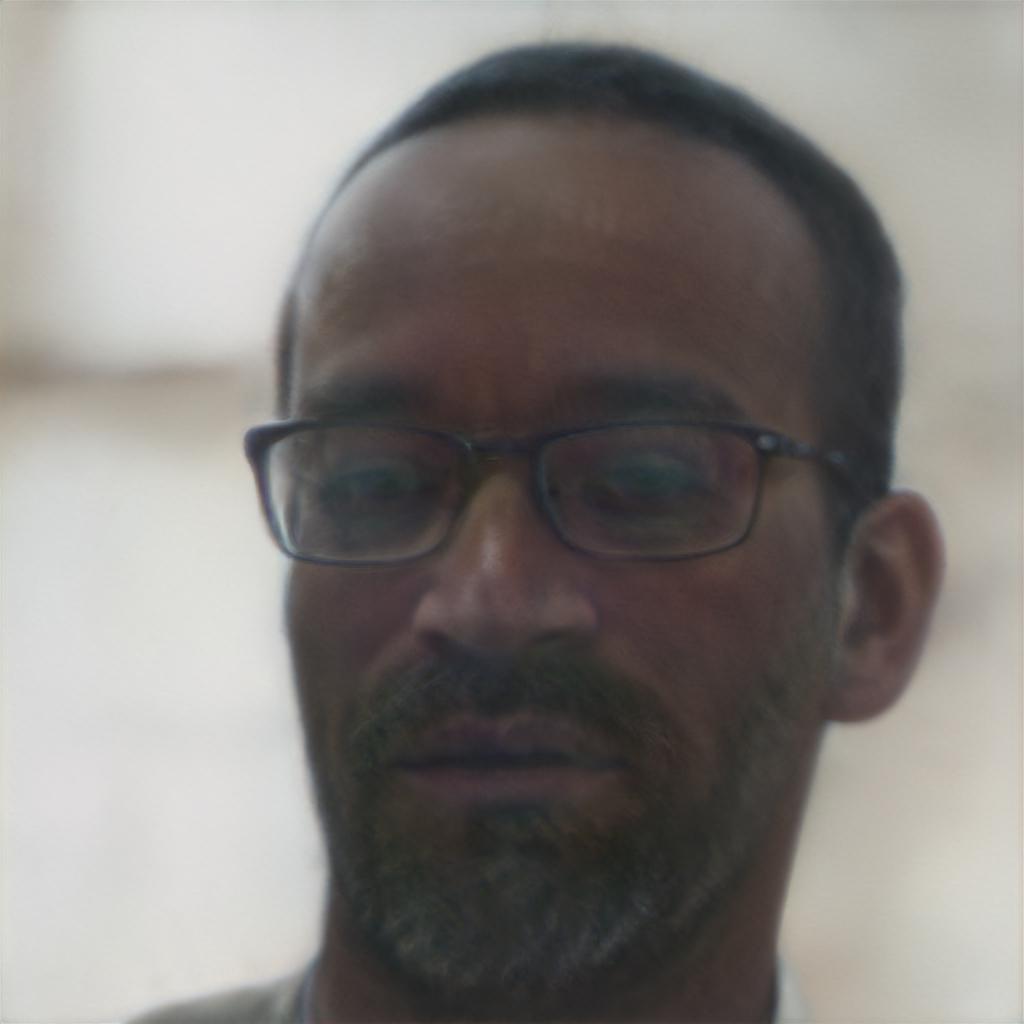}  & 
\includegraphics[width=0.16\linewidth]{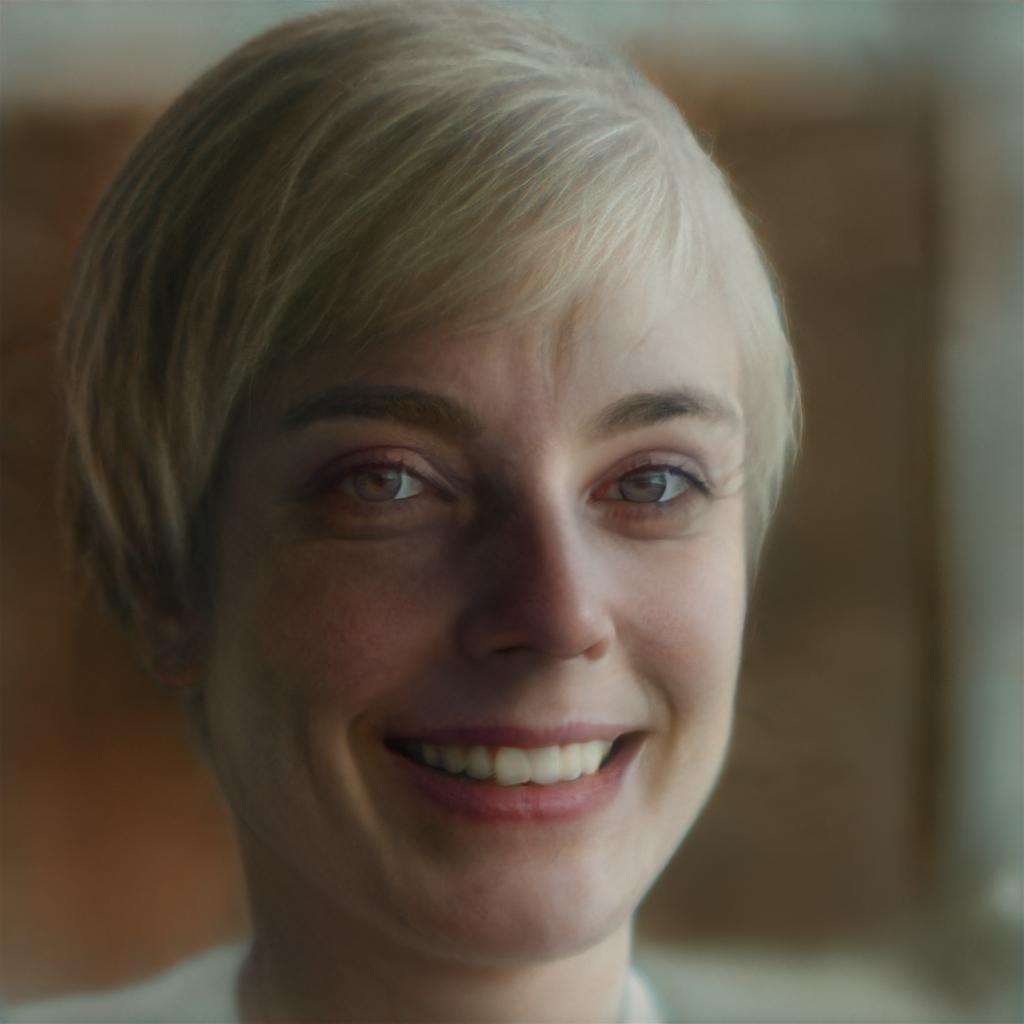} &
\includegraphics[width=0.16\linewidth]{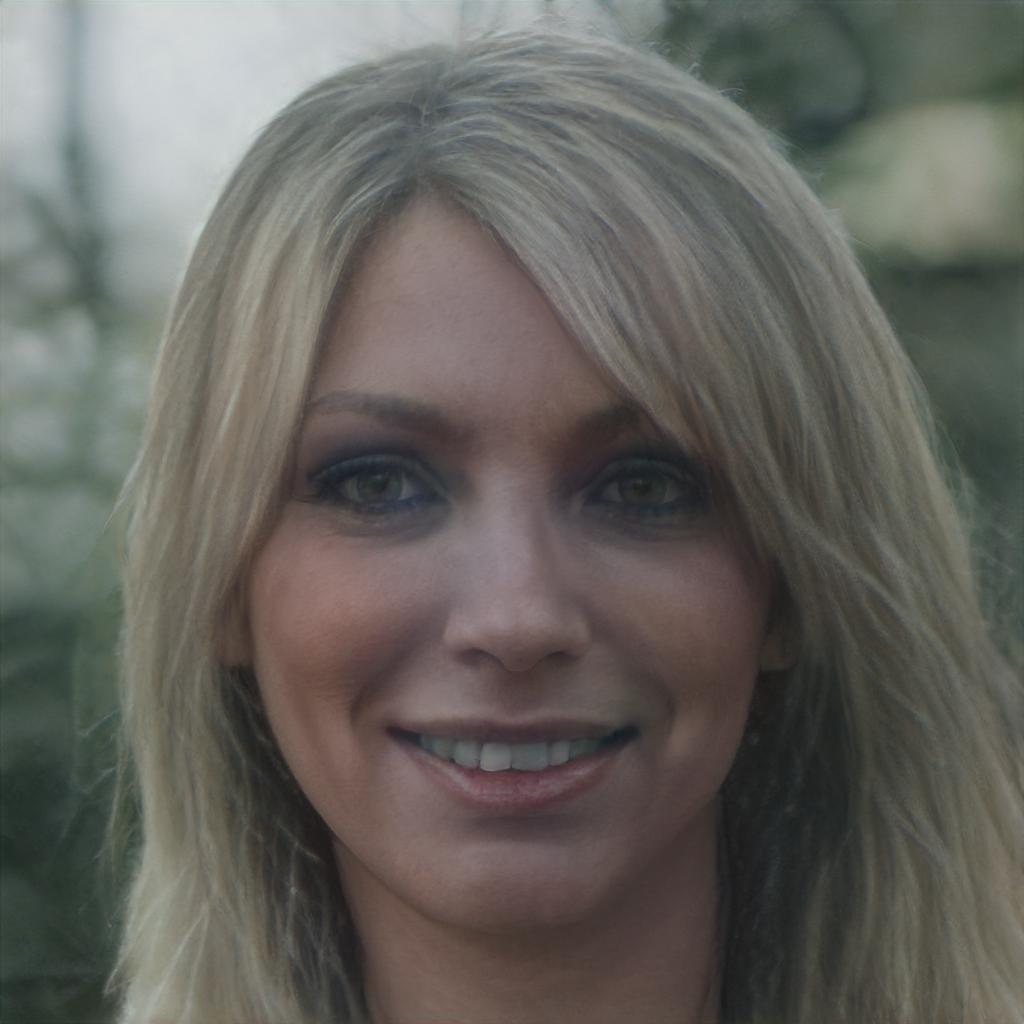} &
\includegraphics[width=0.16\linewidth]{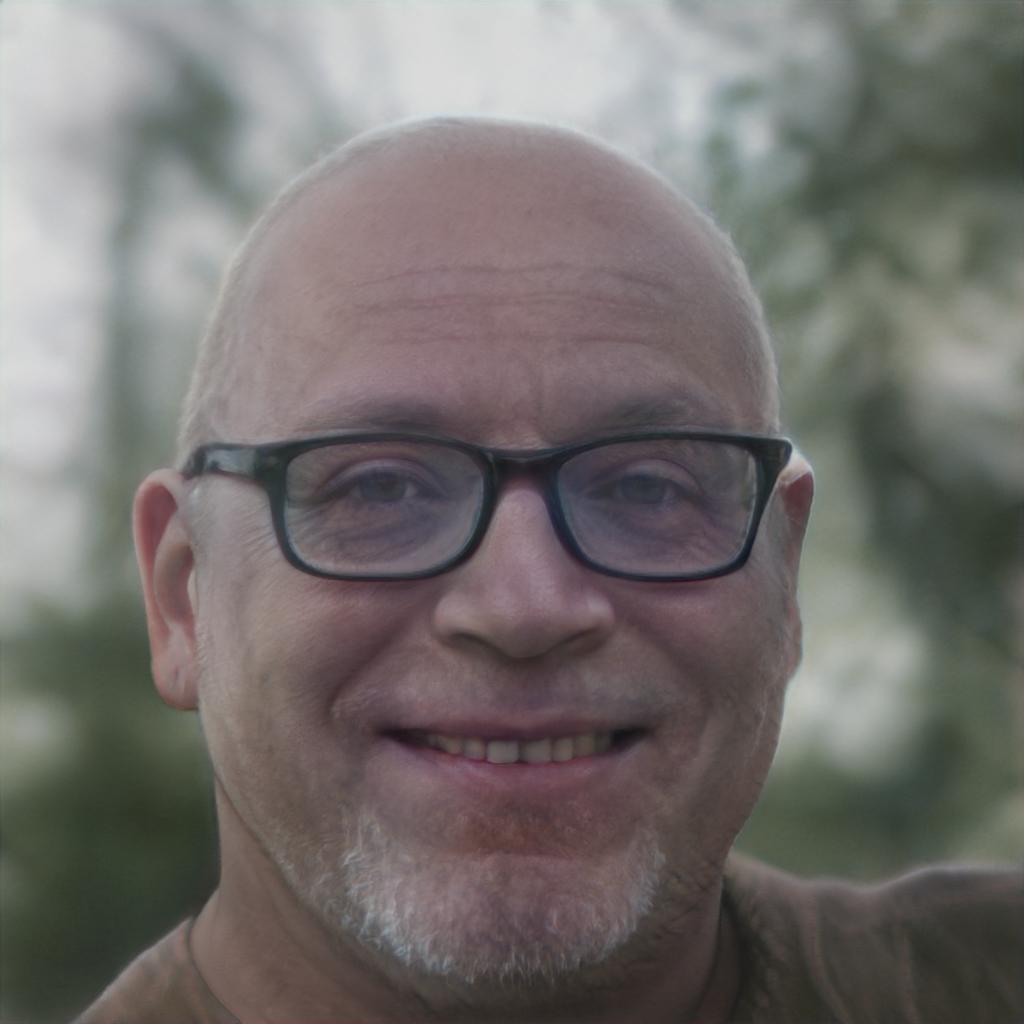} \\ 

 \begin{turn}{90}\hspace{0.6cm} LPIPS. D \end{turn} &
\begin{turn}{90}\hspace{0.6cm} BPP=0.029 \end{turn} &
 \includegraphics[width=0.16\linewidth]{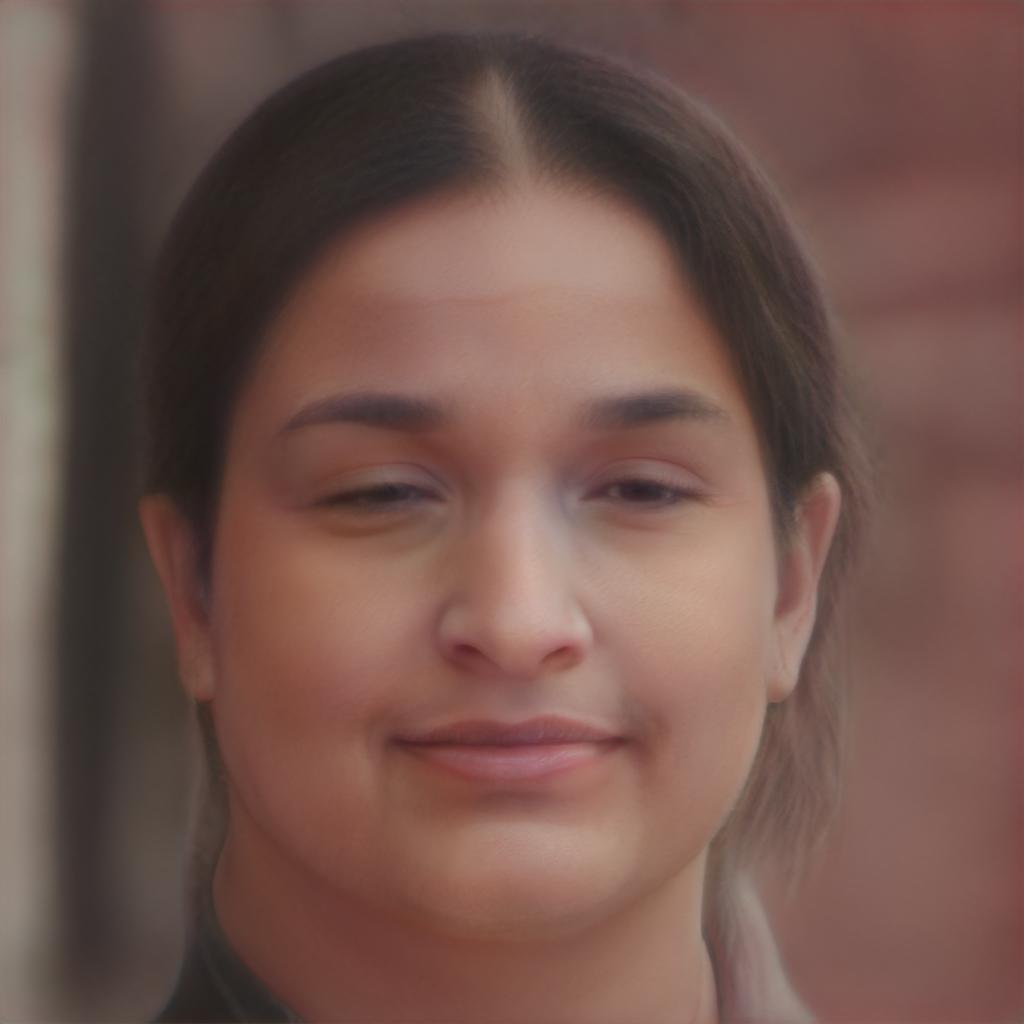}  & 
\includegraphics[width=0.16\linewidth]{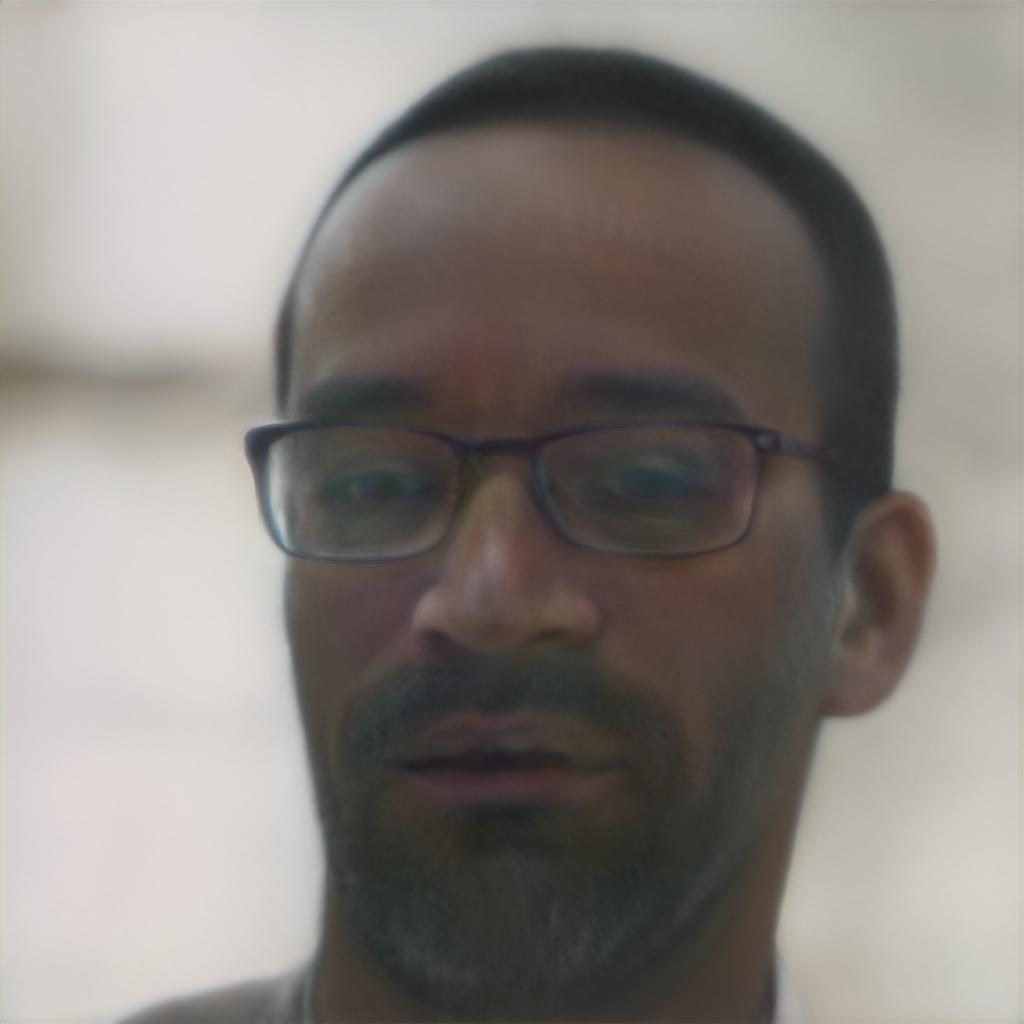}  & 
\includegraphics[width=0.16\linewidth]{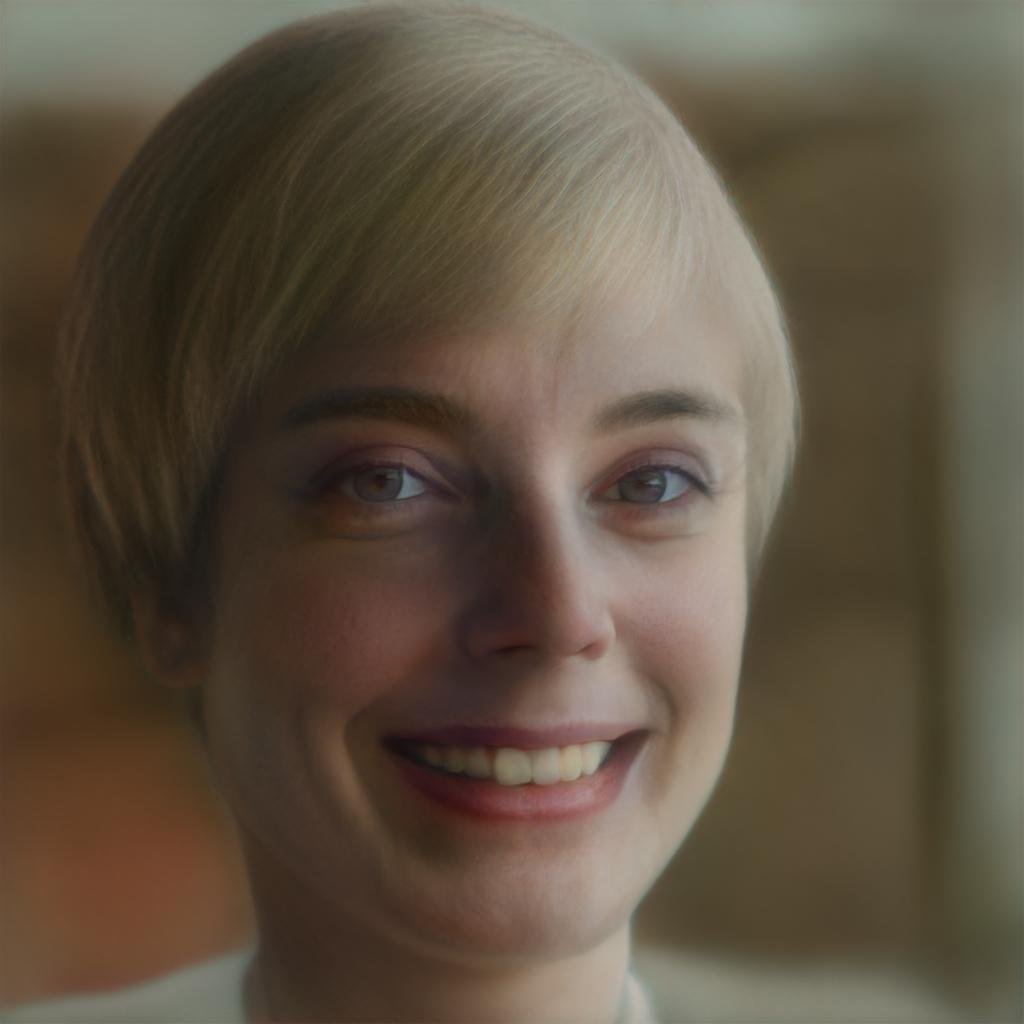} &
\includegraphics[width=0.16\linewidth]{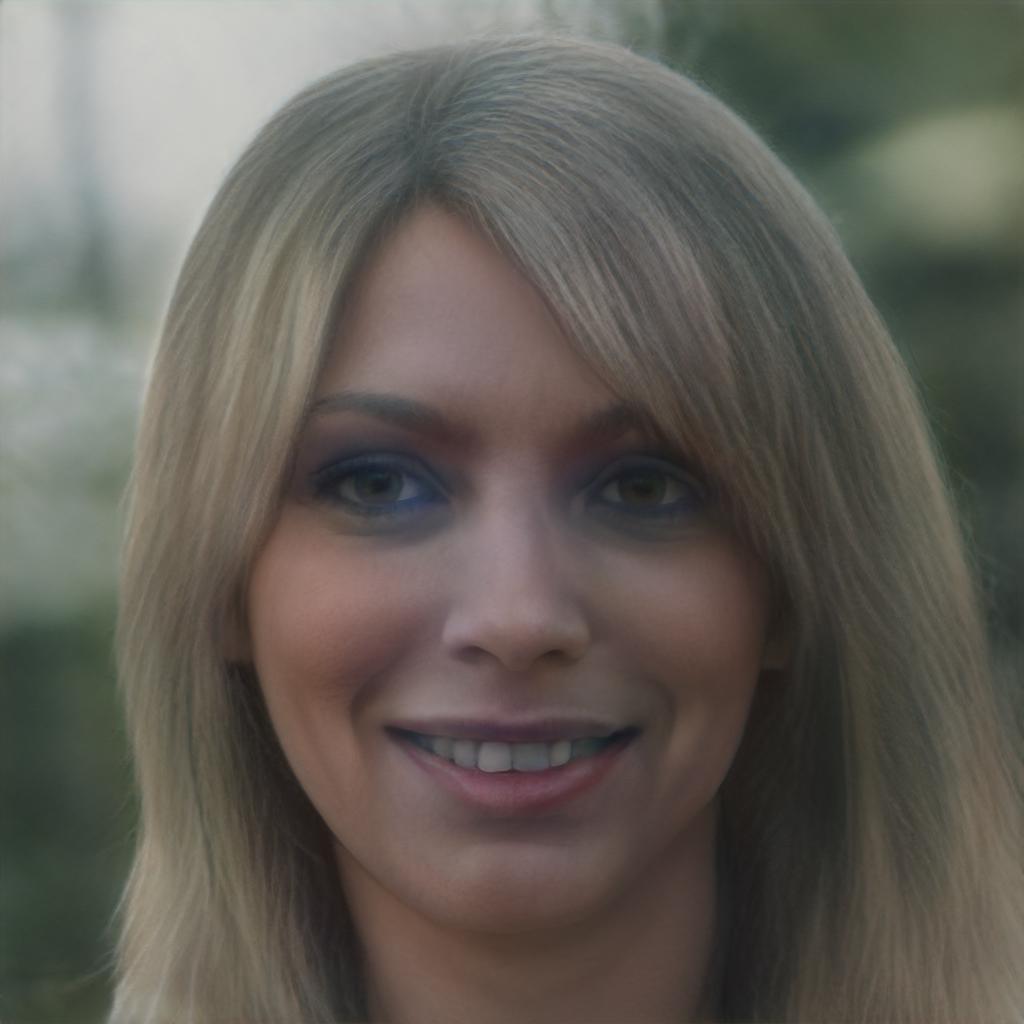} &
\includegraphics[width=0.16\linewidth]{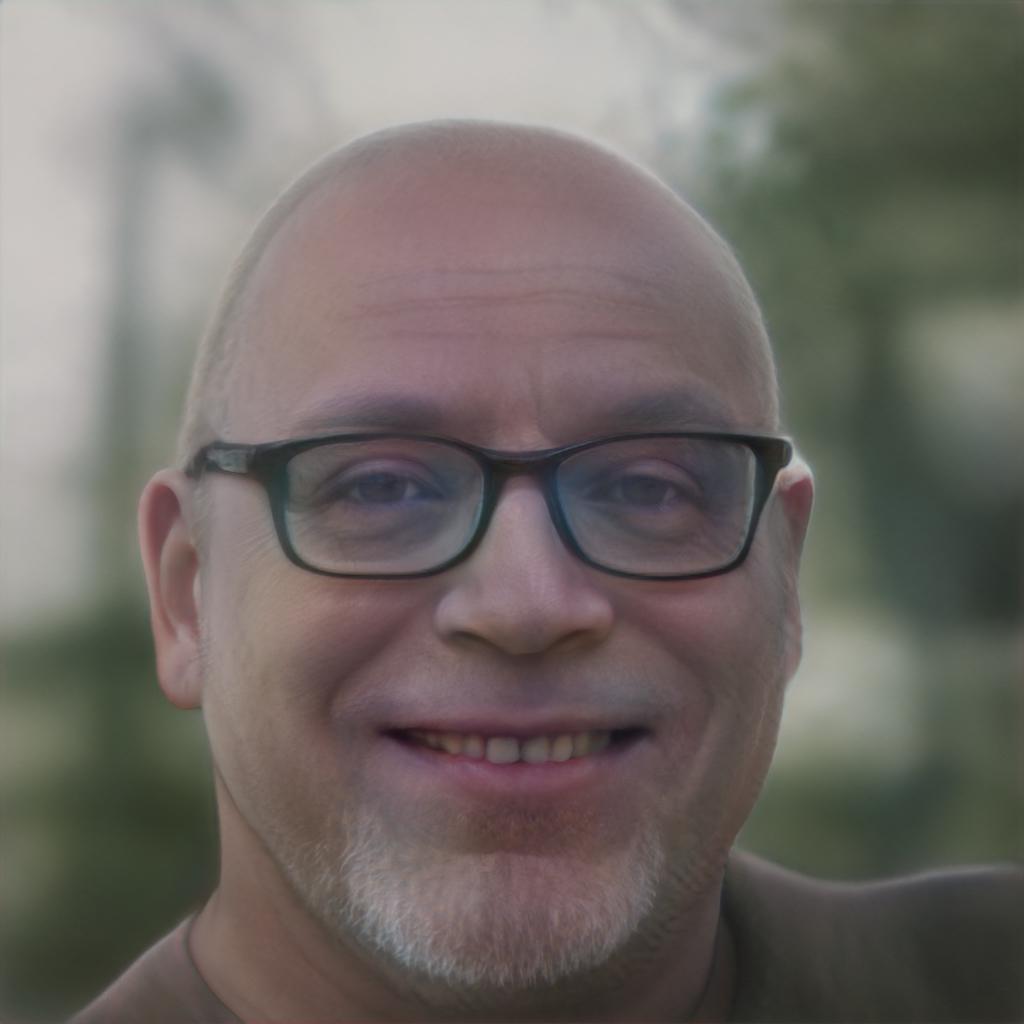} \\

 \begin{turn}{90}\hspace{0.6cm} LPIPS.-Img. D \end{turn} &
\begin{turn}{90}\hspace{0.6cm} BPP=0.031 \end{turn} &
 \includegraphics[width=0.16\linewidth]{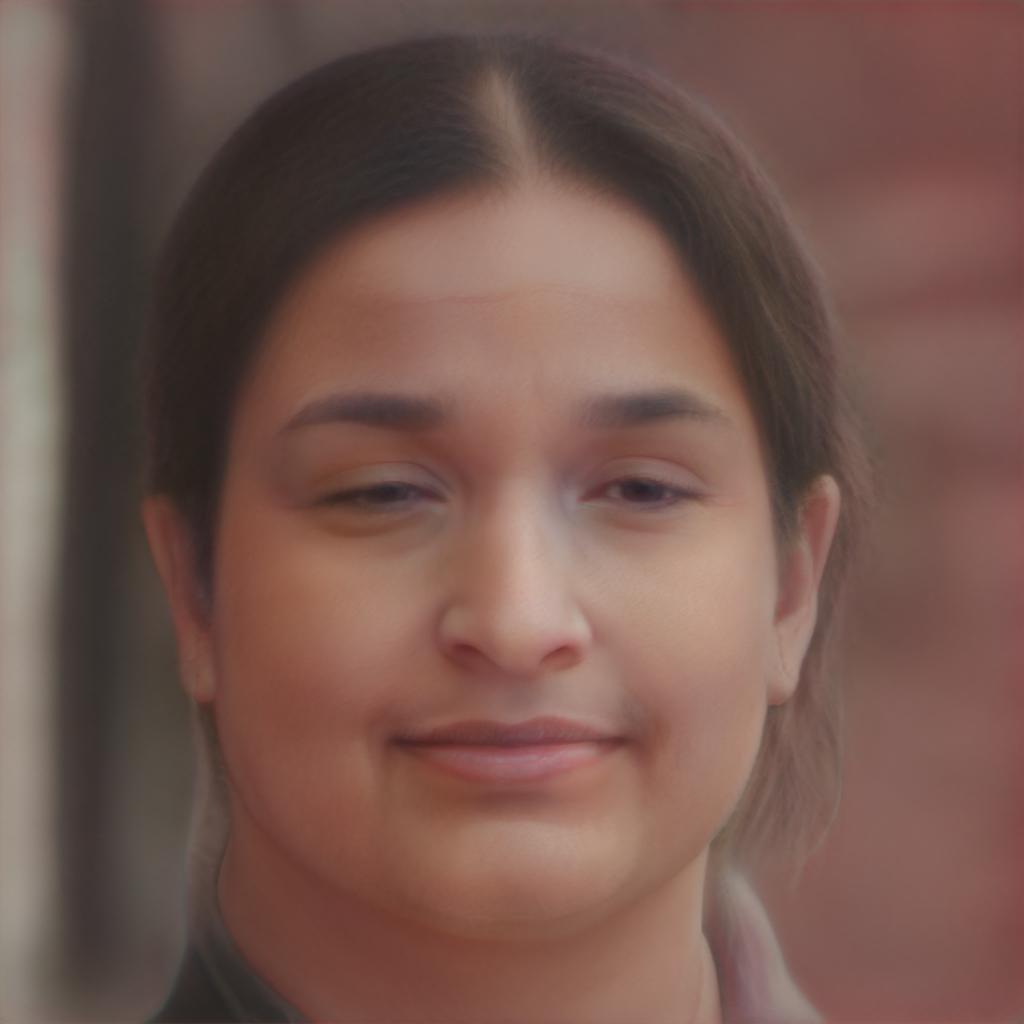}  & 
\includegraphics[width=0.16\linewidth]{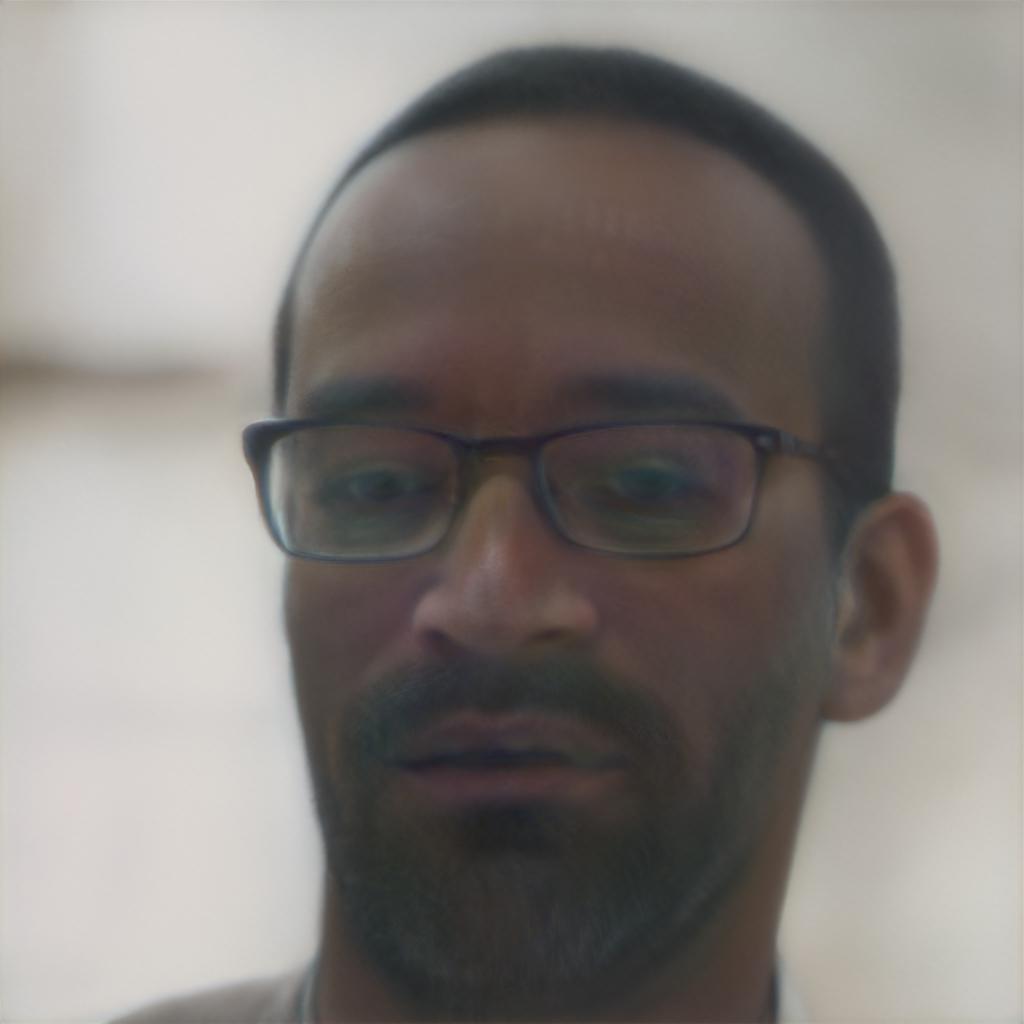}  & 
\includegraphics[width=0.16\linewidth]{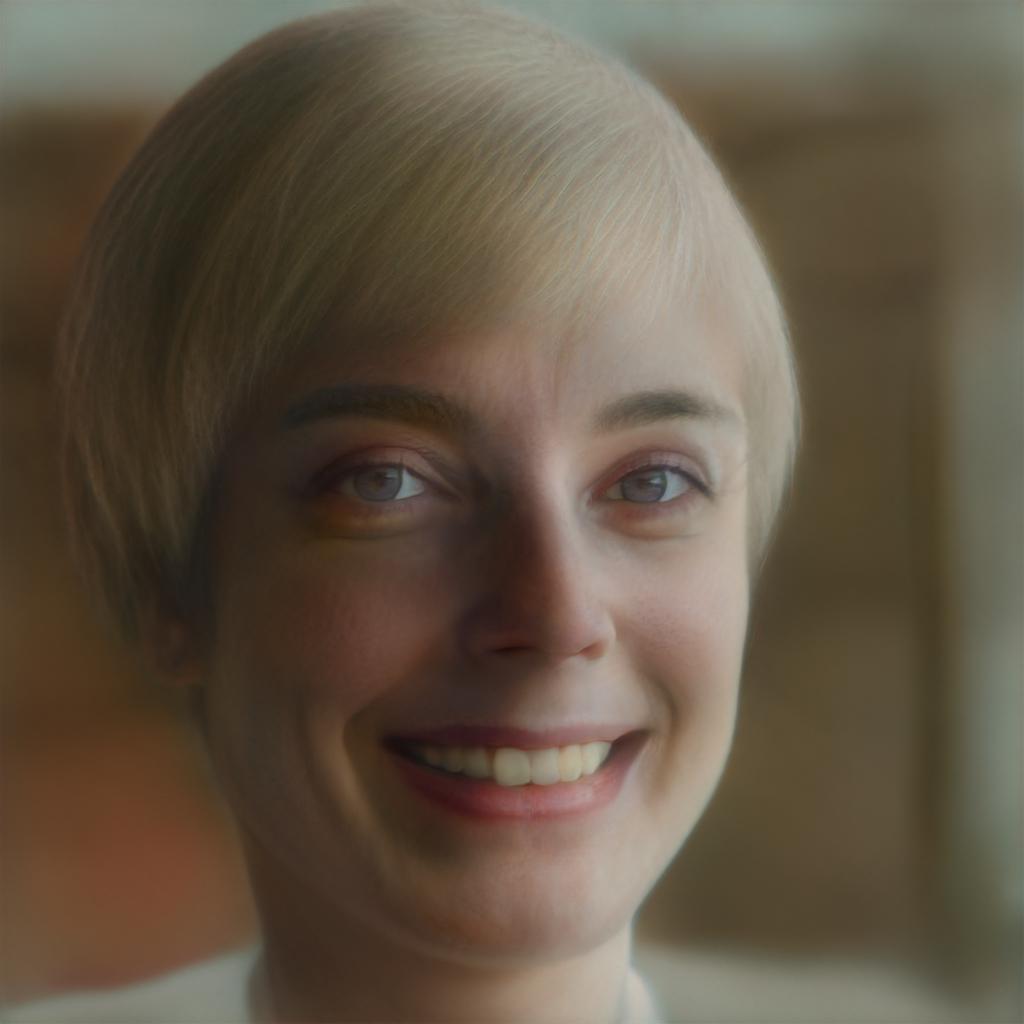} &
\includegraphics[width=0.16\linewidth]{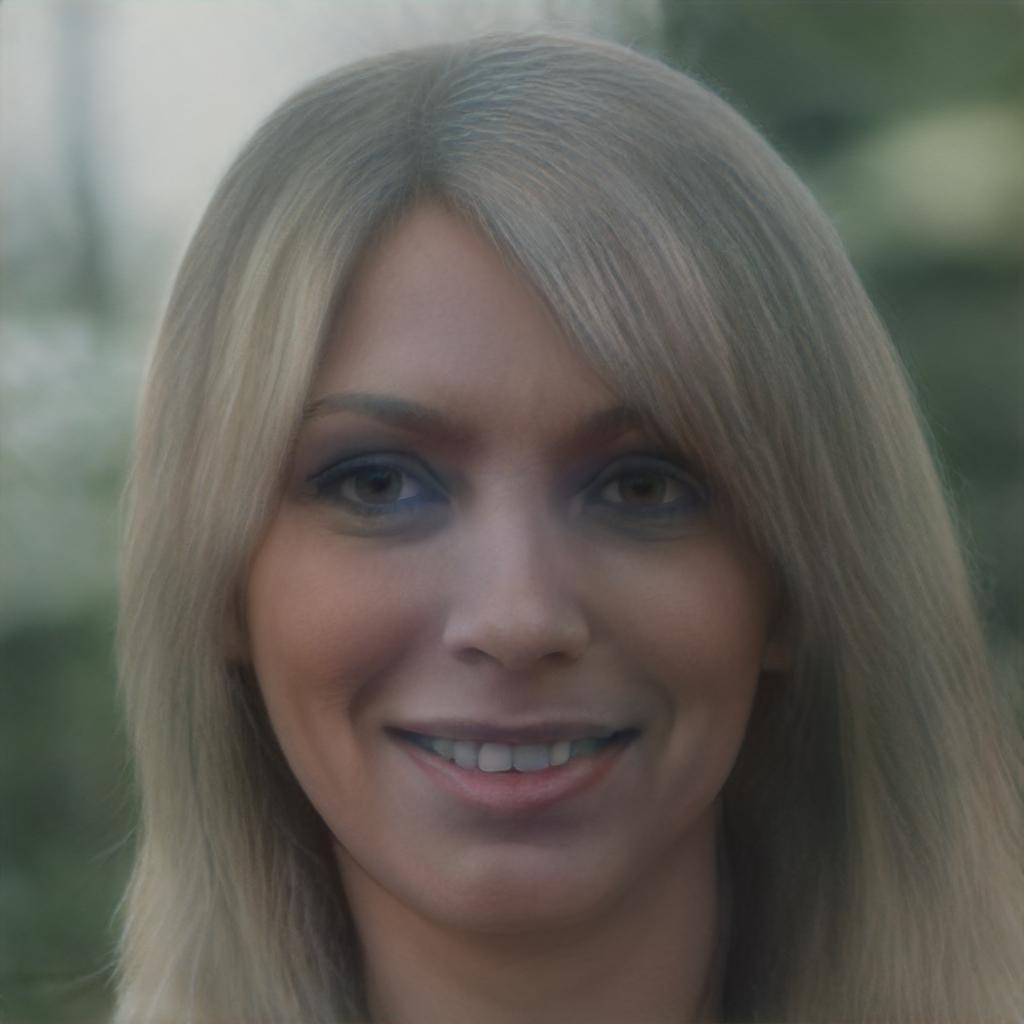} &
\includegraphics[width=0.16\linewidth]{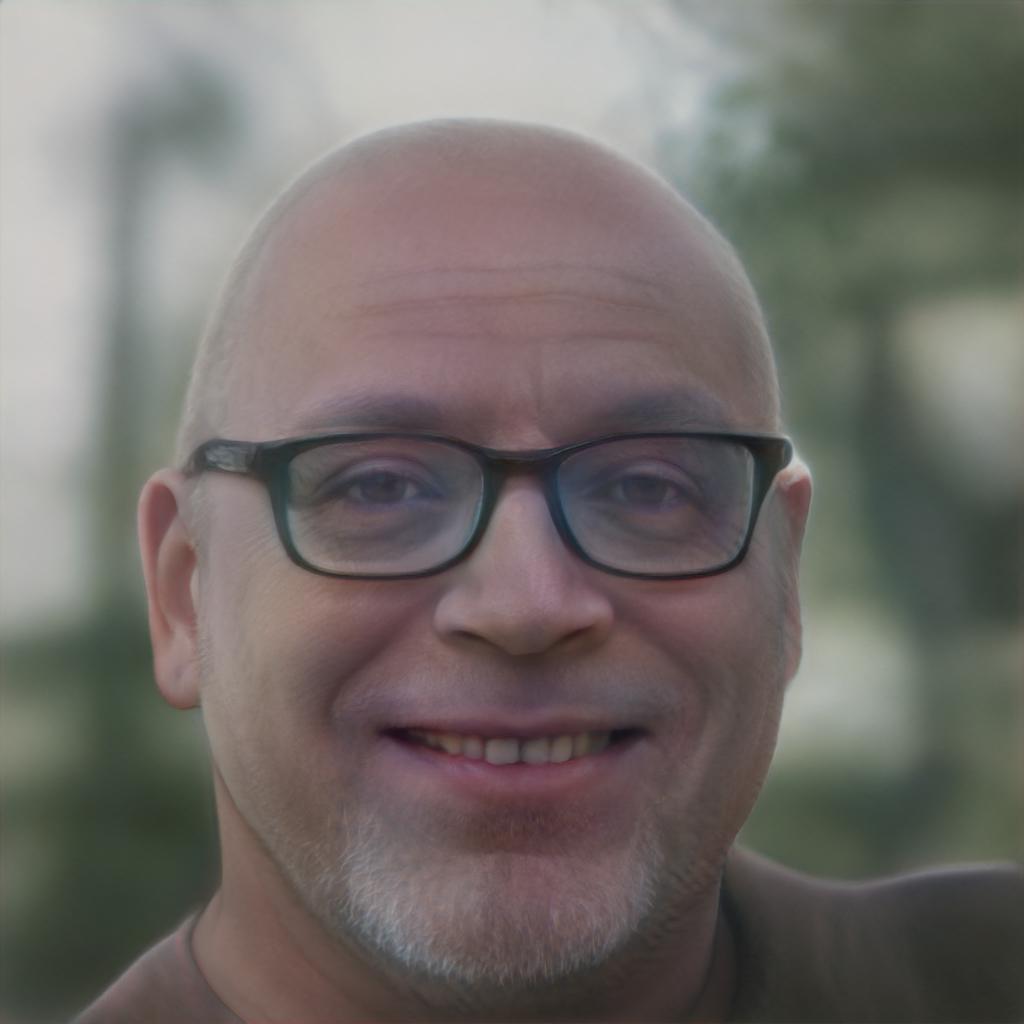} \\ 
\bottomrule

\end{tabular}
\caption{Ablation study for the choice of the distortion loss. We compare the effect of taking the MSE loss in; the latent space (Lat. D, ours), in the image space (Img. D), both in the latent and image space (Lat.-Img. D) and in the image space with the LPIPS perceptual distortion (LPIPS.-Img. D). The loss in the image space favors some artifacts such as white patches and the images are blurred. In addition, the LPIPS loss produce blurred images ((LPIPS.-Img. D and LPIPS. D). Our loss produce artifacts free images. There is no benefit of using additionally the image space loss. The bpp is the taken as the average over the 5 images in each row}
\label{fig:ablation_disto_loss}
\end{figure*}

From Figures \ref{fig:distortion_ablation_MEAD_intra}, \ref{fig:ablation_disto_loss}, we can notice that our loss (Lat. D) outperforms the the MSE and the LPIPS distortion in the image space. Moreover, the loss in the image space produces some artifacts and images are blurred while the loss in the latent space is almost artifacts free. There is no benefit of using the loss in the image space in addition to ours. Note that, the training using only the latent loss is faster than others (\emph{e.g.}, the training took; 5 days (LPIPS.-Img. D), 4 days (Lat.-Img. D) and 6 hours (Lat. D)) and occupies smaller space in GPU (\emph{e.g.}, 24 GB with batch size=4 for LPIPS.-Img. D, 10 GB with batch size =4 for Img. D and Lat.-Img. D and less than 2 GB with batch size =8 for Lat. D). 

\subsubsection{Video Compression (Inter Coding: Explicit sparsification)}
\label{sec:l1_reg}
Having only few dimensions that change between two consecutive latent codes is efficient for entropy coding. Thus, we have investigated how to explicitly sparsify these differences and add an $L1$ regularization on the latent codes differences during training. The loss becomes:
\begin{equation}
\label{eq:l1_reg}
    \mathcal{L}_{IC-L1} = \mathcal{L}_{IC} + \lambda_{L_1}| \bm{w^}{*}_{t-1} - \bm{w^}{*}_{t} |
\end{equation}
Note that, as the transformation $T$ and the entropy model are trained jointly, we expect that the latent codes in \wstarc to be transformed to fit efficiently the entropy model. In the following, we assess whether explicit sparsification brings additional improvement (Section \ref{sec:app_ablation_inter_coding}).

\subsubsection{Video Compression (Inter Coding)}
\label{sec:app_ablation_inter_coding}
\begin{figure*}[h]
     \centering
     \begin{subfigure}[b]{0.32\linewidth}
         \centering
         \caption{\tiny{PSNR}}
         \includegraphics[width=\linewidth, height=5cm]{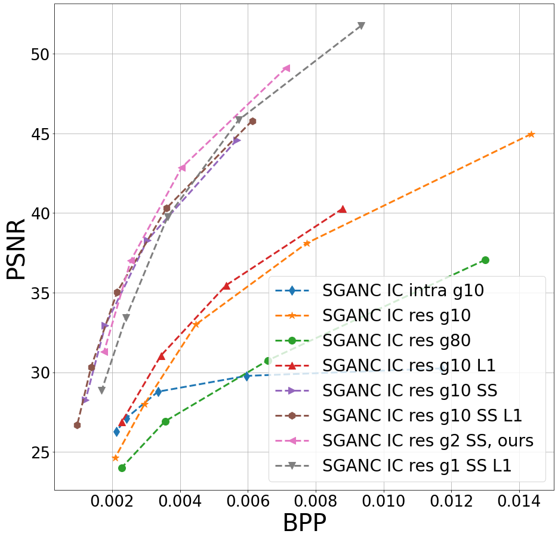}
     \end{subfigure}
     \hfill
     \begin{subfigure}[b]{0.32\linewidth}
         \centering
         \caption{\tiny{MS-SSIM}}
         \includegraphics[width=\linewidth, height=5cm]{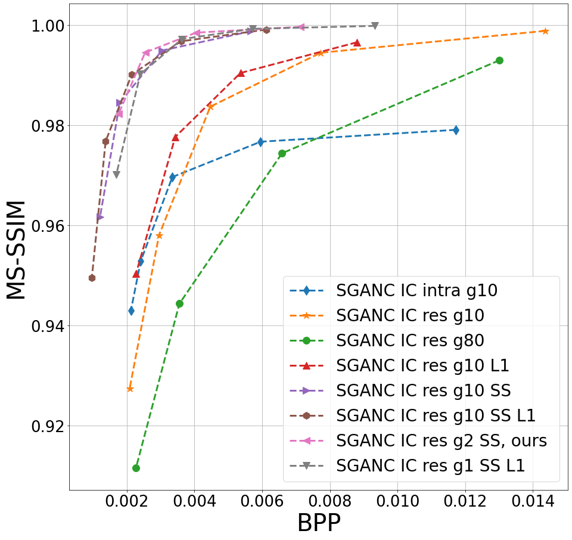}
     \end{subfigure}
     \hfill
     \begin{subfigure}[b]{0.32\linewidth}
         \centering
         \caption{\tiny{LPIPS \cite{zhang2018unreasonable}}}
         \includegraphics[width=\linewidth, height=5cm]{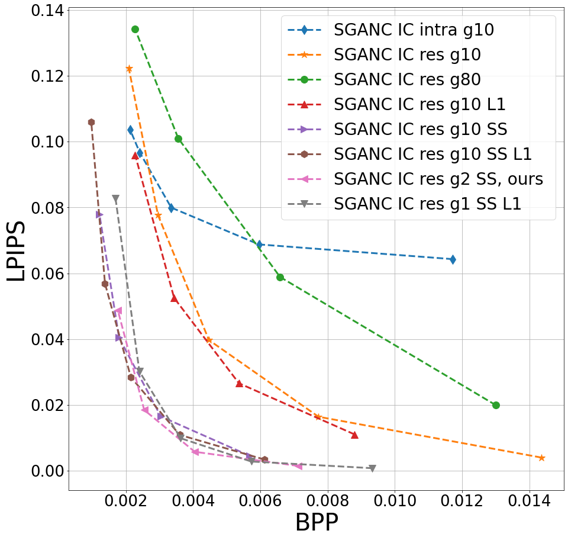}
     \end{subfigure}
        \caption{Ablation study for video compression using inter coding with residuals on MEAD inter dataset. res: do residual coding each $g$ frames, intra: do intra coding each $g$ frames, SS: stage specific entropy model and transformation $T$, L1: add L1 regularization during training (section \ref{sec:l1_reg}), EM: entropy model}
        \label{fig:ablation_MEAD_inter_inter_coding}
\end{figure*}

In this section we present the ablation study for SGANC IC, in which we investigate the effect of the following parameters; parameter $g$ for the residual coding (g), L1 regularization \eqref{eq:l1_reg}, residual coding (res) vs intra coding (intra) each $g$, stage specific entropy model; using different entropy models and distortion $\lambda$ for the different stages of StyleGAN2 (SS). We thus compare the following variants:

\begin{itemize}
    \item SGANC IC intra g10: replacing the residual coding at each $g=10$ by intra coding using one of the image processing models that we train as in section 3.2 in the main paper.
    \item SGANC IC res g10: doing residual coding each $g=10$.
    \item SGANC IC res g80: doing residual coding each $g=80$.
    \item SGANC IC res g10 L1: Adding the L1 regularization \eqref{eq:l1_reg} during training and doing residual coding each $g=10$.
    \item SGANC res g10 SS: Using 3 entropy models and 3 $T$ models for each stage of the StyleGAN2 (1-7, 7-13, 13-18). In addition, training with layer specific distortion lambda; $\lambda = w\lambda$ (where $w=1$ for the first stage and decrease from $1$ to $0.01$ for the second and third stage.
    \item SGANC res g10 SS L1: residual coding each $g=10$, stage specific entropy models and L1 regularization.
    \item SGANC res g2 SS: residual coding each $g=2$, stage specific entropy.
    \item SGANC res g1 SS L1: Same as before but doing a residual coding each frame ($g=1$).
\end{itemize}
From Figure \ref{fig:ablation_MEAD_inter_inter_coding}, we can notice the following:

\begin{itemize}
    \item Decreasing $g$ leads to better results, at the expense of increased BPP especially for $g=1$.
    \item Significant improvements is obtained by performing residual coding (SGANC IC intra g10 vs SGANC IC res g10) and using stage specific entropy models (SGANC IC res g10 vs SGANC IC res g10 SS).
    \item Only a marginal improvement was obtained by adding the L1 regularization during training (SGANC res g10 vs SGANC res g10 L1). In addition, the improvement becomes negligible when using SS entropy models (SGANC res g10 SS vs SGANC res g10 SS L1).
\end{itemize}

\subsection{More results}
\label{sec:app_more_results_img_vid}
In this section we show more quantitative and qualitative results for image and video compression.
\subsubsection{Image Compression}
\begin{figure*}[h]
\setlength\tabcolsep{2pt}%
\centering
\small
\begin{tabular}{p{0.7cm}cccc}
\toprule
Model &
Original &
Projected & 
SGANC &
MeanHP  \\
\hline
 &
\includegraphics[width=0.23\linewidth]{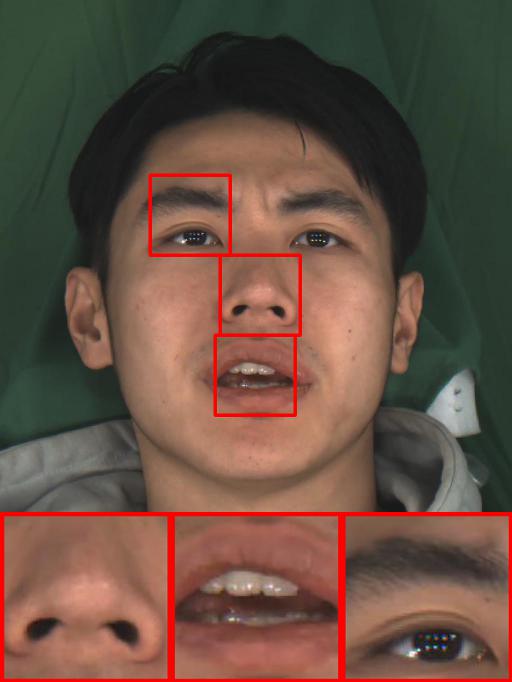}  & 
\includegraphics[width=0.23\linewidth]{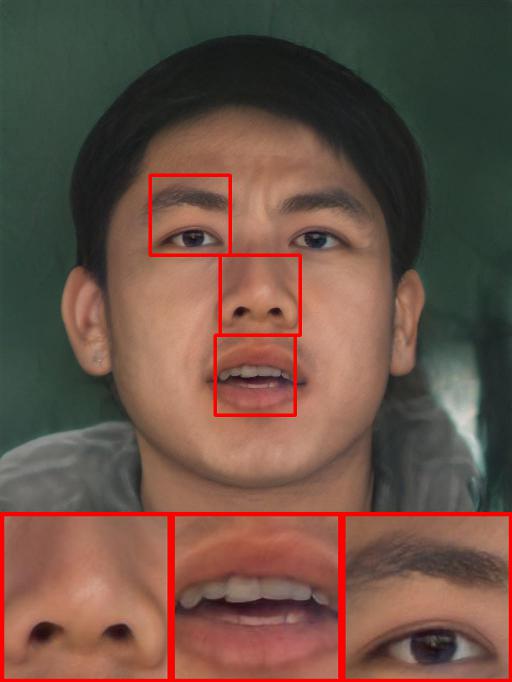} &
\includegraphics[width=0.23\linewidth]{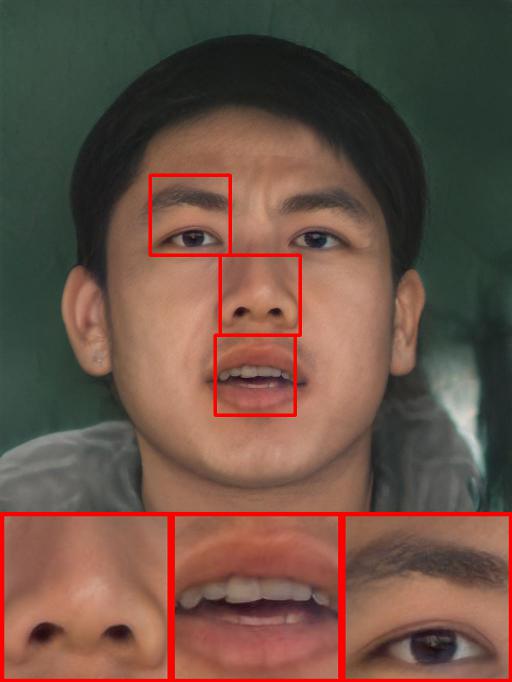} &
\includegraphics[width=0.23\linewidth]{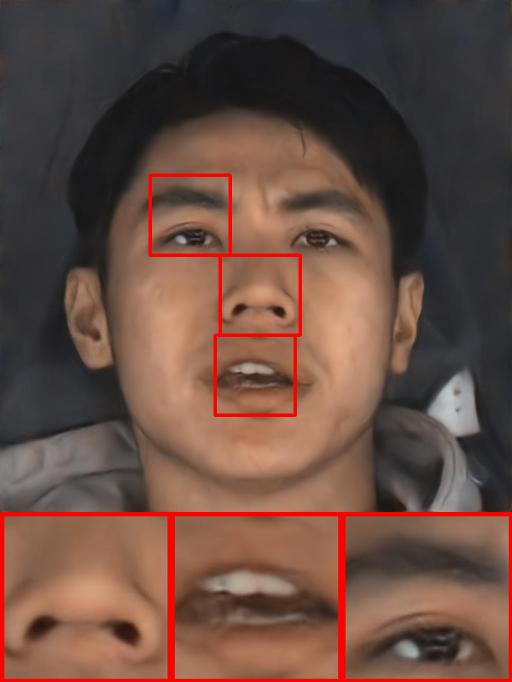}  \\  \cline{4-5} 
BPP &
 & 
 &
0.014  &
0.011   \\ \midrule
Model &
&
VTM &
HP & 
\textit{Cheng et. al}\cite{cheng2020learned}  \\
\hline
 &
 &
\includegraphics[width=0.23\linewidth]{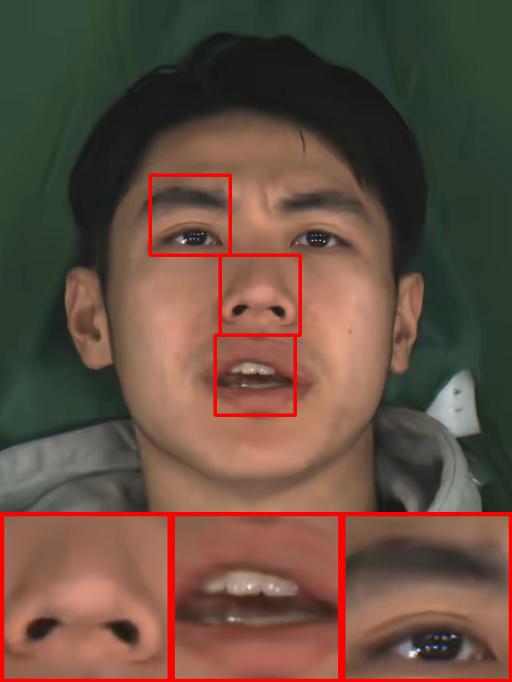} &
\includegraphics[width=0.23\linewidth]{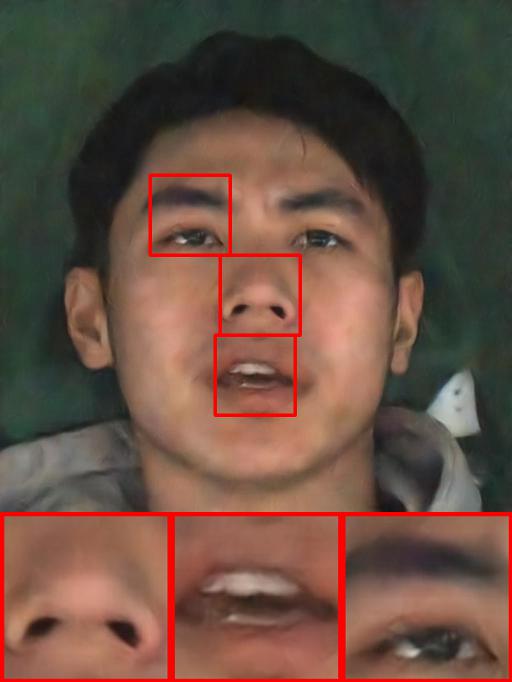} &
\includegraphics[width=0.23\linewidth]{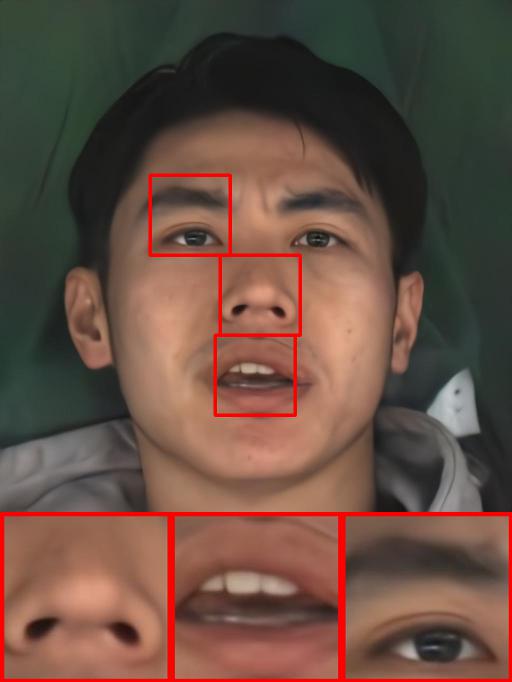} \\  \cline{3-5} 
BPP &
 & 
0.017  &
0.031  & 
0.013 \\ 
\bottomrule
\end{tabular}
\caption{Qualitative results for image compression: Other methods introduce blocking artifacts and blurring. Our method (SGANC) leads to high quality reconstruction  and perceptually lower distortion. Images are better seen zoomed.}
\label{fig:qual_results_low_mead_intra_4_main}

\end{figure*}

\begin{figure*}[h]
\setlength\tabcolsep{2pt}%
\centering
\small
\begin{tabular}{p{1.4cm}cccc}
\toprule
Model &
Original   &
Projected &
VTM   & 
\begin{tabular}{@{}c@{}} SGANC \\ l=1e-5  \end{tabular}  \\
\hline
 &
\includegraphics[width=0.22\linewidth]{images/same_bpp_draw_new/FP006734MD02/real_nvp_compression_lam1e_3_old_abl_only_nf_8_coupl_hard_lat_disto_fixopt_compai/orig_img_119.jpg}  & 
\includegraphics[width=0.22\linewidth]{images/same_bpp_draw_new/FP006734MD02/real_nvp_compression_lam1e_3_old_abl_only_nf_8_coupl_hard_lat_disto_fixopt_compai/projected_img_119.jpg} &
\includegraphics[width=0.22\linewidth]{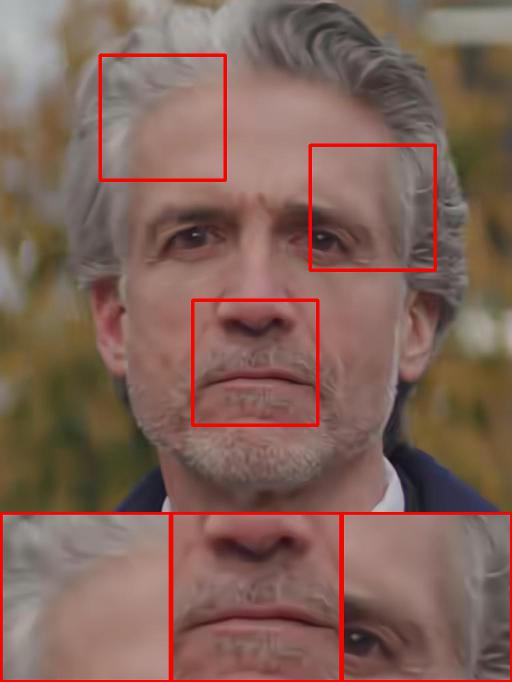} &
\includegraphics[width=0.22\linewidth]{images/same_bpp_draw_new/FP006734MD02/real_nvp_compression_lam1e_5_old_abl_only_nf_8_coupl_hard_lat_disto_fixopt_compai/rec_img_119.jpg}   \\
 \cline{4-5} 
PSNR &
 & 
 &
35.46 &
29.31  \\

MS-SSIM &
 & 
 &
0.933 &
0.970  \\
BPP &
 & 
 &
0.0157 &
0.0146  \\
LPIPS &
 & 
 &
0.274 &
0.062  \\\midrule
Model &
AV1   &
\begin{tabular}{c} MeanHP   \end{tabular}   & 
\begin{tabular}{c} HP   \end{tabular}  & 
\textit{Cheng et. al}\cite{chen2020perceptually}  \\
\hline
 &
\includegraphics[width=0.22\linewidth]{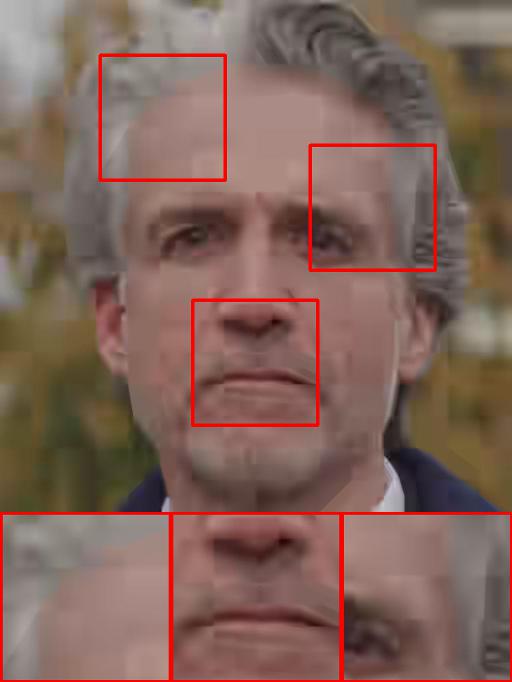} &
\includegraphics[width=0.22\linewidth]{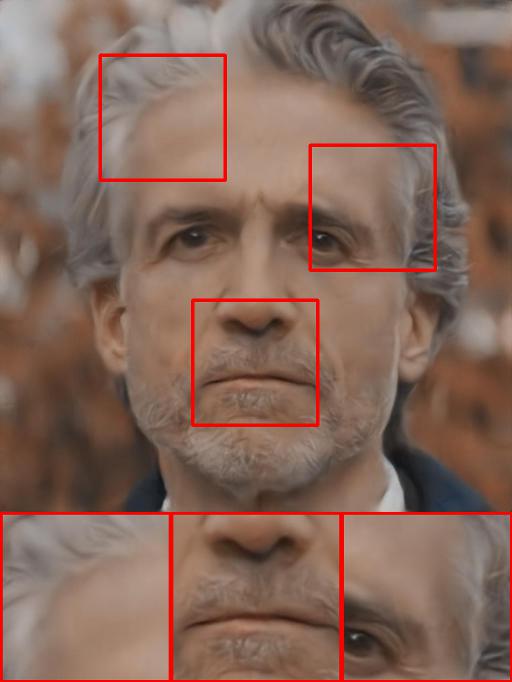} & 
\includegraphics[width=0.22\linewidth]{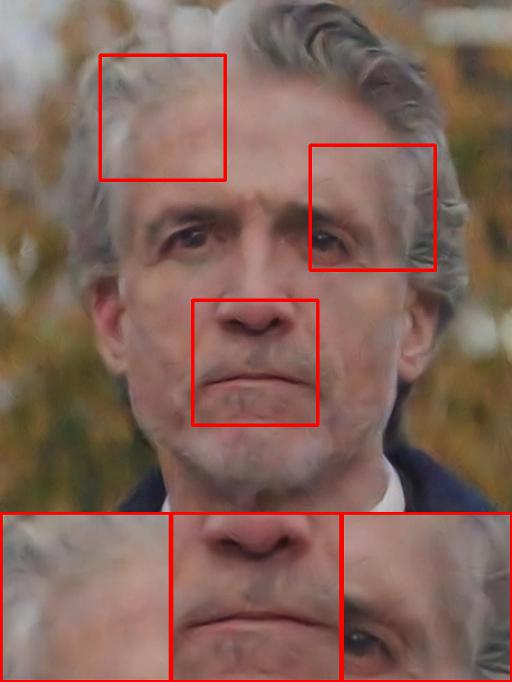} & 
\includegraphics[width=0.22\linewidth]{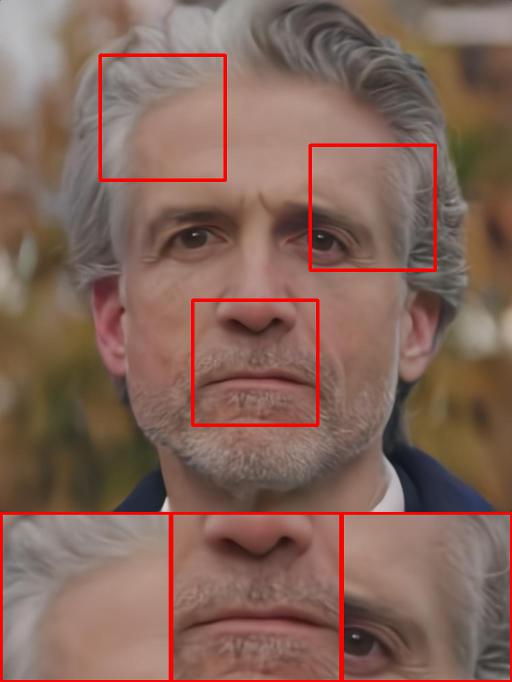} \\
\cline{2-5} 
PSNR &
32.79 &
31.64  &
33.50 &
34.79  \\

MS-SSIM &
0.894 &
0.945 &
0.915  & 
0.959  \\
BPP &
0.0127 &
0.0129 & 
0.0217 & 
0.0152  \\
LPIPS &
0.360 &
0.344 & 
0.320  & 
0.281   \\\midrule
Model &
&
\begin{tabular}{@{}c@{}} SGANC \\ l=1e-4  \end{tabular}  & 
\begin{tabular}{@{}c@{}} SGANC \\ l=1e-6  \end{tabular}  &
\\
\hline
 &
 &
\includegraphics[width=0.22\linewidth]{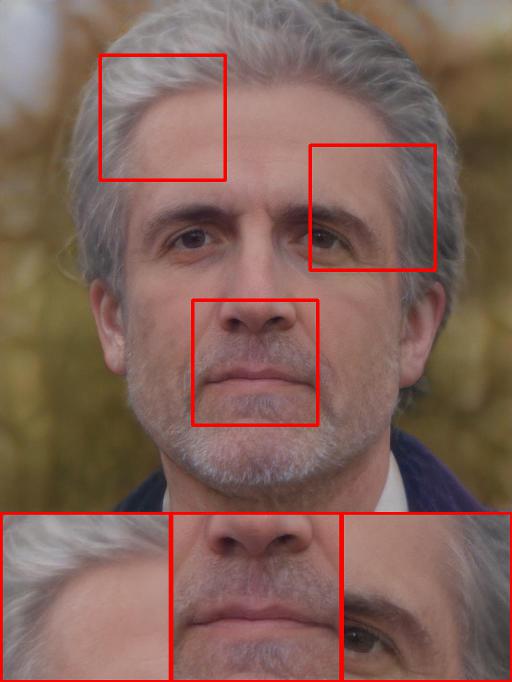} &
\includegraphics[width=0.22\linewidth]{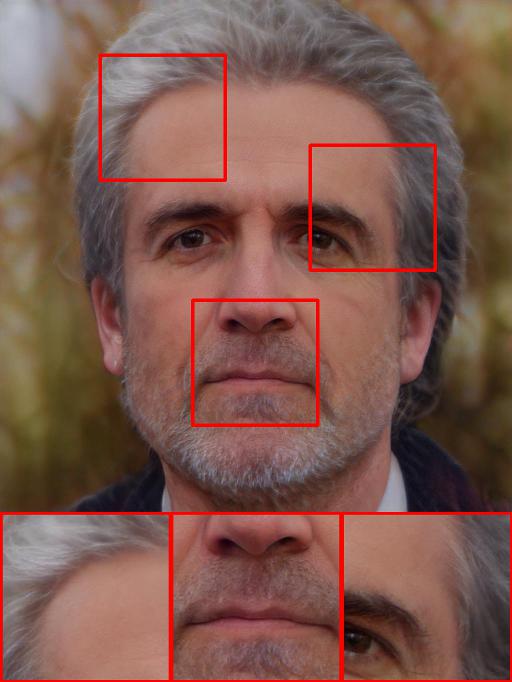} &
\\
\cline{3-4} 
PSNR &
 &
38.85 & 
18.71 &
\\

MS-SSIM &
  & 
0.995 &
0.834 &
\\
BPP &
 & 
0.0280 &
0.0066 &
\\
LPIPS &
  & 
0.010 &
0.250 &
\\
\bottomrule
\end{tabular}
\caption{Qualitative results for image compression: Other methods introduce blocking artifacts and blurring. Our method (SGANC) leads to high quality reconstruction  and perceptually lower distortion. Images are better seen zoomed.}
\label{fig:app_qual_results_FP006734MD02}

\end{figure*}

\begin{figure*}[h]
\setlength\tabcolsep{2pt}%
\centering
\small
\begin{tabular}{p{1.4cm}cccc}
\toprule
Model &
Original   &
Projected &
VTM   & 
\begin{tabular}{@{}c@{}} SGANC \\ l=1e-5  \end{tabular}  \\
\hline
 &
\includegraphics[width=0.22\linewidth]{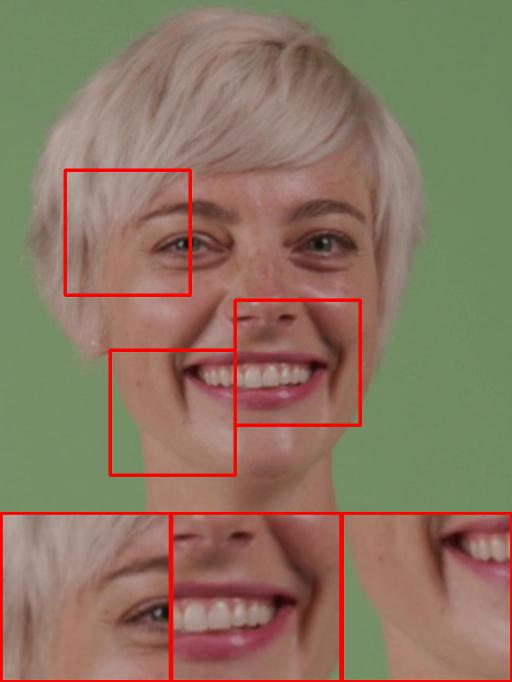}  & 
\includegraphics[width=0.22\linewidth]{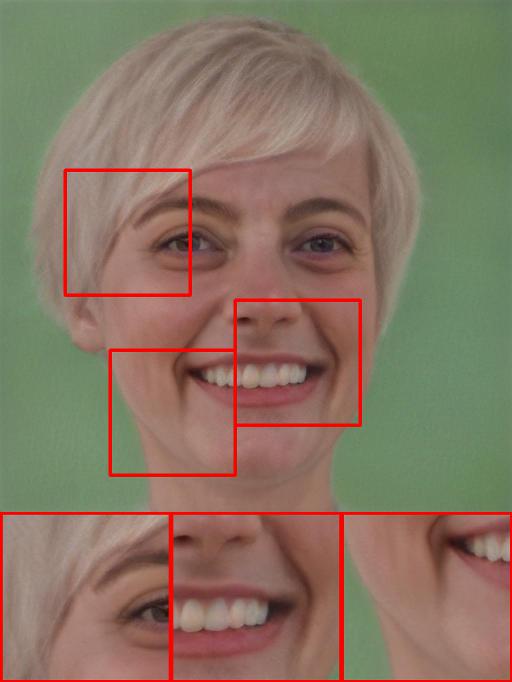} &
\includegraphics[width=0.22\linewidth]{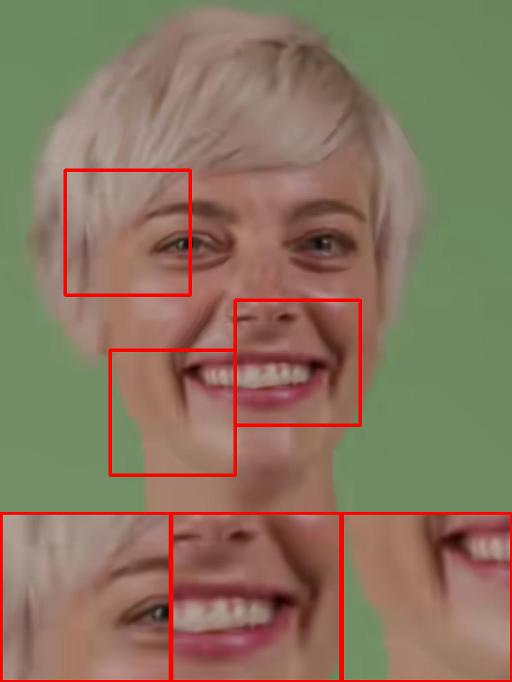} &
\includegraphics[width=0.22\linewidth]{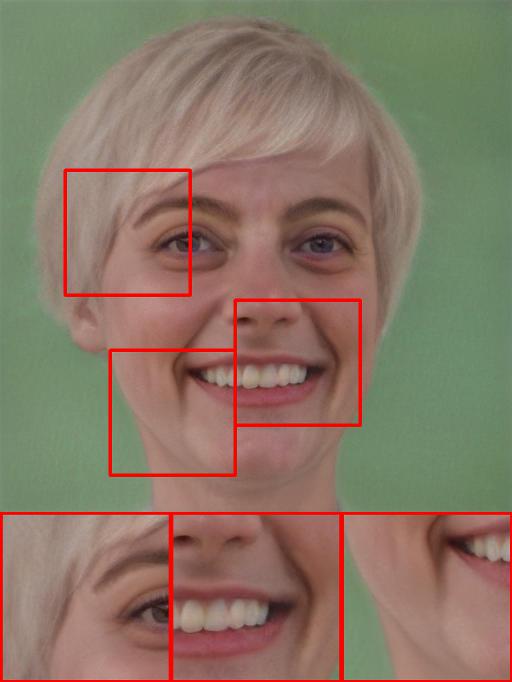}   \\
 \cline{4-5} 
PSNR &
 & 
 &
38.73 &
30.56  \\

MS-SSIM &
 & 
 &
0.966 &
0.967  \\
BPP &
 & 
 &
0.0.0102 &
0.0149  \\
LPIPS &
 & 
 &
0.1758 &
0.061  \\\midrule
Model &
AV1   &
\begin{tabular}{c} MeanHP   \end{tabular}   & 
\begin{tabular}{c} HP   \end{tabular}  & 
\textit{Cheng et. al}\cite{chen2020perceptually}  \\
\hline
 &
\includegraphics[width=0.22\linewidth]{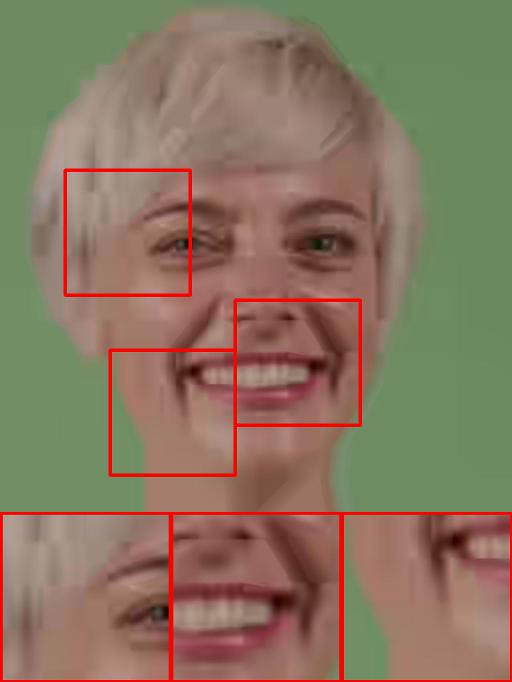} &
\includegraphics[width=0.22\linewidth]{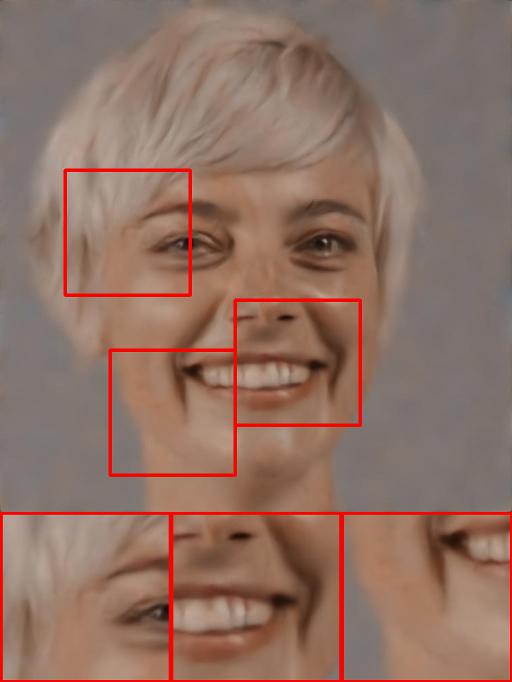} & 
\includegraphics[width=0.22\linewidth]{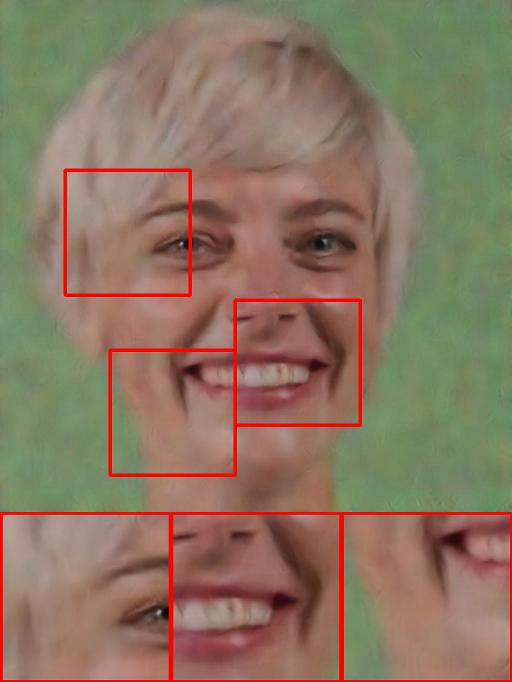} & 
\includegraphics[width=0.22\linewidth]{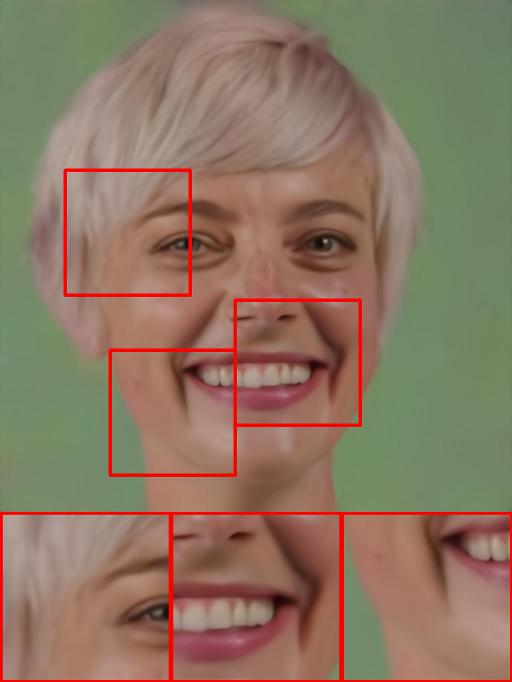} \\
\cline{2-5} 
PSNR &
35.51 &
27.15  &
34.98  &
36.89   \\

MS-SSIM &
0.940 &
0.953 &
0.938   & 
0.978   \\
BPP &
0.0095 &
0.0098 & 
0.0216 & 
0.0089  \\
LPIPS &
0.2201 &
0.4470 & 
0.2498   & 
0.1718    \\\midrule
Model &
&
\begin{tabular}{@{}c@{}} SGANC \\ l=1e-4  \end{tabular}  & 
\begin{tabular}{@{}c@{}} SGANC \\ l=1e-6  \end{tabular}  &
\\
\hline
 &
 &
\includegraphics[width=0.22\linewidth]{images/same_bpp_draw_new/FP010363HD03/real_nvp_compression_lam1e_4_old_abl_only_nf_8_coupl_hard_lat_disto_fixopt_compai/projected_img_59.jpg} &
\includegraphics[width=0.22\linewidth]{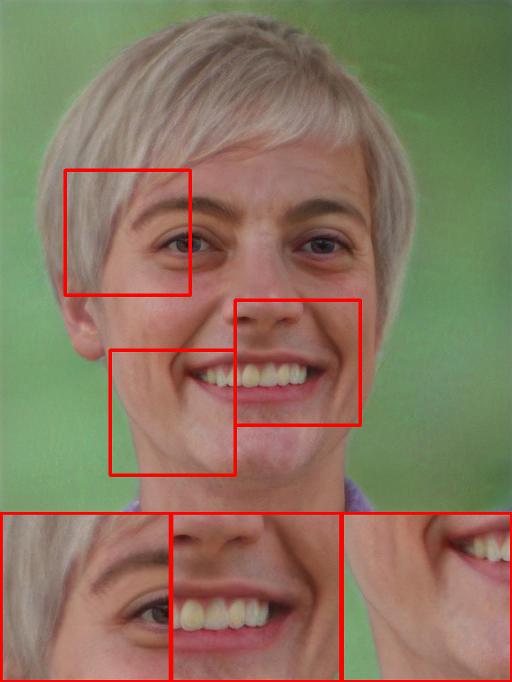} &
\\
\cline{3-4} 
PSNR &
 &
41.43 & 
17.97 &
\\

MS-SSIM &
  & 
0.995 &
0.716 &
\\
BPP &
 & 
0.0289 &
0.0068 &
\\
LPIPS &
  & 
0.008 &
0.317 &
\\
\bottomrule
\end{tabular}
\caption{Qualitative results: Other methods introduce blocking artifacts and blurring. Our method (SGANC) leads to high quality reconstruction  and perceptually lower distortion. Images are better seen zoomed}
\label{fig:qual_results_FP010363HD03_2}

\end{figure*}

In this section we show the results for image compression (Figures \ref{fig:qual_results_low_mead_intra_4_main}, \ref{fig:qual_results_FP010363HD03_2}, \ref{fig:app_qual_results_FP006734MD02}). Contrary to the results presented in the main paper, we here present for sake of completeness the results by comparing to the original image (except for our method).

\begin{figure*}[h]
     \centering
     \begin{subfigure}[b]{0.32\linewidth}
         \centering
         \caption{\tiny{PSNR}}
         \includegraphics[width=\linewidth, height=5cm]{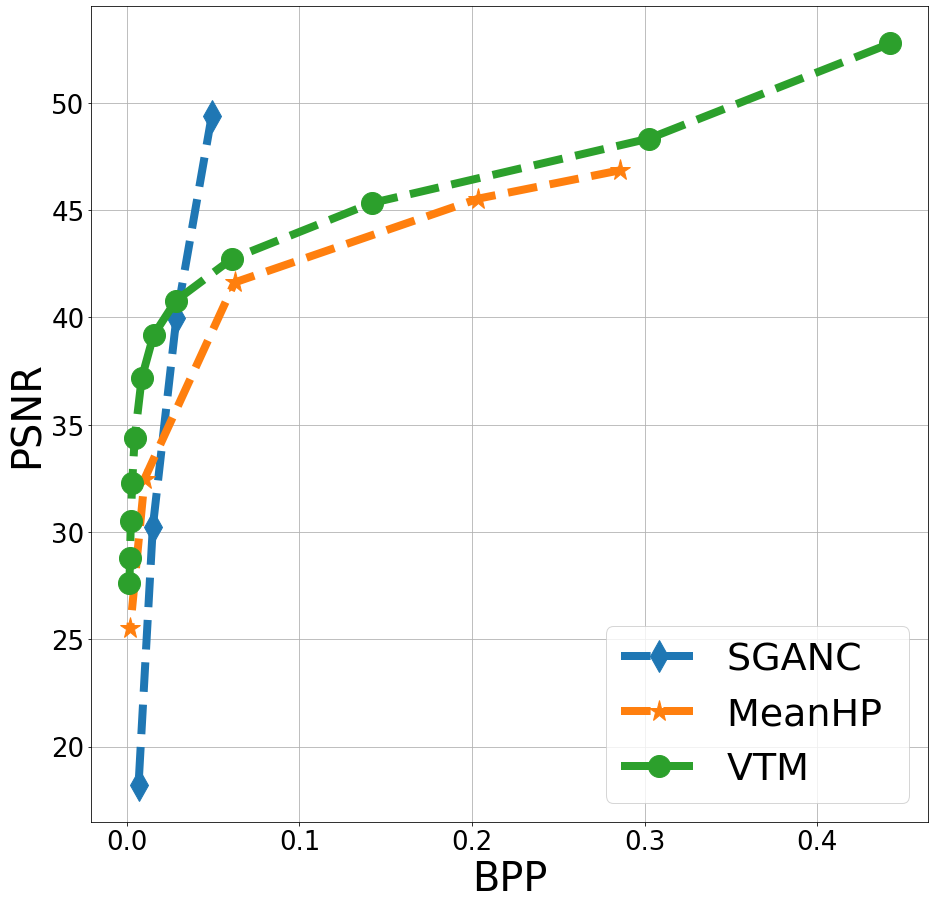}
     \end{subfigure}
     \hfill
     \begin{subfigure}[b]{0.32\linewidth}
         \centering
         \caption{\tiny{MS-SSIM}}
         \includegraphics[width=\linewidth, height=5cm]{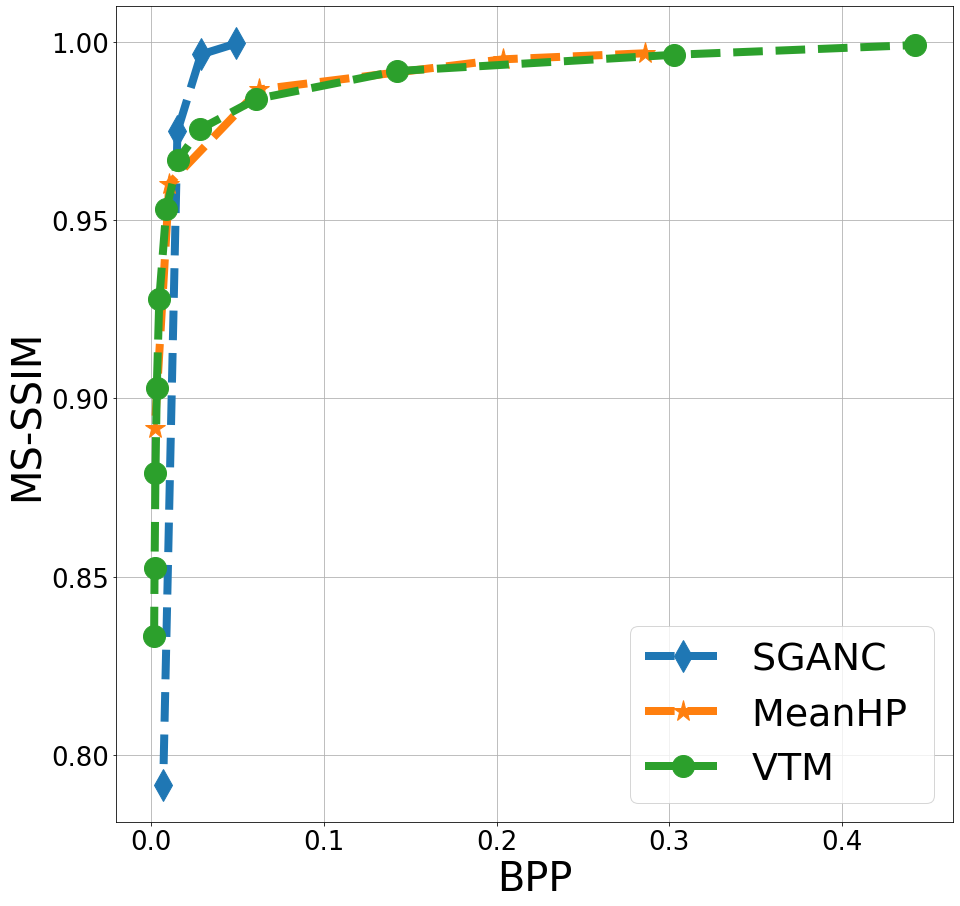}
     \end{subfigure}
     \hfill
     \begin{subfigure}[b]{0.32\linewidth}
         \centering
         \caption{\tiny{LPIPS \cite{zhang2018unreasonable}}}
         \includegraphics[width=\linewidth, height=5cm]{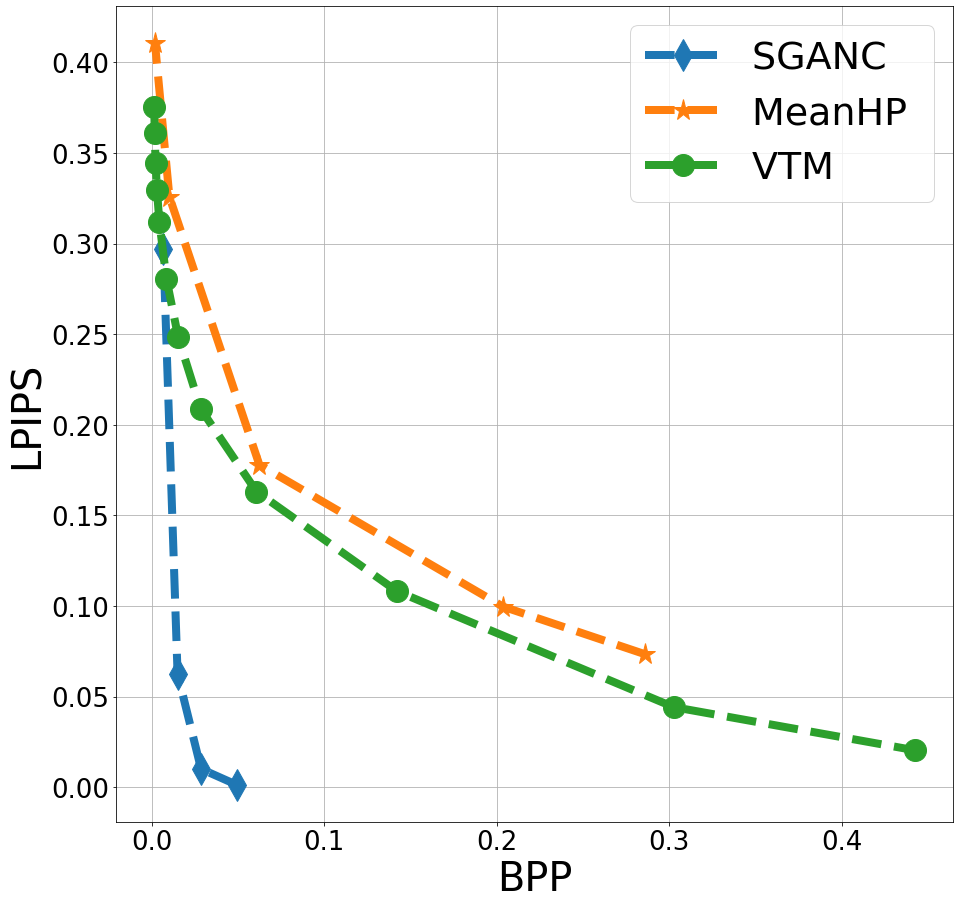}
     \end{subfigure}
        \caption{Rate distortion curves on FILMPAC dataset: for medium and large BPP, Our method (Blue) is better in terms of LPIPS and MS-SSIM than VTM (Green), MeanHP (Orange). For high BPP, our method is better in terms of PSNR.}
        \label{fig:results_filmpac_avg}
\end{figure*}

\begin{figure*}[h]
     \centering
     \begin{subfigure}[b]{0.32\linewidth}
         \centering
         \caption{\tiny{PSNR}}
         \includegraphics[width=\linewidth, height=5cm]{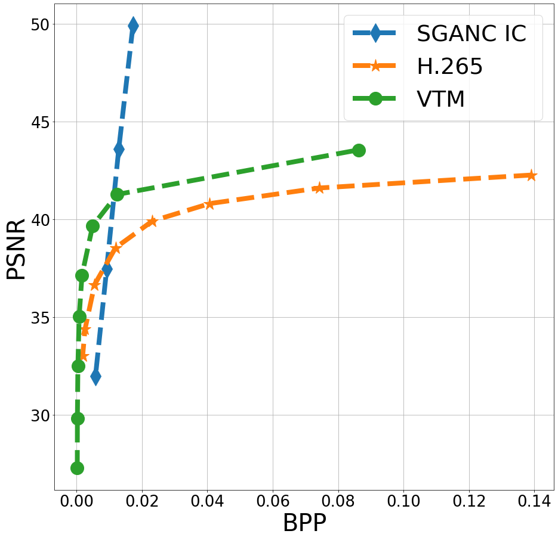}
     \end{subfigure}
     \hfill
     \begin{subfigure}[b]{0.32\linewidth}
         \centering
         \caption{\tiny{MS-SSIM}}
         \includegraphics[width=\linewidth, height=5cm]{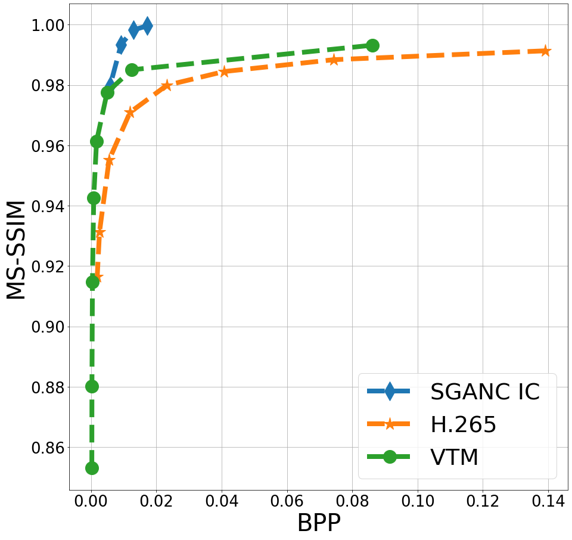}
     \end{subfigure}
     \hfill
     \begin{subfigure}[b]{0.32\linewidth}
         \centering
         \caption{\tiny{LPIPS \cite{zhang2018unreasonable}}}
         \includegraphics[width=\linewidth, height=5cm]{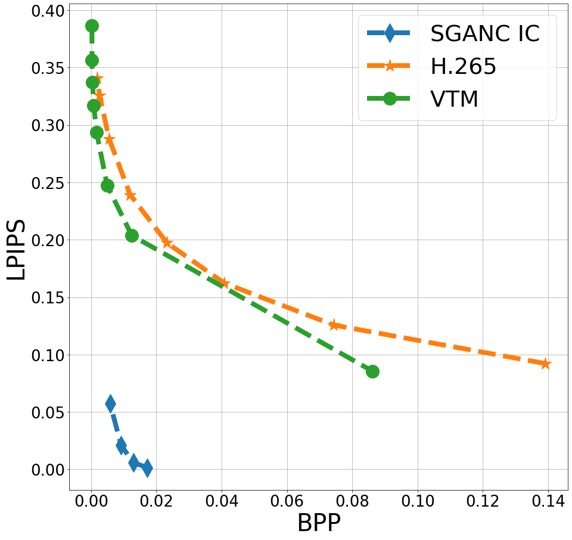}
     \end{subfigure}
        \caption{Rate distortion curves for video compression on Filmpac dataset. The perceptual metric LPIPS clearly shows a perceptual improvement of our method.}
        \label{fig:results_filmpac_inter}
\end{figure*}

\subsubsection{Video Compression}
\label{sec:app_more_results_video}
In this section, we show quantitative and qualitative results for video compression. Contrary to our main results in the paper, we here compress the original video and compute the metrics with original frames (except for our approach still compared to the projected frames).

\begin{figure*}[!h]
\setlength\tabcolsep{2pt}%
\centering
\small
\begin{tabular}{p{0.7cm}ccc}
\toprule
Model &
Original &
Projected & 
SGANC IC \\
\hline
 &
\includegraphics[width=0.31\linewidth]{images/inter_qualitative_draw_new/W016/orig/frame0002.jpg}  & 
\includegraphics[width=0.31\linewidth]{images/inter_qualitative_draw_new/W016/projected/frame0002.jpg} &
\includegraphics[width=0.31\linewidth]{images/inter_qualitative_draw_new/W016/sganc_ic/l1_6/frame0001.jpg}  \\ \cline{4-4} 
BPP &
 & 
 &
0.0017 \\ \midrule
Model &
& 
VTM & 
H.265 \\
\hline
&
&
\includegraphics[width=0.31\linewidth]{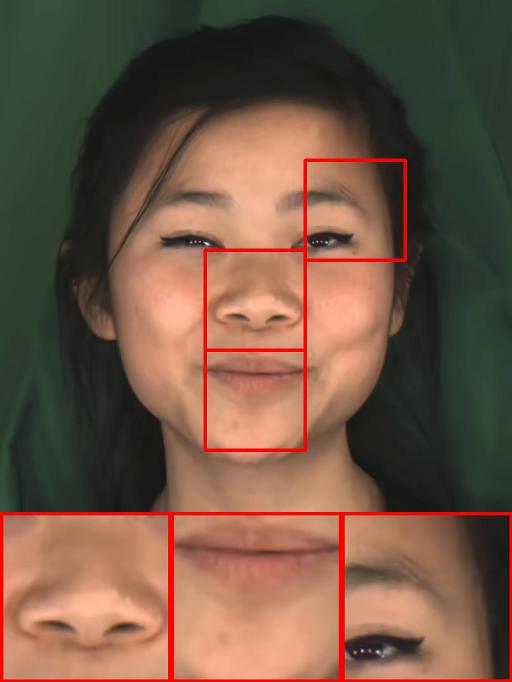}  &
\includegraphics[width=0.31\linewidth]{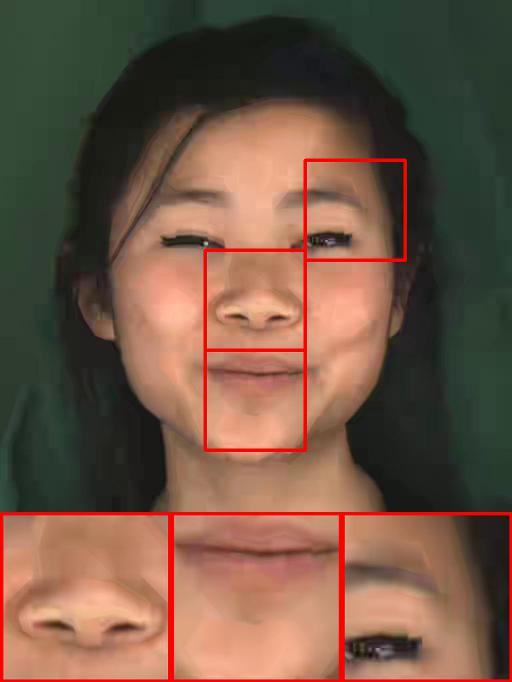} \\\cline{3-4} 
BPP &
 & 
0.0025 &
0.0023 \\ 
\bottomrule
\end{tabular}
\caption{Qualitative results for video compression: Other methods introduce blocking artifacts and blurring. Our method (SGANC IC) leads to high quality reconstruction  and perceptually lower distortion. Images are better seen zoomed}
\label{fig:app_qual_results_low_mead_inter_4_main}

\end{figure*}

\begin{figure*}[h]
\setlength\tabcolsep{2pt}%
\centering
\begin{tabular}{p{0.5cm}cccc}
\toprule
 \begin{turn}{90} \hspace{1.2cm} Original \end{turn}  &
 &
 \includegraphics[width=0.22\linewidth]{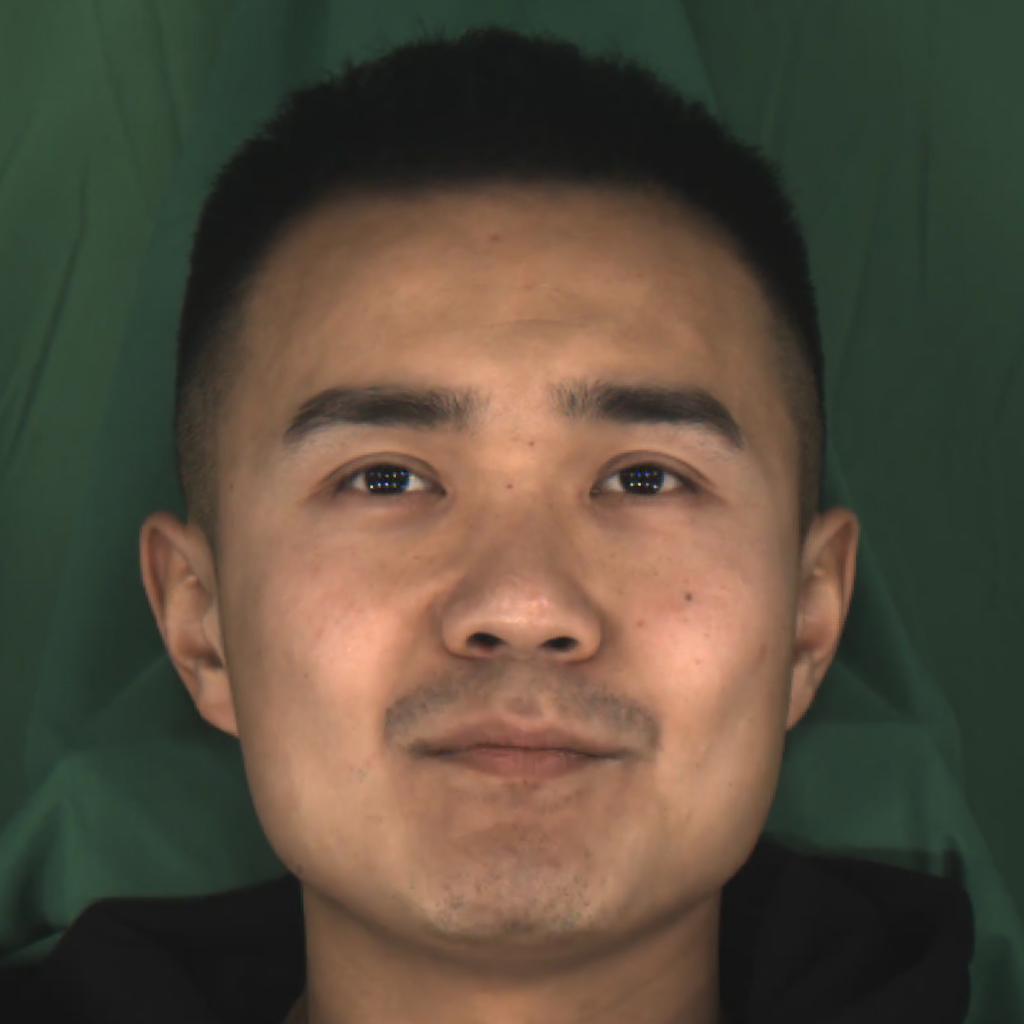}  & 
 \includegraphics[width=0.22\linewidth]{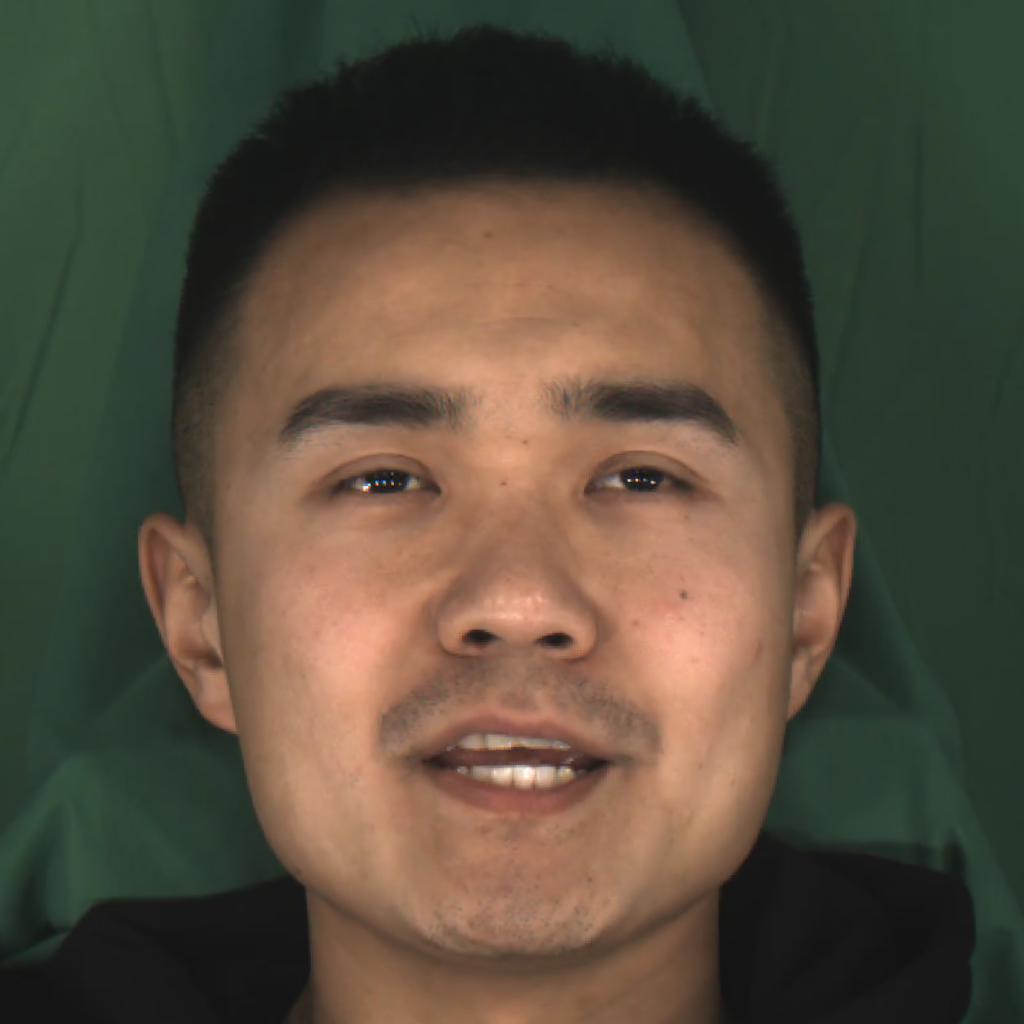}  & 
 \includegraphics[width=0.22\linewidth]{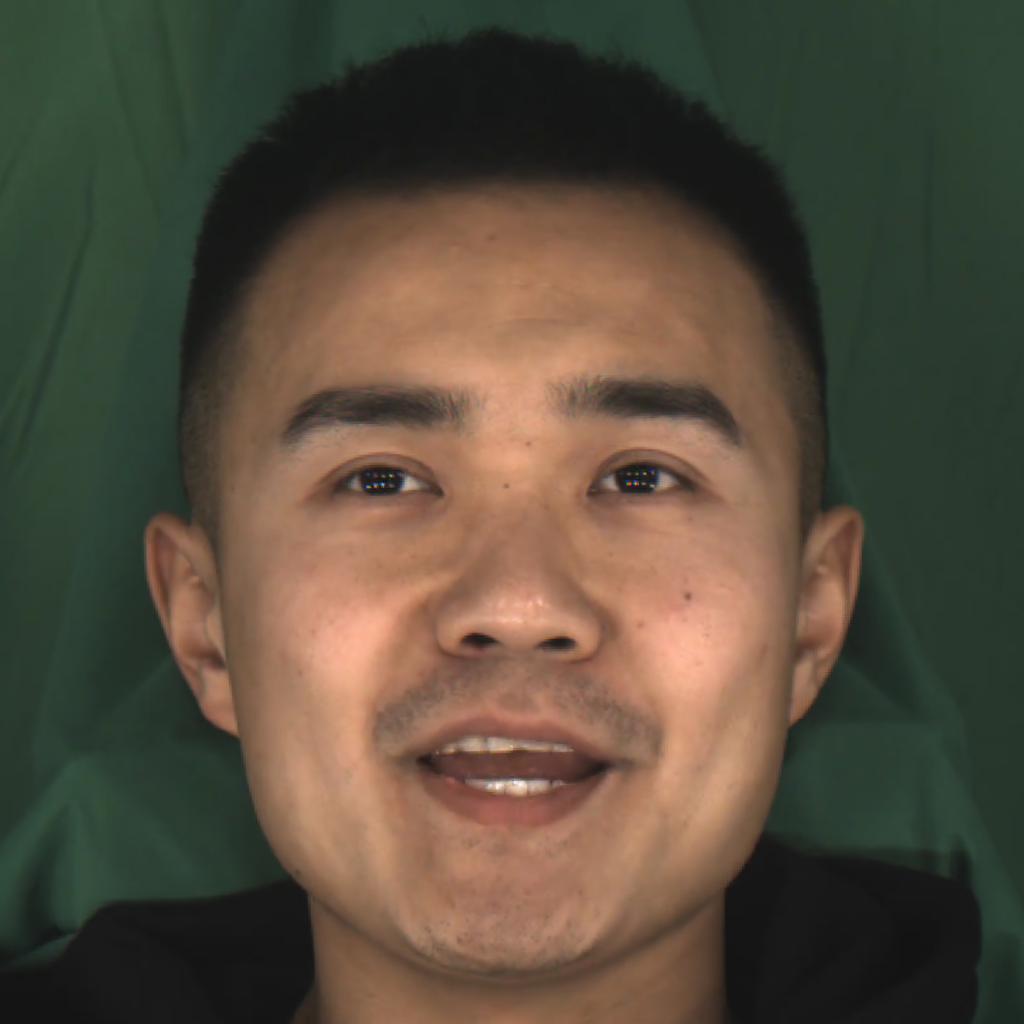}  \\ 

 \begin{turn}{90}\hspace{1.2cm} Projected\end{turn} &
 &
 \includegraphics[width=0.22\linewidth]{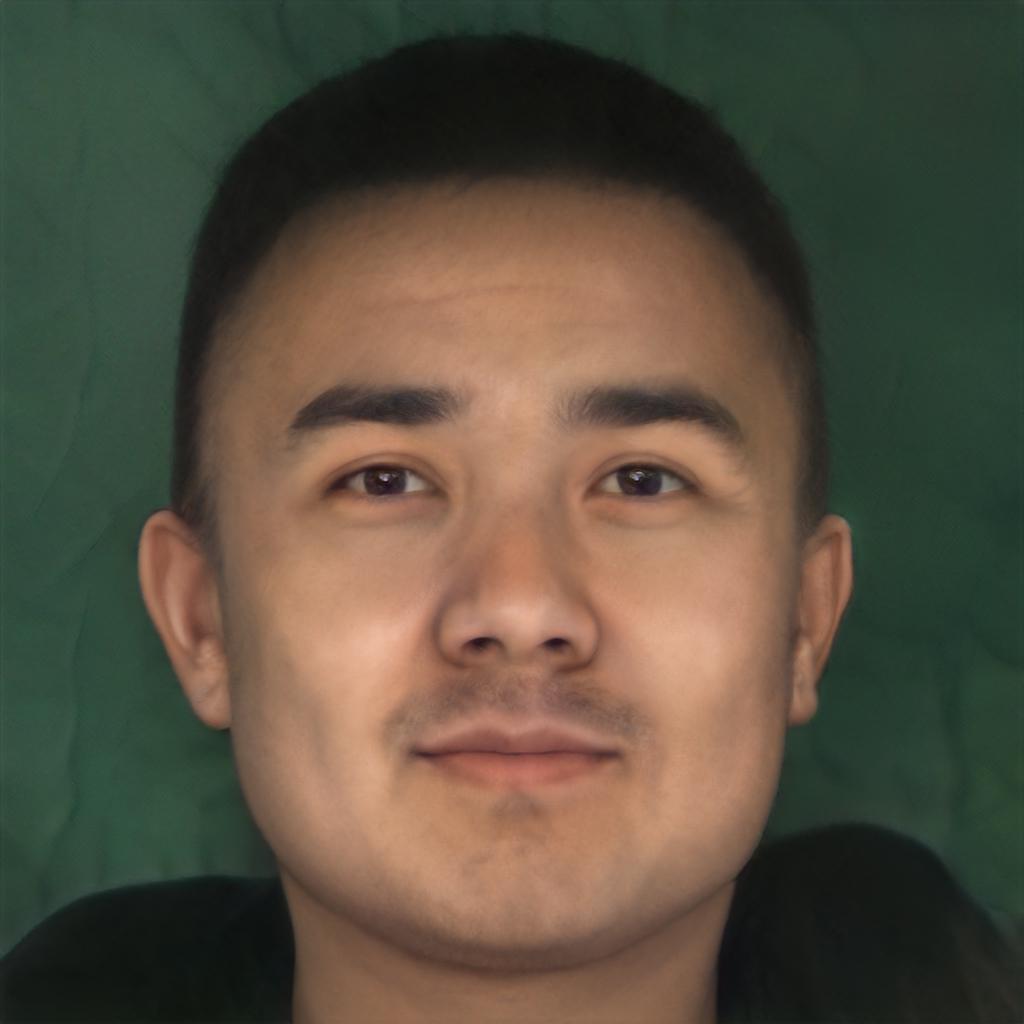}  & 
 \includegraphics[width=0.22\linewidth]{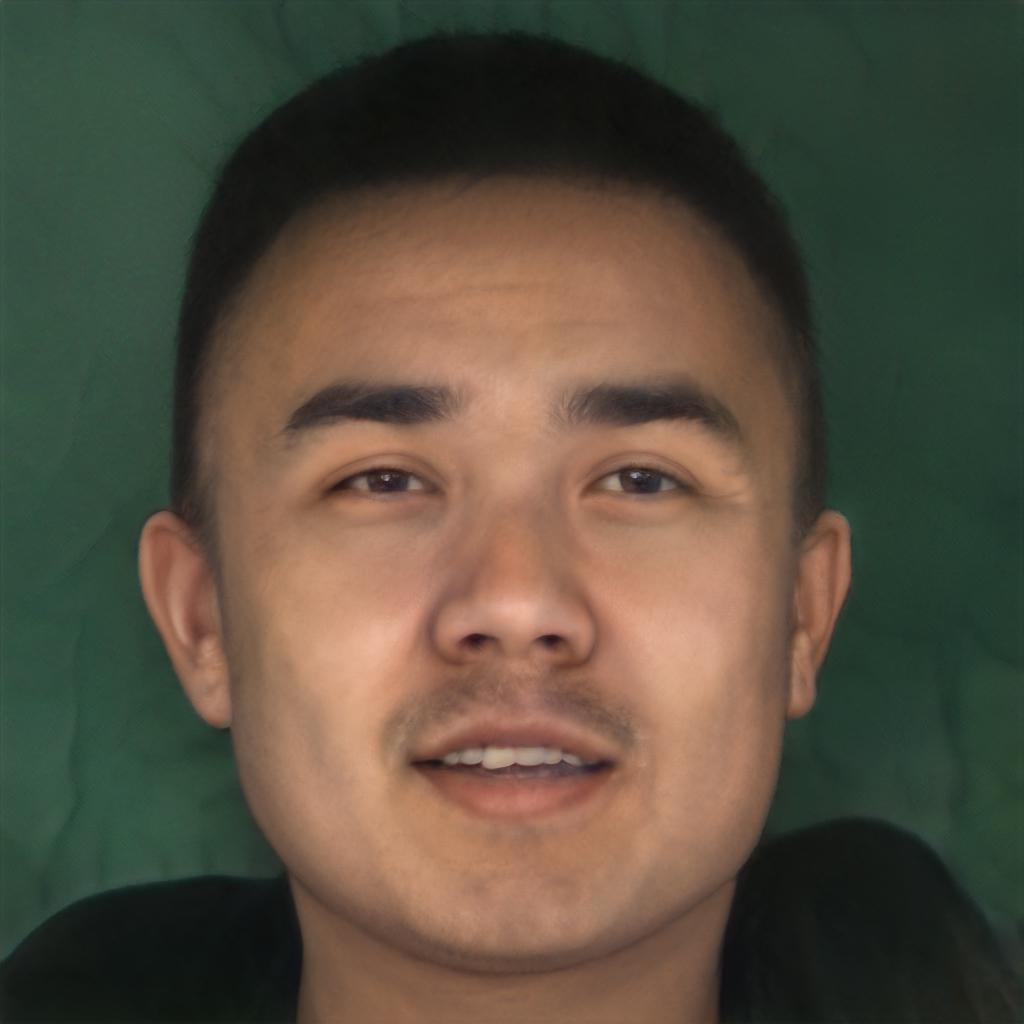}  & 
 \includegraphics[width=0.22\linewidth]{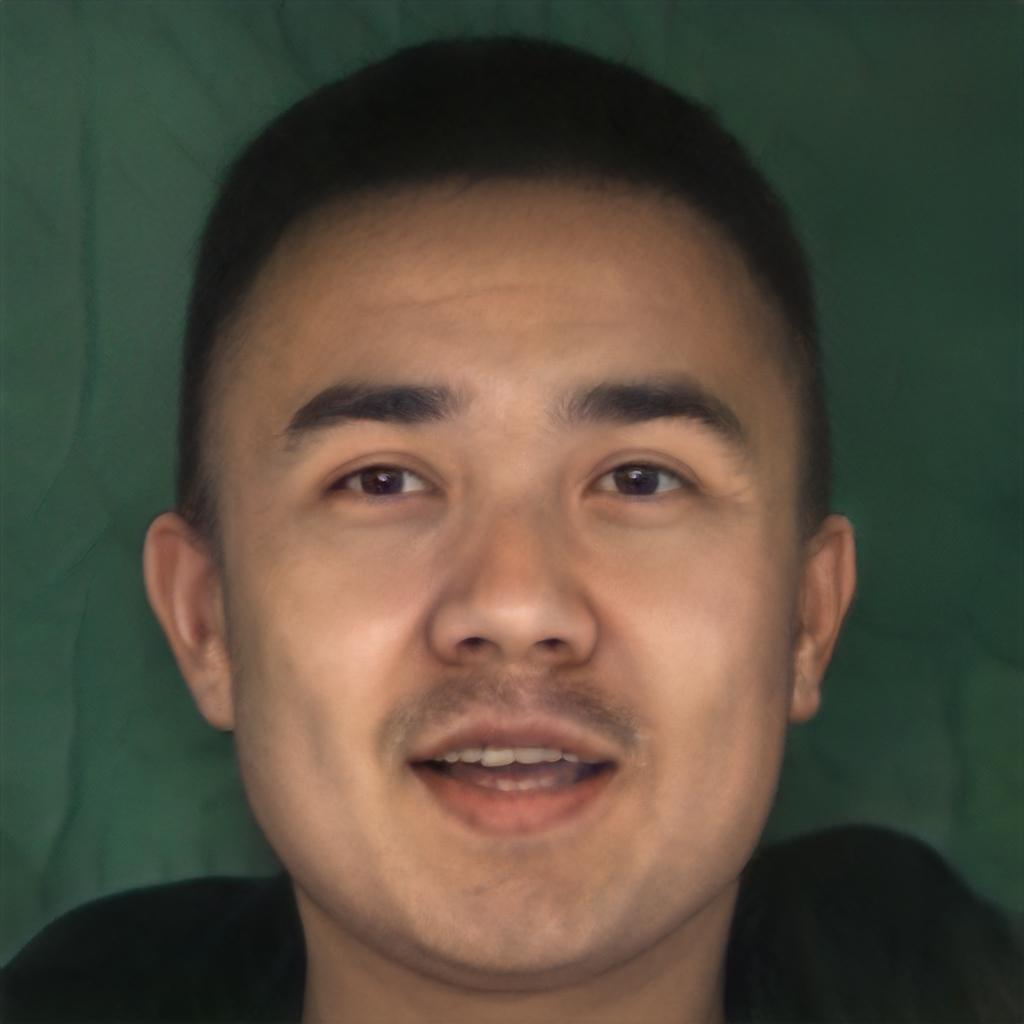}  \\

\begin{turn}{90}\hspace{0.5cm} SGANC IC \end{turn} &
\begin{turn}{90}\hspace{0.5cm} BPP=0.00156 \end{turn} &
 \includegraphics[width=0.22\linewidth]{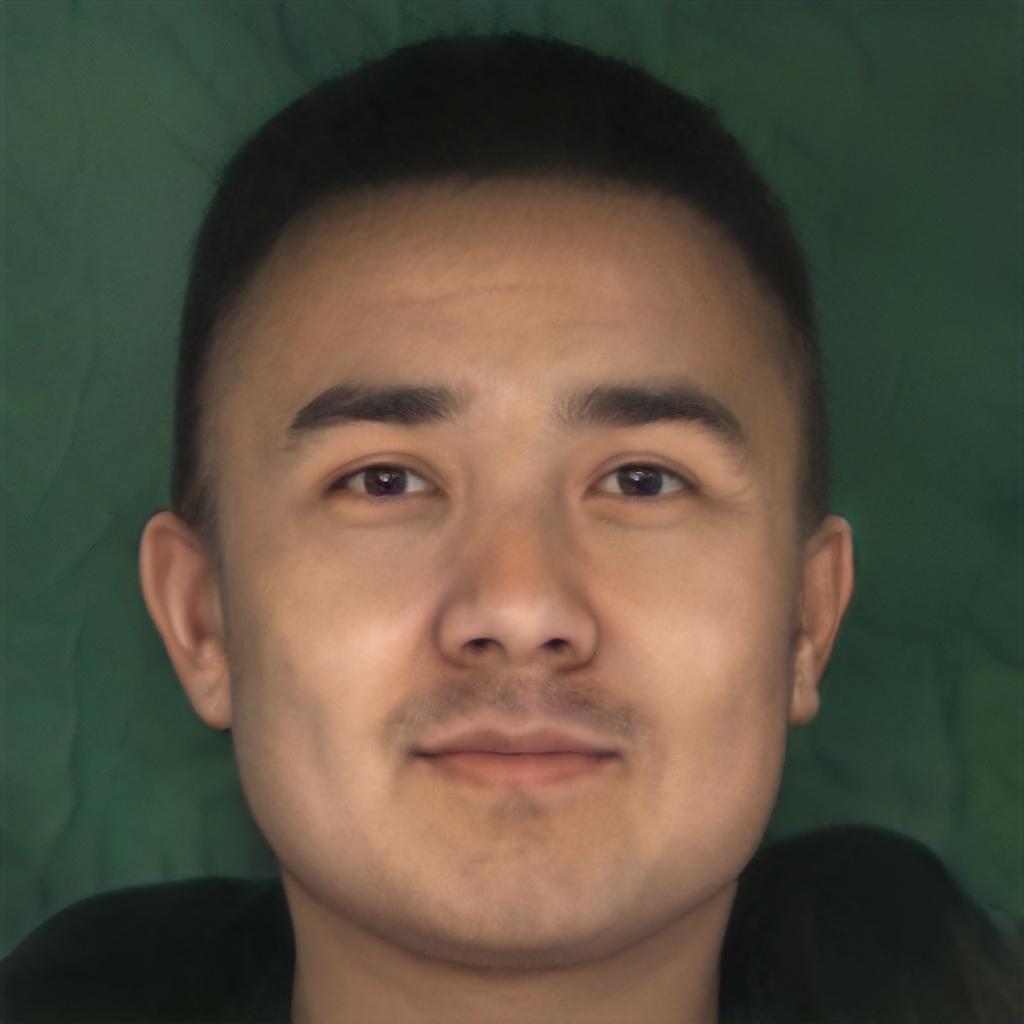}  & 
 \includegraphics[width=0.22\linewidth]{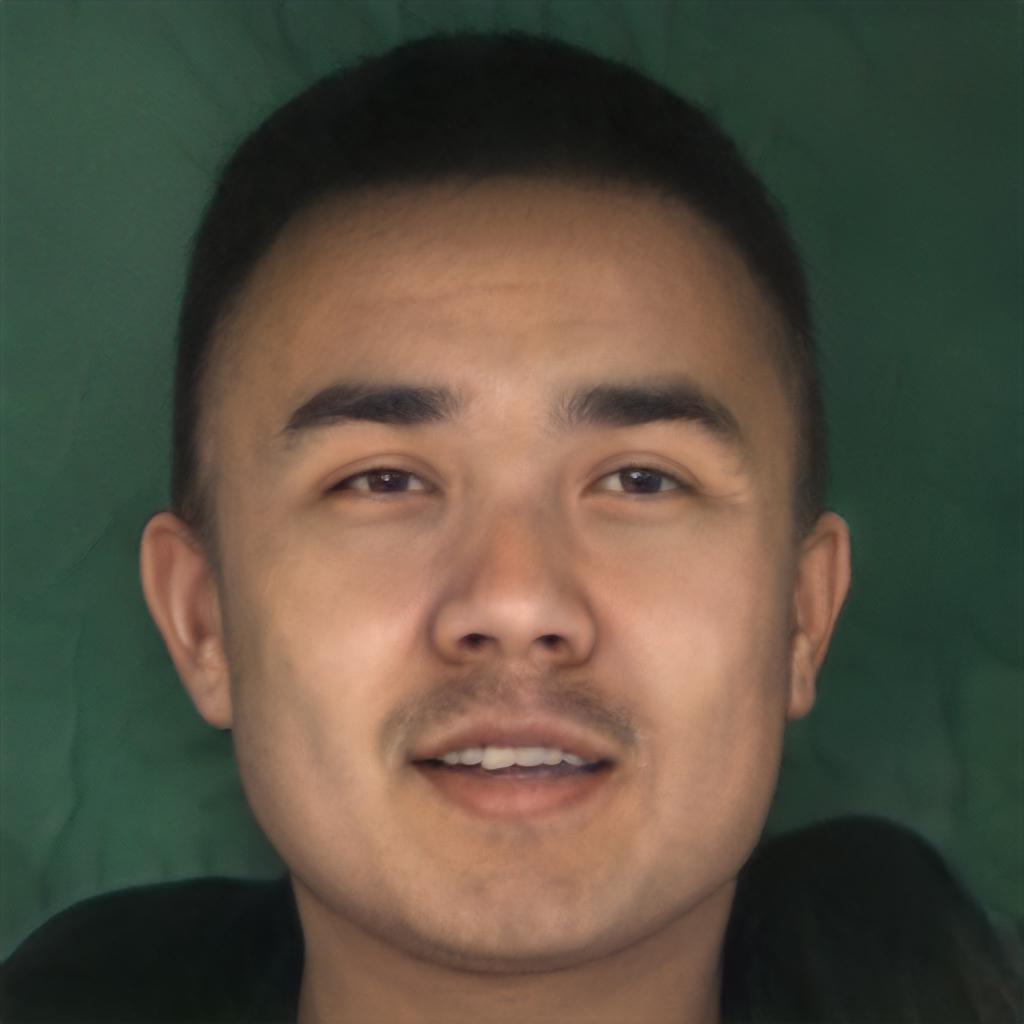}   &
 \includegraphics[width=0.22\linewidth]{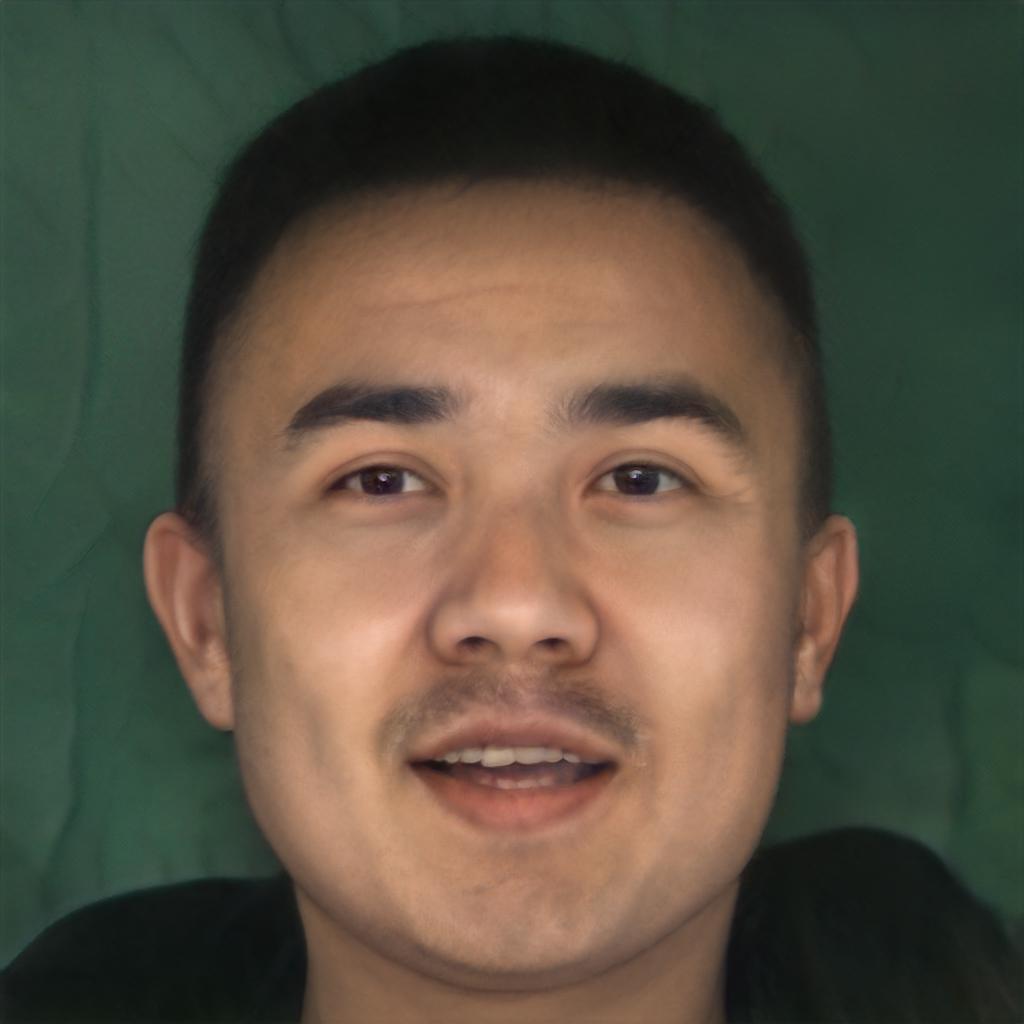}  \\
 
 \begin{turn}{90}\hspace{1cm} VTM  \end{turn}&
 \begin{turn}{90}\hspace{0.5cm} BPP=0.00189\end{turn}&
 \includegraphics[width=0.22\linewidth]{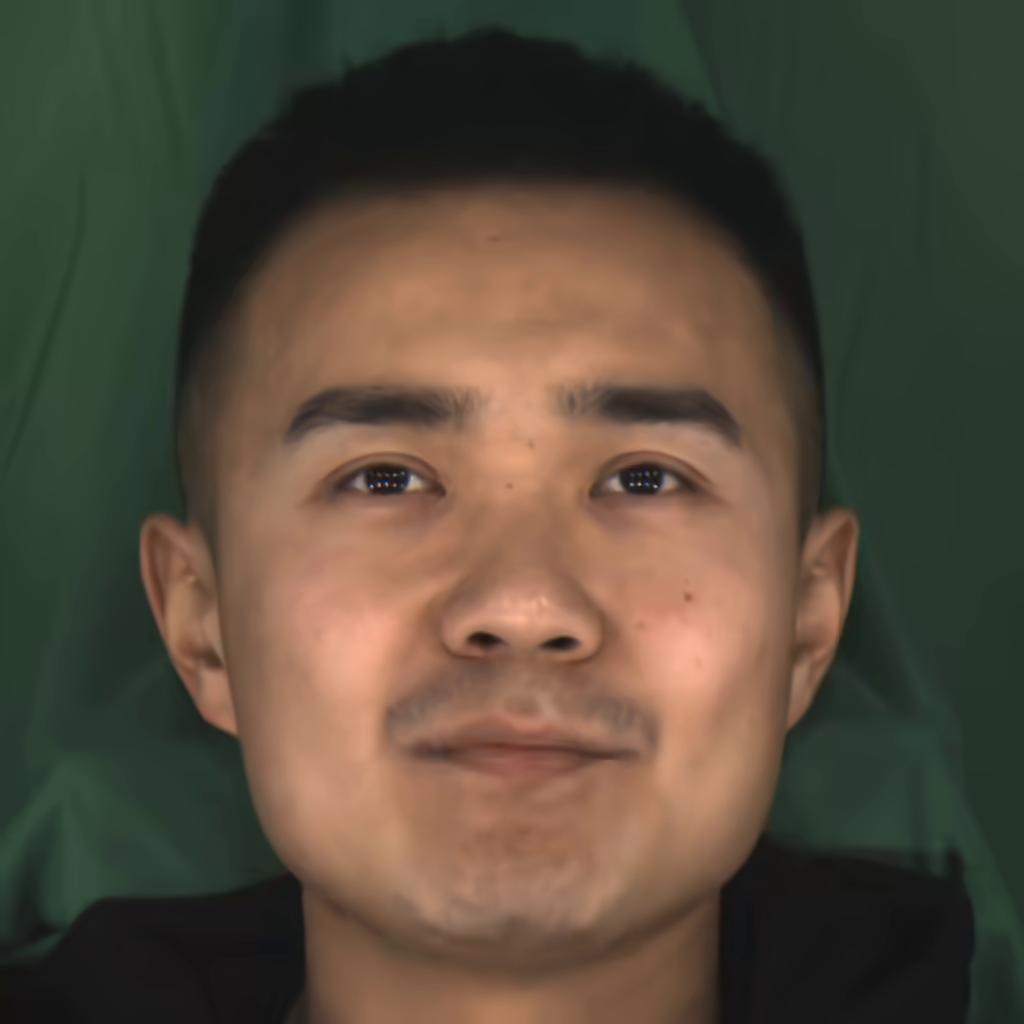}  & 
 \includegraphics[width=0.22\linewidth]{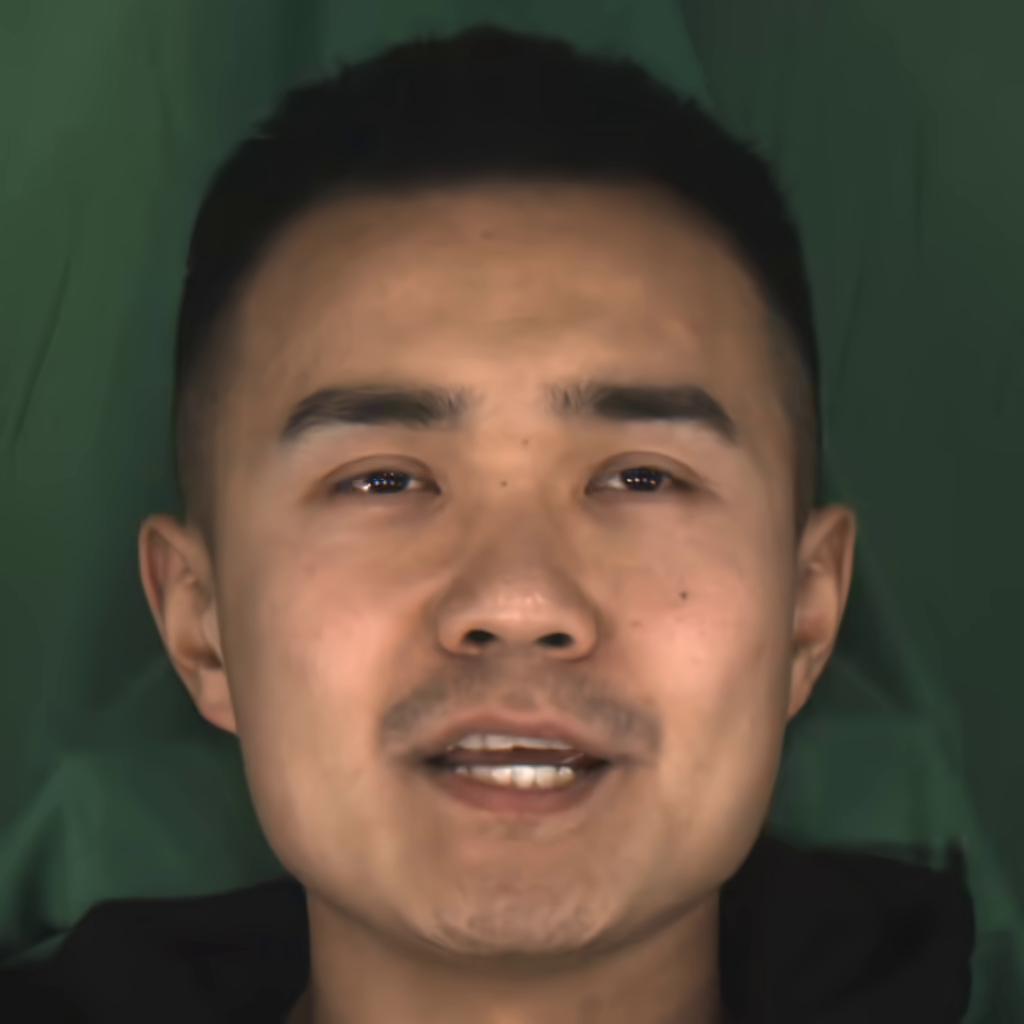}  & 
 \includegraphics[width=0.22\linewidth]{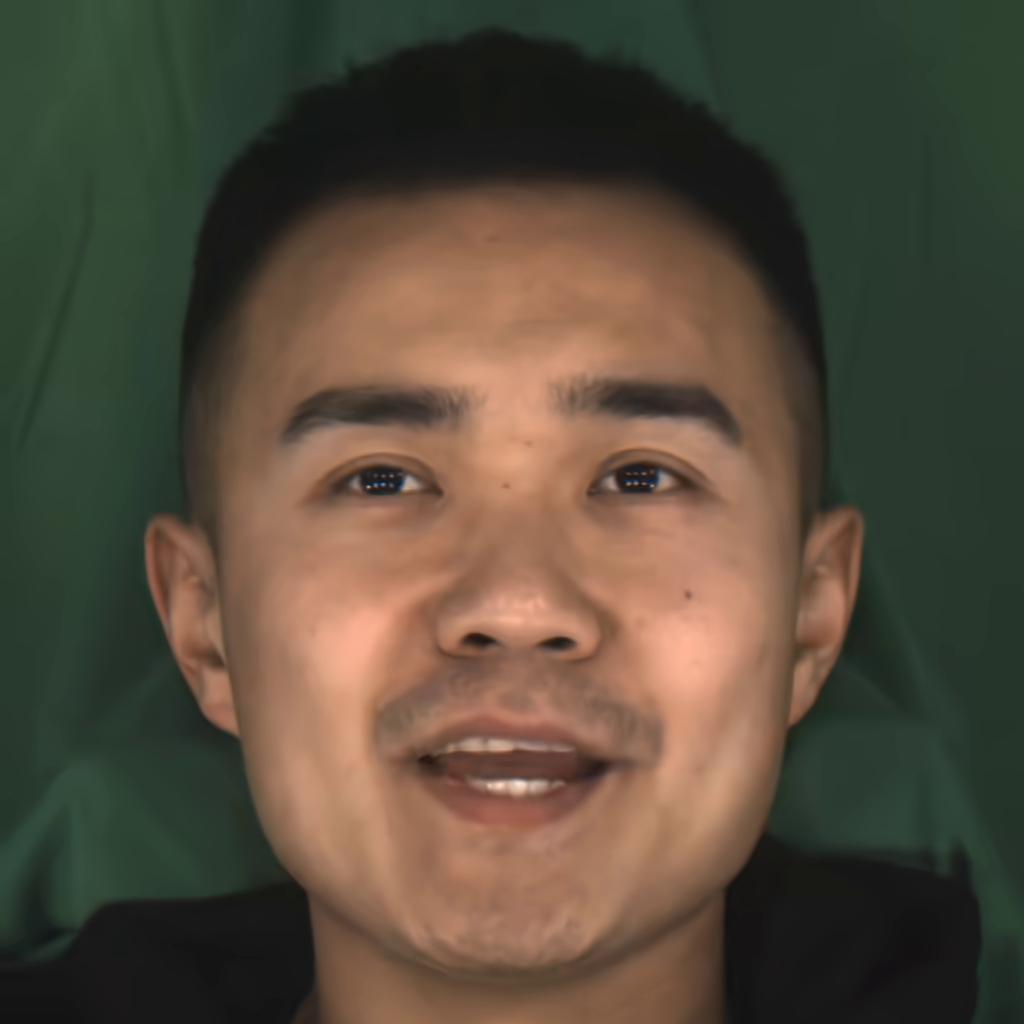}  \\ 

 \begin{turn}{90}\hspace{1cm} H.265 \end{turn} &
 \begin{turn}{90}\hspace{0.5cm} BPP=0.00186\end{turn}&
 \includegraphics[width=0.22\linewidth]{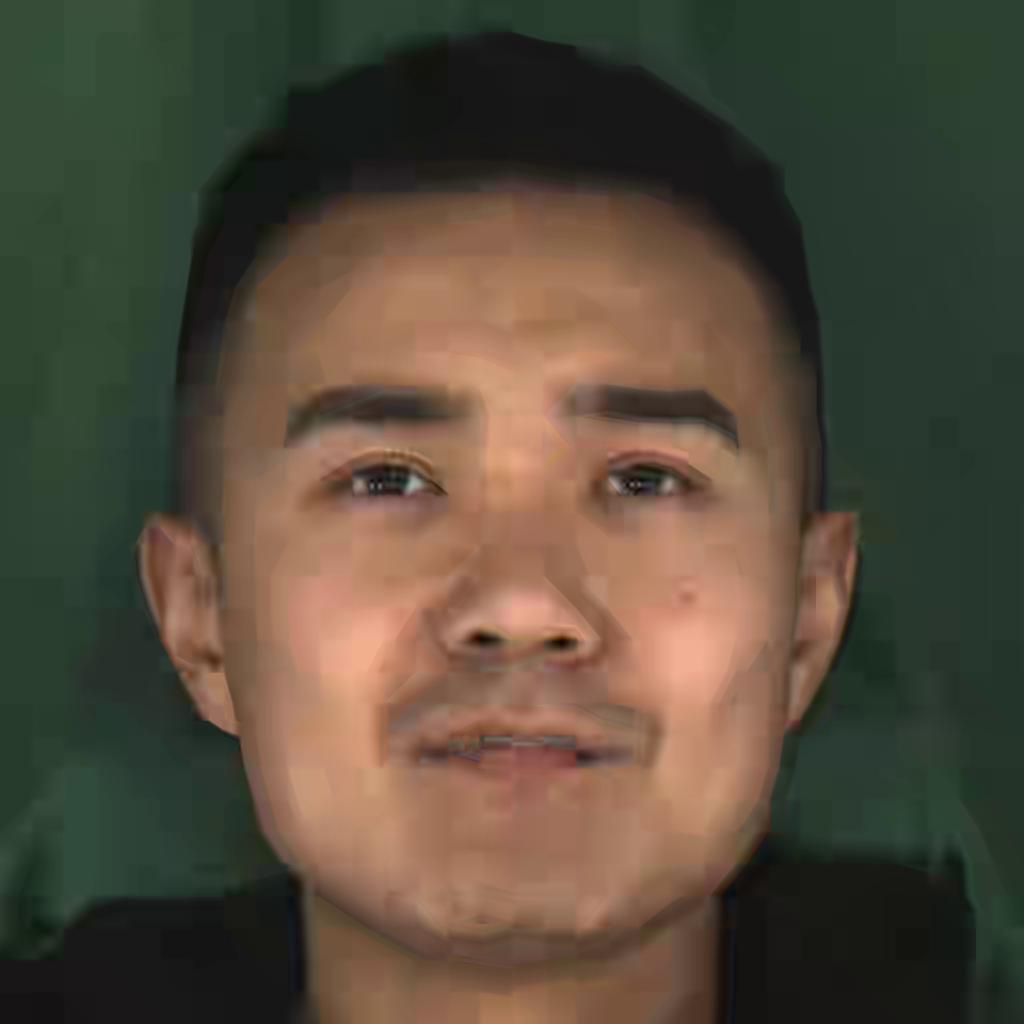}  & 
 \includegraphics[width=0.22\linewidth]{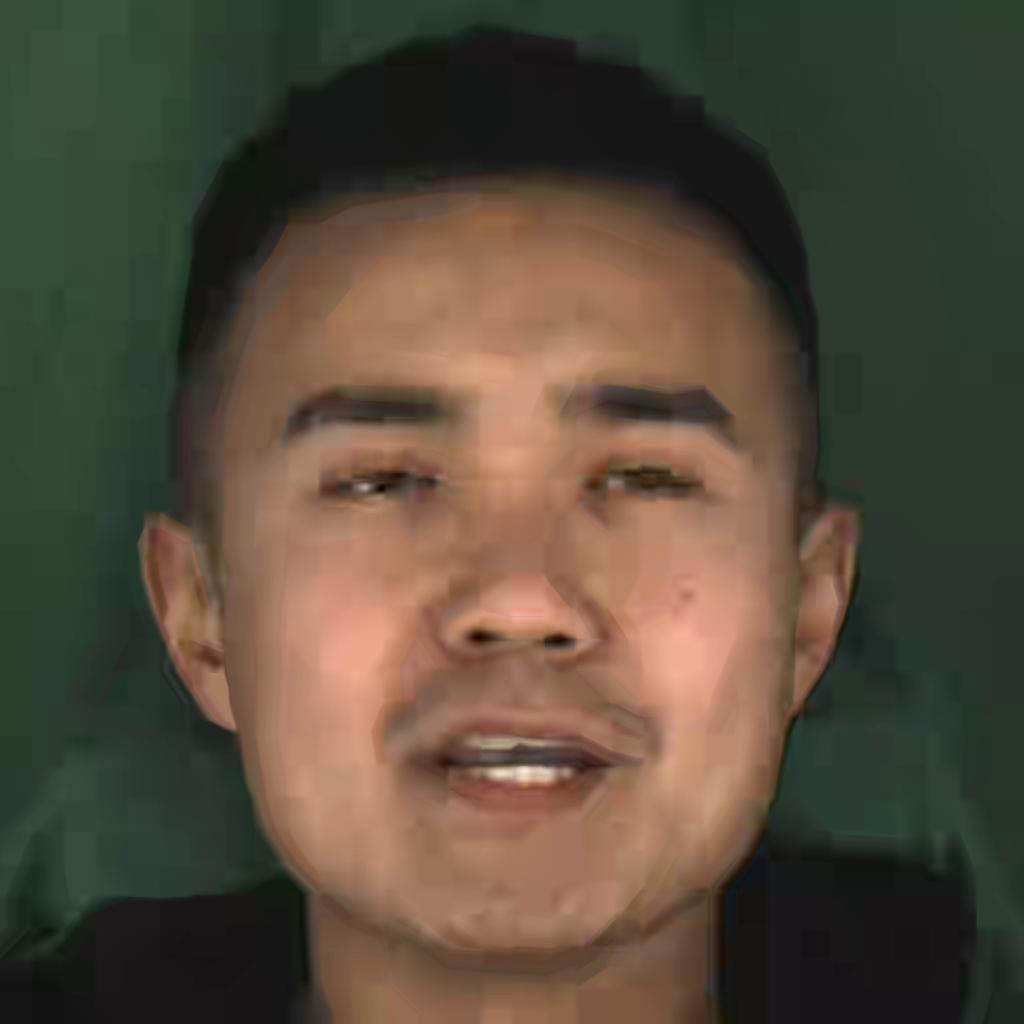}  & 
 \includegraphics[width=0.22\linewidth]{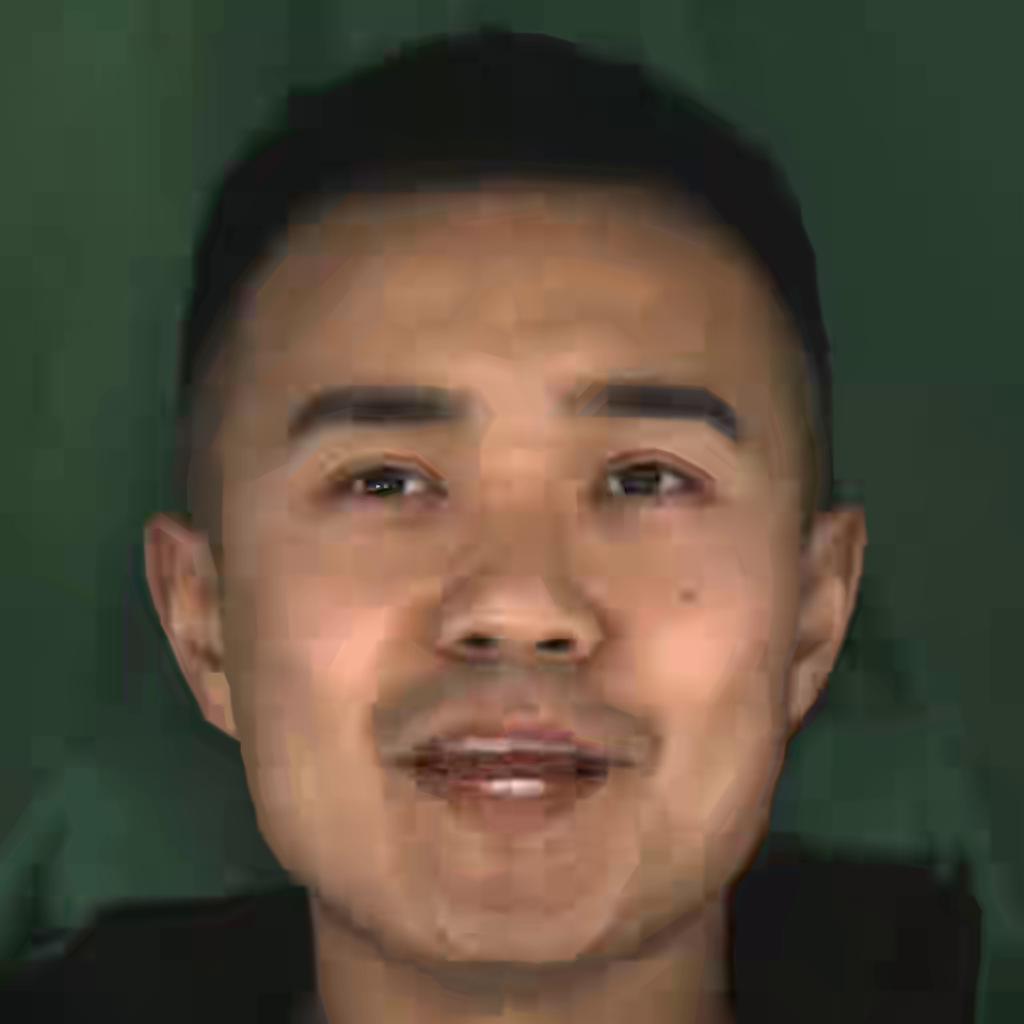}  \\ 
 \\
\bottomrule
\end{tabular}
\caption{Qualitative results on MEAD inter dataset for extreme BPP, computed for three video frames. H.265 shows blocking artifacts and blurring, VTM shows blurring especially at the edges of the face and the hair while our method (SGANC) is almost artifacts free and with high quality images.}
\label{fig:qual_results_low_mead_inter_1}
\end{figure*}

\begin{figure*}[h]
\setlength\tabcolsep{2pt}%
\centering
\begin{tabular}{p{0.5cm}cccc}
\toprule
 \begin{turn}{90} \hspace{1.2cm} Original \end{turn}  &
 &
 \includegraphics[width=0.22\linewidth]{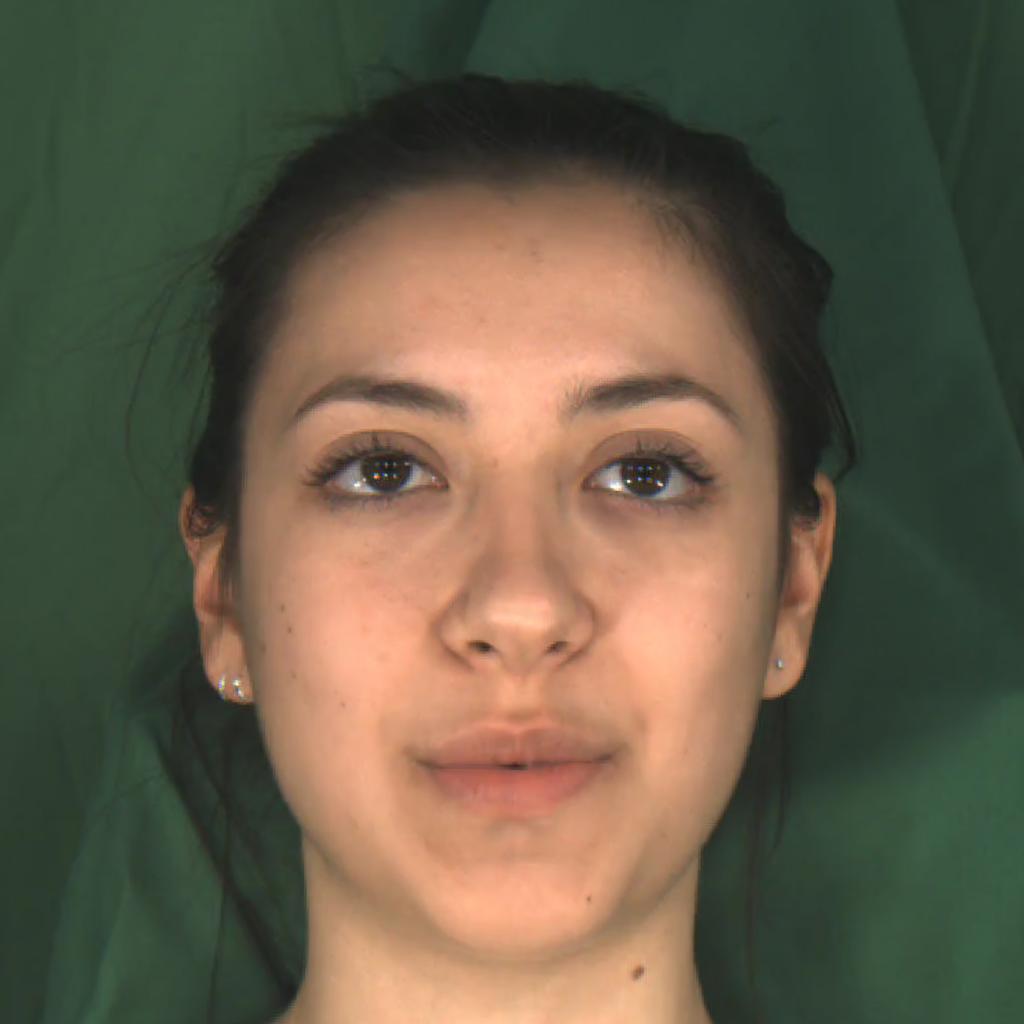}  & 
 \includegraphics[width=0.22\linewidth]{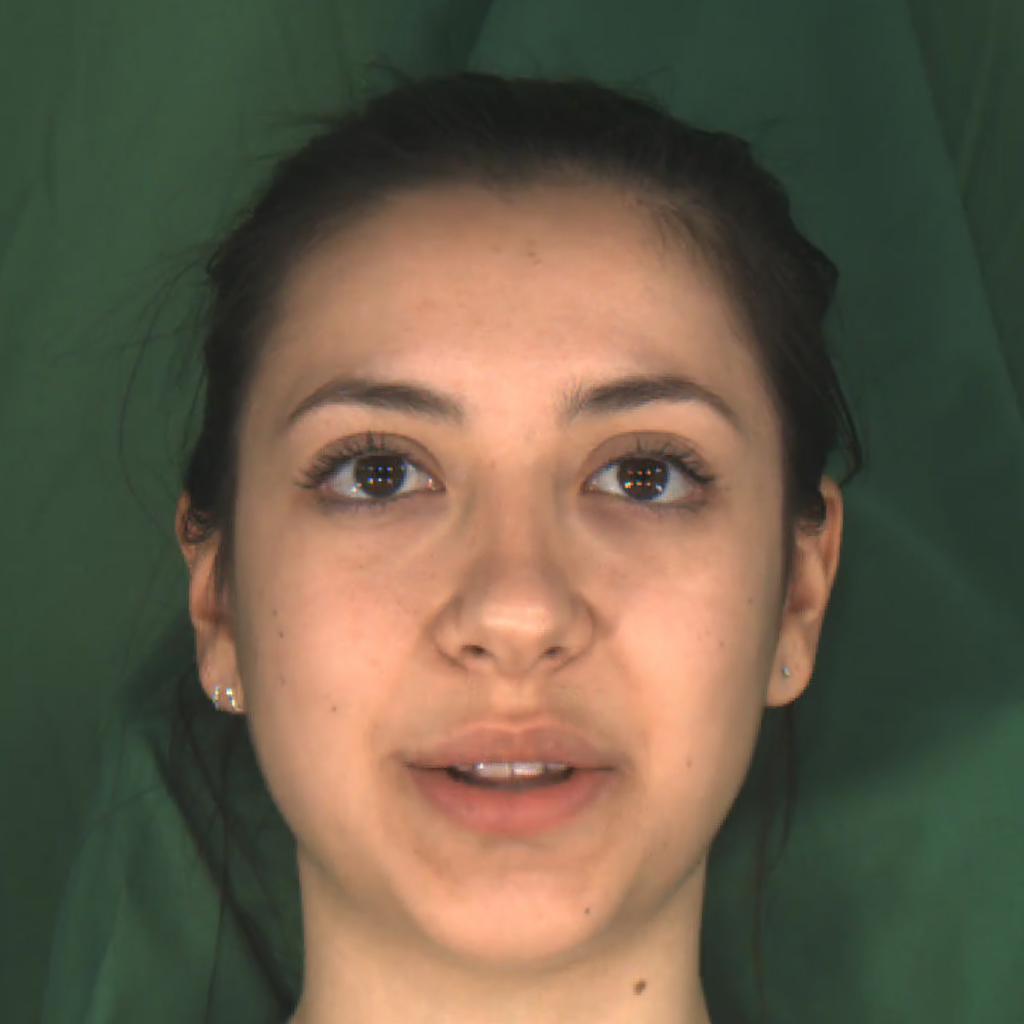}  & 
 \includegraphics[width=0.22\linewidth]{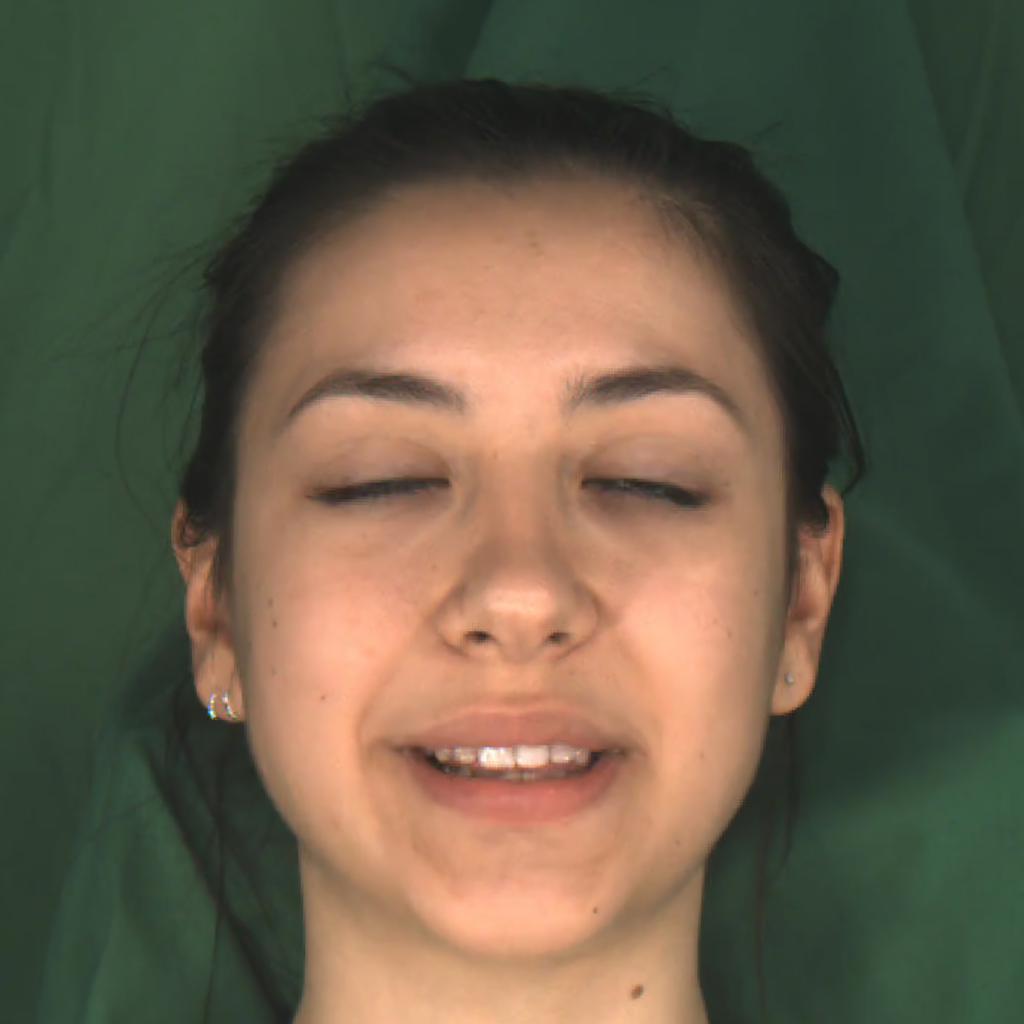}  \\

 \begin{turn}{90}\hspace{1.2cm} Projected\end{turn} &
 &
 \includegraphics[width=0.22\linewidth]{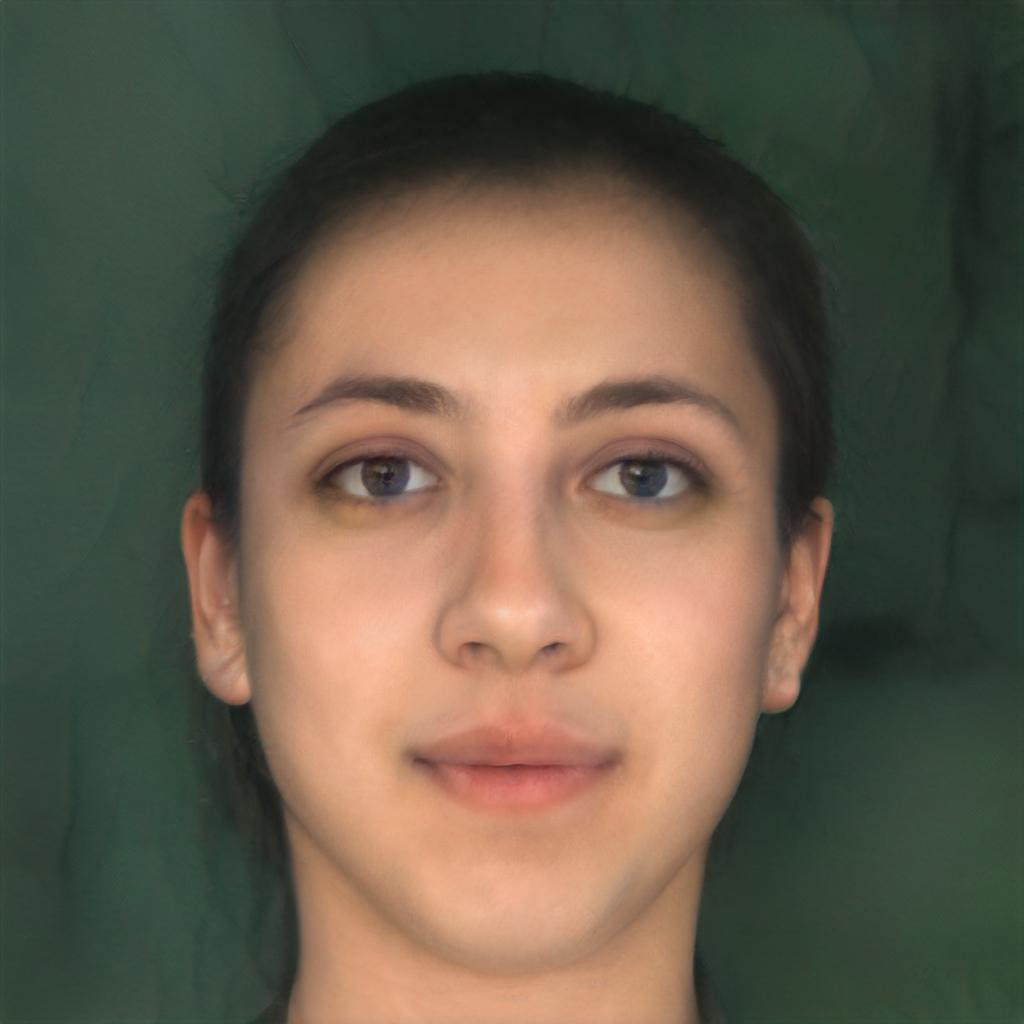}  & 
 \includegraphics[width=0.22\linewidth]{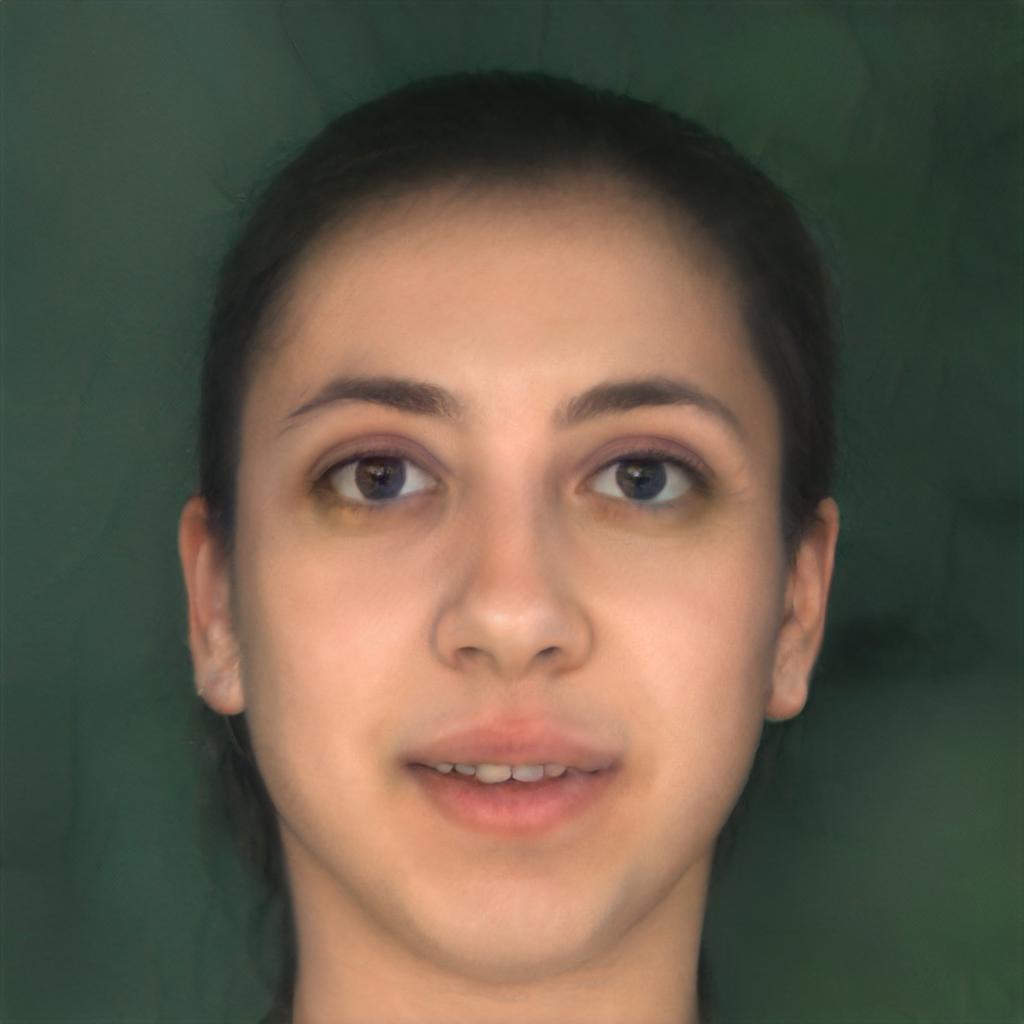}  & 
 \includegraphics[width=0.22\linewidth]{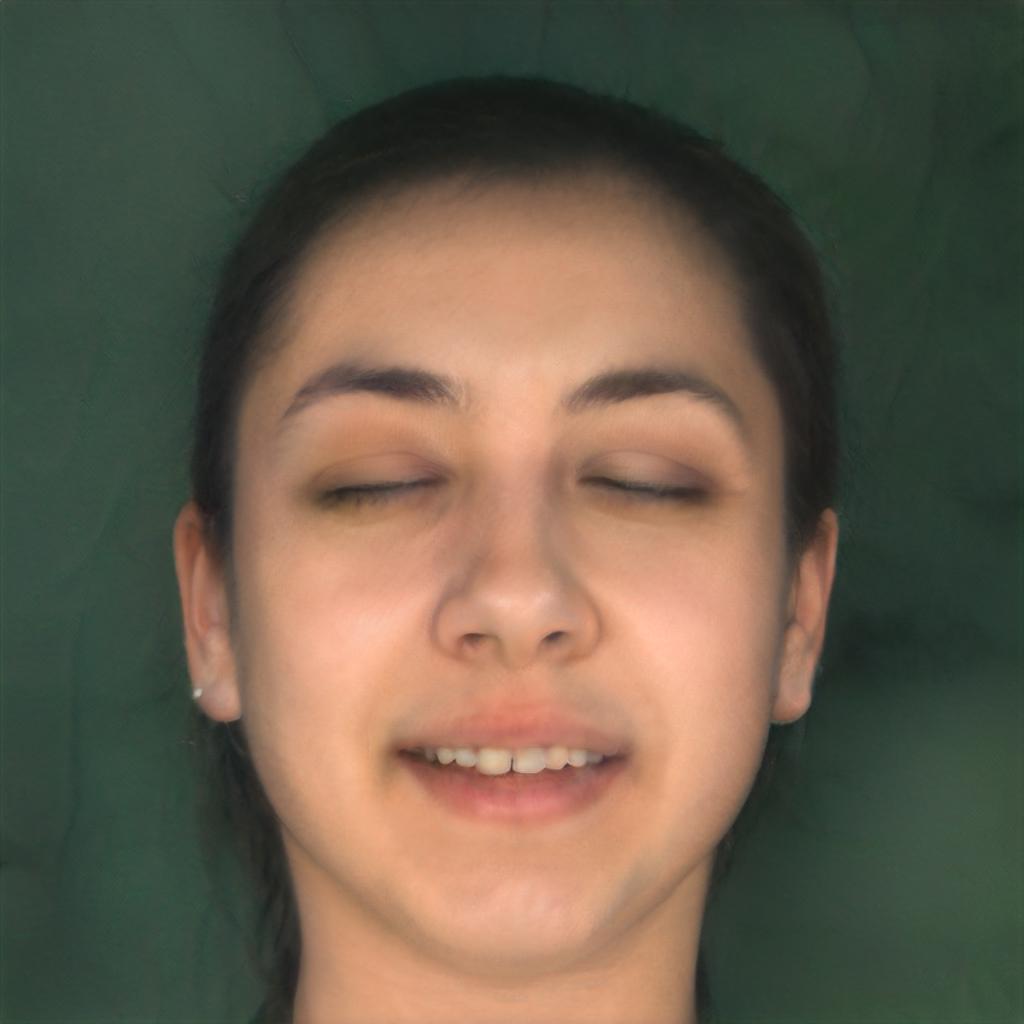}  \\

\begin{turn}{90}\hspace{0.5cm} SGANC IC \end{turn} &
\begin{turn}{90}\hspace{0.5cm} BPP=0.00204 \end{turn} &
 \includegraphics[width=0.22\linewidth]{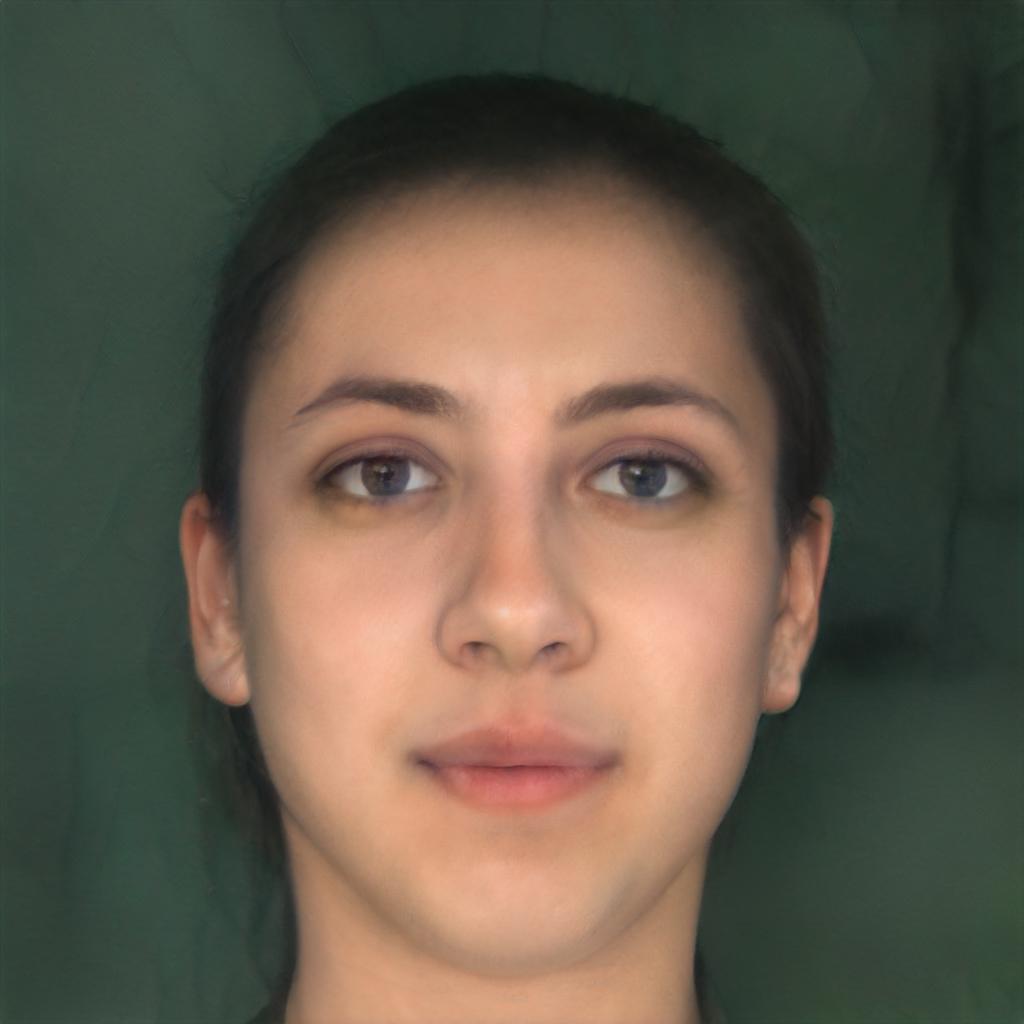}  & 
 \includegraphics[width=0.22\linewidth]{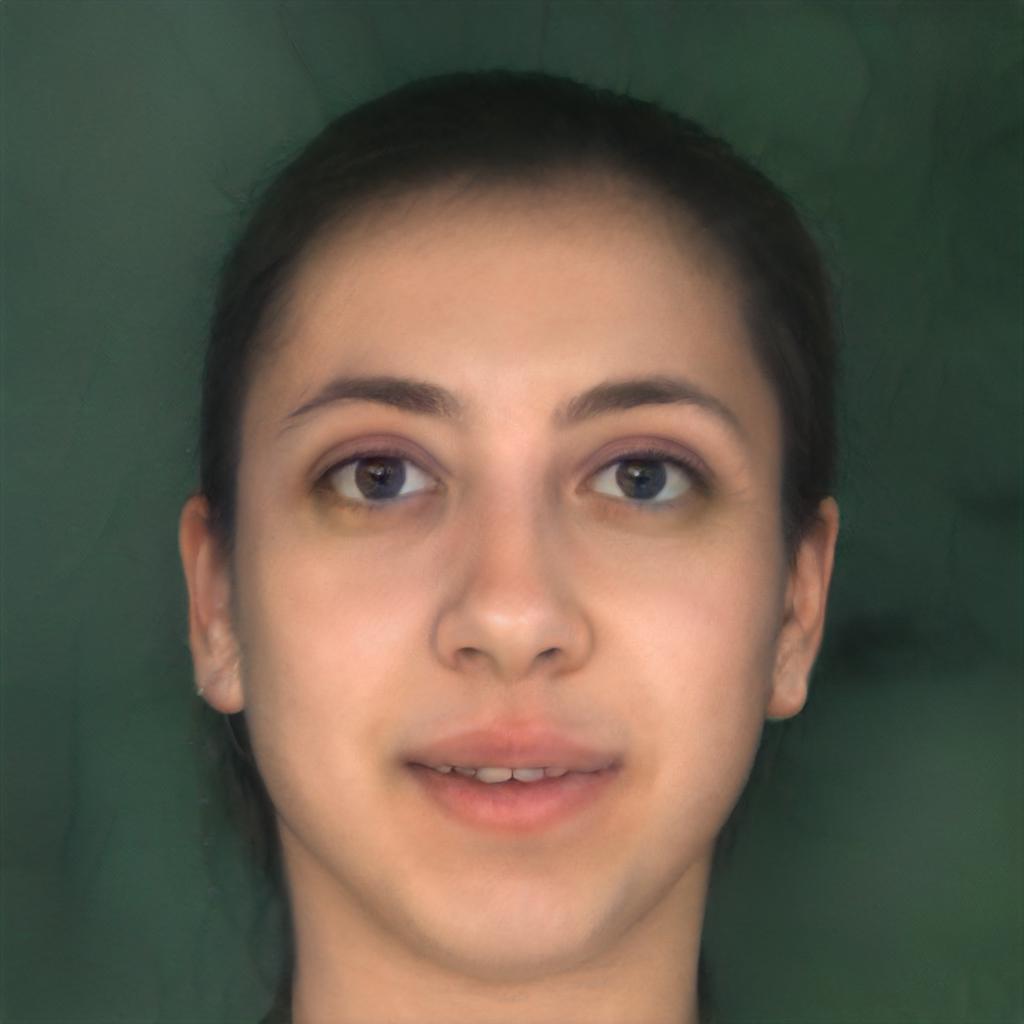}  & 
 \includegraphics[width=0.22\linewidth]{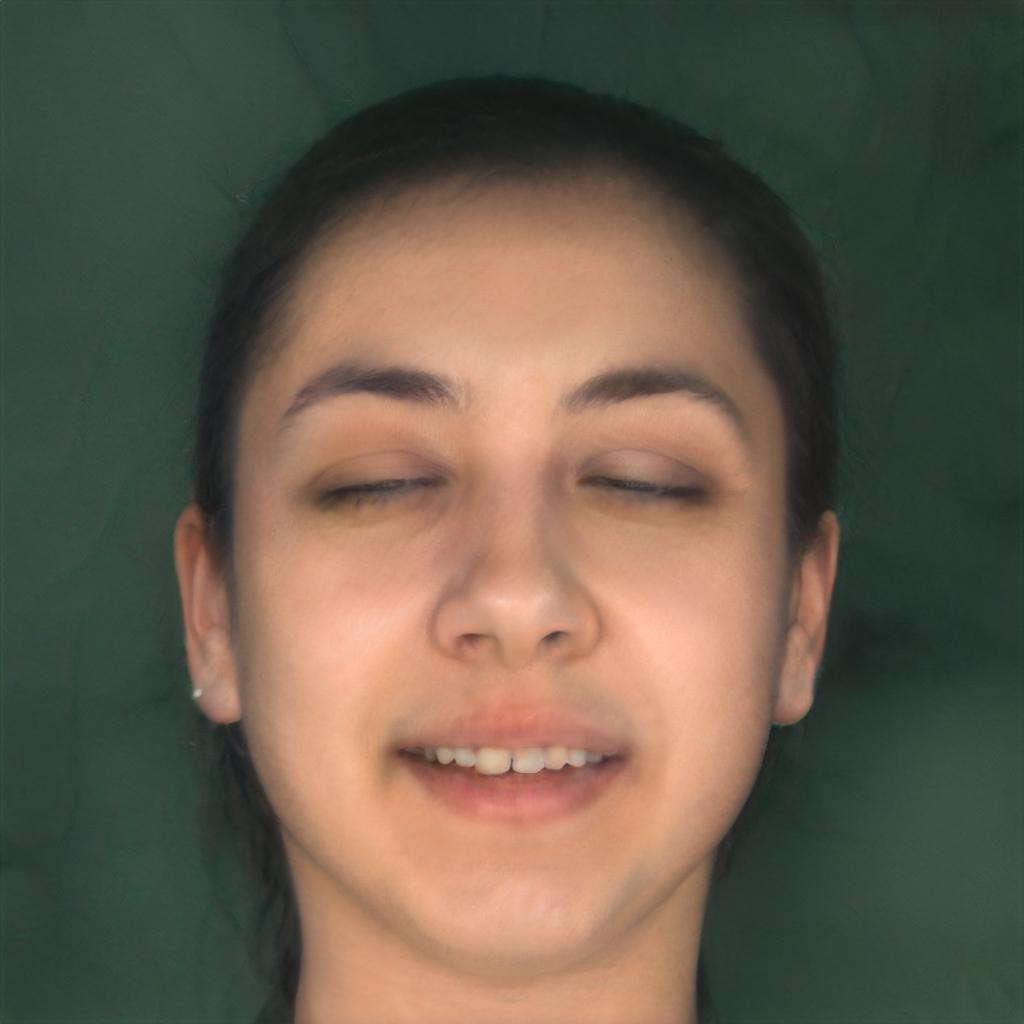}  \\ 
 
 \begin{turn}{90}\hspace{1cm} VTM  \end{turn}&
 \begin{turn}{90}\hspace{0.5cm} BPP=0.00184\end{turn}&
 \includegraphics[width=0.22\linewidth]{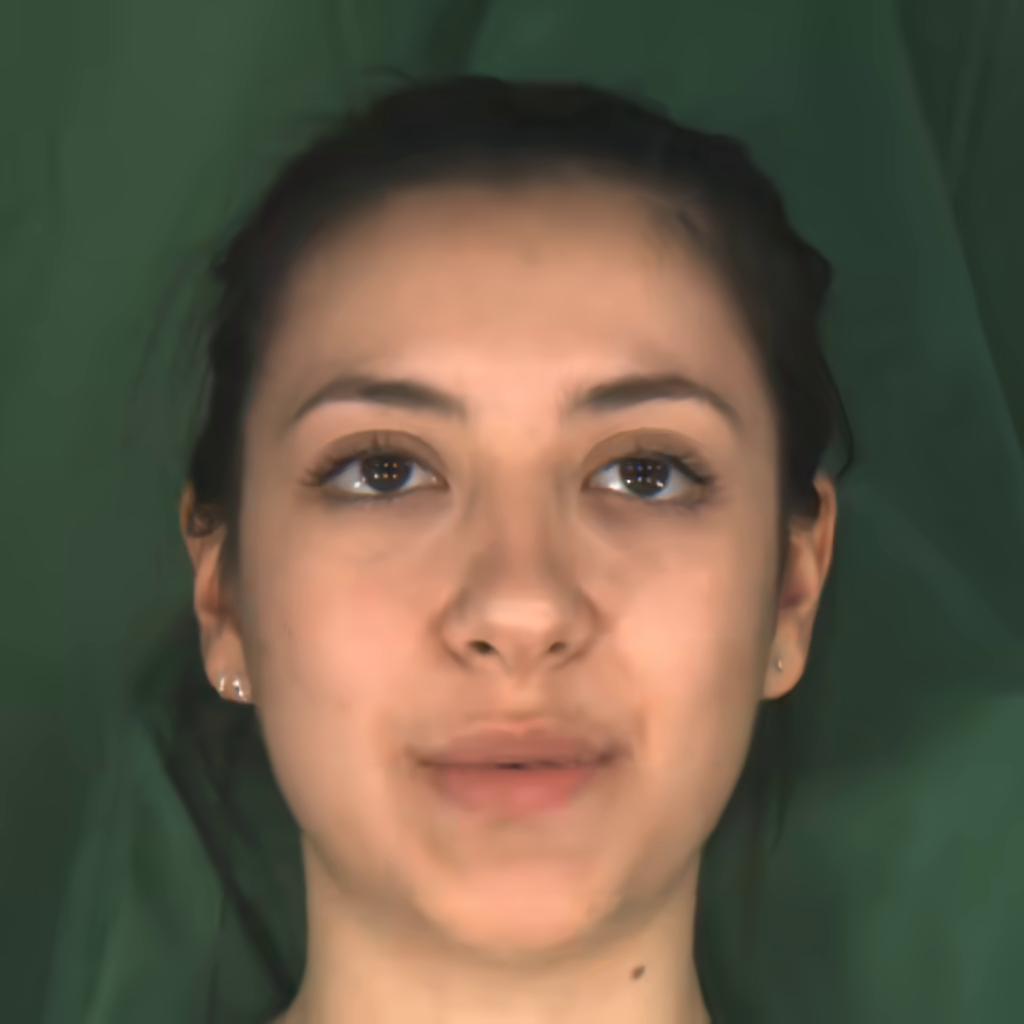}  & 
 \includegraphics[width=0.22\linewidth]{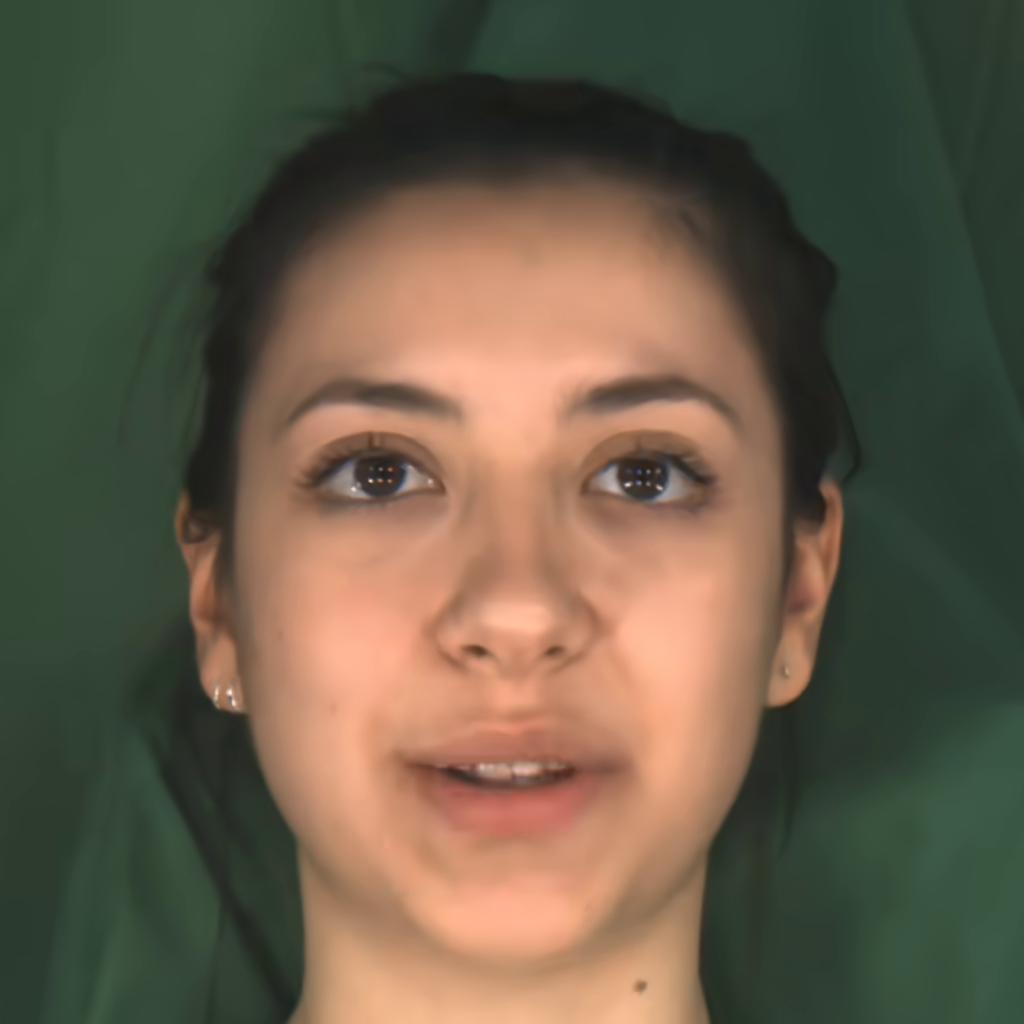}  & 
 \includegraphics[width=0.22\linewidth]{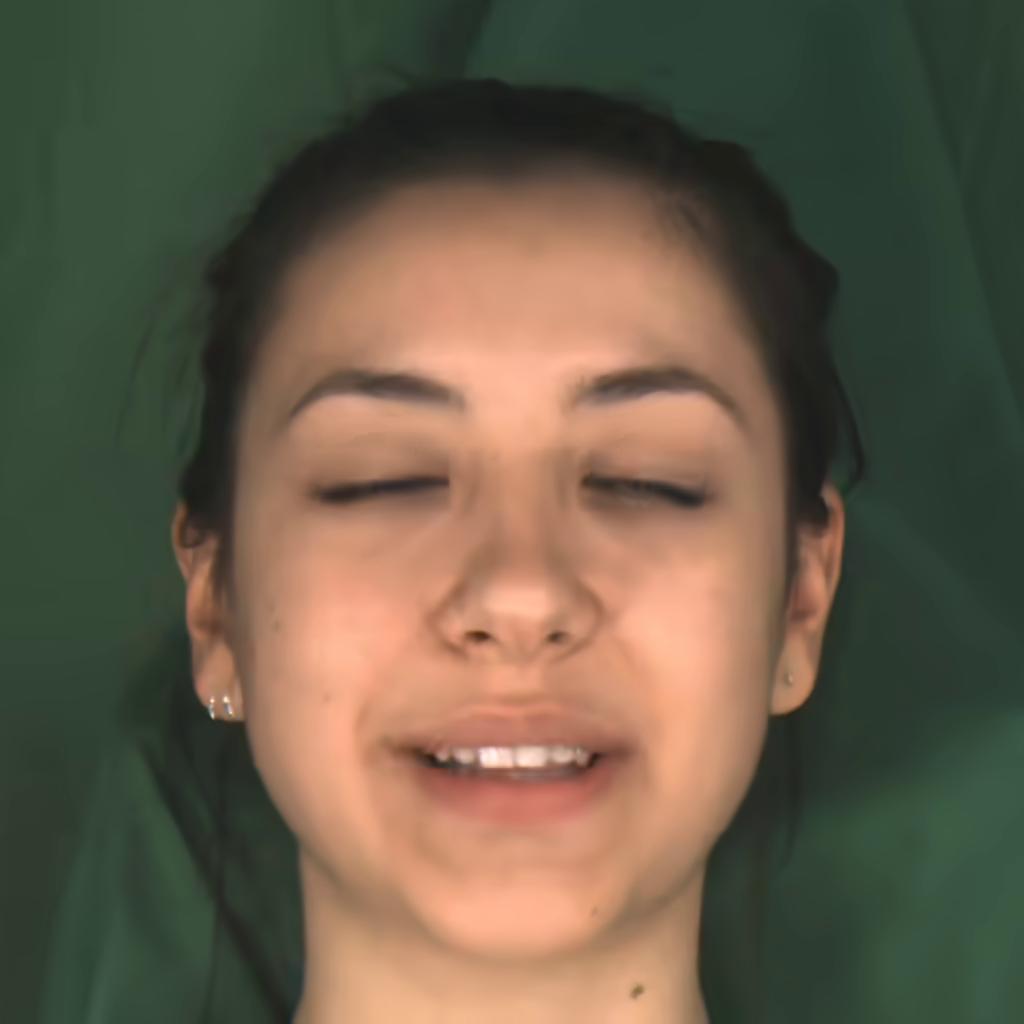}  \\ 

 \begin{turn}{90}\hspace{1cm} H.265 \end{turn} &
 \begin{turn}{90}\hspace{0.5cm} BPP=0.00205\end{turn}&
 \includegraphics[width=0.22\linewidth]{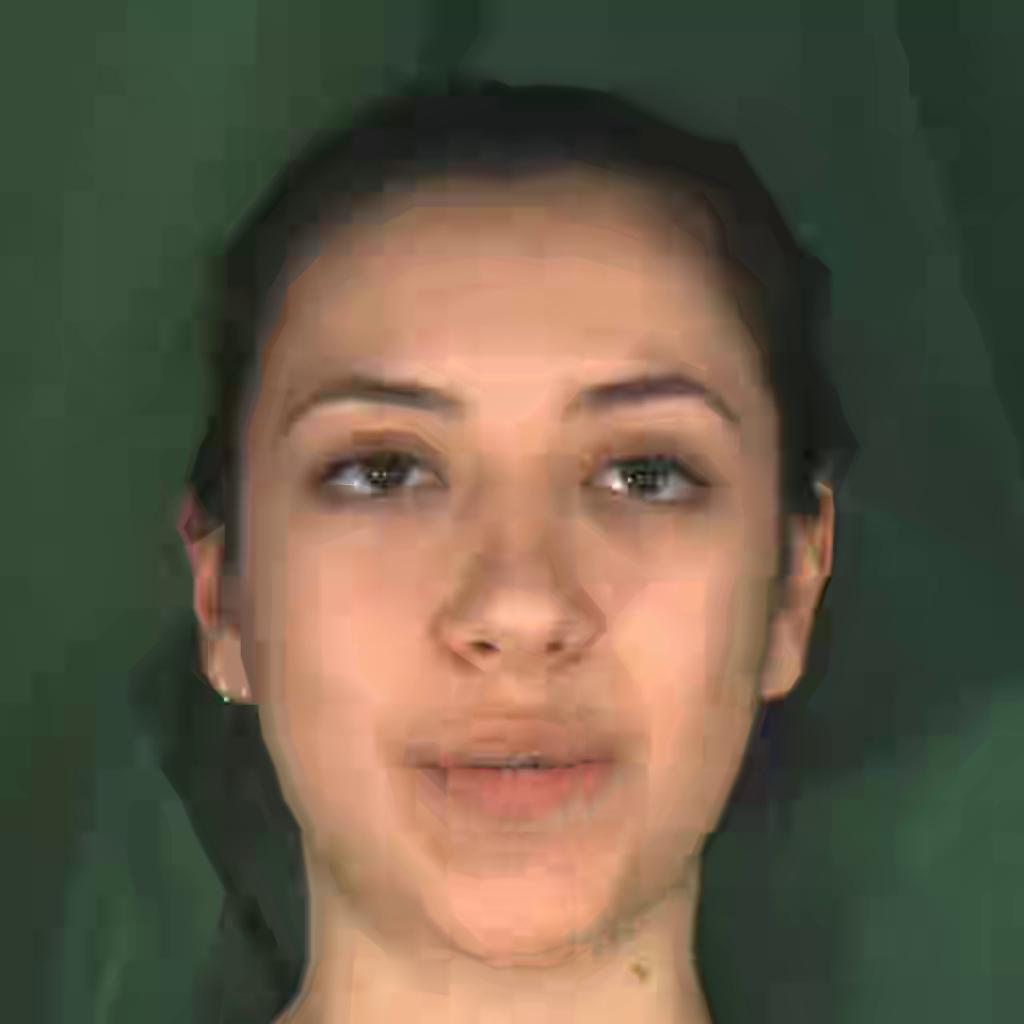}  & 
 \includegraphics[width=0.22\linewidth]{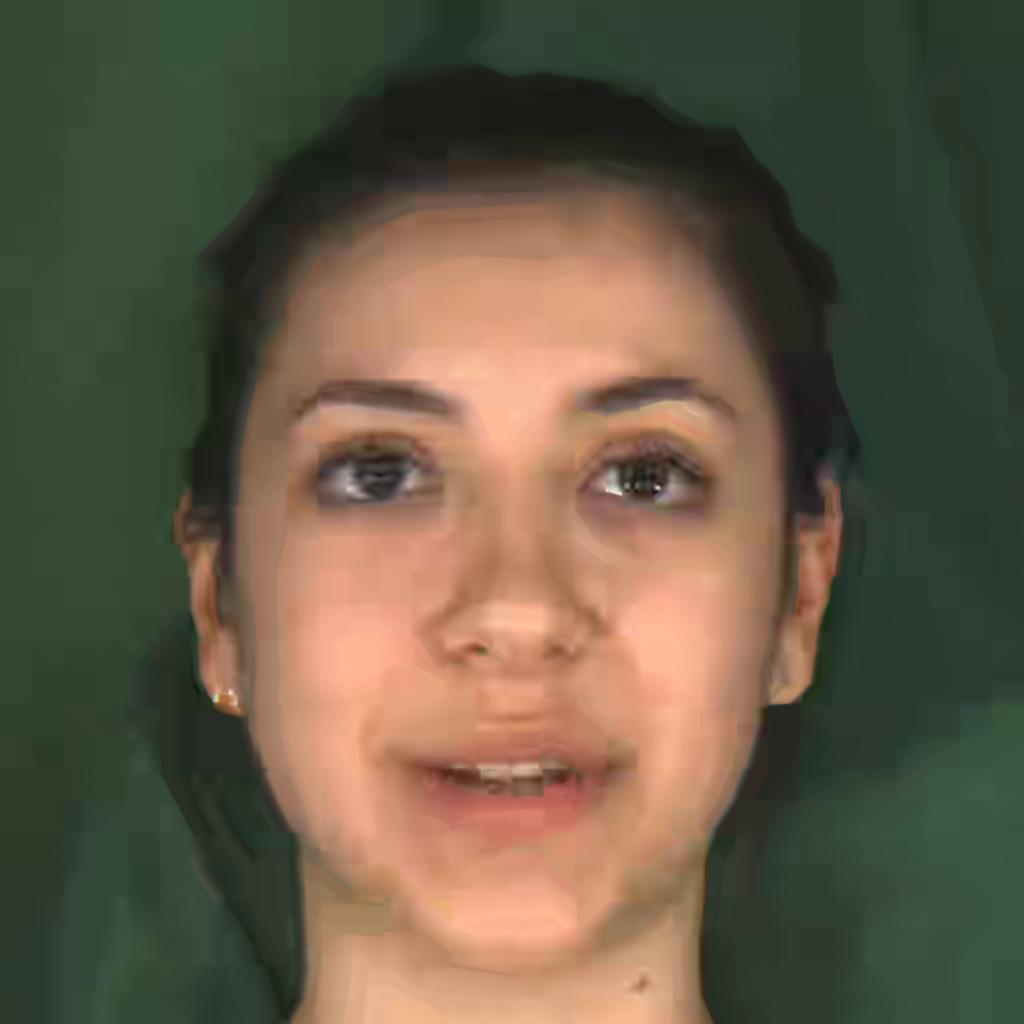}  & 
 \includegraphics[width=0.22\linewidth]{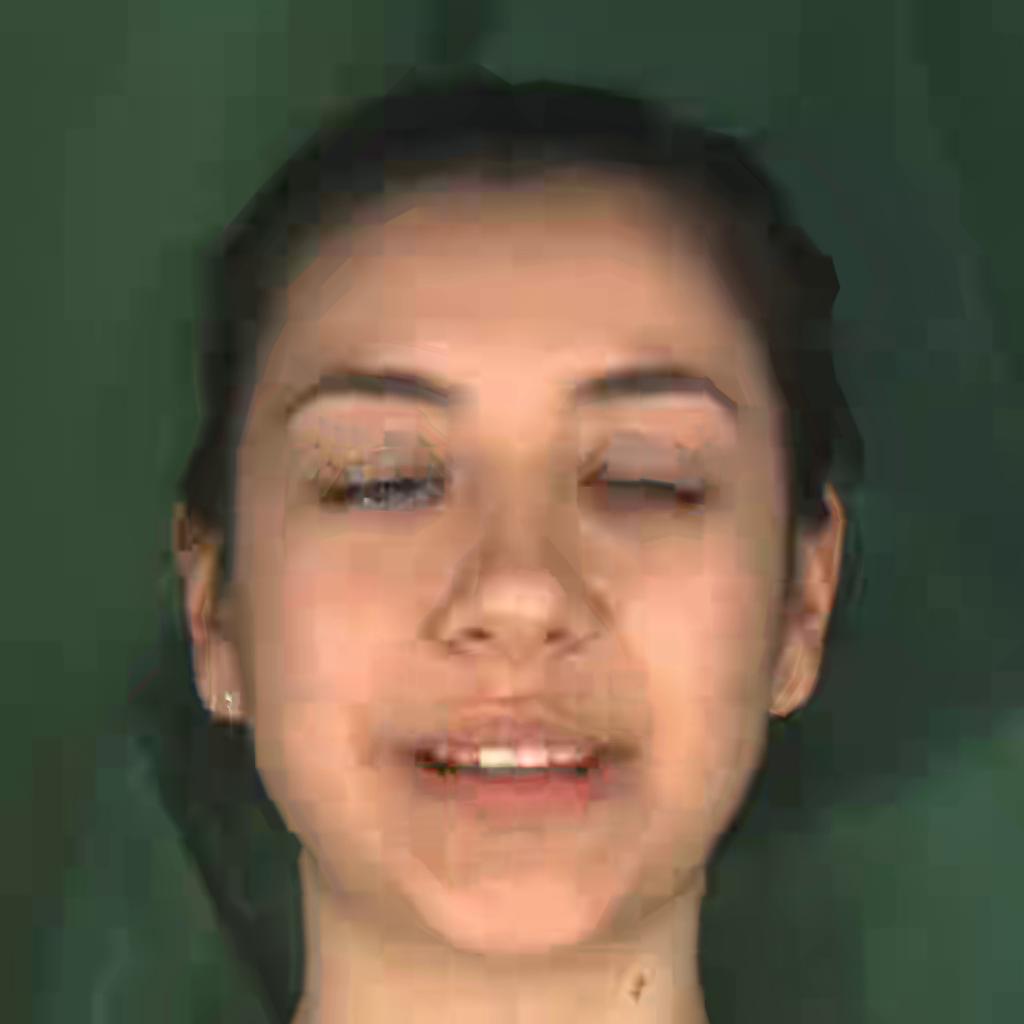}  \\ 
 \\
\bottomrule
\end{tabular}
\caption{Qualitative results on MEAD inter dataset for extreme BPP, computed for three video frames (columns). H.265 shows blocking artifacts and blurring, VTM shows blurring especially at the edges of the face and the hair while our method (SGANC) is almost artifacts free and with high quality images.}
\label{fig:qual_results_low_mead_inter_2}
\end{figure*}

\begin{figure*}[h]
\setlength\tabcolsep{2pt}%
\centering
\begin{tabular}{p{0.5cm}cccc}
\toprule
 \begin{turn}{90} \hspace{1.2cm} Original \end{turn}  &
 &
 \includegraphics[width=0.22\linewidth]{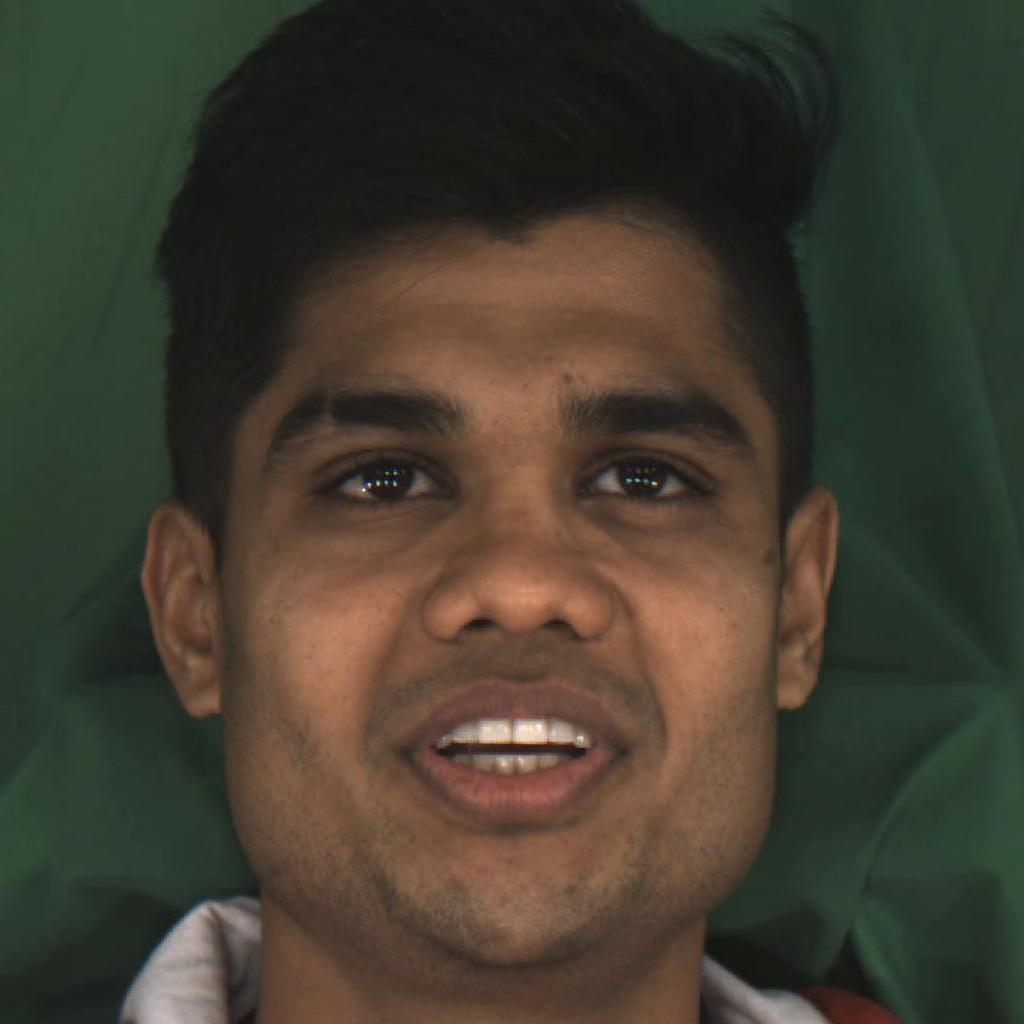}  & 
 \includegraphics[width=0.22\linewidth]{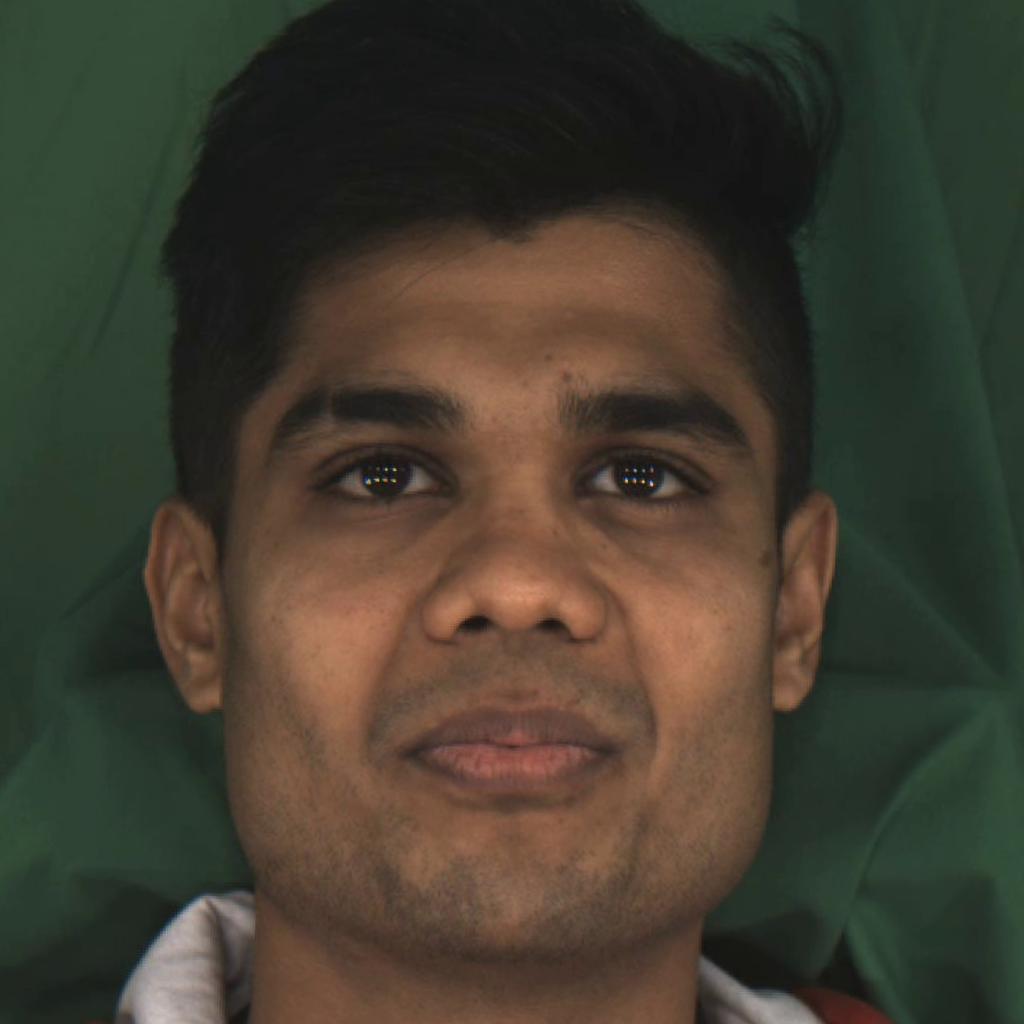}  &
 \includegraphics[width=0.22\linewidth]{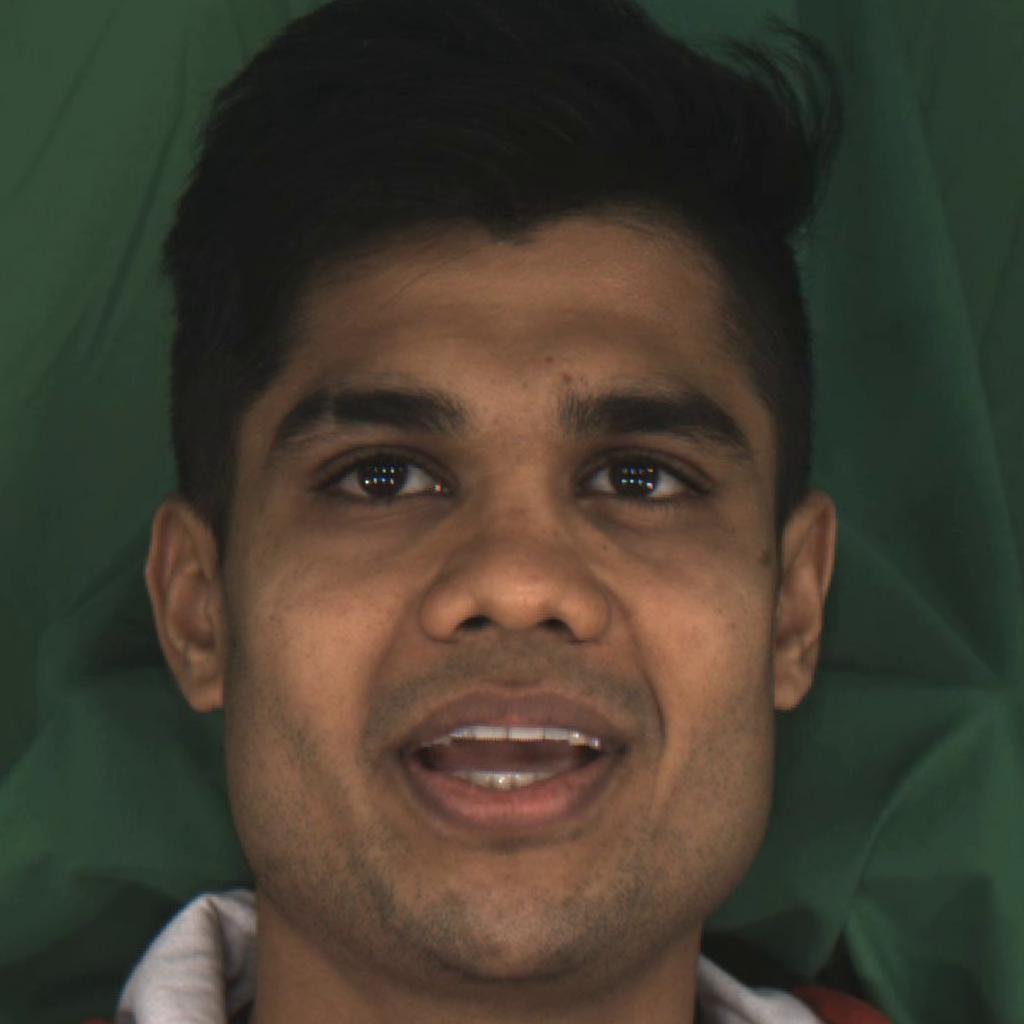}  \\

 \begin{turn}{90}\hspace{1.2cm} Projected\end{turn} &
 &
 \includegraphics[width=0.22\linewidth]{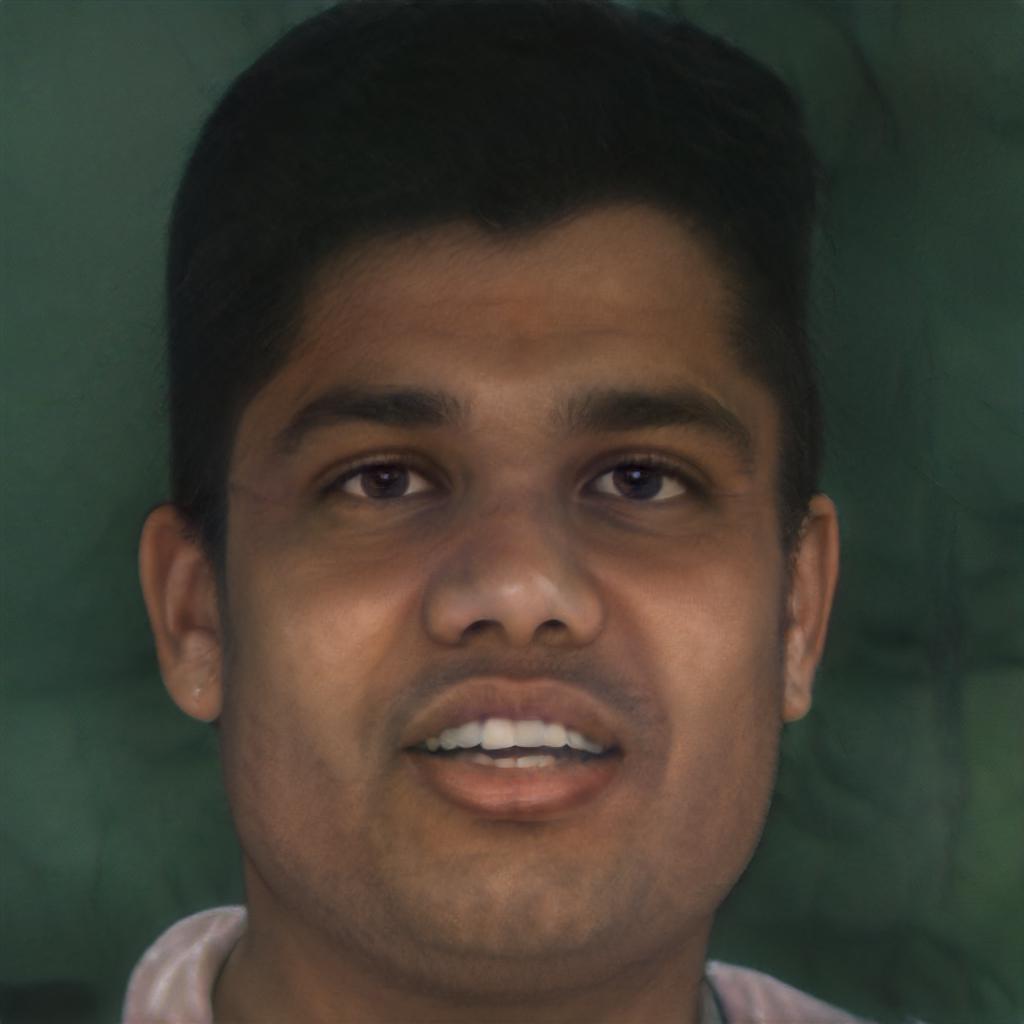}  & 
 \includegraphics[width=0.22\linewidth]{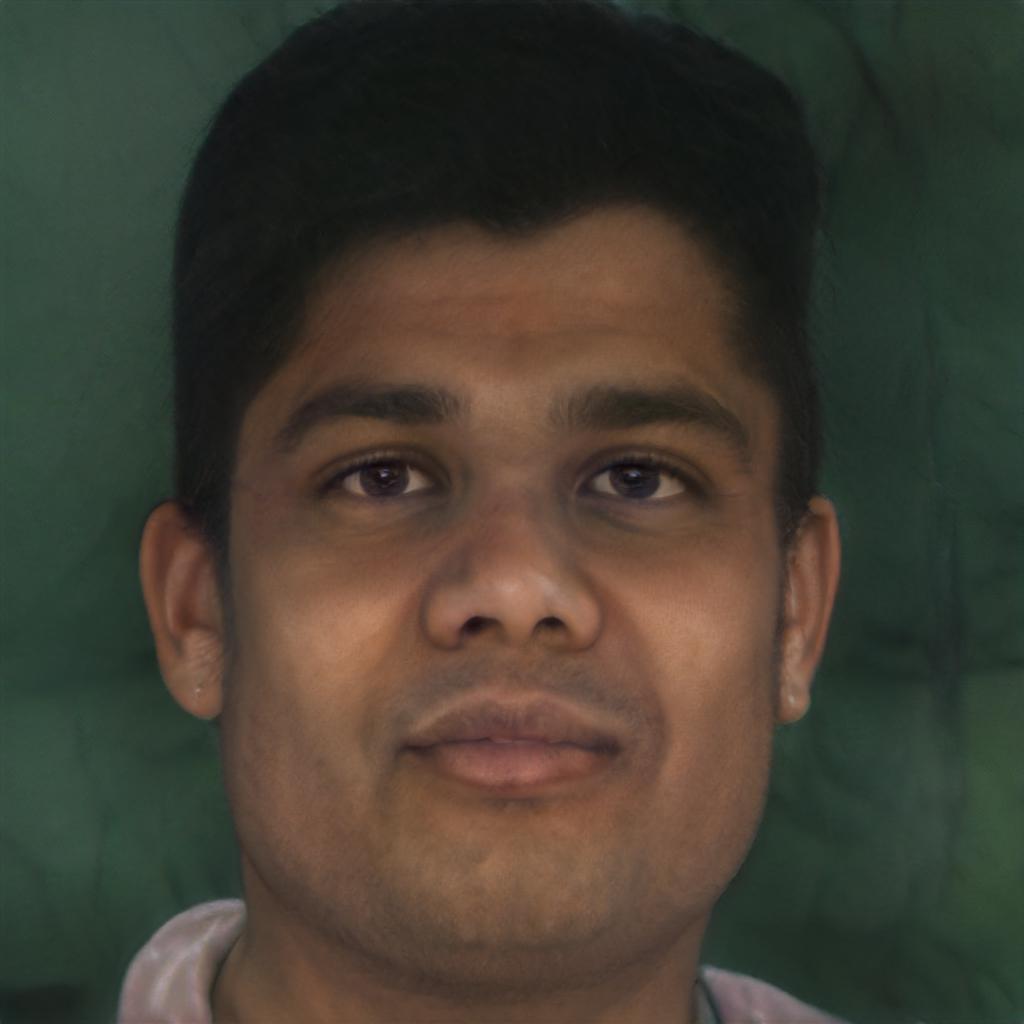}  & 
 \includegraphics[width=0.22\linewidth]{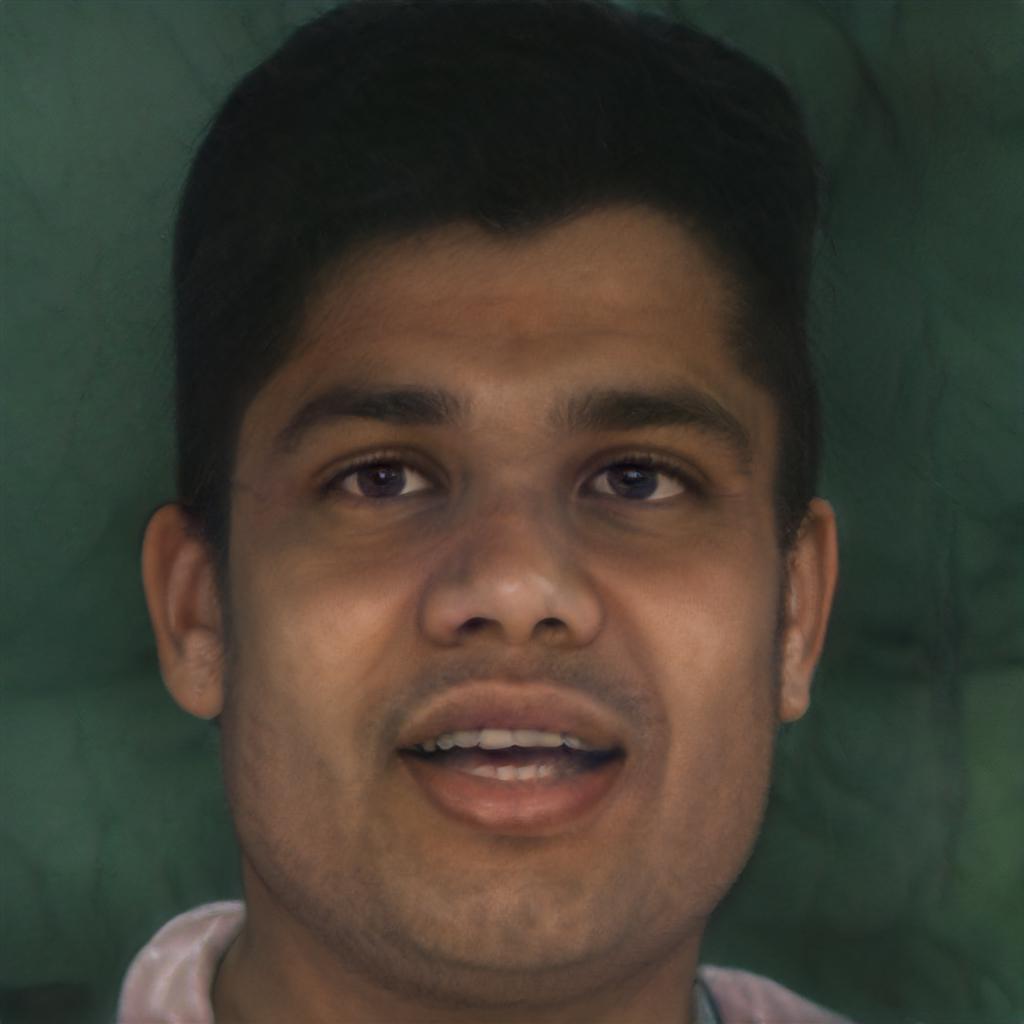}  \\

\begin{turn}{90}\hspace{0.5cm} SGANC IC \end{turn} &
\begin{turn}{90}\hspace{0.5cm} BPP=0.00176 \end{turn} &
 \includegraphics[width=0.22\linewidth]{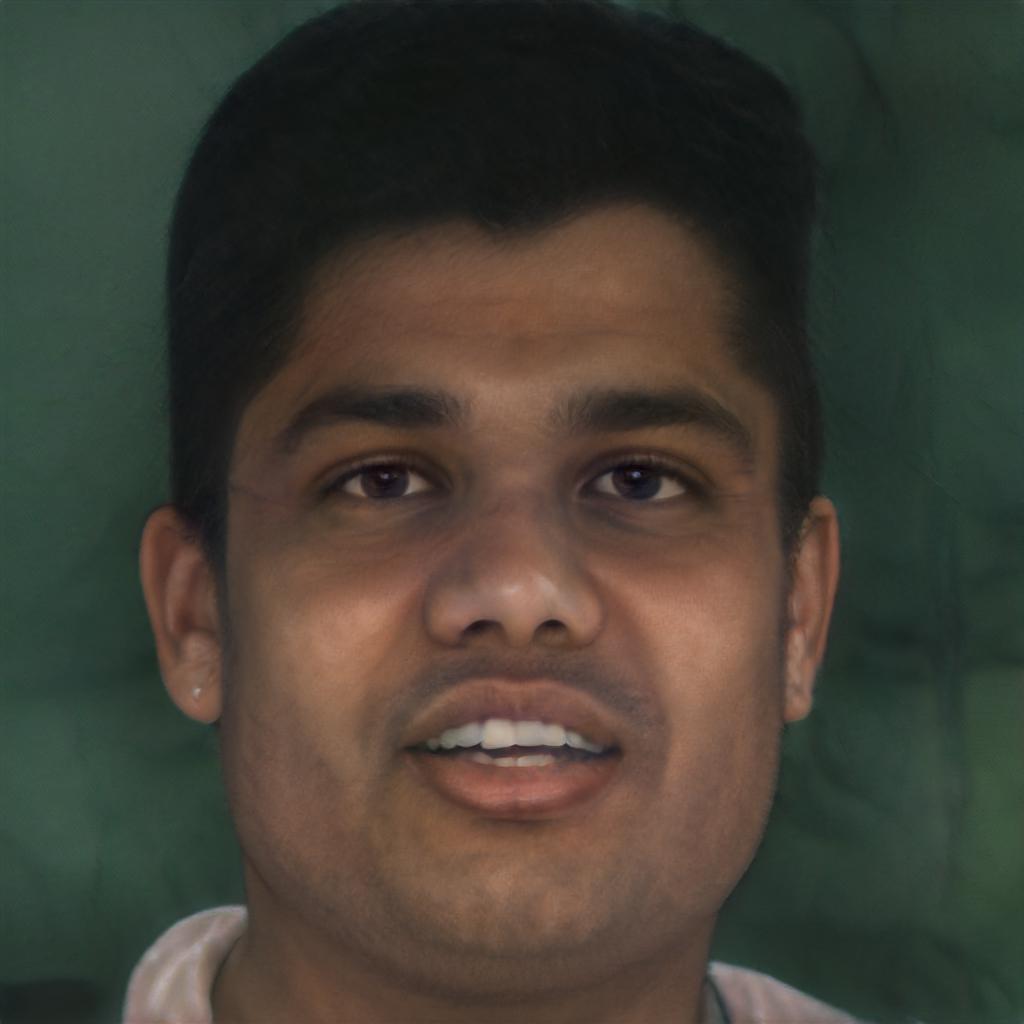}  & 
 \includegraphics[width=0.22\linewidth]{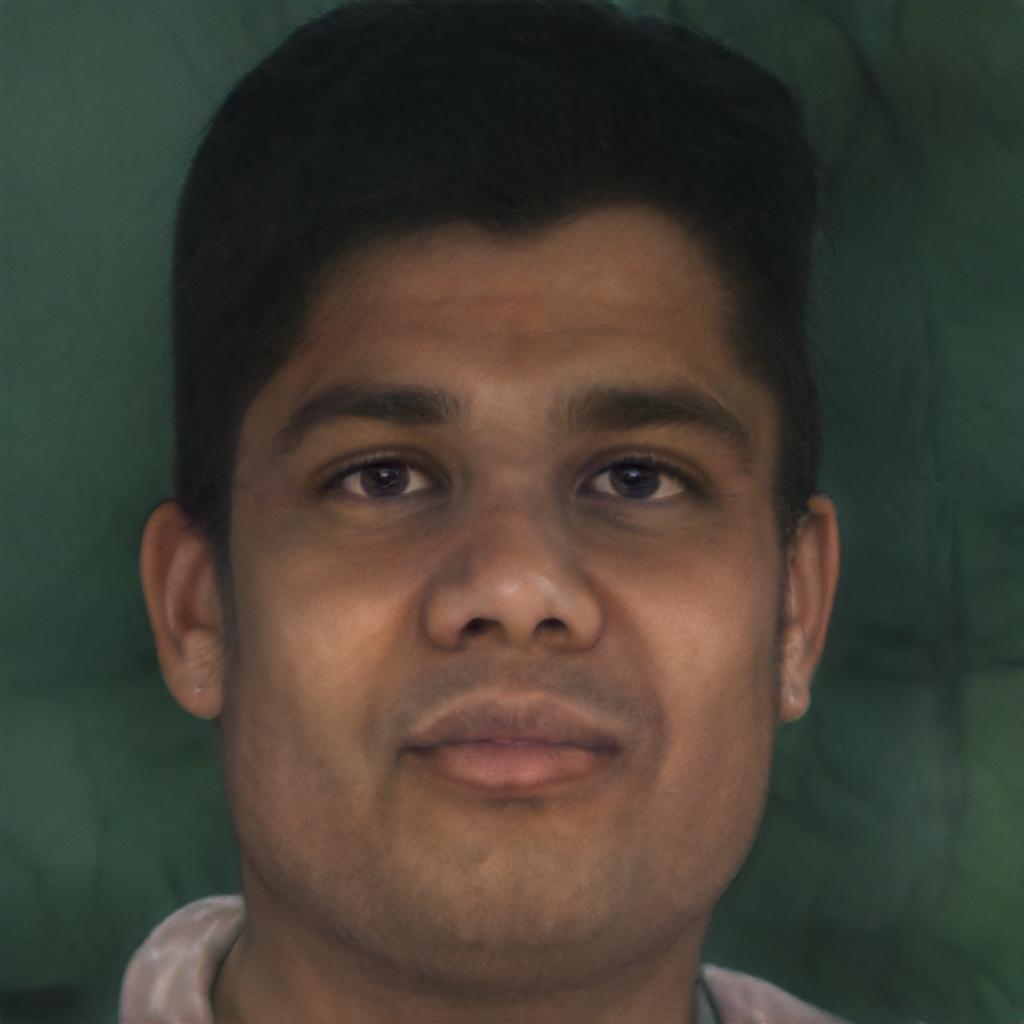}  & 
 \includegraphics[width=0.22\linewidth]{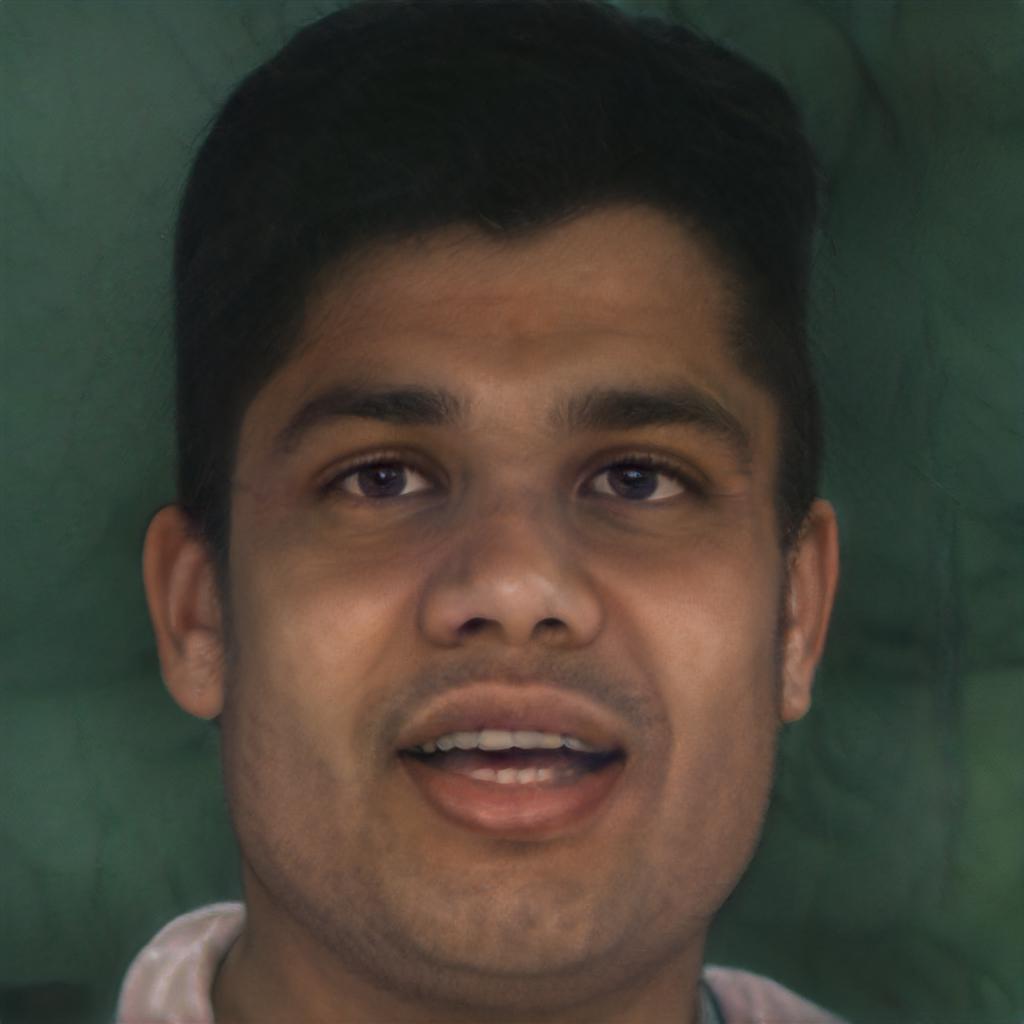}  \\ 
 
 \begin{turn}{90}\hspace{1cm} VTM  \end{turn}&
 \begin{turn}{90}\hspace{0.5cm} BPP=0.00190\end{turn}&
 \includegraphics[width=0.22\linewidth]{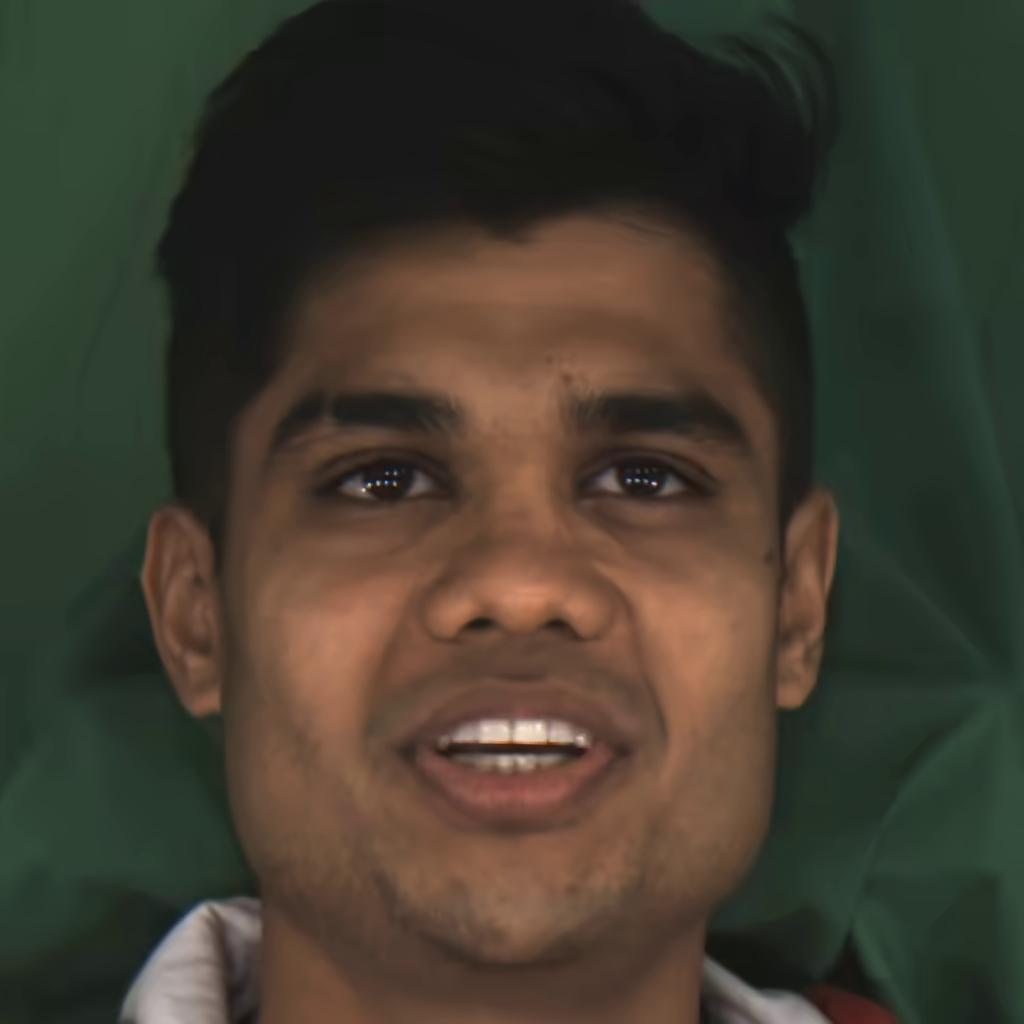}  & 
 \includegraphics[width=0.22\linewidth]{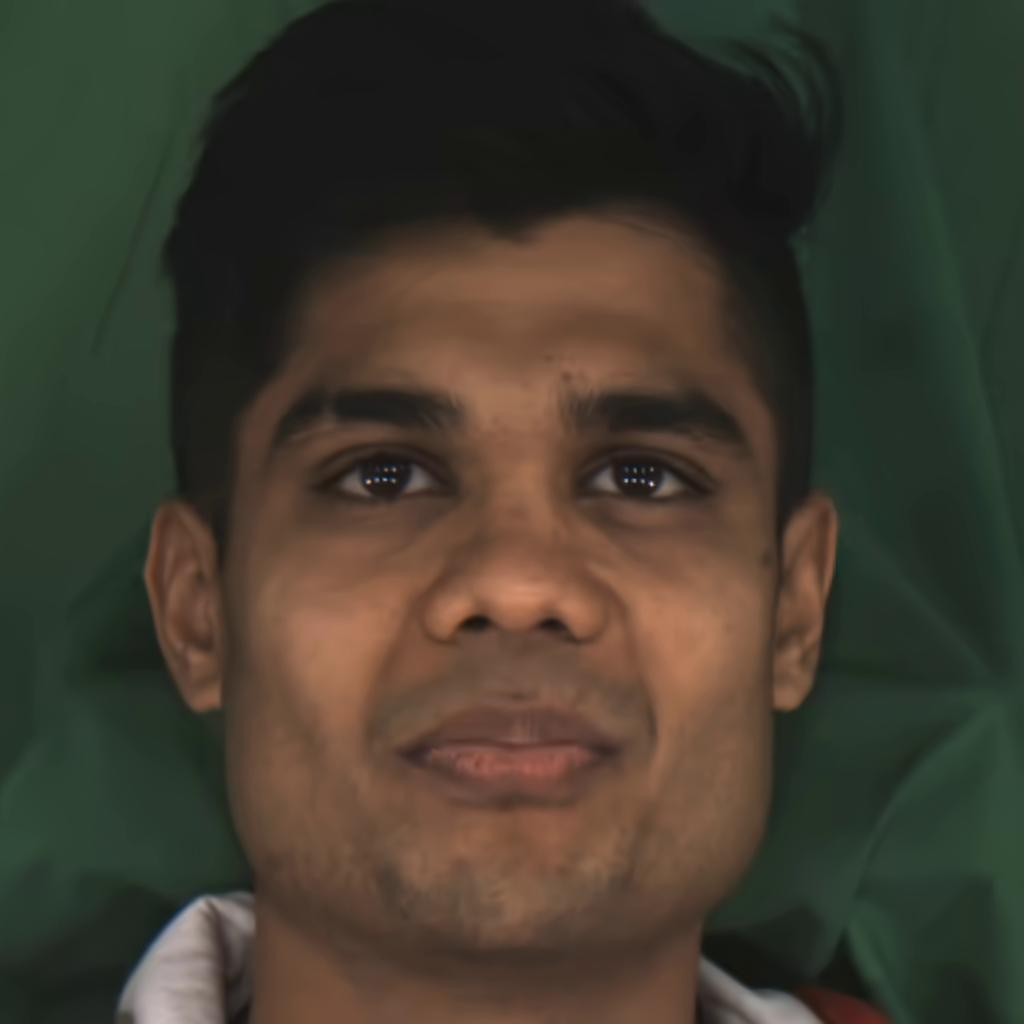}  &
 \includegraphics[width=0.22\linewidth]{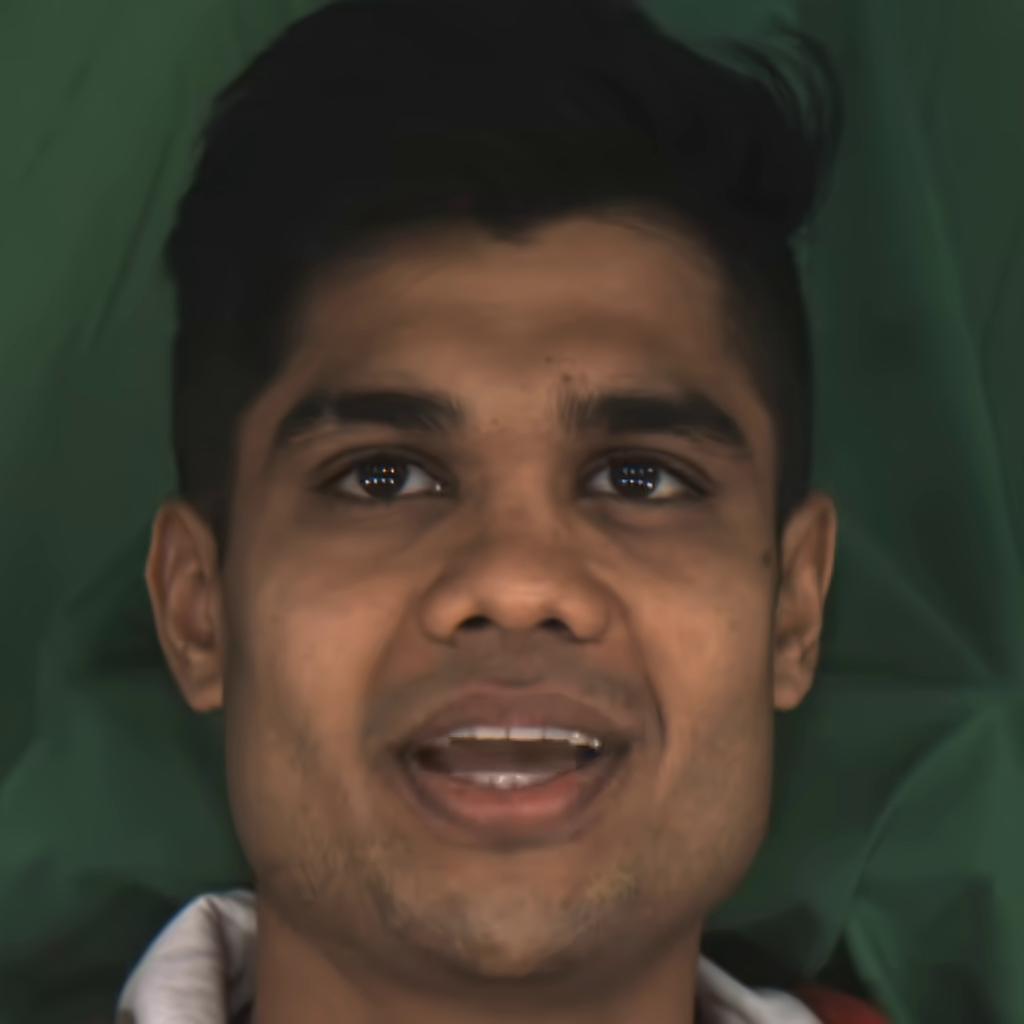}  \\

 \begin{turn}{90}\hspace{1cm} H.265 \end{turn} &
 \begin{turn}{90}\hspace{0.5cm} BPP=0.00199\end{turn} &
 \includegraphics[width=0.22\linewidth]{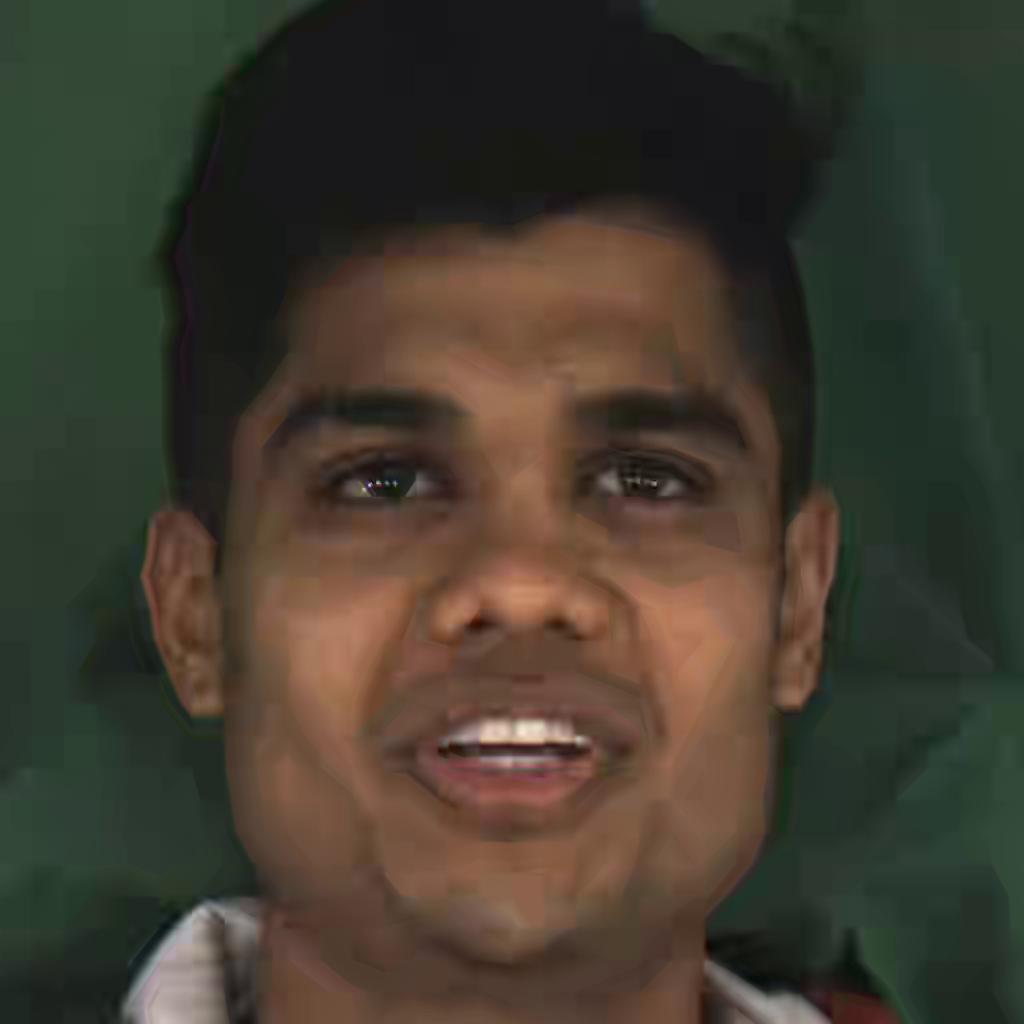}  & 
 \includegraphics[width=0.22\linewidth]{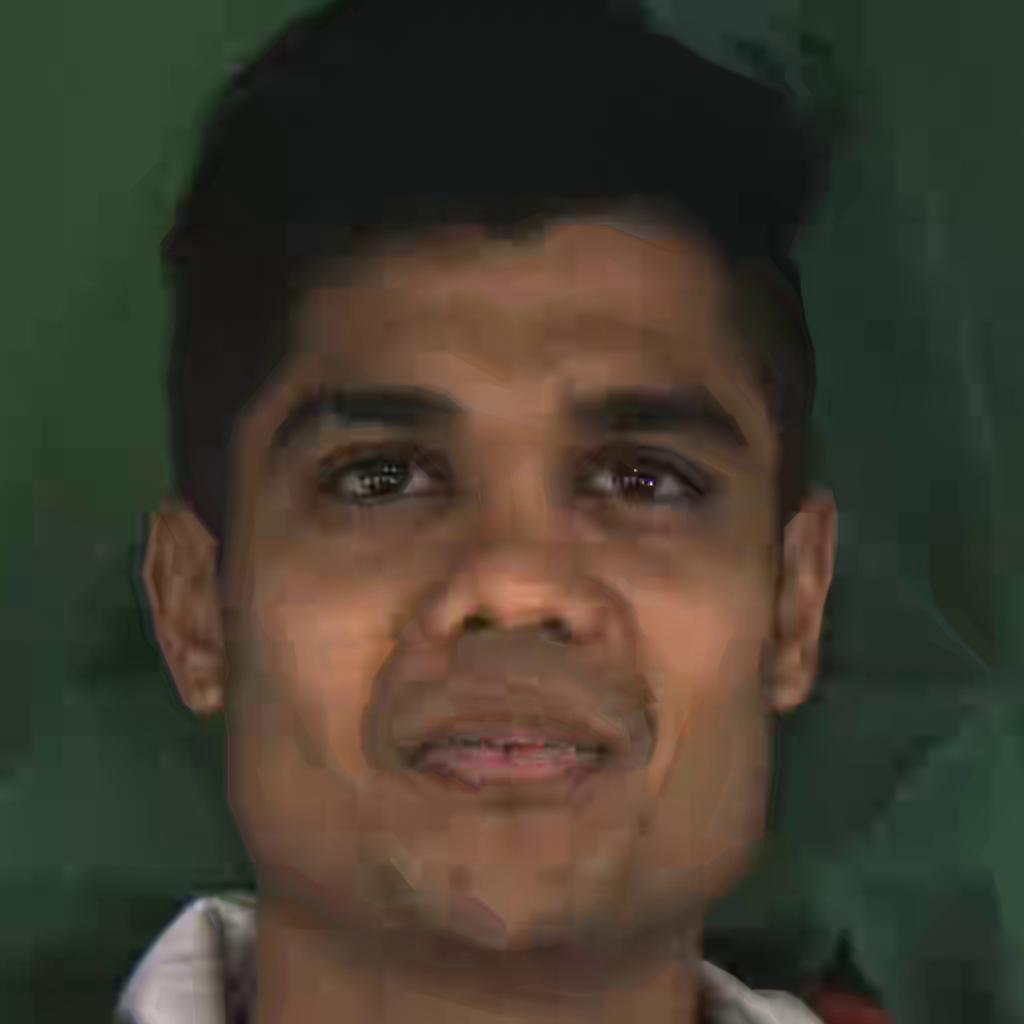}  &
 \includegraphics[width=0.22\linewidth]{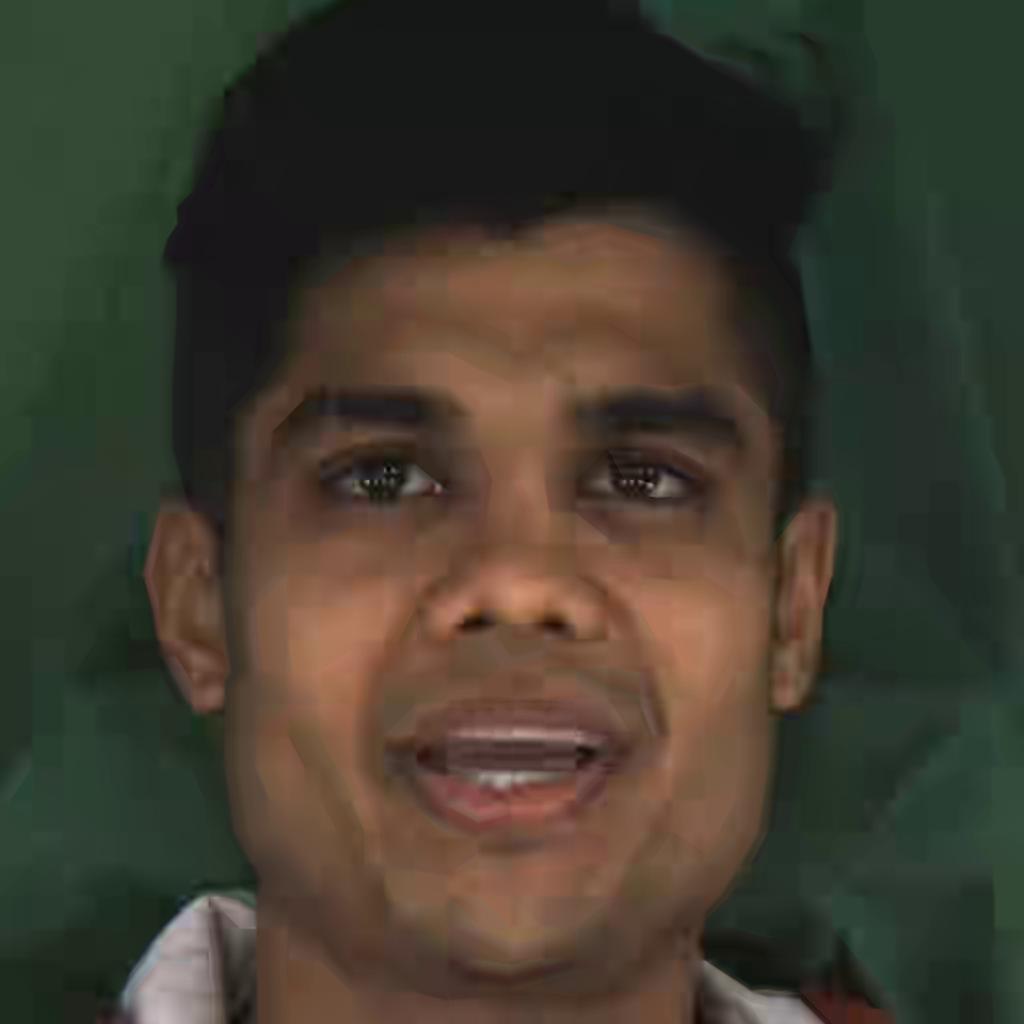}  \\ 
 \\
\bottomrule
\end{tabular}
\caption{Qualitative results on MEAD inter dataset for extreme BPP, computed for three video frames. H.265 shows blocking artifacts and blurring, VTM shows blurring especially at the edges of the face and the hair while our method (SGANC) is almost artifacts free and with high quality images.}
\label{fig:qual_results_low_mead_inter_3}
\end{figure*}

\begin{figure*}[h]
\setlength\tabcolsep{2pt}%
\centering
\begin{tabular}{p{0.5cm}cccc}
\toprule
 \begin{turn}{90} \hspace{1.2cm} Original \end{turn}  &
 &
 \includegraphics[width=0.22\linewidth]{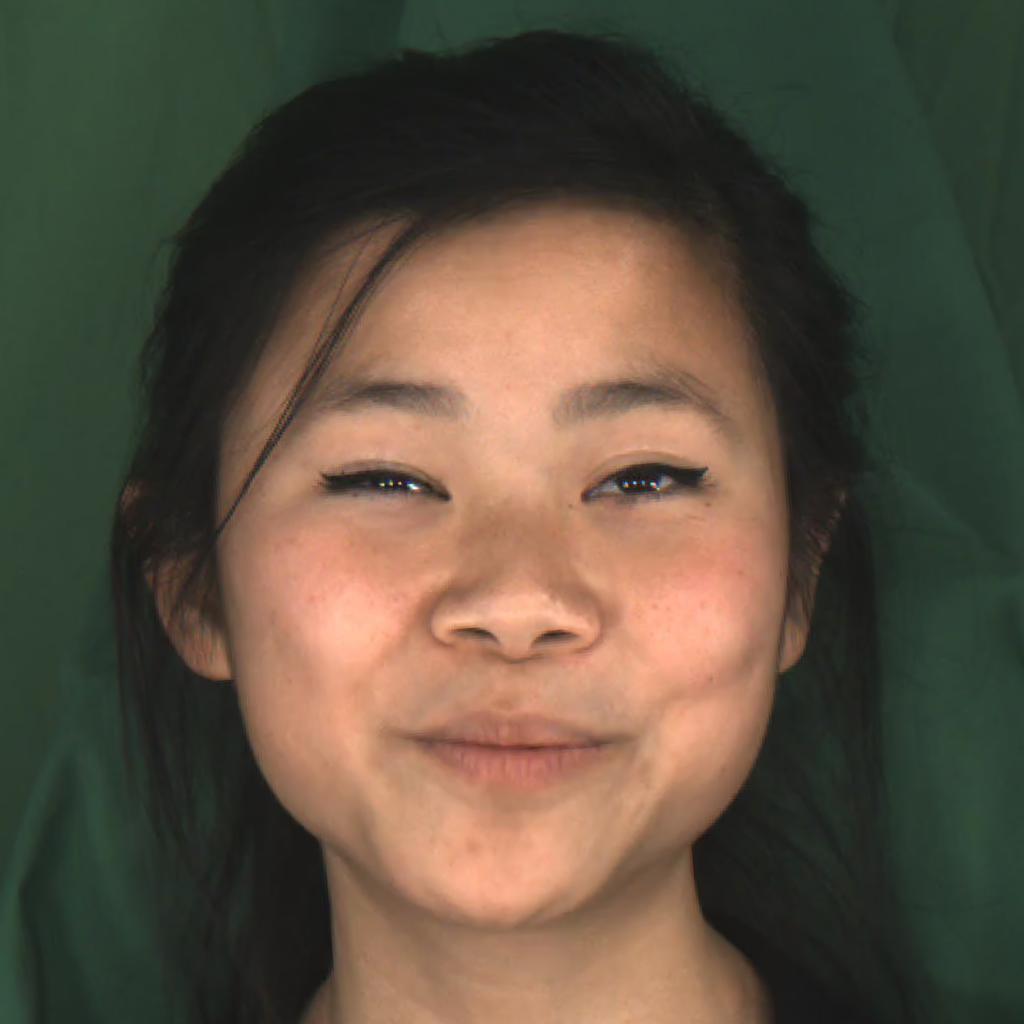}  & 
 \includegraphics[width=0.22\linewidth]{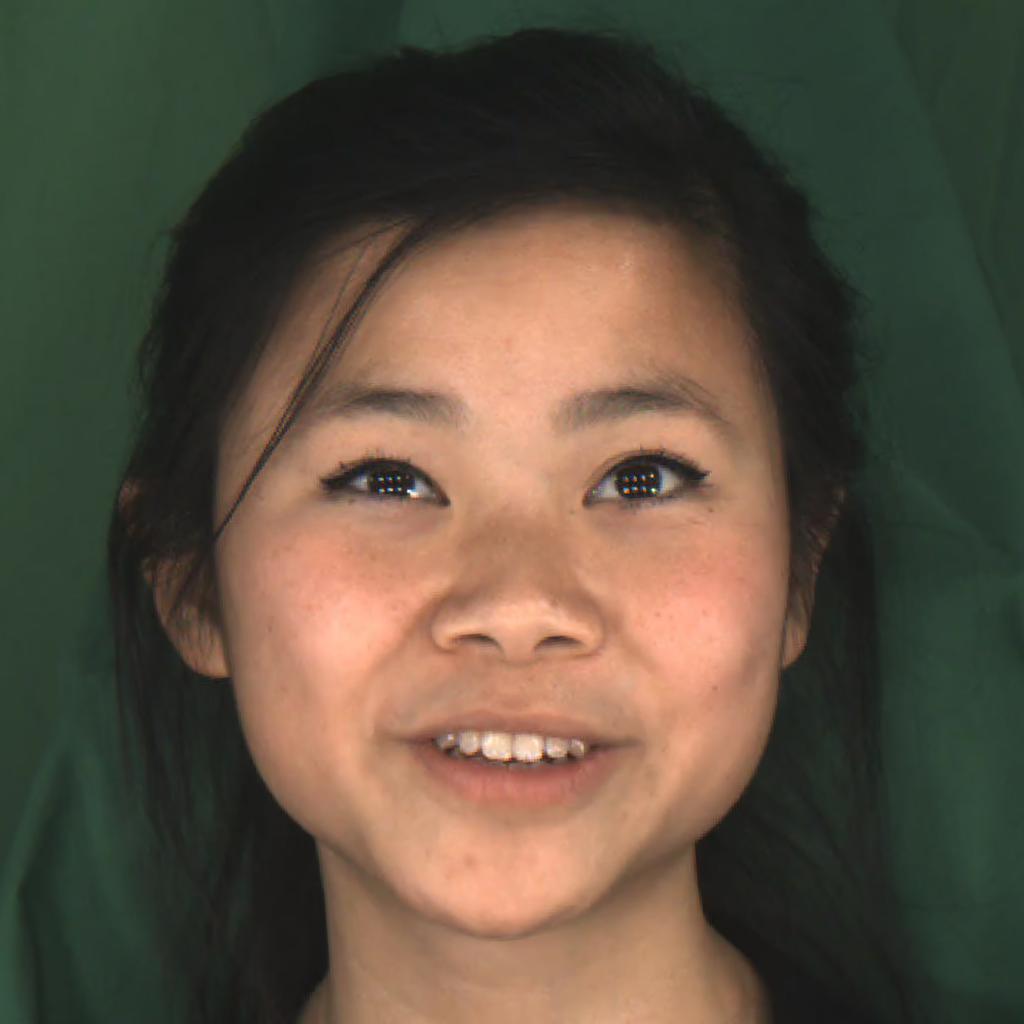}  & 
 \includegraphics[width=0.22\linewidth]{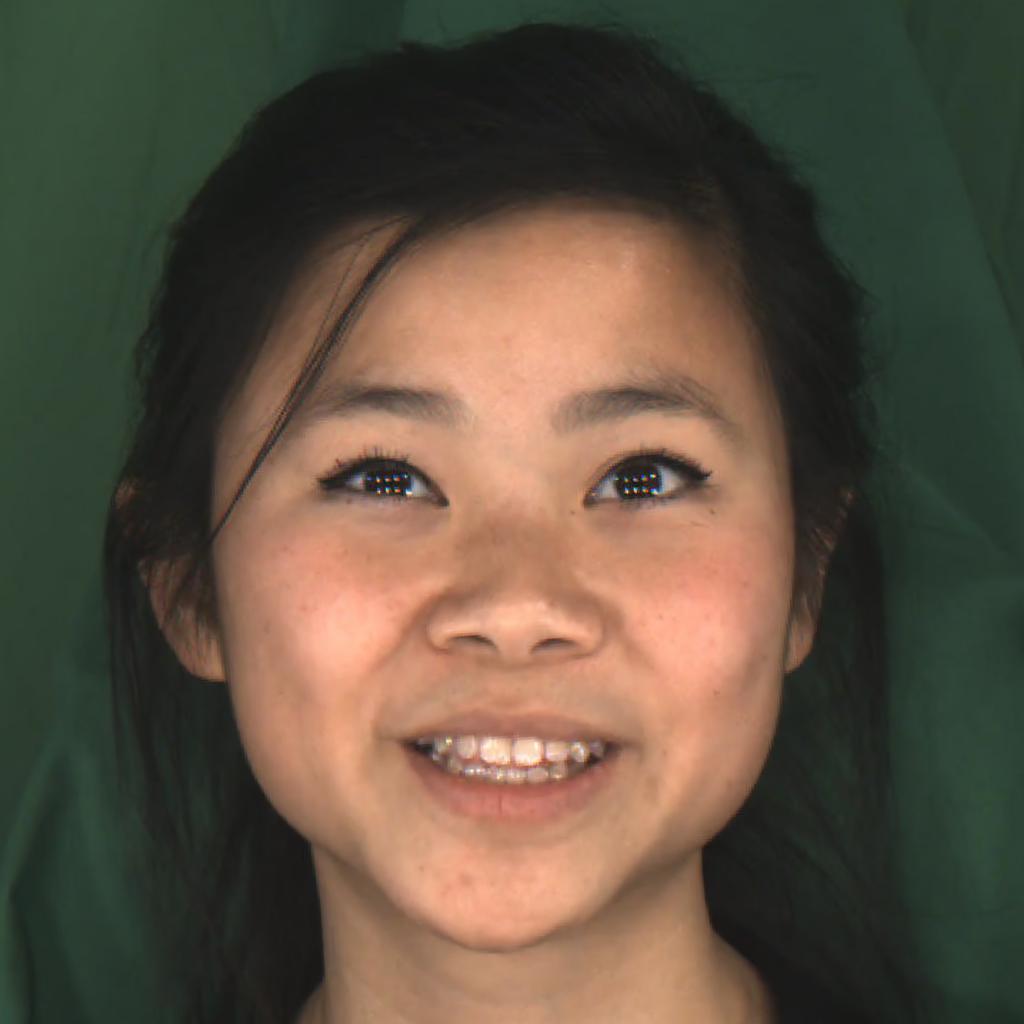}  \\ 

 \begin{turn}{90}\hspace{1.2cm} Projected\end{turn} &
 &
 \includegraphics[width=0.22\linewidth]{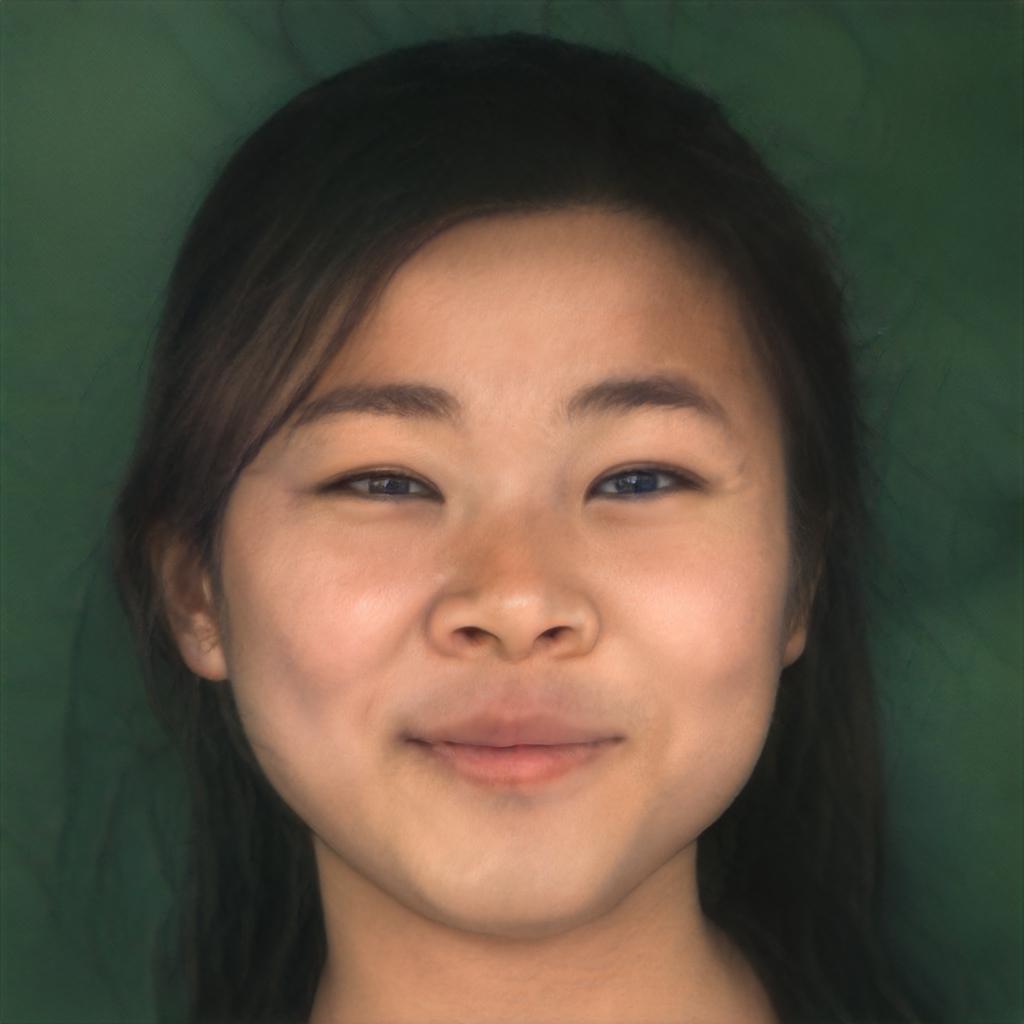}  & 
 \includegraphics[width=0.22\linewidth]{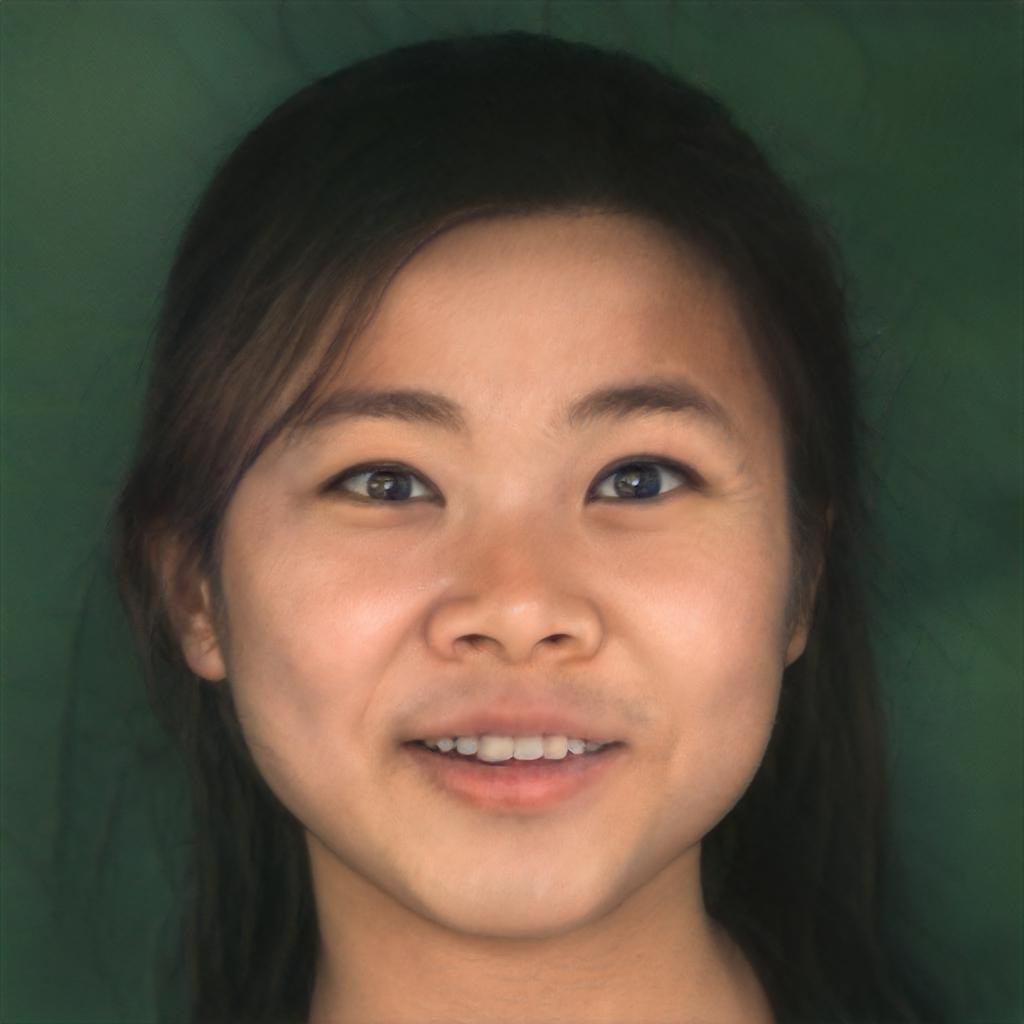}  & 
 \includegraphics[width=0.22\linewidth]{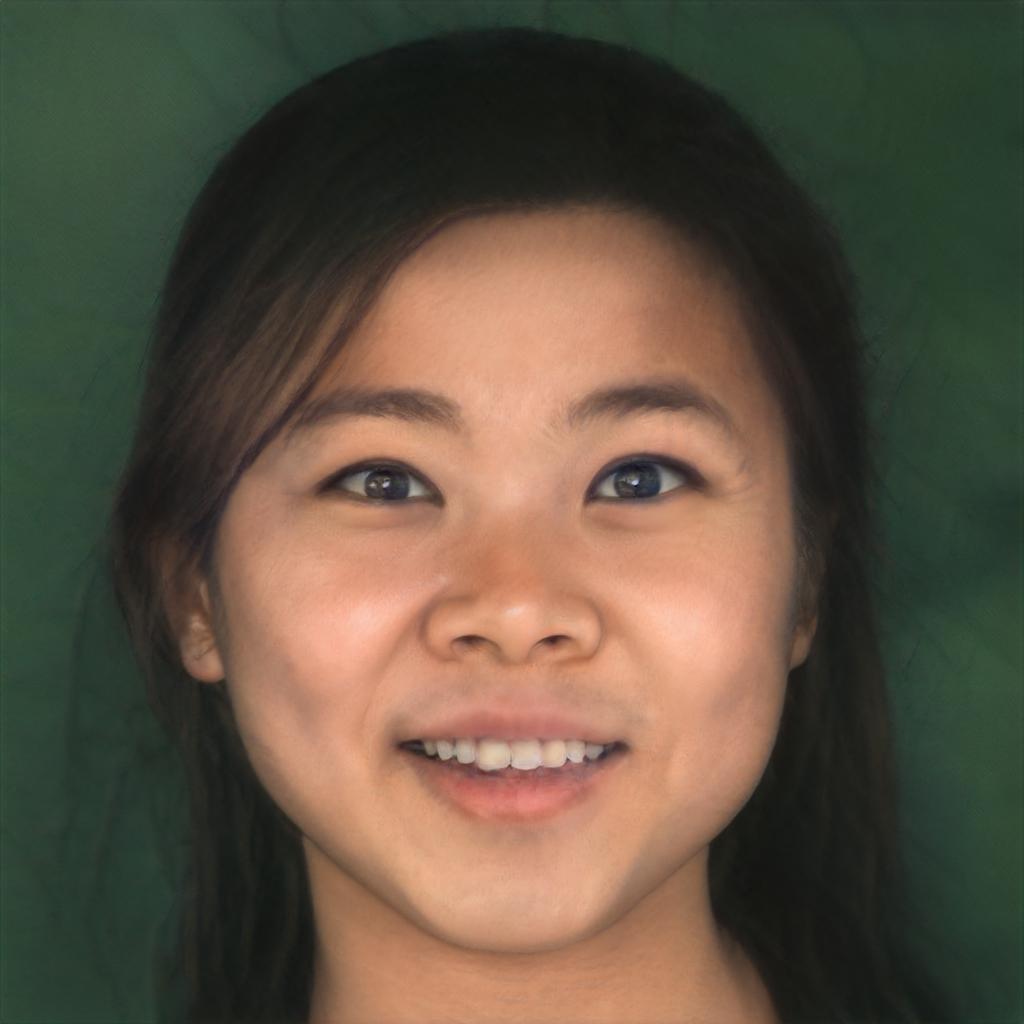}  \\ 

\begin{turn}{90}\hspace{0.5cm} SGANC IC \end{turn} &
\begin{turn}{90}\hspace{0.5cm} BPP=0.0017 \end{turn} &
\includegraphics[width=0.22\linewidth]{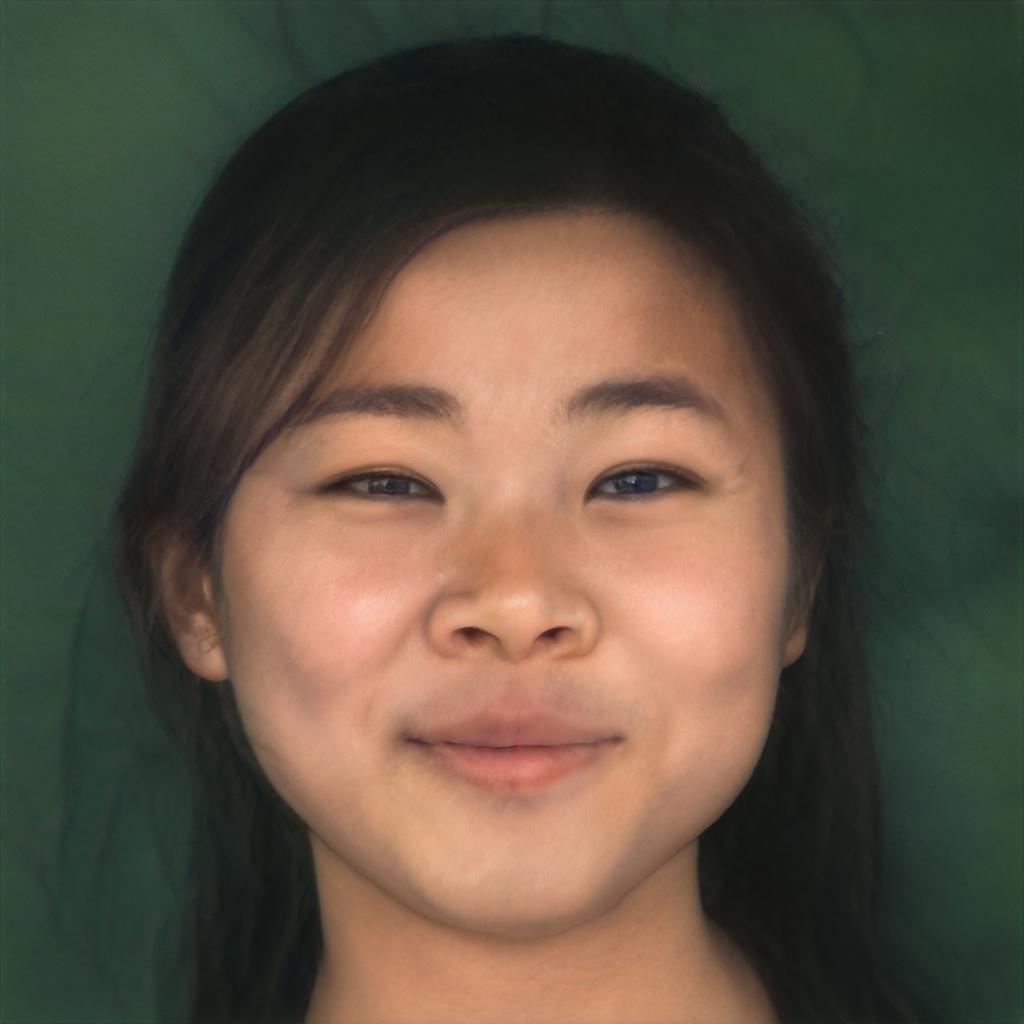} & 
 \includegraphics[width=0.22\linewidth]{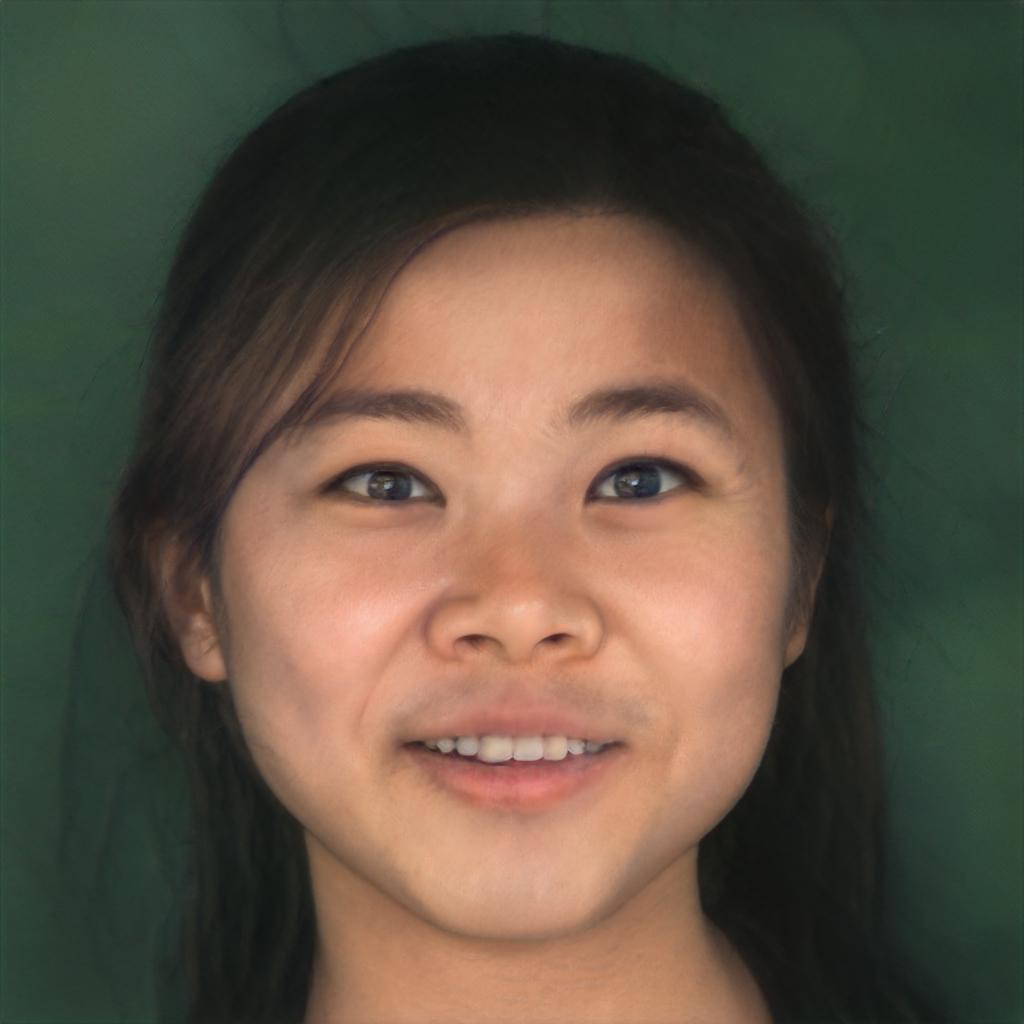}  & 
 \includegraphics[width=0.22\linewidth]{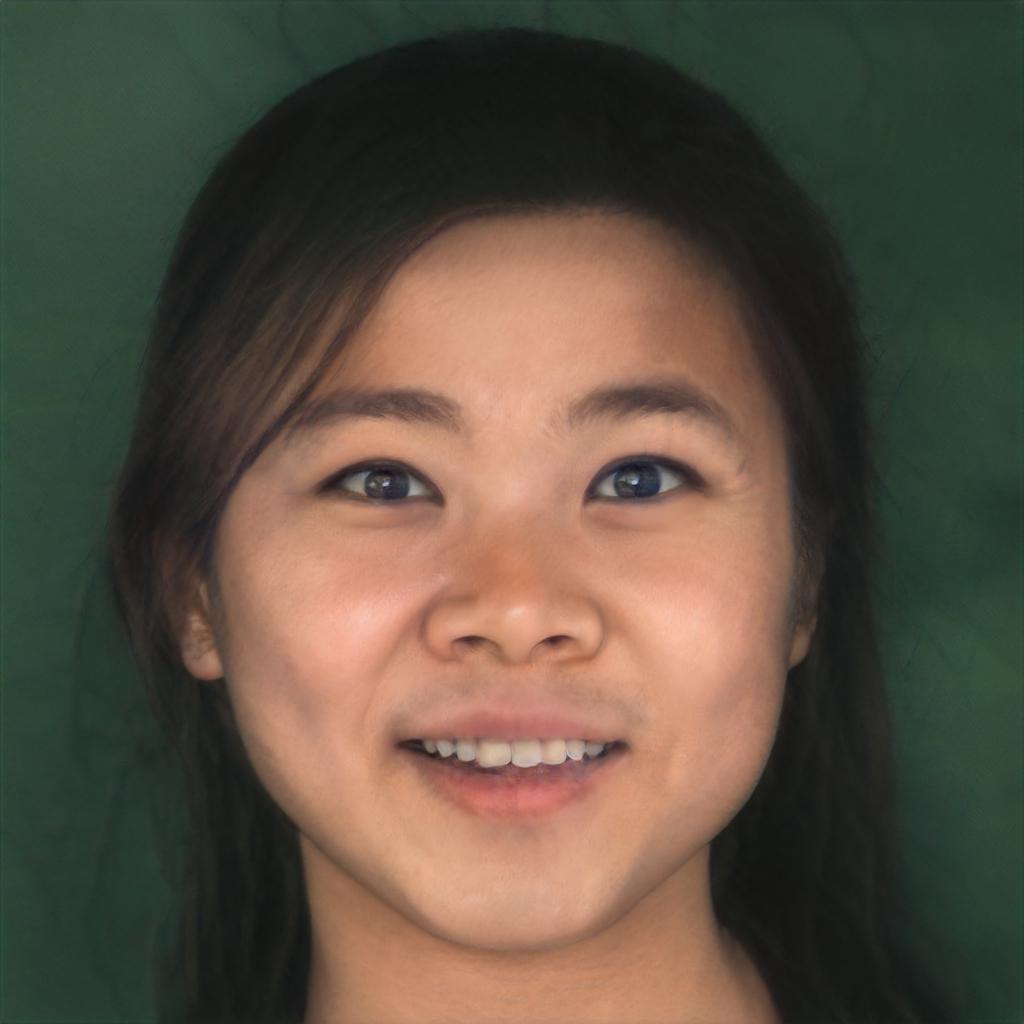}  \\
 
 \begin{turn}{90}\hspace{1cm} VTM  \end{turn}&
 \begin{turn}{90}\hspace{0.5cm} BPP=0.00250\end{turn}&
 \includegraphics[width=0.22\linewidth]{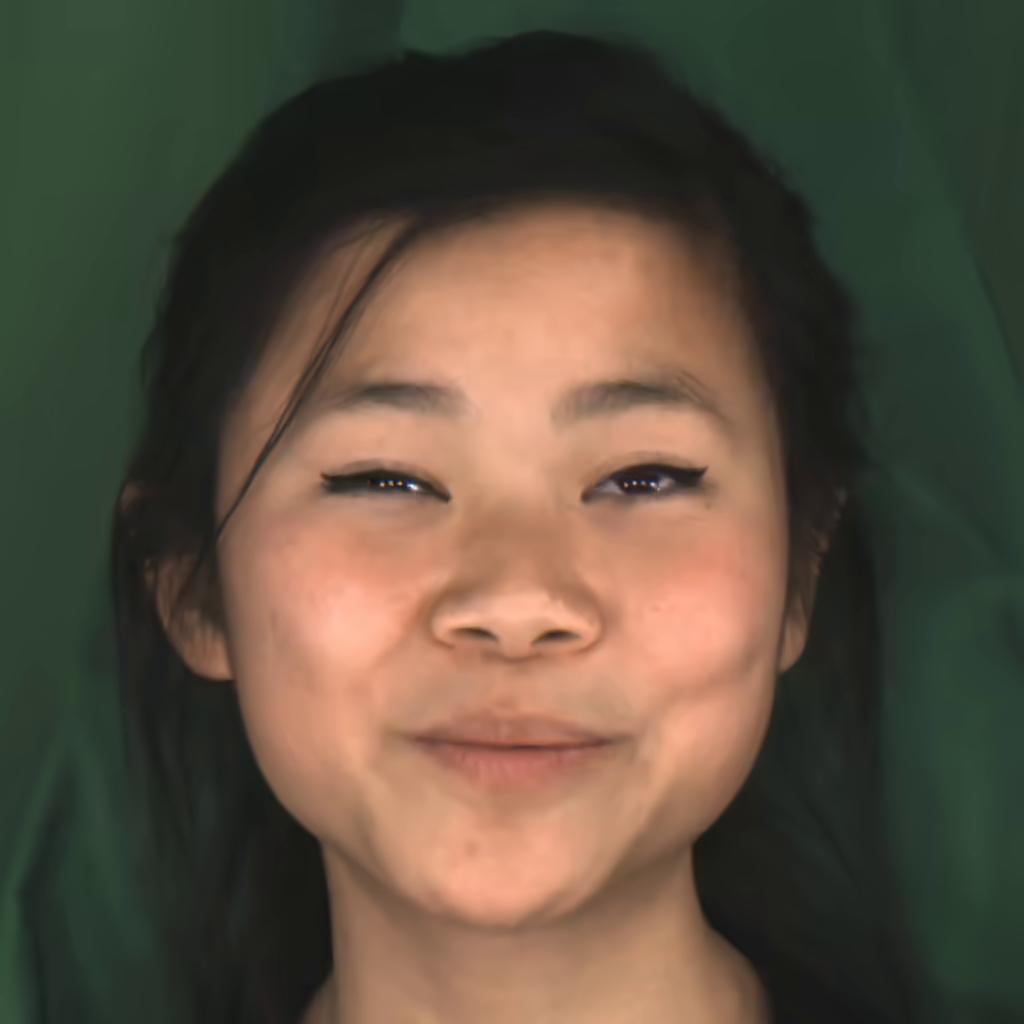}  & 
 \includegraphics[width=0.22\linewidth]{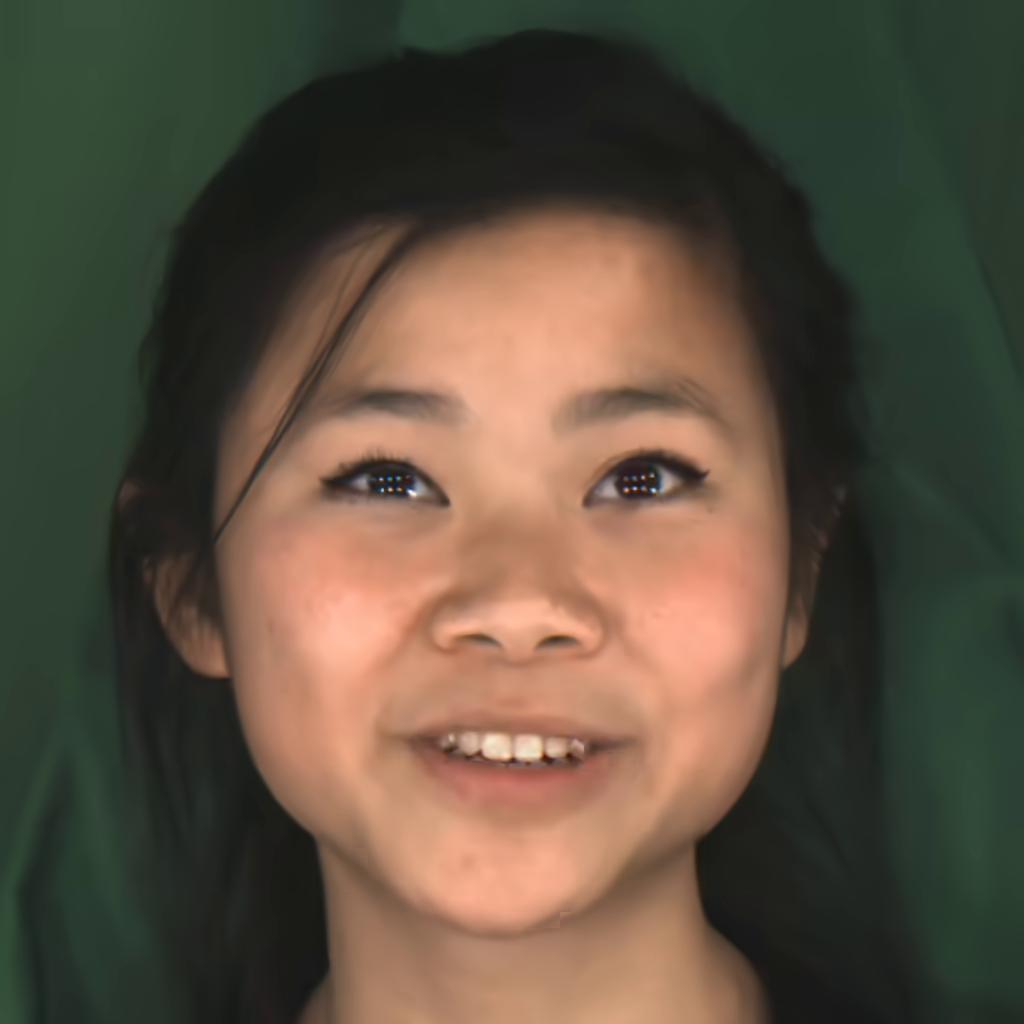}  & 
 \includegraphics[width=0.22\linewidth]{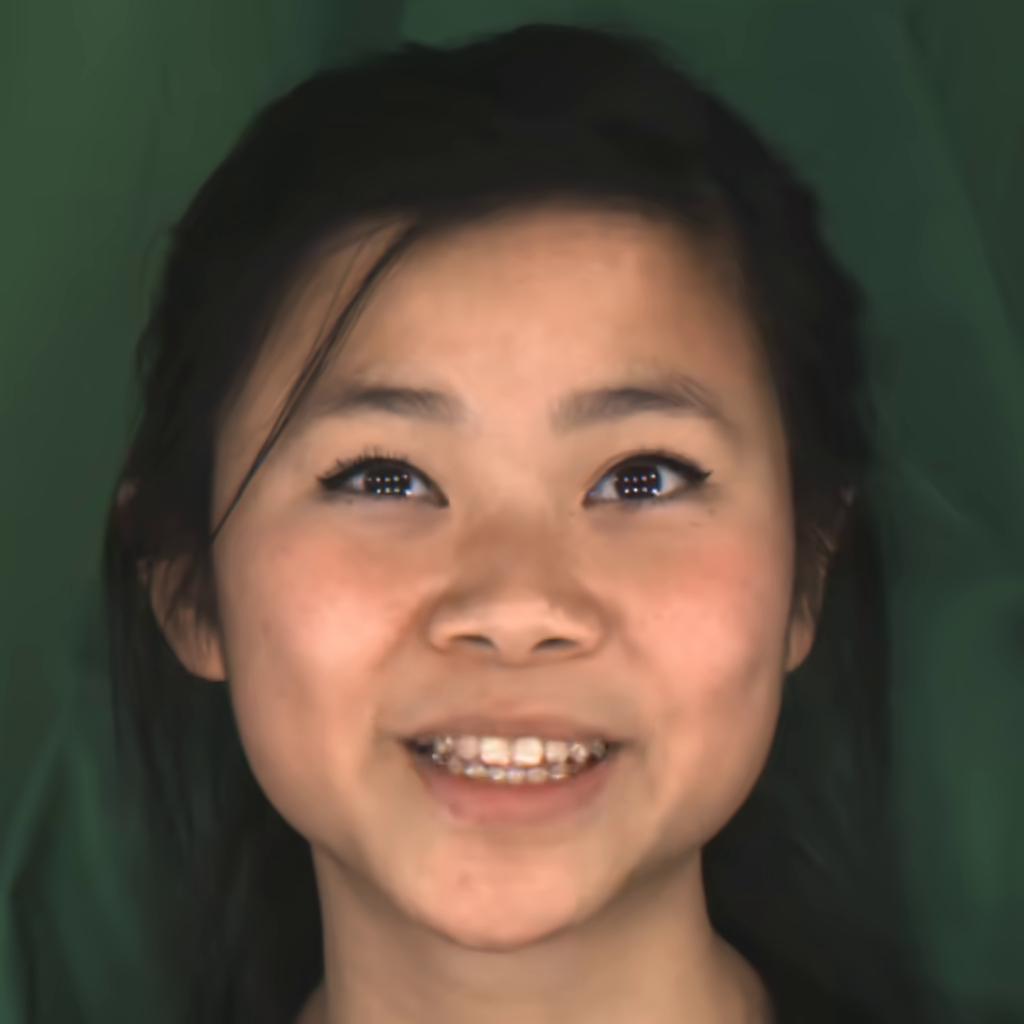}  \\

 \begin{turn}{90}\hspace{1cm} H.265 \end{turn} &
 \begin{turn}{90}\hspace{0.5cm} BPP=0.00237\end{turn}&
 \includegraphics[width=0.22\linewidth]{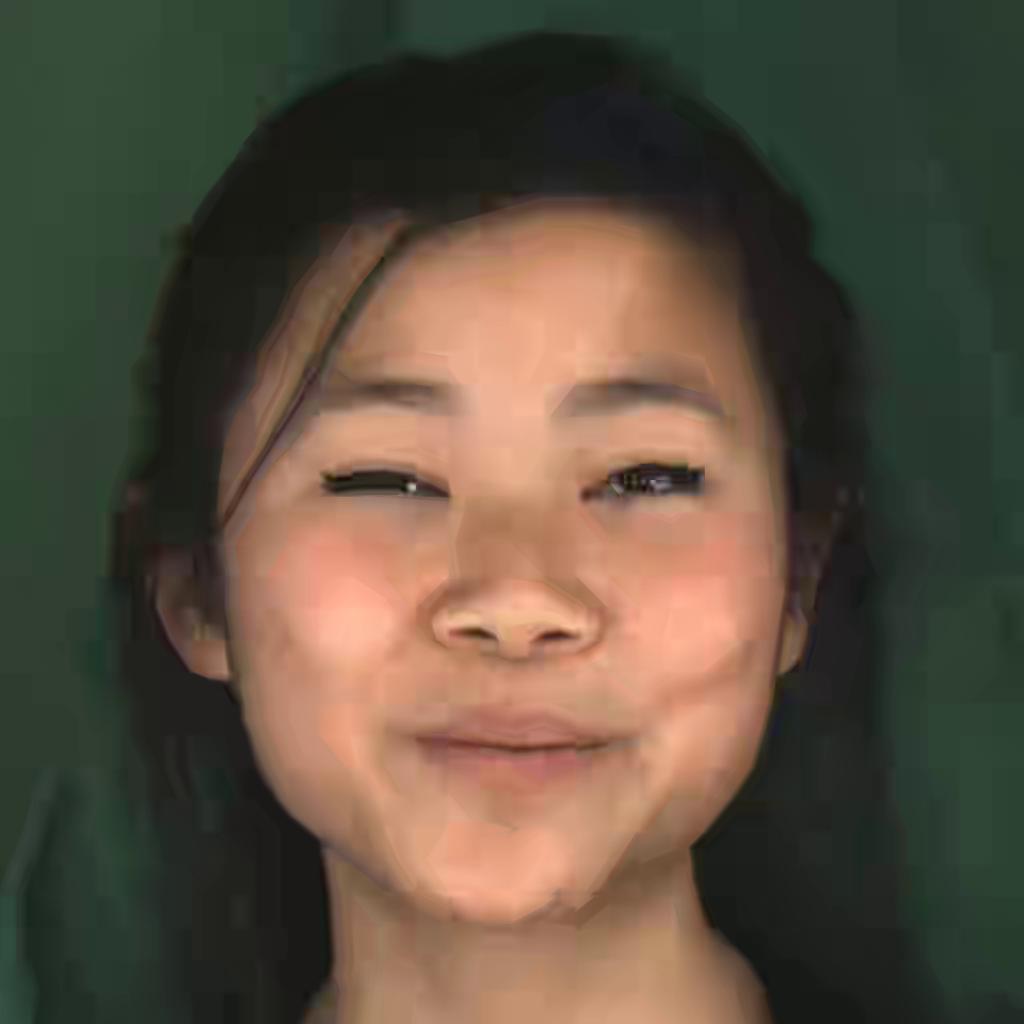}  & 
 \includegraphics[width=0.22\linewidth]{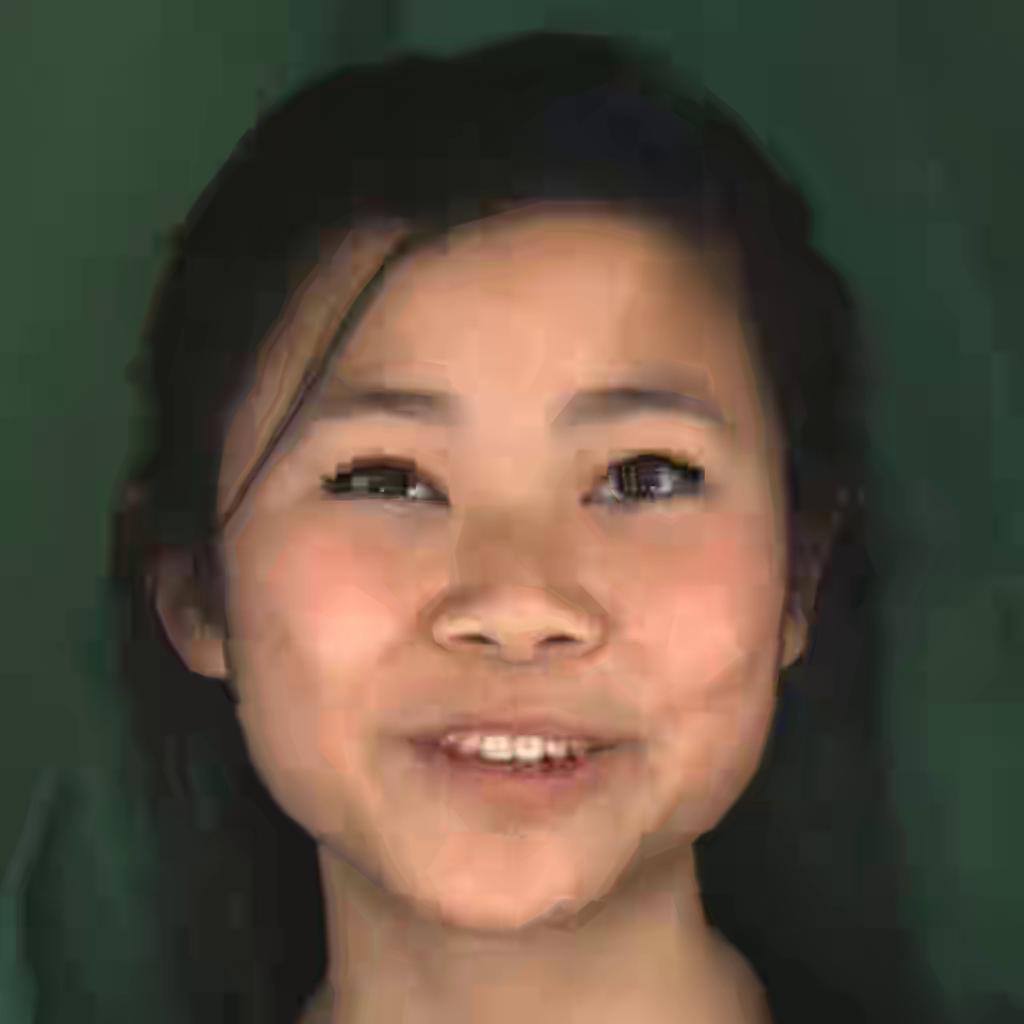}  & 
 \includegraphics[width=0.22\linewidth]{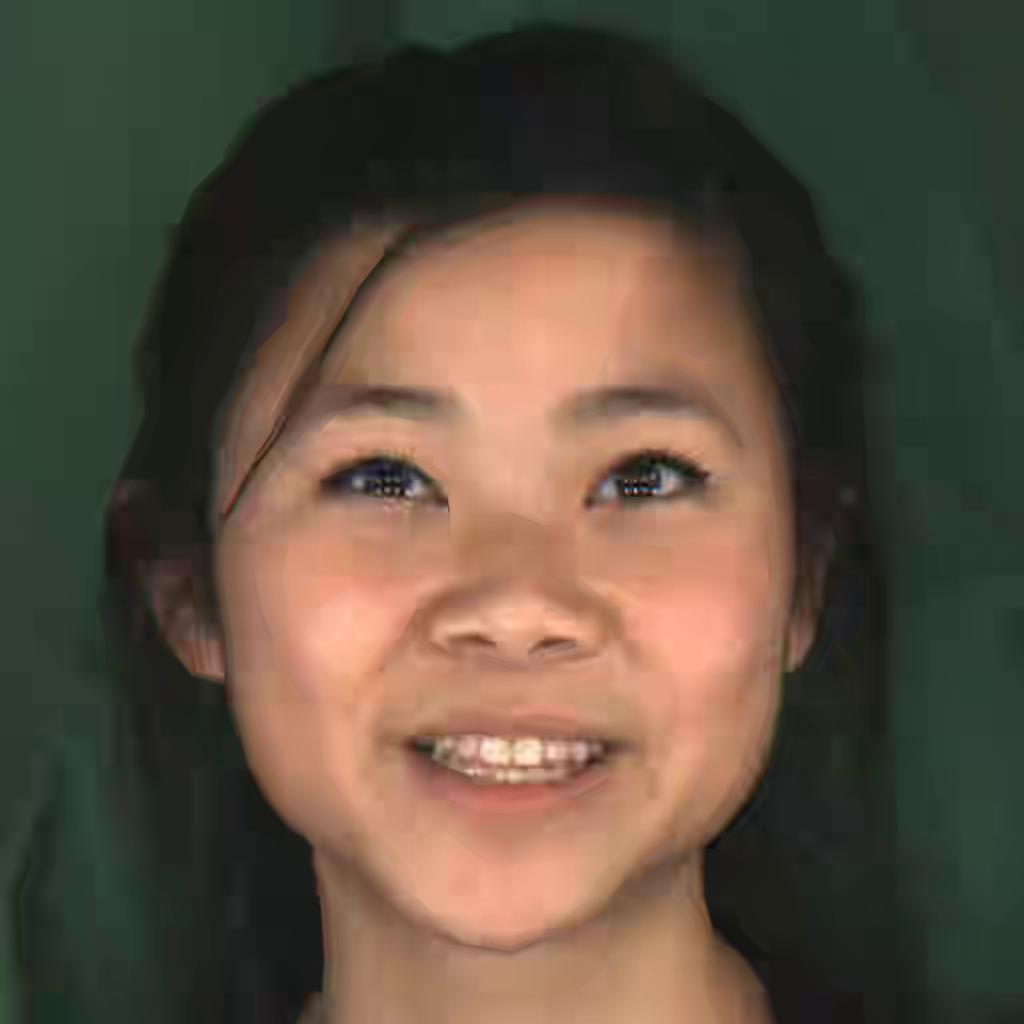}  \\ 
 \\
\bottomrule
\end{tabular}
\caption{Qualitative results on MEAD inter dataset for extreme BPP: H.265 shows blocking artifacts and blurring, VTM shows blurring especially at the edges of the face and the hair while our method (SGANC) is almost artifacts free and with high quality images.}
\label{fig:qual_results_low_mead_inter_4}
\end{figure*}

\begin{figure*}[h]
\setlength\tabcolsep{2pt}%
\centering
\begin{tabular}{p{0.5cm}cccc}
\toprule
 \begin{turn}{90} \hspace{1.2cm} Original \end{turn}  &
 &
 \includegraphics[width=0.22\linewidth]{images/inter_qualitative/M019/orig/frame0041.jpg}  & 
 \includegraphics[width=0.22\linewidth]{images/inter_qualitative/M019/orig/frame0113.jpg}  & 
 \includegraphics[width=0.22\linewidth]{images/inter_qualitative/M019/orig/frame0175.jpg}  \\ 

 \begin{turn}{90}\hspace{1.2cm} Projected\end{turn} &
 &
 \includegraphics[width=0.22\linewidth]{images/inter_qualitative/M019/projected/frame0041.jpg}  & 
 \includegraphics[width=0.22\linewidth]{images/inter_qualitative/M019/projected/frame0113.jpg}  & 
 \includegraphics[width=0.22\linewidth]{images/inter_qualitative/M019/projected/frame0175.jpg}  \\

\begin{turn}{90}\hspace{0.5cm} SGANC IC \end{turn} &
\begin{turn}{90}\hspace{0.5cm} BPP=0.00237 \end{turn} &
 \includegraphics[width=0.22\linewidth]{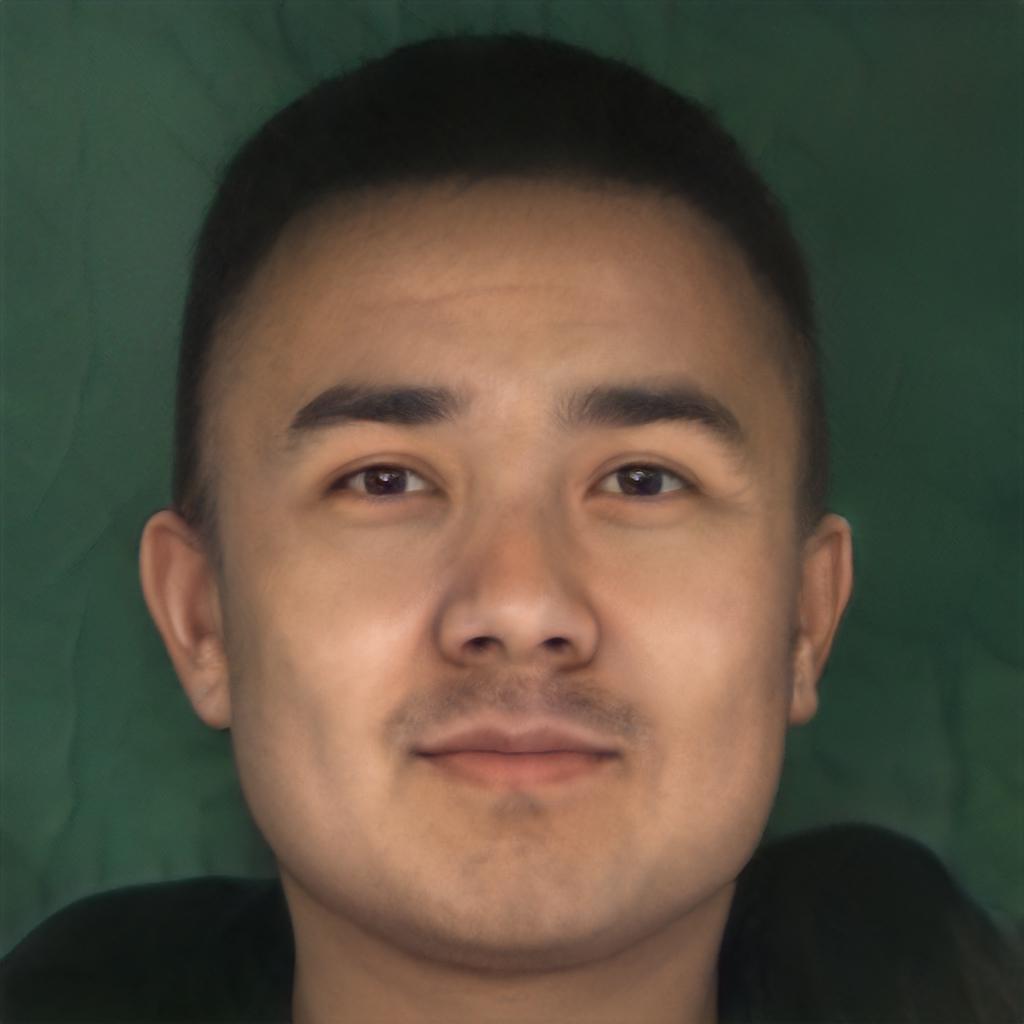}  & 
 \includegraphics[width=0.22\linewidth]{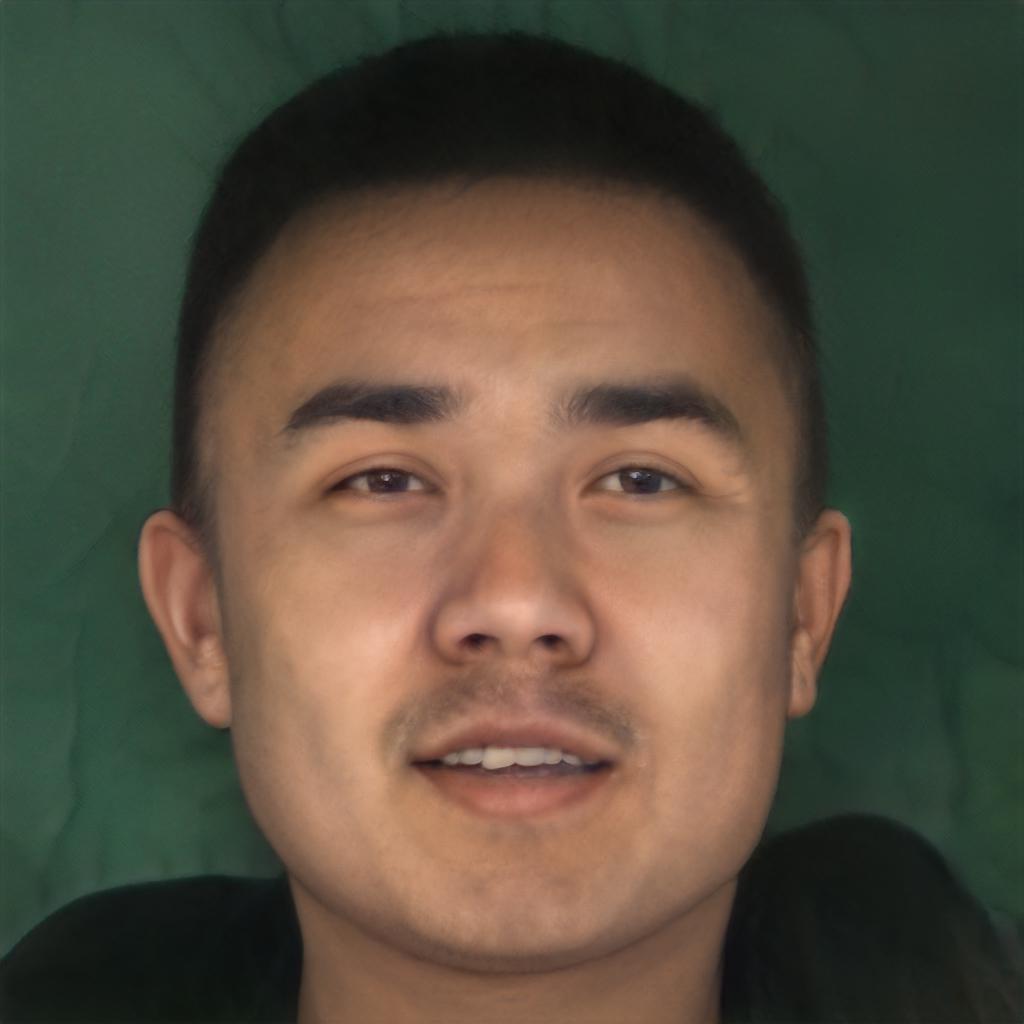}   &
 \includegraphics[width=0.22\linewidth]{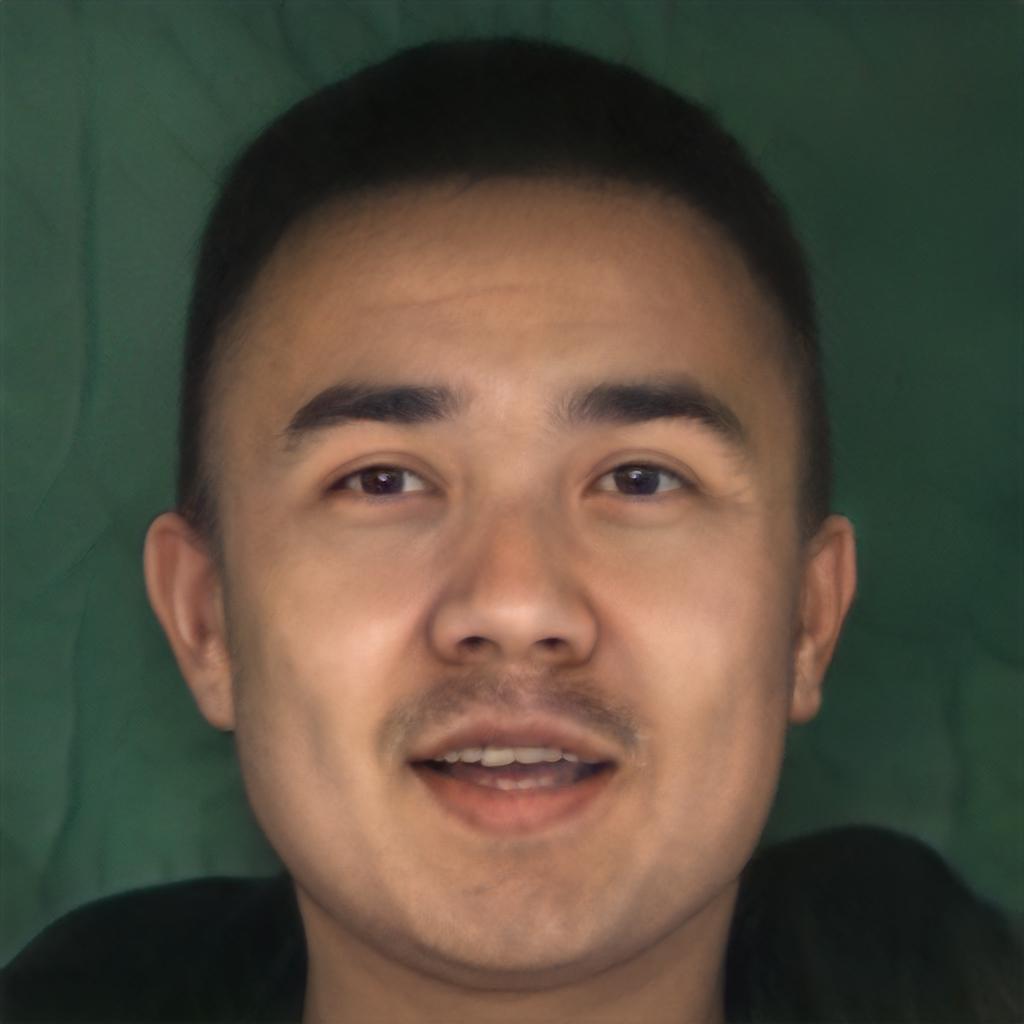}  \\
 
 \begin{turn}{90}\hspace{1cm} VTM  \end{turn}&
 \begin{turn}{90}\hspace{0.5cm} BPP=0.00317 \end{turn}&
 \includegraphics[width=0.22\linewidth]{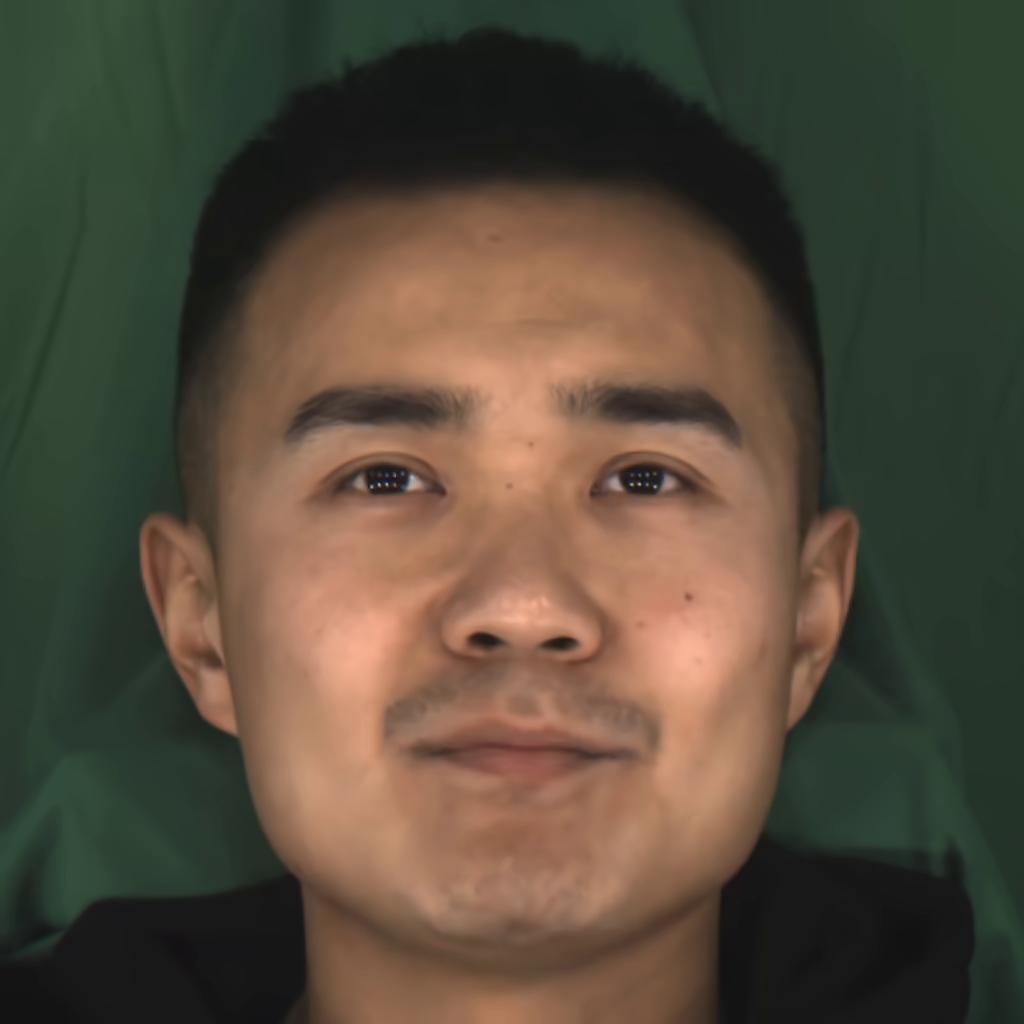}  & 
 \includegraphics[width=0.22\linewidth]{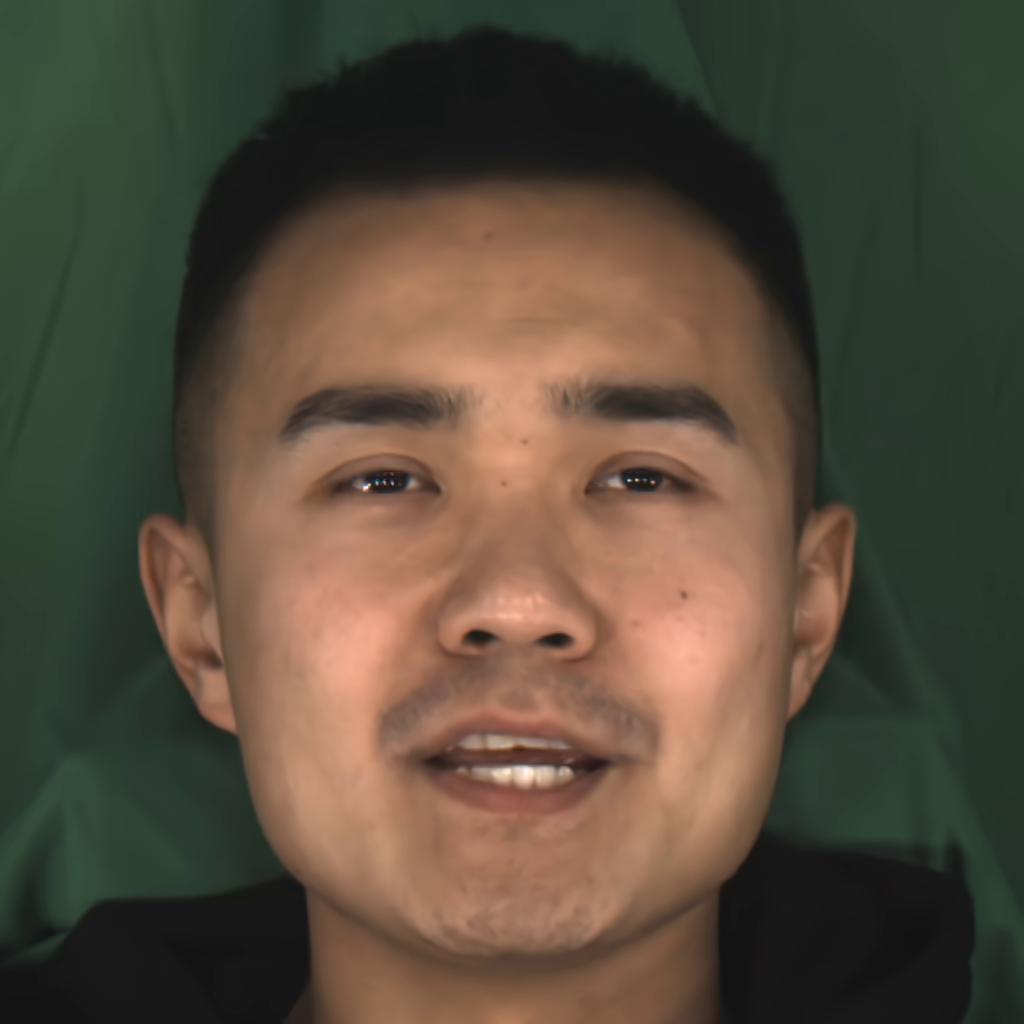}  & 
 \includegraphics[width=0.22\linewidth]{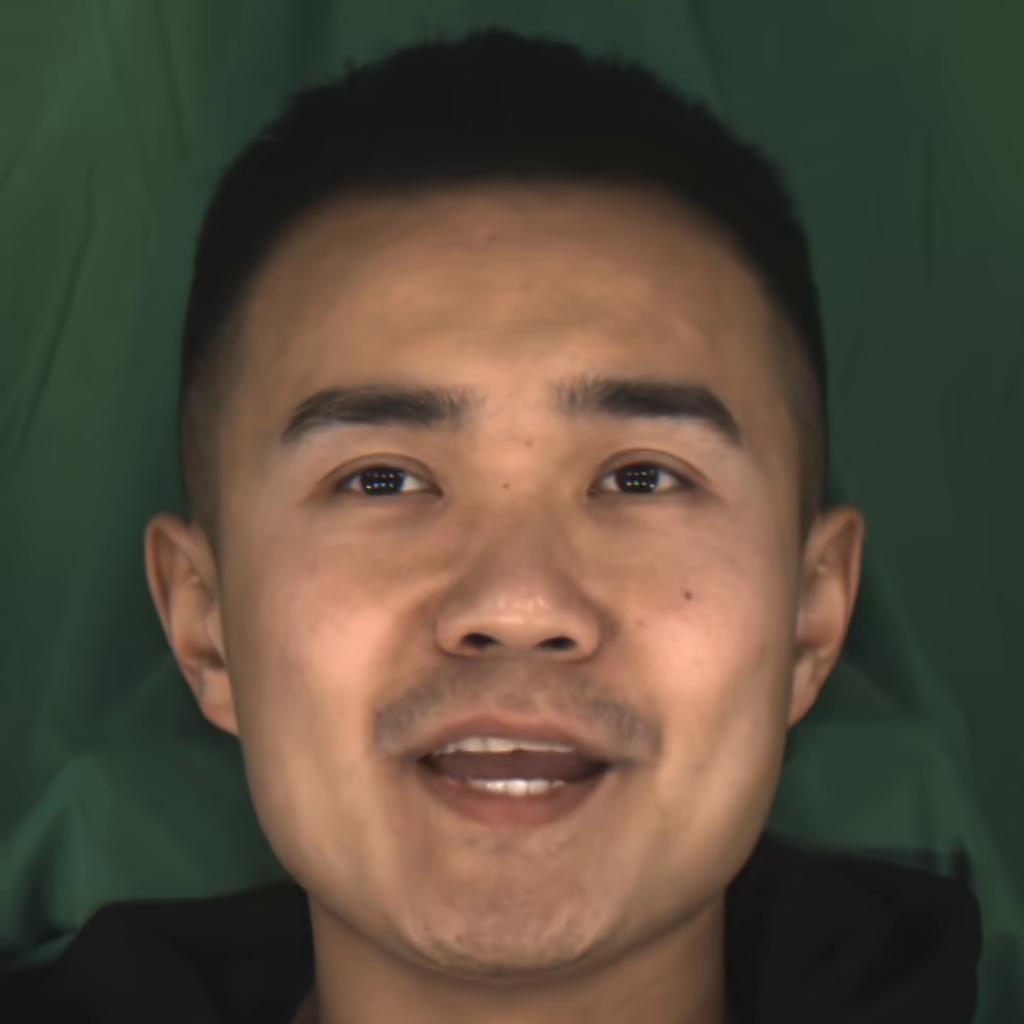}  \\ 

 \begin{turn}{90}\hspace{1cm} H.265 \end{turn} &
 \begin{turn}{90}\hspace{0.5cm} BPP=0.00355 \end{turn}&
 \includegraphics[width=0.22\linewidth]{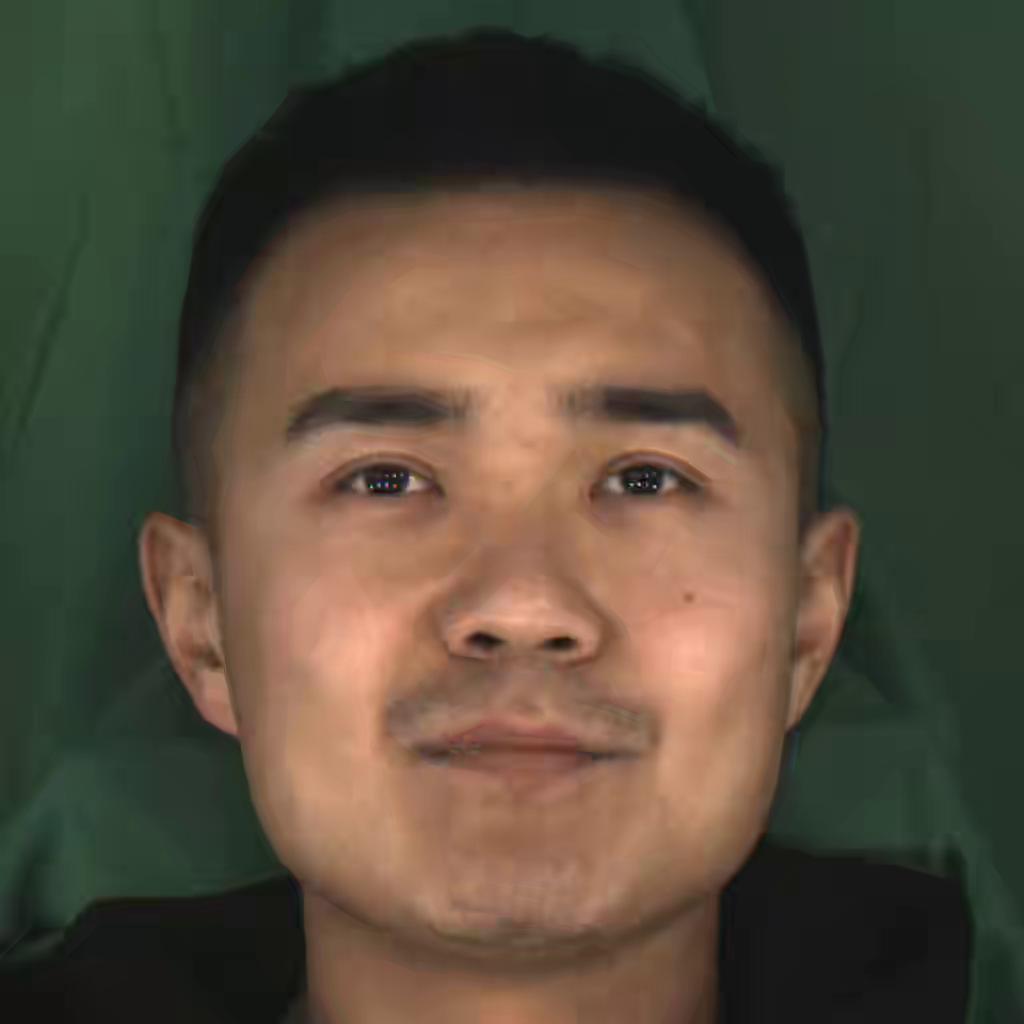}  & 
 \includegraphics[width=0.22\linewidth]{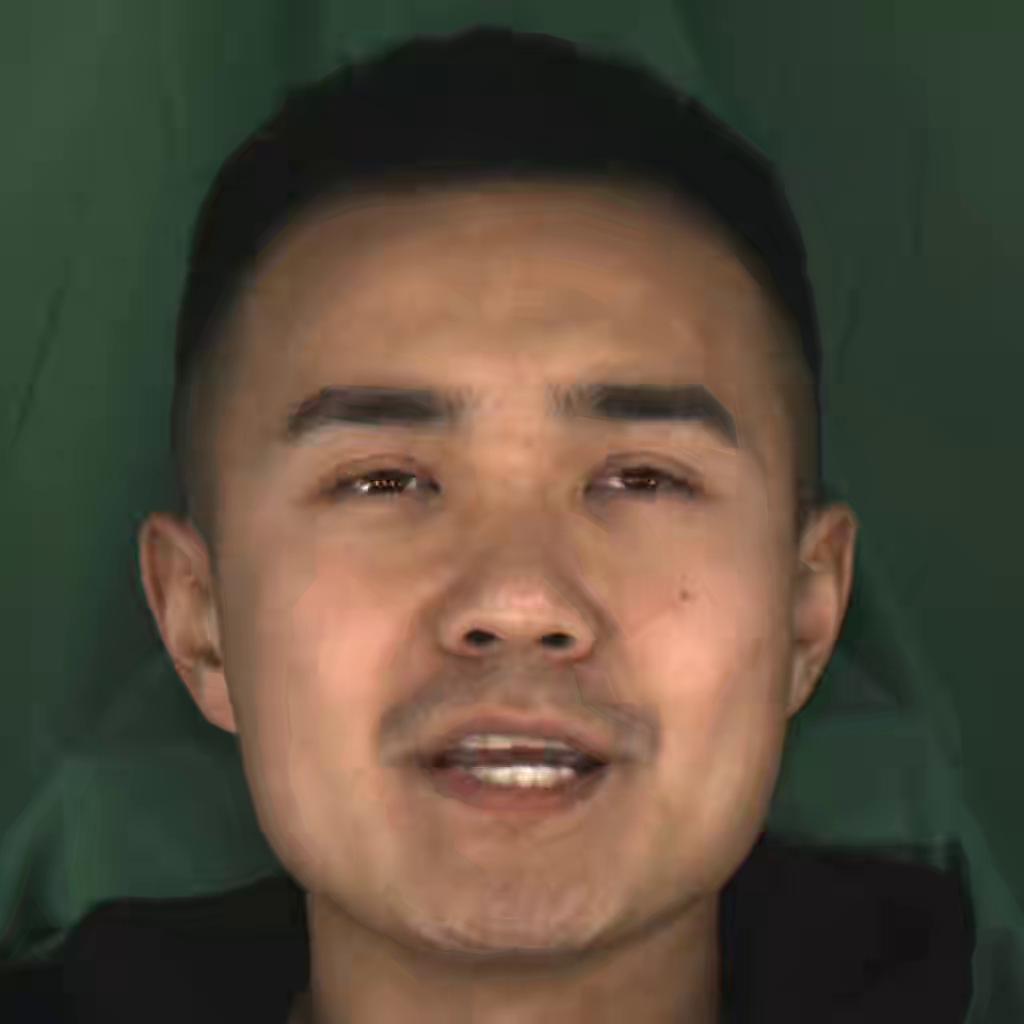}  & 
 \includegraphics[width=0.22\linewidth]{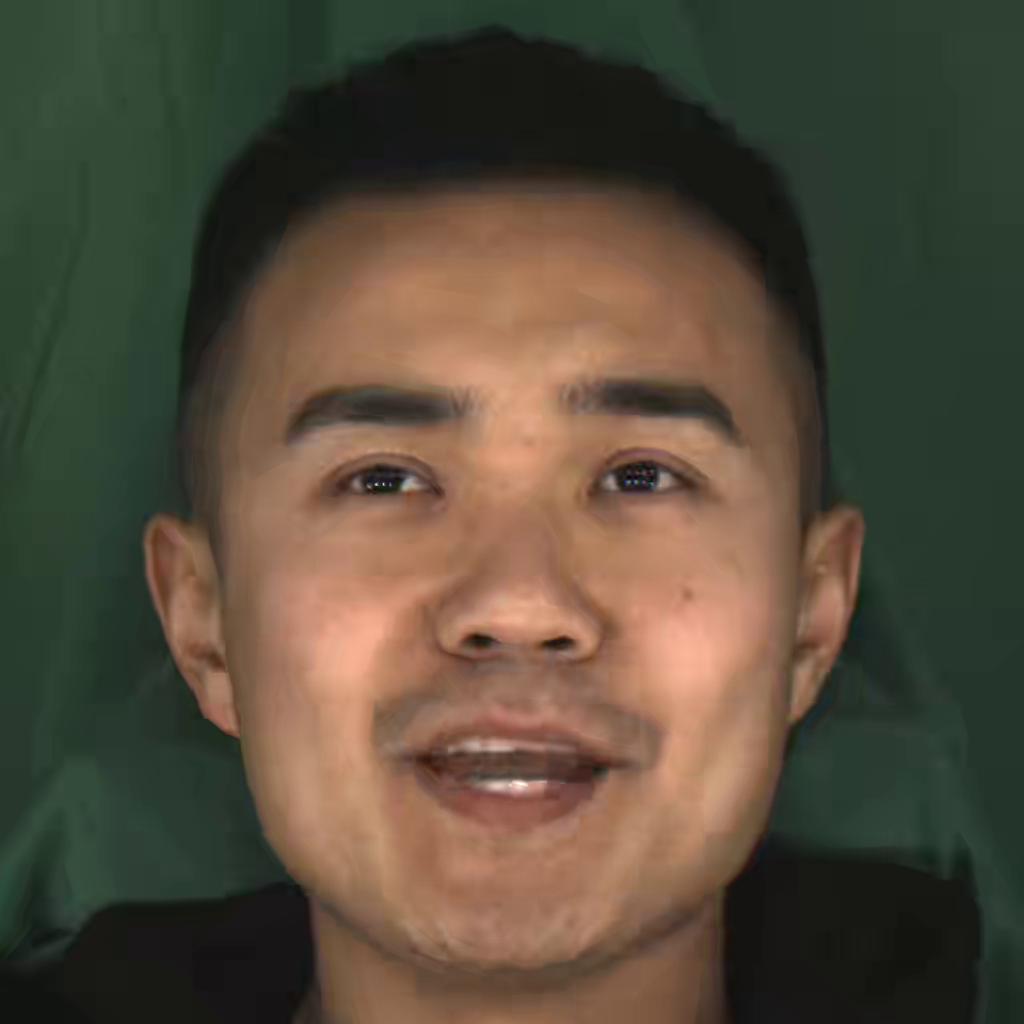}  \\ 
 \\
\bottomrule
\end{tabular}
\caption{Qualitative results on MEAD inter dataset: H.265 shows blocking artifacts and blurring, VTM shows blurring especially at the edges of the face and the hair while our method (SGANC) is almost artifacts free and with high quality images.}
\label{fig:qual_results_medium_mead_inter_1}
\end{figure*}

\begin{figure*}[h]
\setlength\tabcolsep{2pt}%
\centering
\begin{tabular}{p{0.5cm}cccc}
\toprule
 \begin{turn}{90} \hspace{1.2cm} Original \end{turn}  &
 &
 \includegraphics[width=0.22\linewidth]{images/inter_qualitative/W015/orig/frame0004.jpg}  & 
 \includegraphics[width=0.22\linewidth]{images/inter_qualitative/W015/orig/frame0018.jpg}  & 
 \includegraphics[width=0.22\linewidth]{images/inter_qualitative/W015/orig/frame0029.jpg}  \\

 \begin{turn}{90}\hspace{1.2cm} Projected\end{turn} &
 &
 \includegraphics[width=0.22\linewidth]{images/inter_qualitative/W015/projected/frame0004.jpg}  & 
 \includegraphics[width=0.22\linewidth]{images/inter_qualitative/W015/projected/frame0018.jpg}  & 
 \includegraphics[width=0.22\linewidth]{images/inter_qualitative/W015/projected/frame0029.jpg}  \\

\begin{turn}{90}\hspace{0.5cm} SGANC IC \end{turn} &
\begin{turn}{90}\hspace{0.5cm} BPP=0.00313 \end{turn} &
 \includegraphics[width=0.22\linewidth]{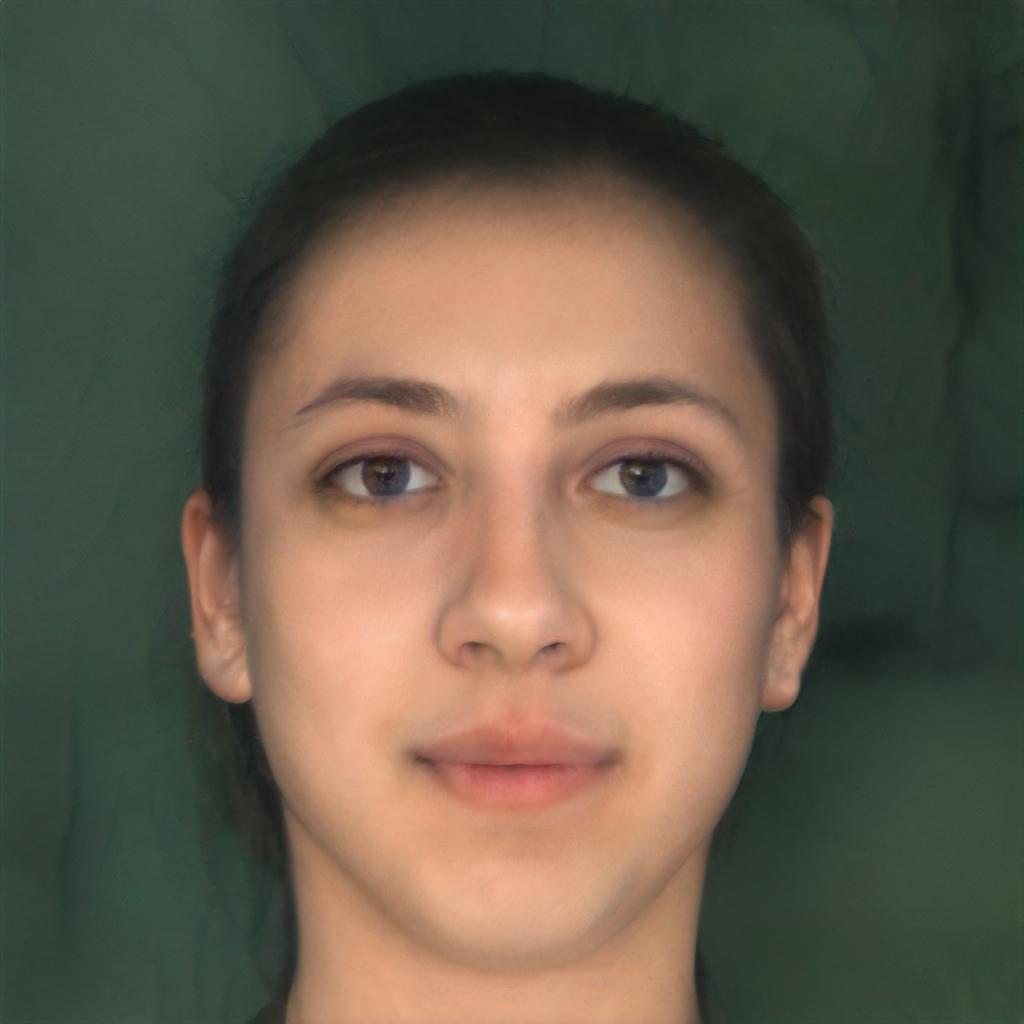}  & 
 \includegraphics[width=0.22\linewidth]{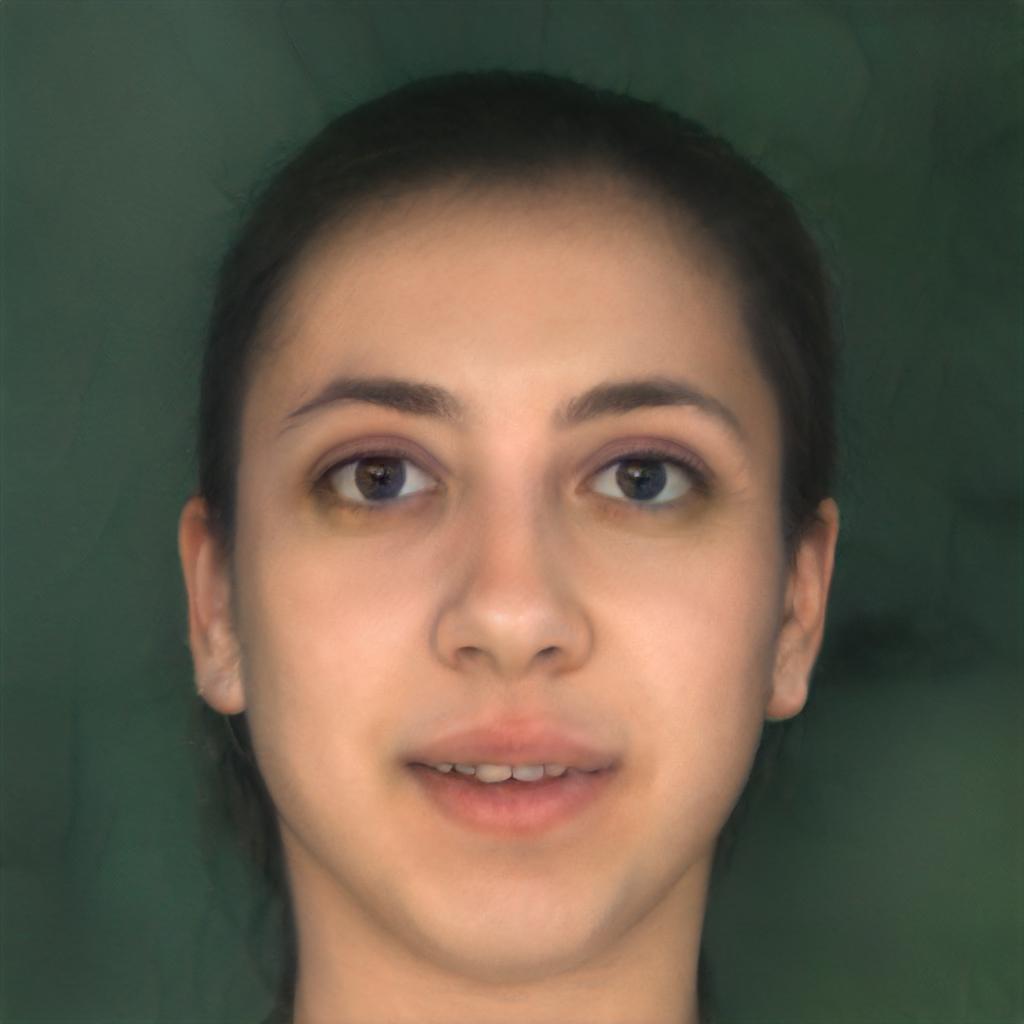}  & 
 \includegraphics[width=0.22\linewidth]{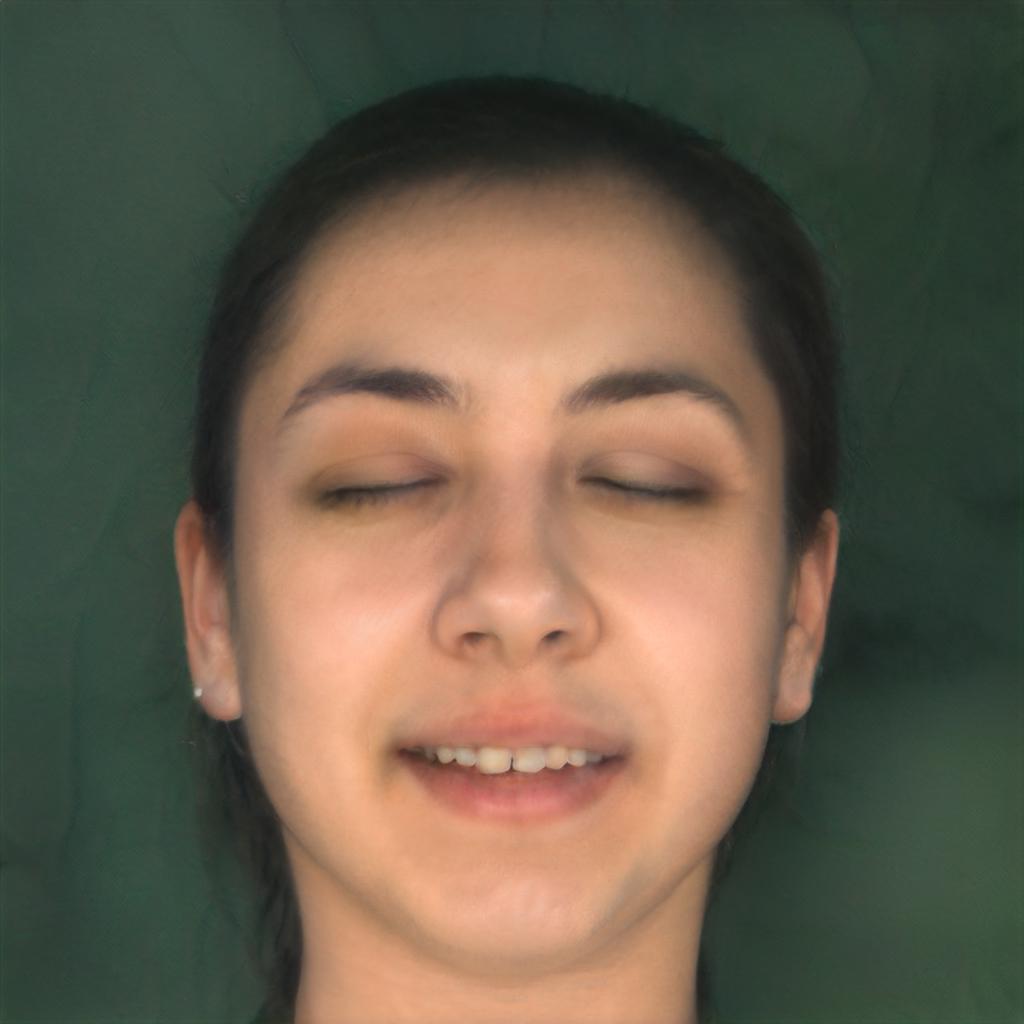}  \\ 
 
 \begin{turn}{90}\hspace{1cm} VTM  \end{turn}&
 \begin{turn}{90}\hspace{0.5cm} BPP=0.0035 \end{turn}&
 \includegraphics[width=0.22\linewidth]{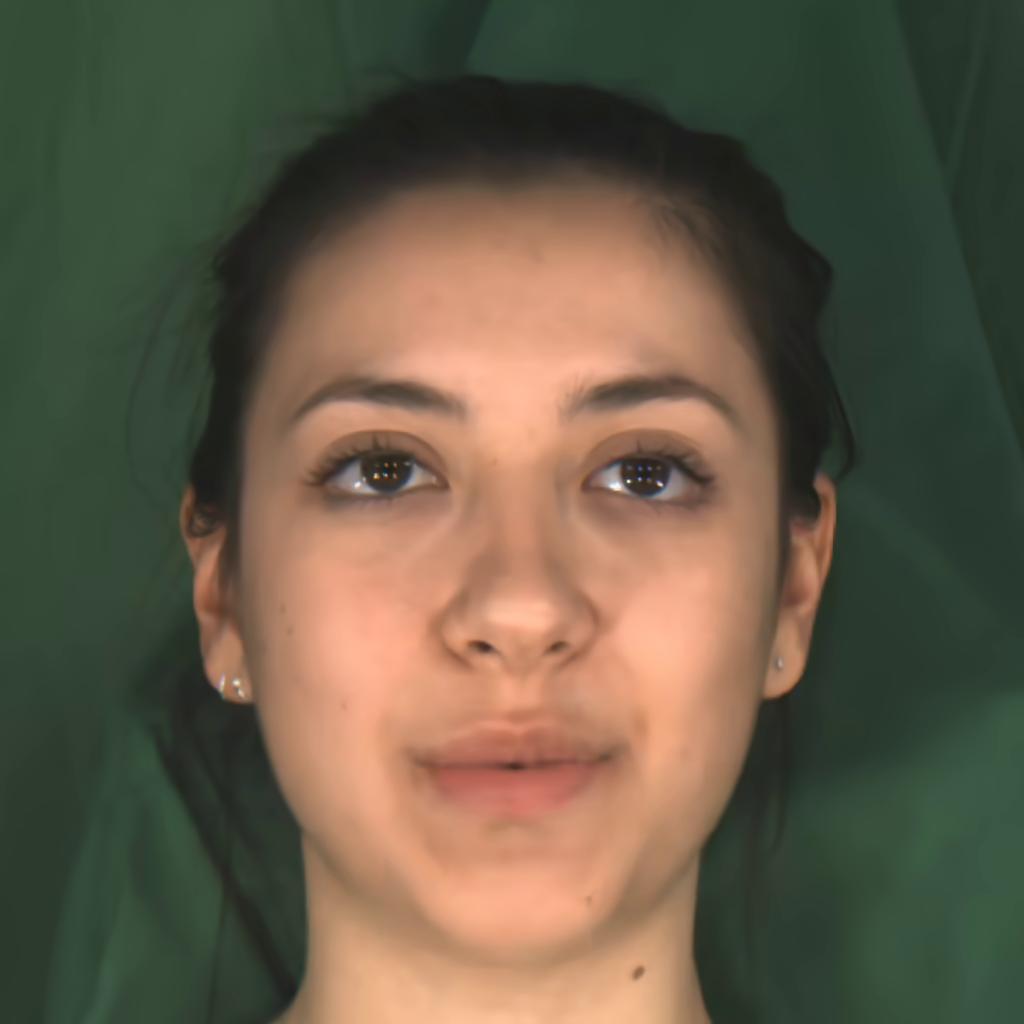}  & 
 \includegraphics[width=0.22\linewidth]{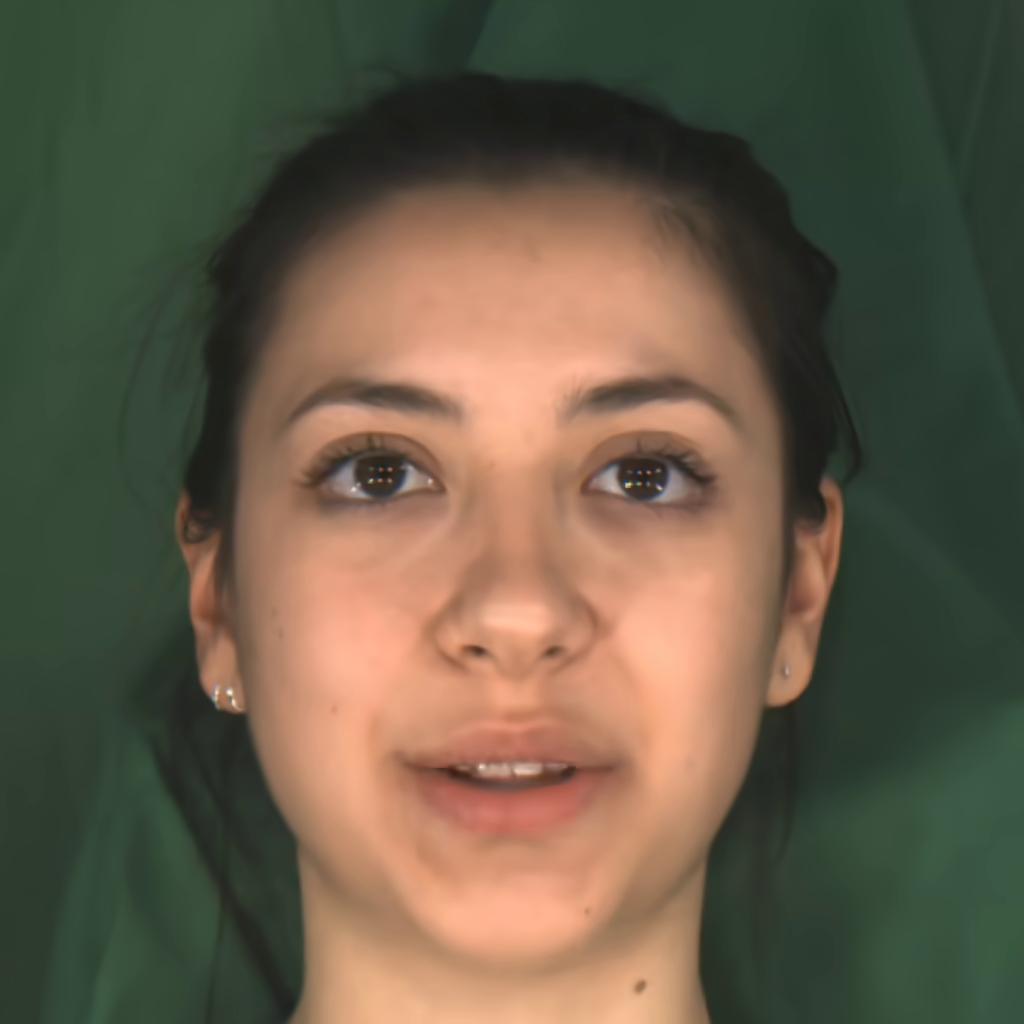}  & 
 \includegraphics[width=0.22\linewidth]{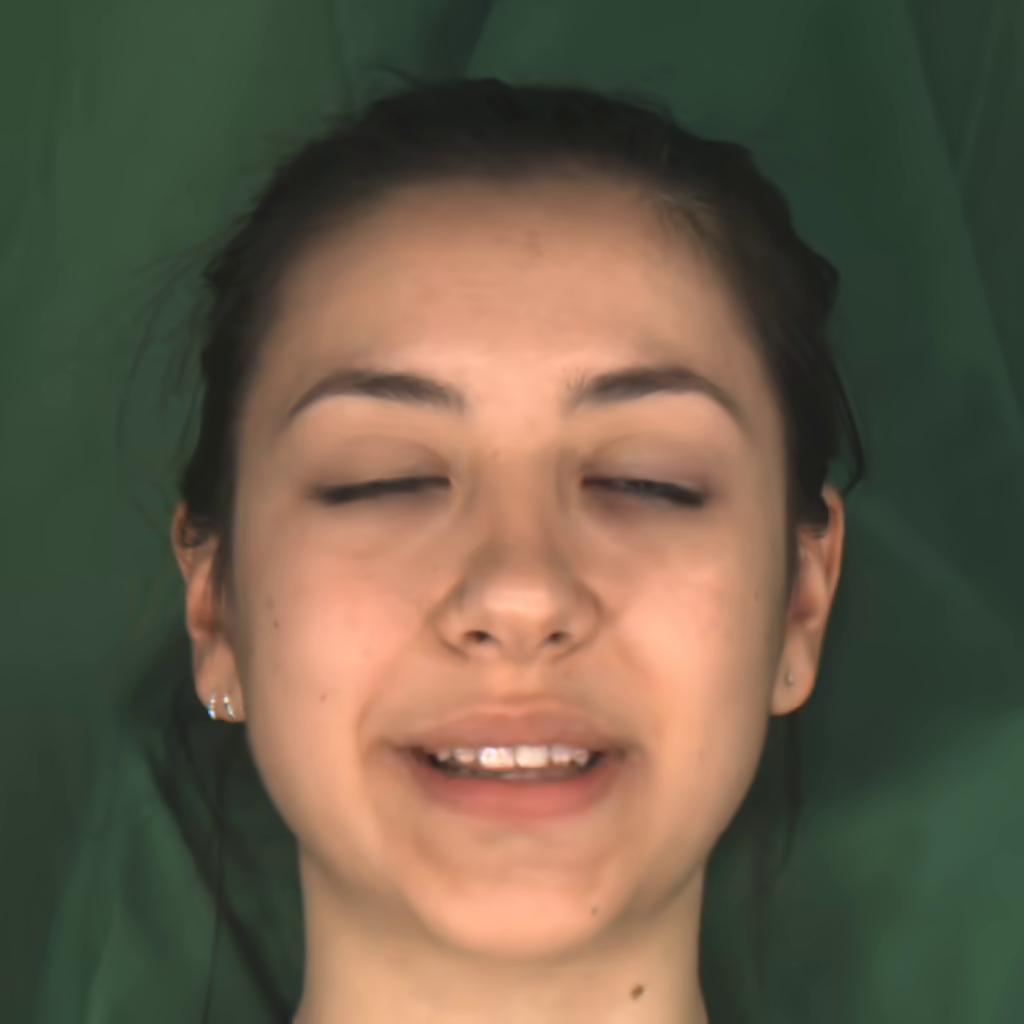}  \\ 

 \begin{turn}{90}\hspace{1cm} H.265 \end{turn} &
 \begin{turn}{90}\hspace{0.5cm} BPP=0.0036\end{turn}&
 \includegraphics[width=0.22\linewidth]{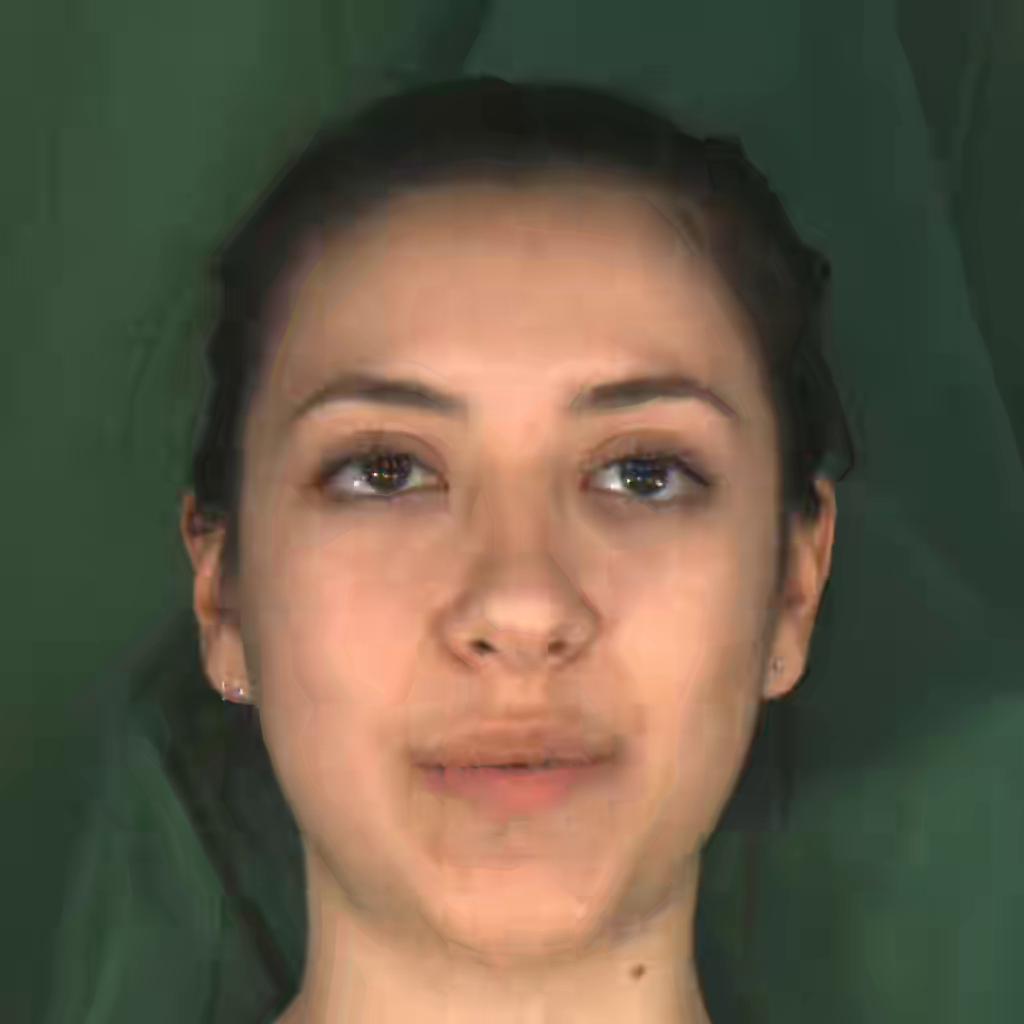}  & 
 \includegraphics[width=0.22\linewidth]{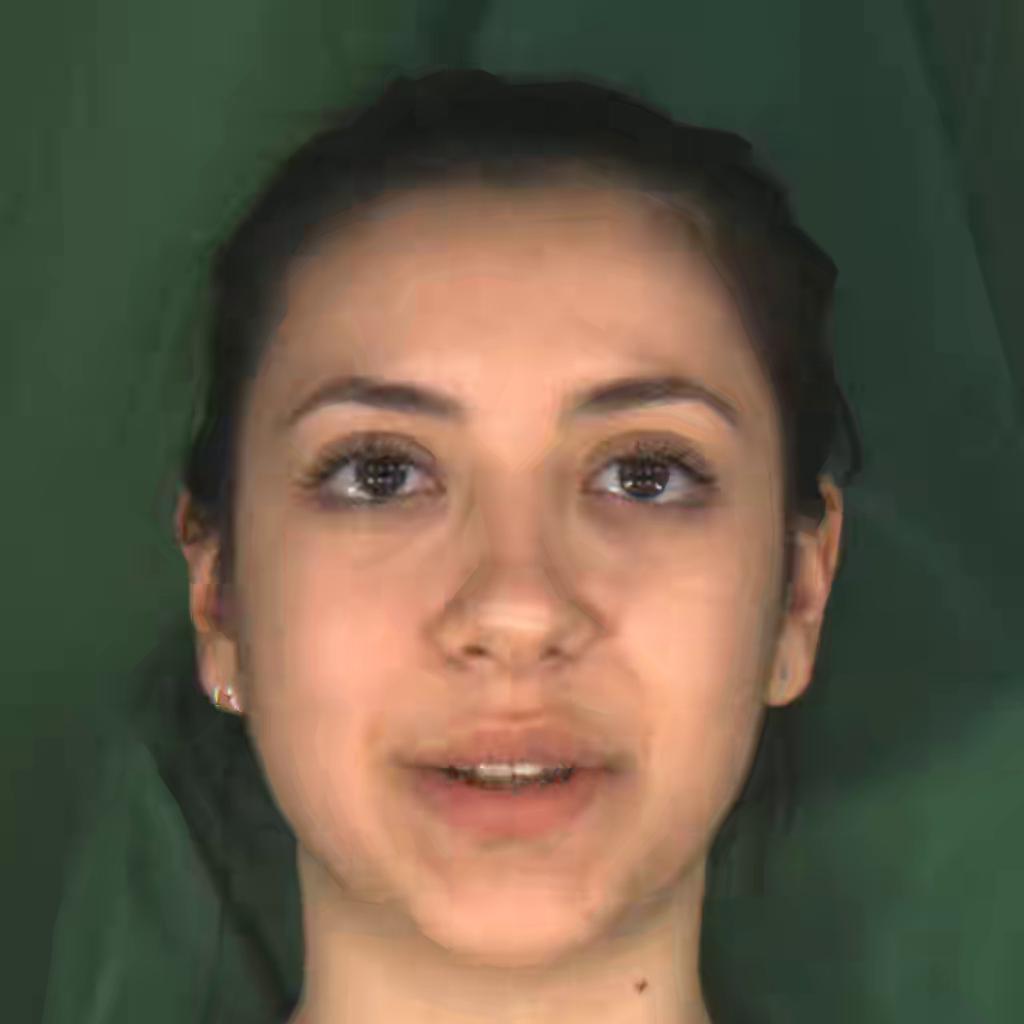}  & 
 \includegraphics[width=0.22\linewidth]{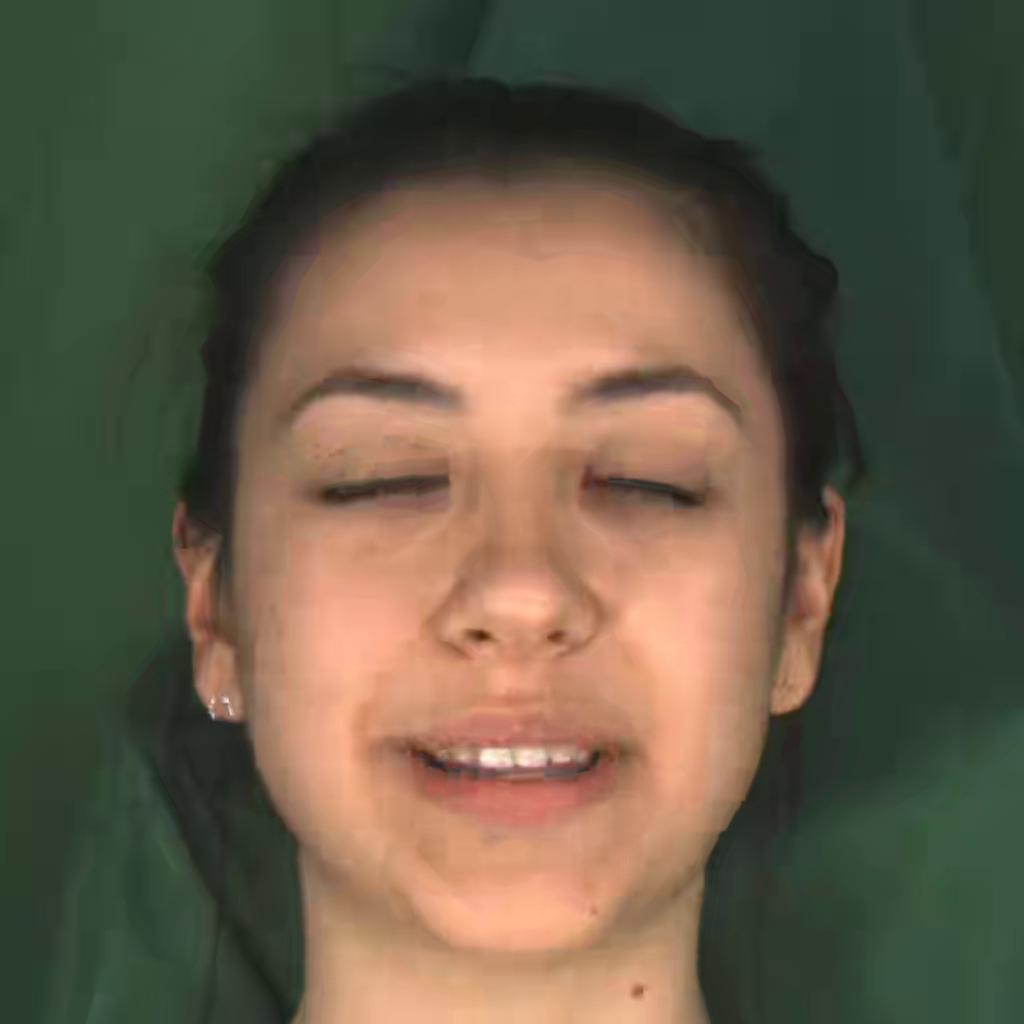}  \\ 
 \\
\bottomrule
\end{tabular}
\caption{Qualitative results on MEAD inter dataset: H.265 shows blocking artifacts and blurring, VTM shows blurring especially at the edges of the face and the hair while our method (SGANC) is almost artifacts free and with high quality images.}
\label{fig:qual_results_medium_mead_inter_2}
\end{figure*}

\begin{figure*}[h]
\setlength\tabcolsep{2pt}%
\centering
\begin{tabular}{p{0.5cm}cccc}
\toprule
 \begin{turn}{90} \hspace{1.2cm} Original \end{turn}  &
 &
 \includegraphics[width=0.22\linewidth]{images/inter_qualitative/M032/orig/frame0002.jpg}  & 
 \includegraphics[width=0.22\linewidth]{images/inter_qualitative/M032/orig/frame0028.jpg}  & 
 \includegraphics[width=0.22\linewidth]{images/inter_qualitative/M032/orig/frame0053.jpg}  \\ 

 \begin{turn}{90}\hspace{1.2cm} Projected\end{turn} &
 &
 \includegraphics[width=0.22\linewidth]{images/inter_qualitative/M032/projected/frame0002.jpg}  & 
 \includegraphics[width=0.22\linewidth]{images/inter_qualitative/M032/projected/frame0028.jpg}  & 
 \includegraphics[width=0.22\linewidth]{images/inter_qualitative/M032/projected/frame0053.jpg}  \\

\begin{turn}{90}\hspace{0.5cm} SGANC IC \end{turn} &
\begin{turn}{90}\hspace{0.5cm} BPP=0.00269 \end{turn} &
 \includegraphics[width=0.22\linewidth]{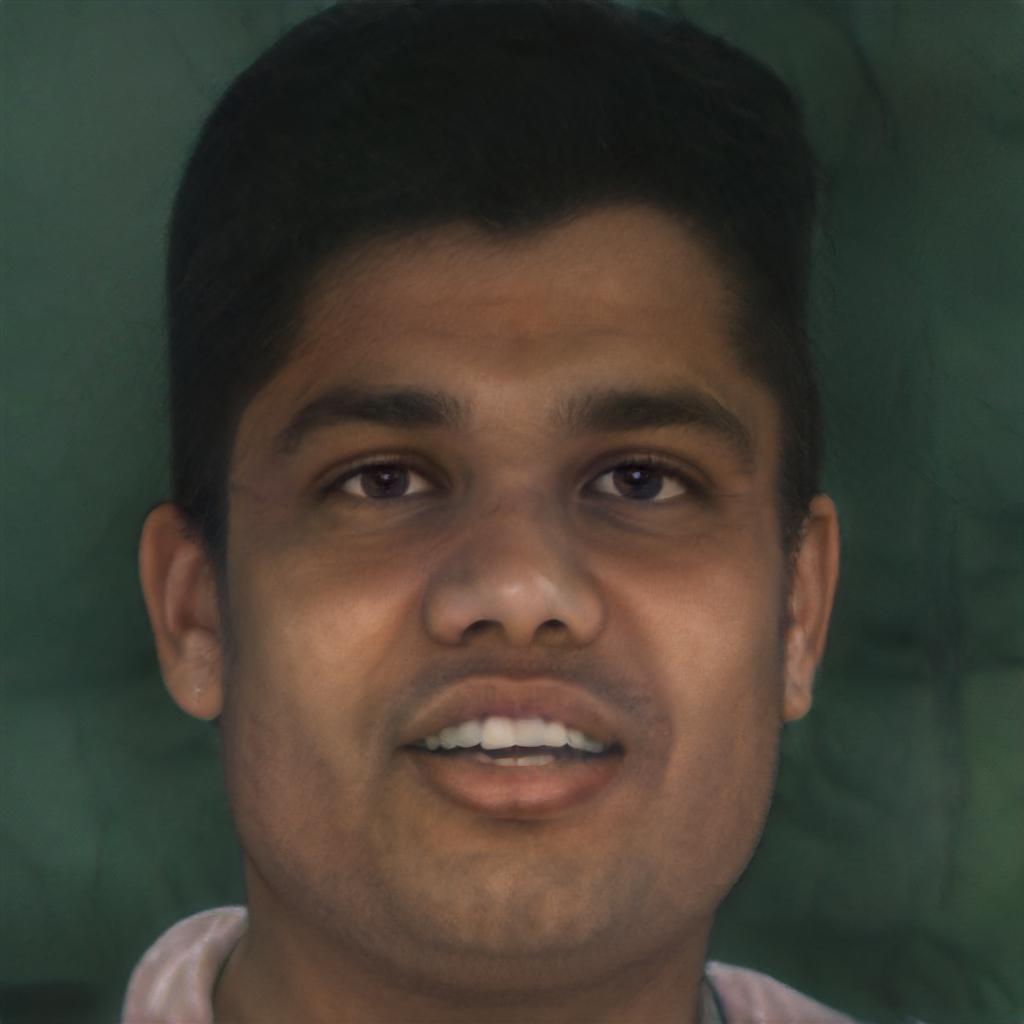}  & 
 \includegraphics[width=0.22\linewidth]{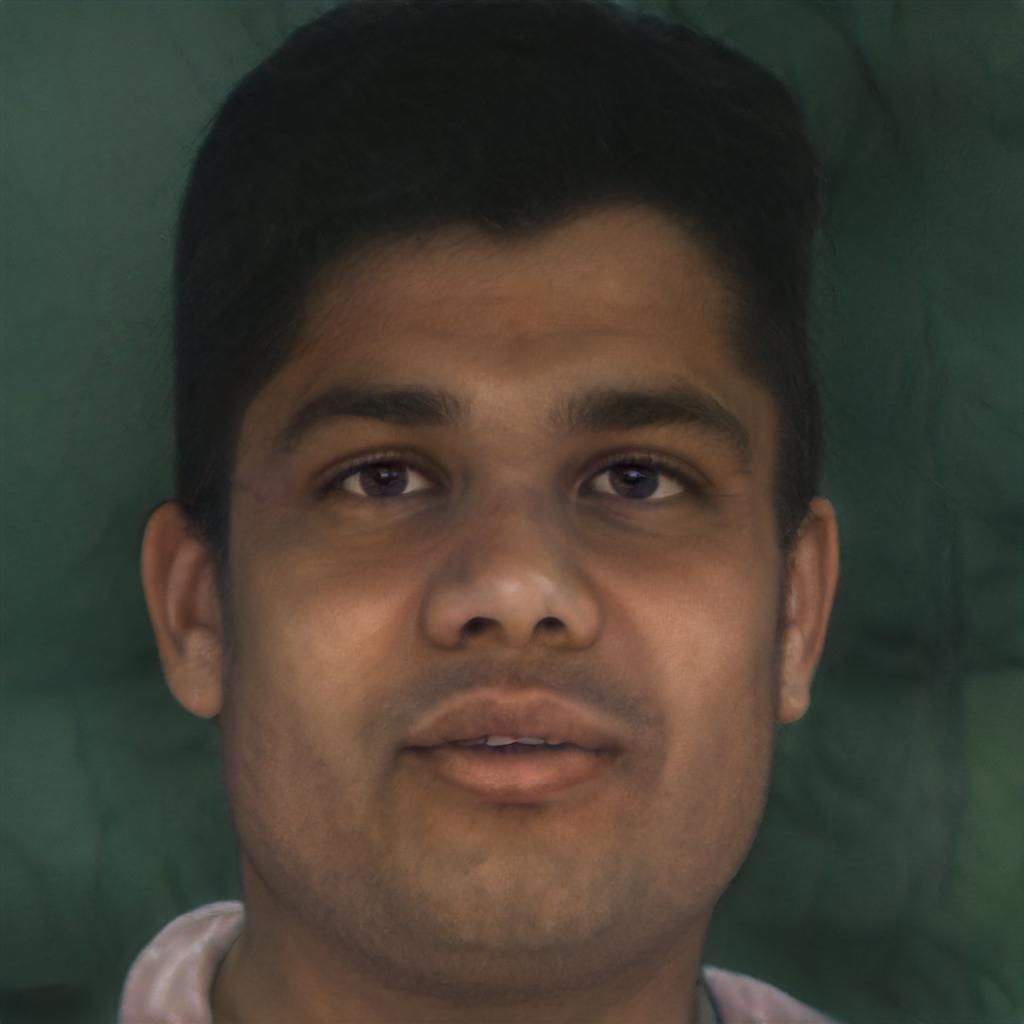}  & 
 \includegraphics[width=0.22\linewidth]{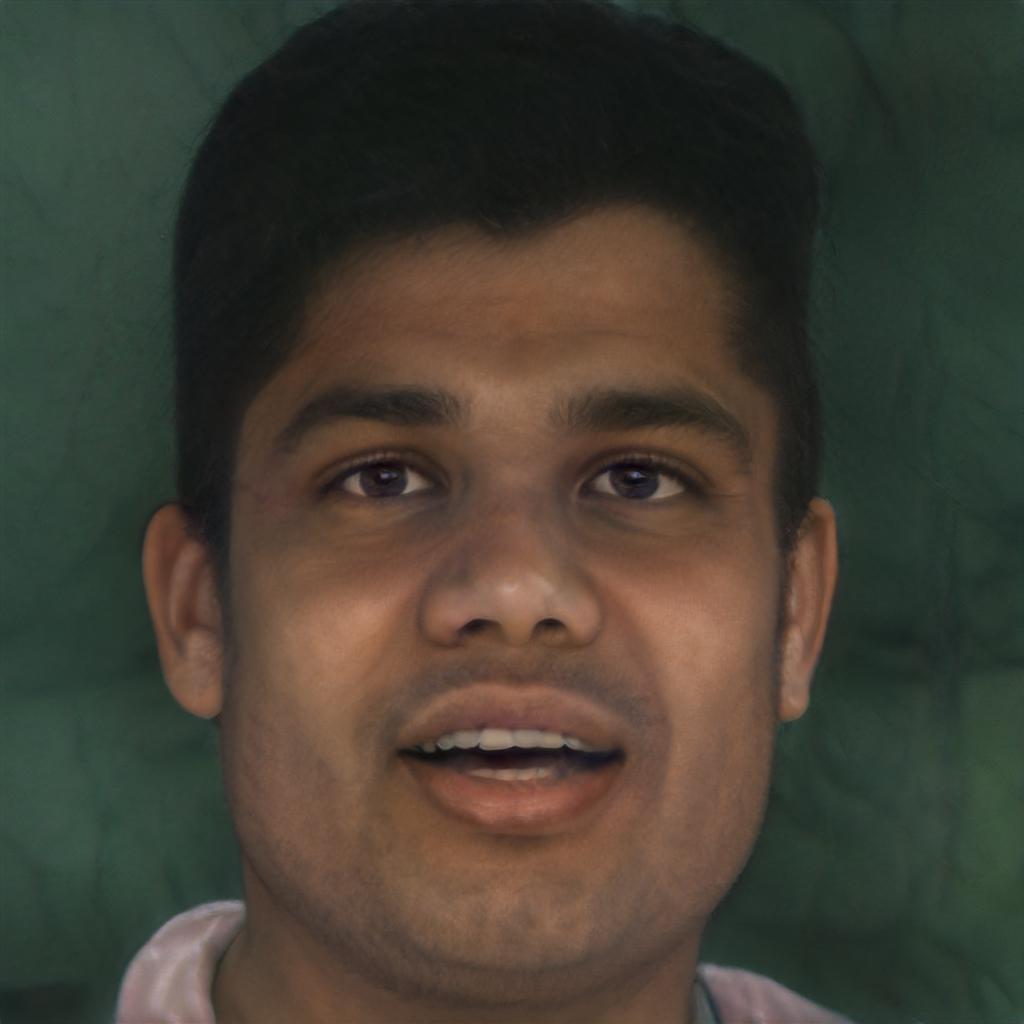}  \\ 
 
 \begin{turn}{90}\hspace{1cm} VTM  \end{turn}&
 \begin{turn}{90}\hspace{0.5cm} BPP=0.00326 \end{turn}&
 \includegraphics[width=0.22\linewidth]{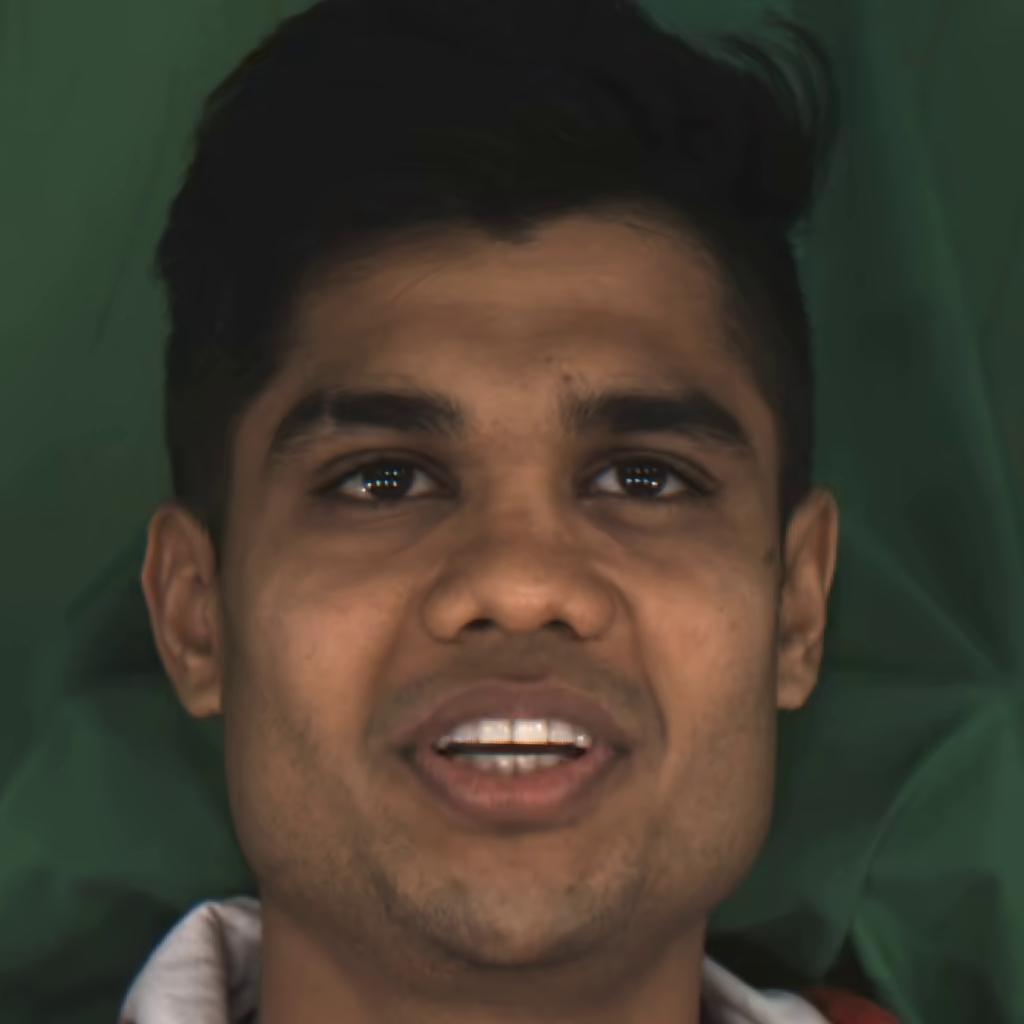}  & 
 \includegraphics[width=0.22\linewidth]{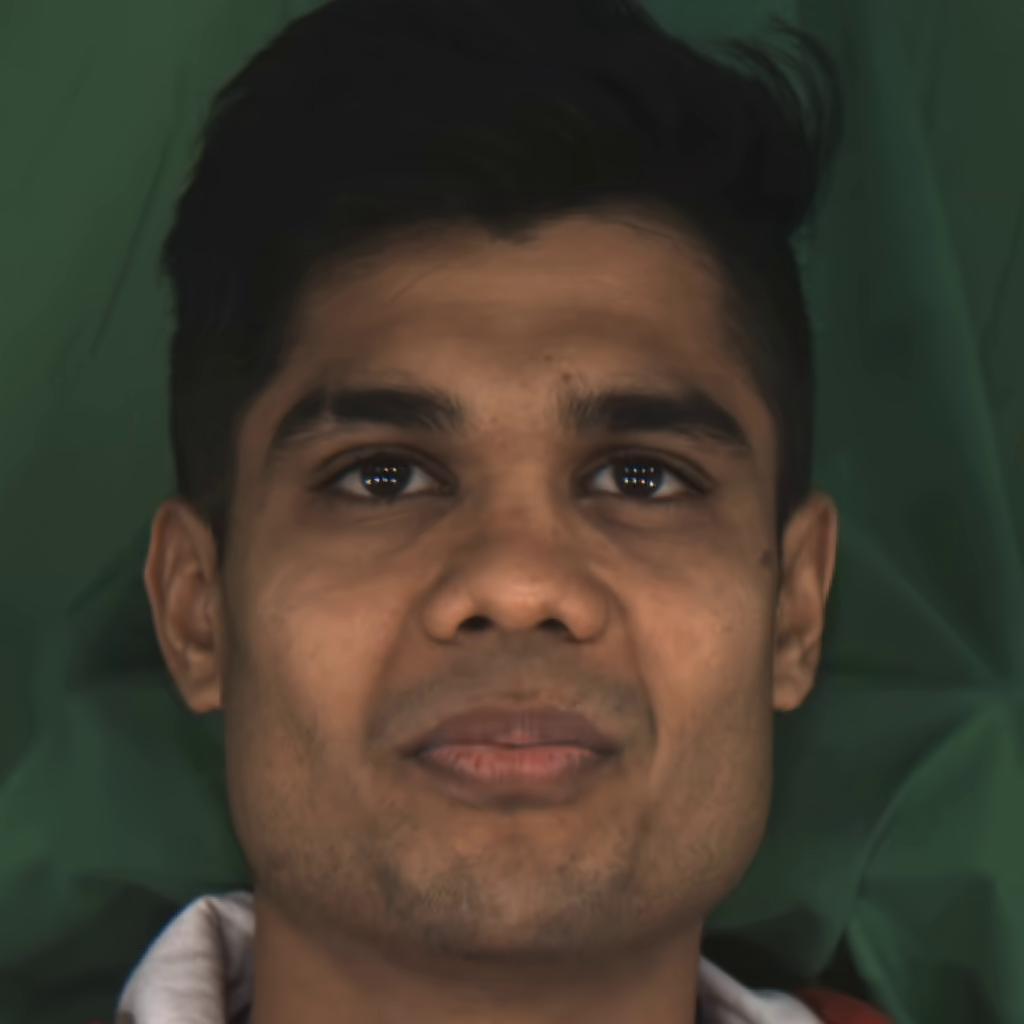}  &
 \includegraphics[width=0.22\linewidth]{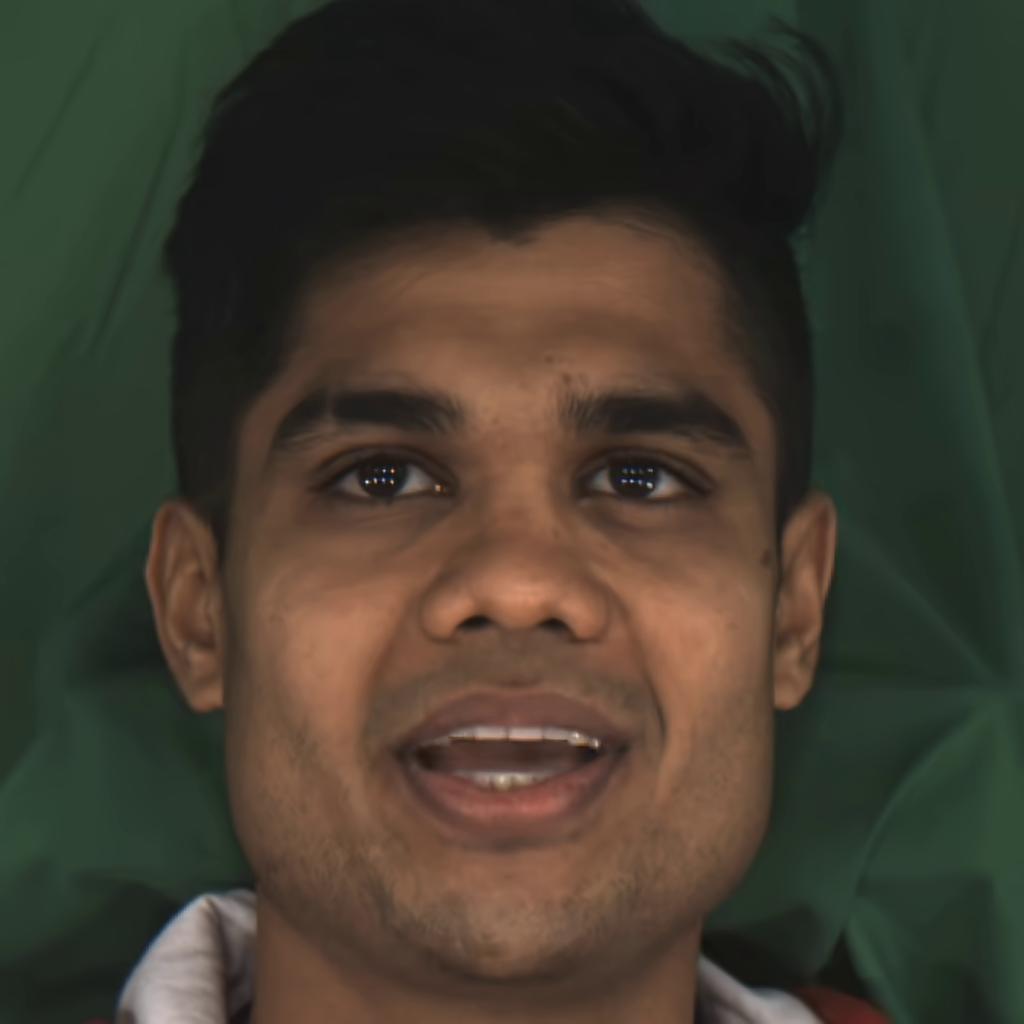}  \\ 

 \begin{turn}{90}\hspace{1cm} H.265 \end{turn} &
 \begin{turn}{90}\hspace{0.5cm} BPP=0.00351\end{turn} &
 \includegraphics[width=0.22\linewidth]{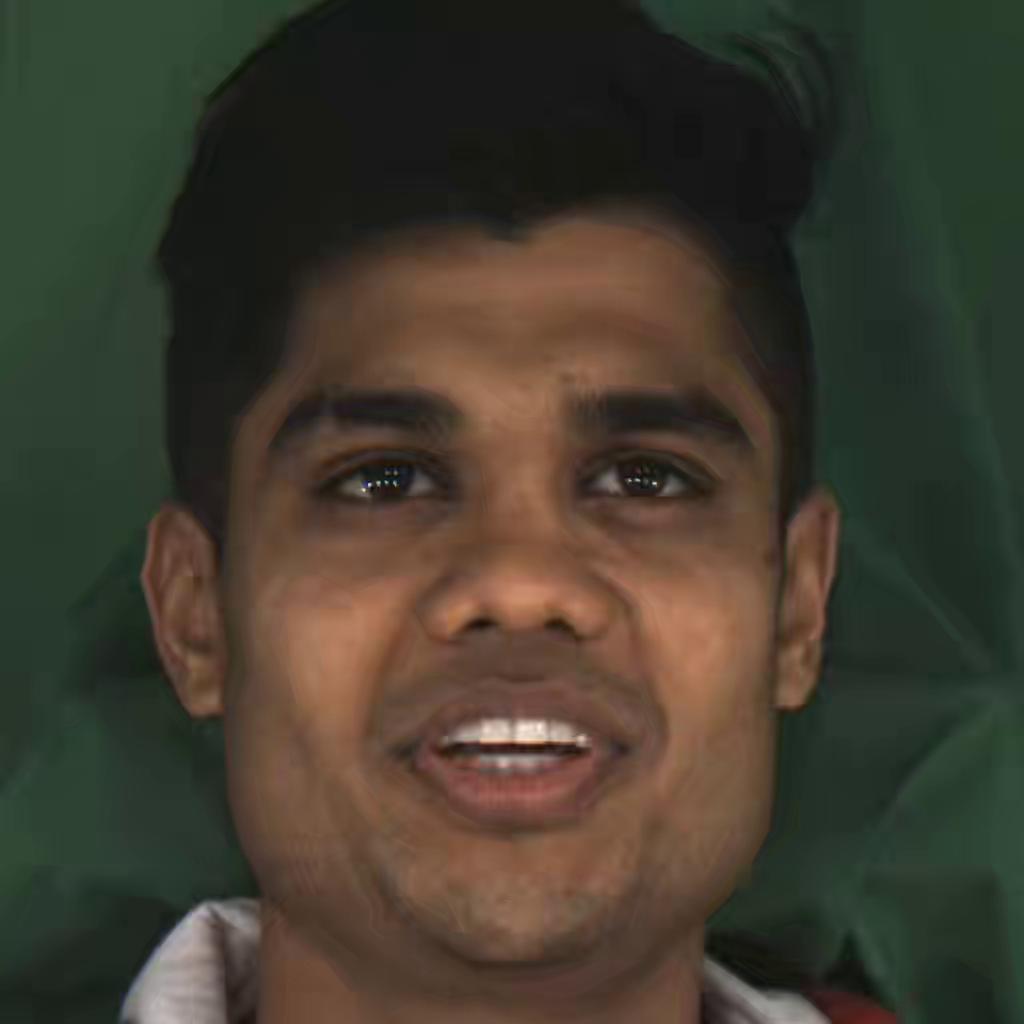}  & 
 \includegraphics[width=0.22\linewidth]{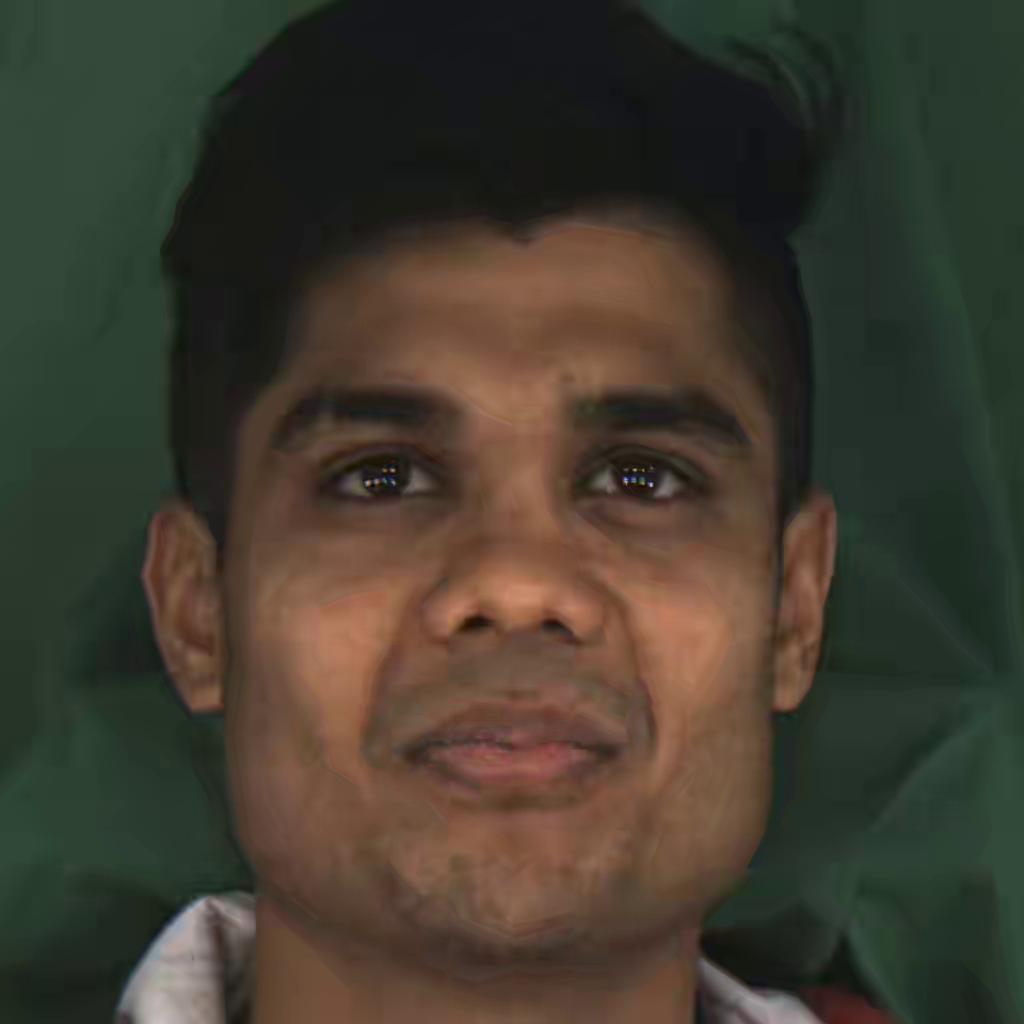}  &
 \includegraphics[width=0.22\linewidth]{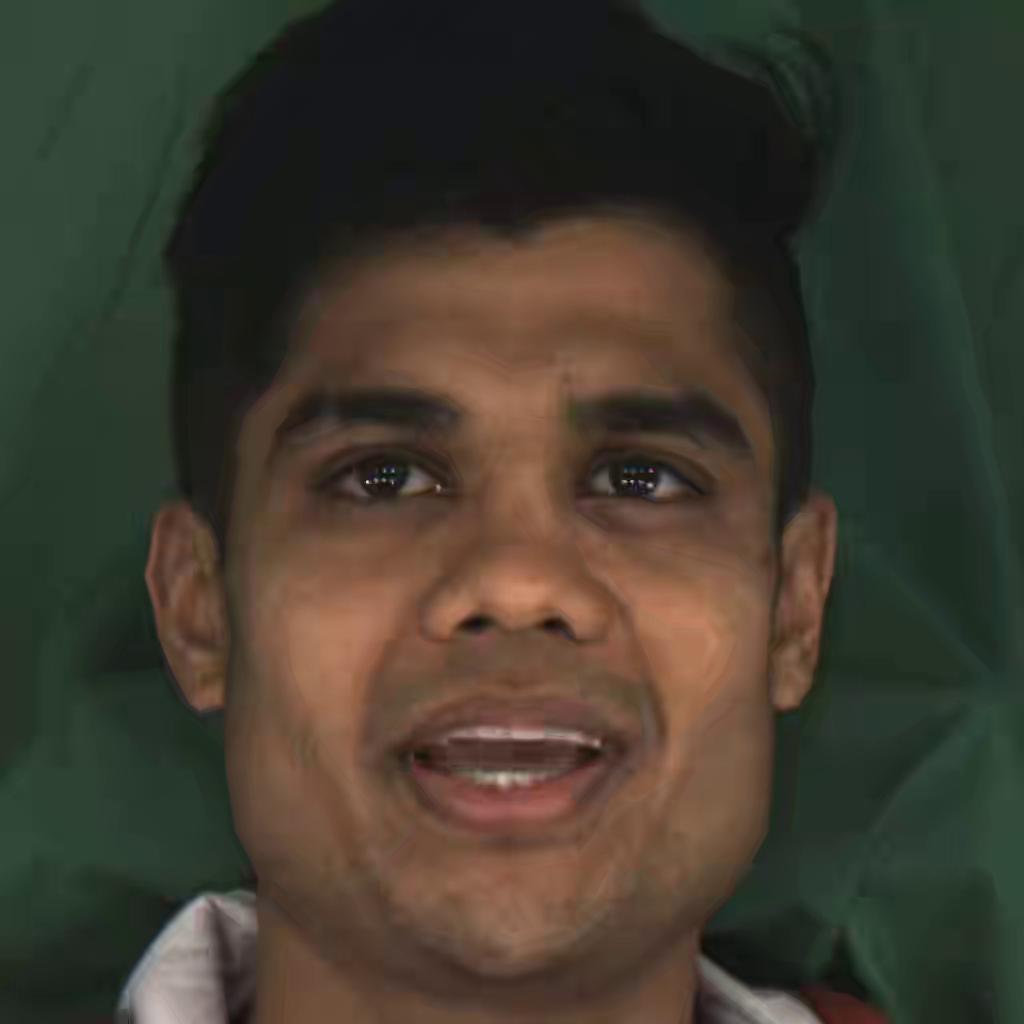}  \\ 
 \\
\bottomrule
\end{tabular}
\caption{Qualitative results on MEAD inter dataset: H.265 shows blocking artifacts and blurring, VTM shows blurring especially at the edges of the face and the hair while our method (SGANC) is almost artifacts free and with high quality images.}
\label{fig:qual_results_medium_mead_inter_3}
\end{figure*}

\begin{figure*}[h]
\setlength\tabcolsep{2pt}%
\centering
\begin{tabular}{p{0.5cm}cccc}
\toprule
 \begin{turn}{90} \hspace{1.2cm} Original \end{turn}  &
 &
 \includegraphics[width=0.22\linewidth]{images/inter_qualitative/W016/orig/frame0002.jpg}  & 
 \includegraphics[width=0.22\linewidth]{images/inter_qualitative/W016/orig/frame0006.jpg}  & 
 \includegraphics[width=0.22\linewidth]{images/inter_qualitative/W016/orig/frame0030.jpg}  \\ 

 \begin{turn}{90}\hspace{1.2cm} Projected\end{turn} &
 &
 \includegraphics[width=0.22\linewidth]{images/inter_qualitative/W016/projected/frame0002.jpg}  & 
 \includegraphics[width=0.22\linewidth]{images/inter_qualitative/W016/projected/frame0006.jpg}  & 
 \includegraphics[width=0.22\linewidth]{images/inter_qualitative/W016/projected/frame0030.jpg}  \\

\begin{turn}{90}\hspace{0.5cm} SGANC IC \end{turn} &
\begin{turn}{90}\hspace{0.5cm} BPP=0.00289 \end{turn} &
 \includegraphics[width=0.22\linewidth]{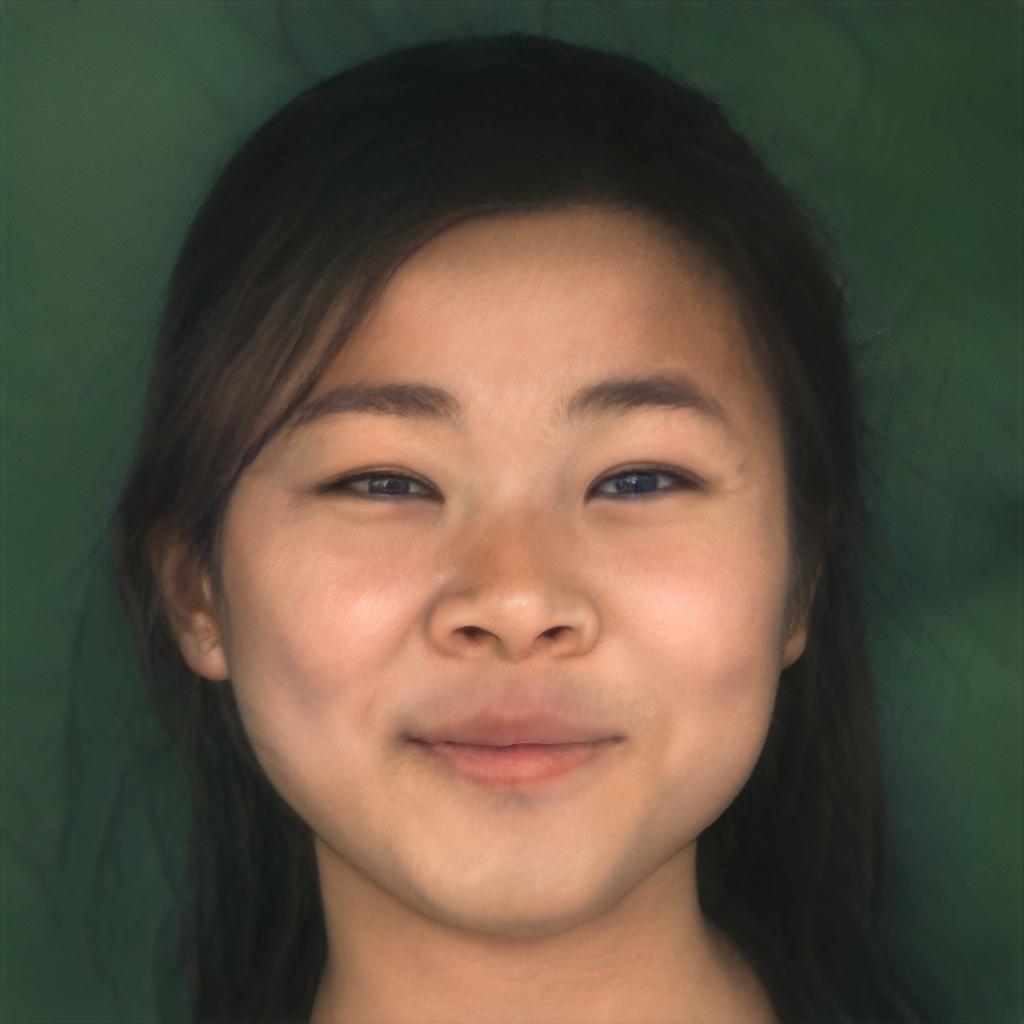}  & 
 \includegraphics[width=0.22\linewidth]{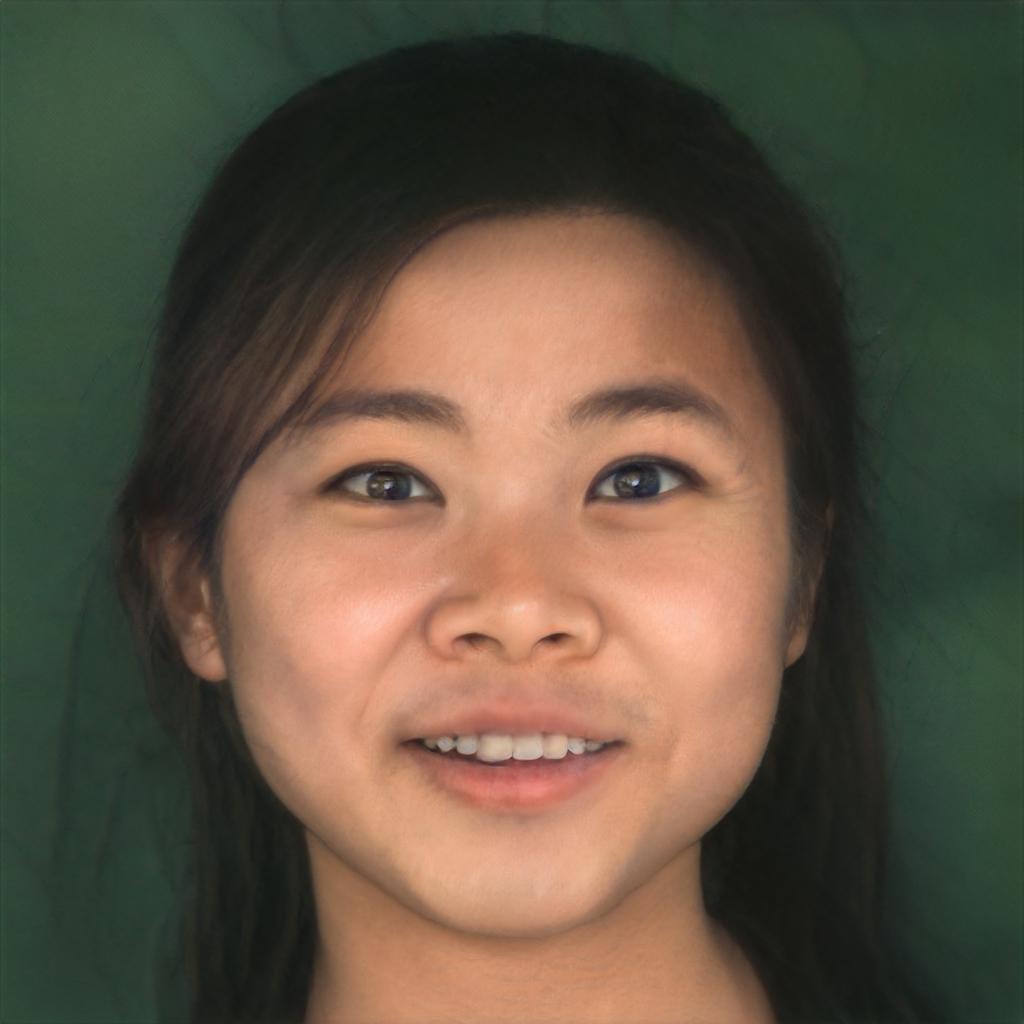}  & 
 \includegraphics[width=0.22\linewidth]{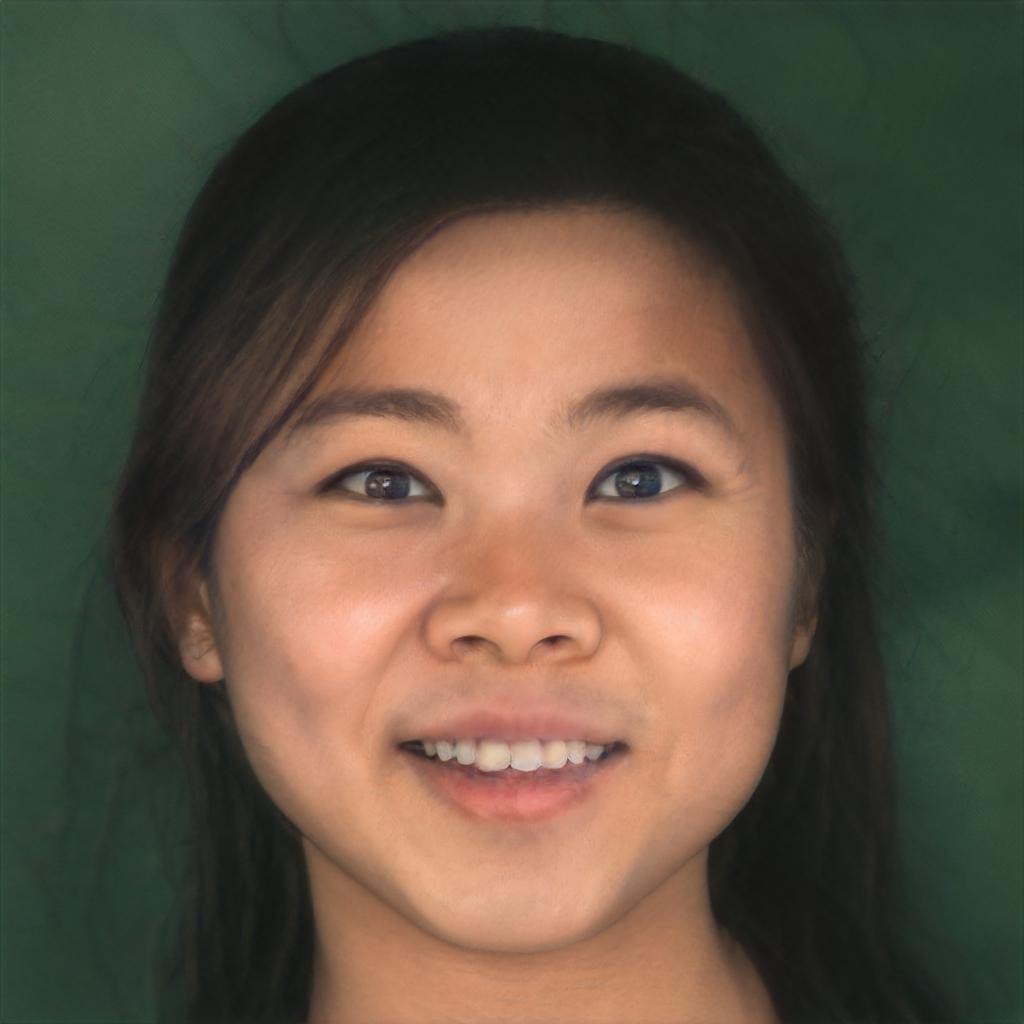}  \\
 
 \begin{turn}{90}\hspace{1cm} VTM  \end{turn}&
 \begin{turn}{90}\hspace{0.5cm} BPP=0.00446 \end{turn}&
 \includegraphics[width=0.22\linewidth]{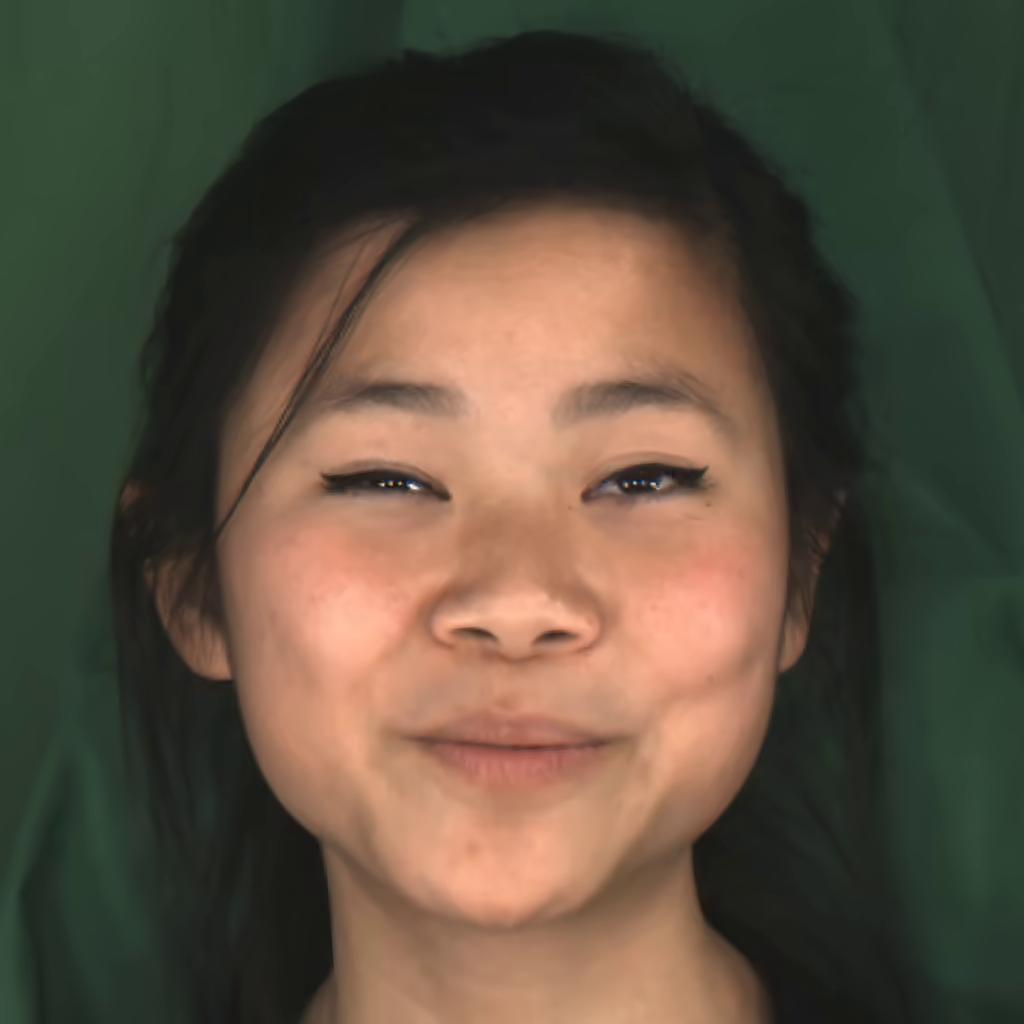}  & 
 \includegraphics[width=0.22\linewidth]{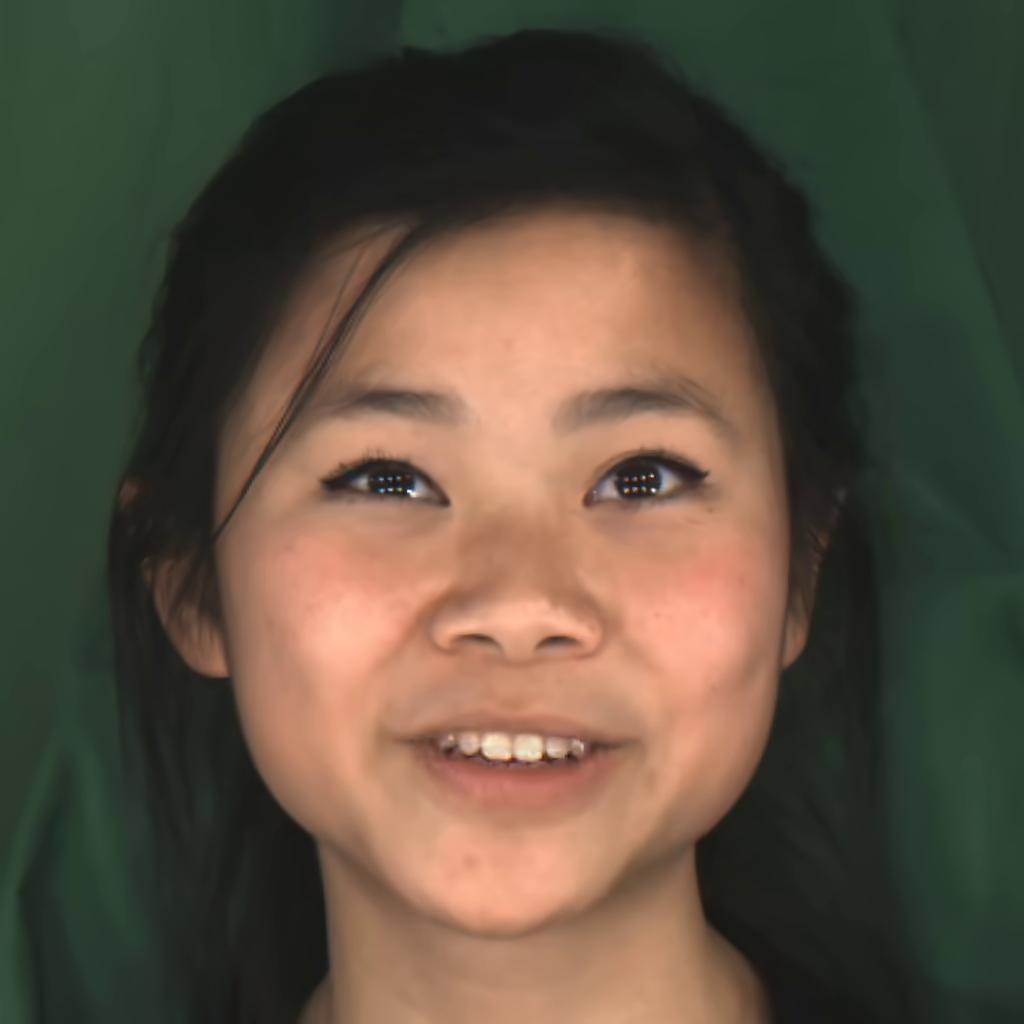}  & 
 \includegraphics[width=0.22\linewidth]{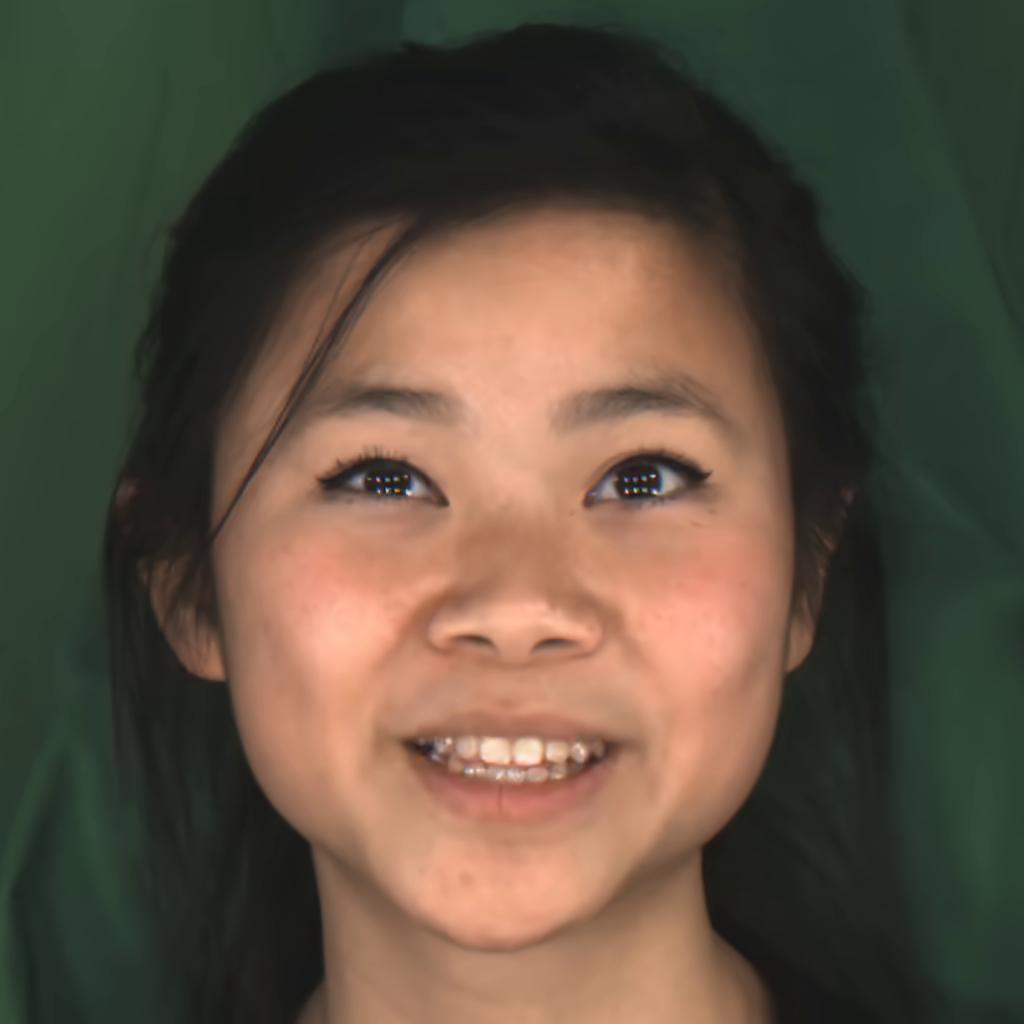}  \\ 

 \begin{turn}{90}\hspace{1cm} H.265 \end{turn} &
 \begin{turn}{90}\hspace{0.5cm} BPP=0.00437\end{turn}&
 \includegraphics[width=0.22\linewidth]{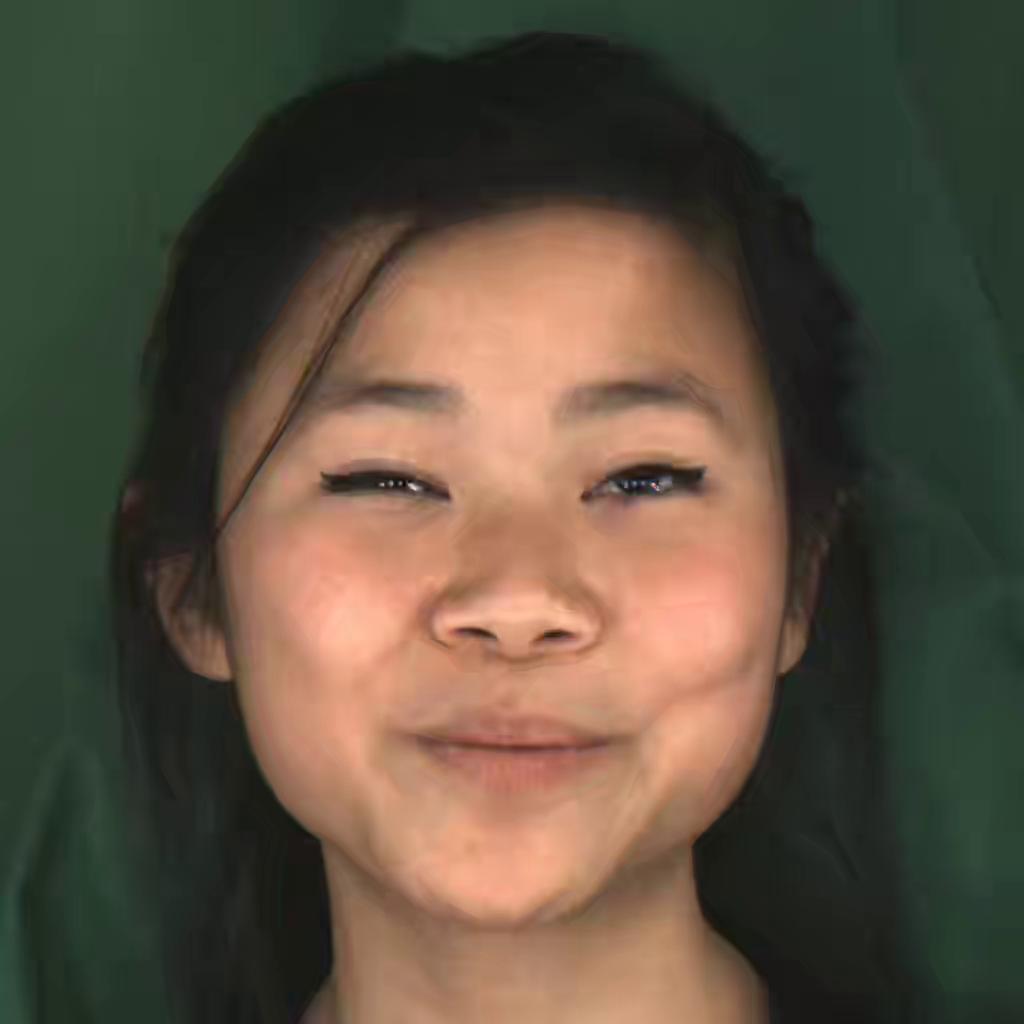}  & 
 \includegraphics[width=0.22\linewidth]{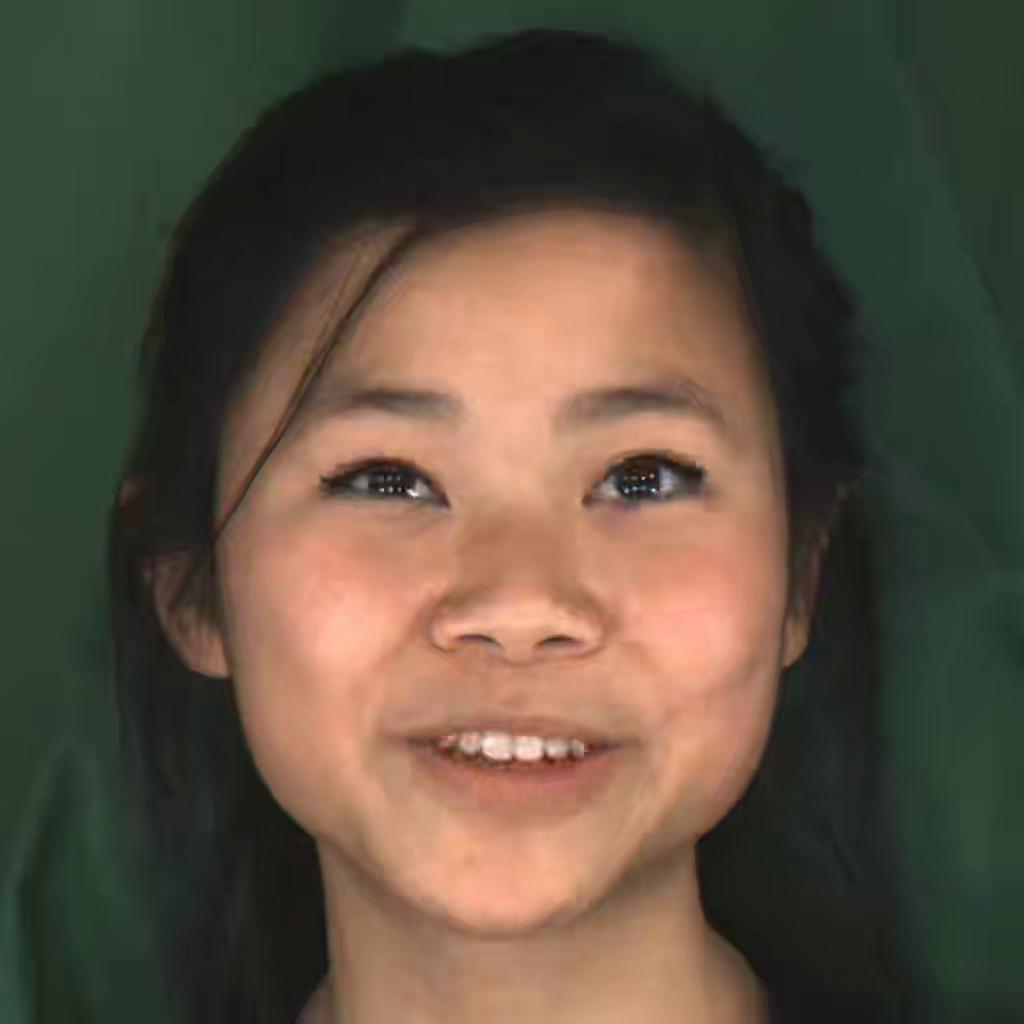}  & 
 \includegraphics[width=0.22\linewidth]{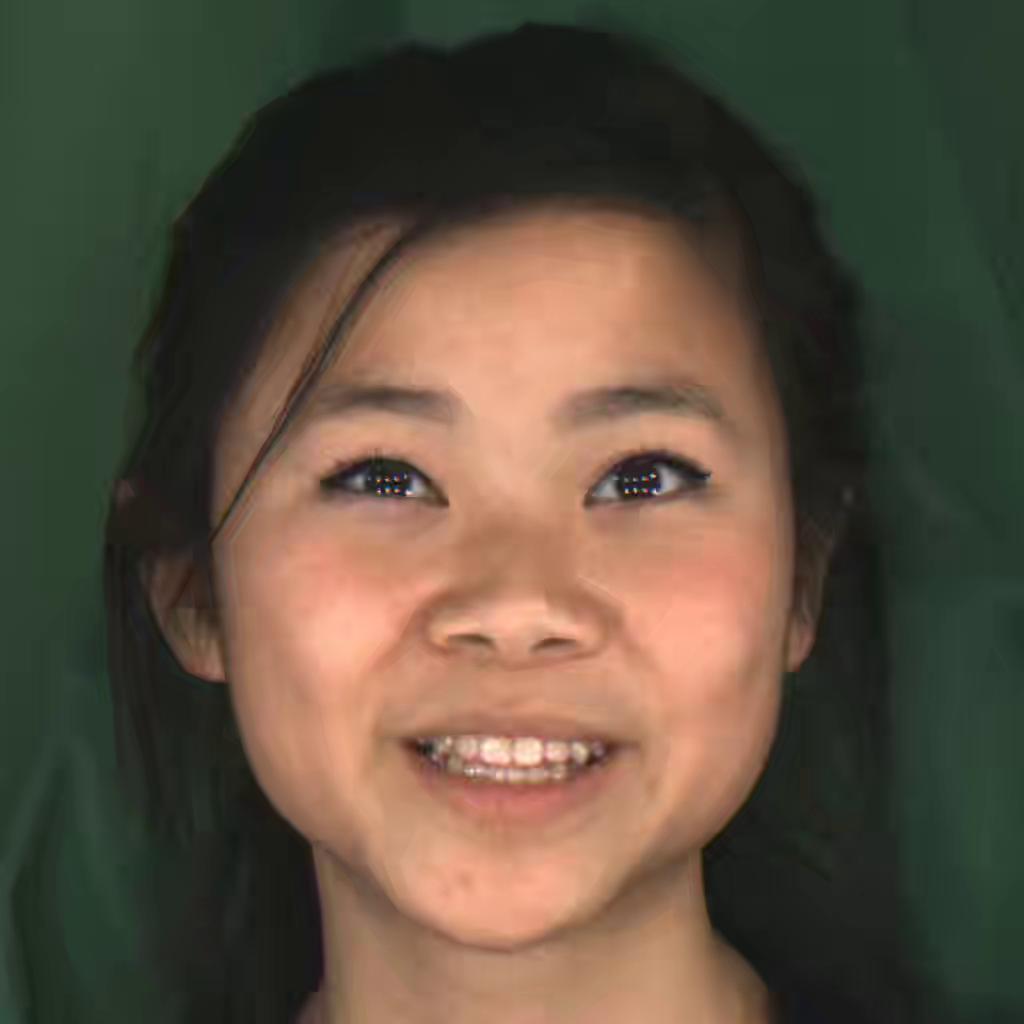}  \\ 
 \\
\bottomrule
\end{tabular}
\caption{Qualitative results on MEAD inter dataset: H.265 shows blocking artifacts and blurring, VTM shows blurring especially at the edges of the face and the hair while our method (SGANC) is almost artifacts free and with high quality images.}
\label{fig:qual_results_medium_mead_inter_4}
\end{figure*}

From Figures \ref{fig:app_qual_results_low_mead_inter_4_main}, \ref{fig:qual_results_low_mead_inter_1}, \ref{fig:qual_results_low_mead_inter_2}, \ref{fig:qual_results_low_mead_inter_3} and \ref{fig:qual_results_low_mead_inter_4}, we can notice that VTM and H.265 introduces blocking and blurring artifacts which is not case for our approach SGANC IC. We can also notice similar observations in Figures \ref{fig:qual_results_medium_mead_inter_1}, \ref{fig:qual_results_medium_mead_inter_2}, \ref{fig:qual_results_medium_mead_inter_3} and \ref{fig:qual_results_medium_mead_inter_4}.\\
From Figure \ref{fig:results_filmpac_inter}, we can notice that our methods are better perceptually (LPIPS and MS-SSIM) than VTM and H.265.



\clearpage

\newpage
\newpage
\newpage
\bibliographystyle{IEEEbib}
\bibliography{main_arxiv}

\end{document}

%% file: misc/user_colors.tex
\usepackage{xcolor}
\usepackage{xparse}

\usepackage[normalem]{ulem}
\usepackage{ifthen}
\newboolean{final}
\setboolean{final}{false}

\newcommand{\FontFormatter}[2]{%
    \textcolor[rgb]{#1}{\textbf{#2}}%
}

\NewDocumentCommand{\TextFormatter}{ommm}{%
    \FontFormatter{#4}{%
        \IfNoValueTF{#1}{%
            $\scriptstyle \overset{#3}{++}$ #2%
        }{%
            \sout{#1} $\scriptstyle \xrightarrow{#3}$ #2%
        }%
    }%
}

\NewDocumentCommand{\TextFormatterFinal}{ommm}{#2}

\NewDocumentCommand{\pierre}{om}{%
    \TextFormatter[#1]{#2}{Pierre}{1.0, 0.0, 0.5}%
}

\NewDocumentCommand{\ms}{om}{%
    \TextFormatter[#1]{#2}{Mustafa}{0.0, 0.5, 0.0}%
}

\NewDocumentCommand{\bharath}{om}{%
\TextFormatter[#1]{#2}{Bharath}{0.0, 0.0, 1.0}%
}

\NewDocumentCommand{\xu}{om}{%
\TextFormatter[#1]{#2}{Xu}{0.5, 0.0, 1.0}%
}

\NewDocumentCommand{\pat}{om}{%
    \TextFormatter[#1]{#2}{Patent}{0.5, 0.5, 0.0}%
}

%% file: main_arxiv.bbl
\begin{thebibliography}{10}

\bibitem{arithm_code}
J.~Rissanen and G.~G. Langdon,
\newblock ``Arithmetic coding,''
\newblock {\em IBM Journal of Research and Development}, vol. 23, no. 2, pp.
  149--162, 1979.

\bibitem{entropy_information}
Robert~M. Gray,
\newblock {\em Entropy and Information Theory},
\newblock Springer Publishing Company, Incorporated, 2nd edition, 2011.

\bibitem{shanon}
C.~E. Shannon,
\newblock ``A mathematical theory of communication,''
\newblock {\em The Bell System Technical Journal}, vol. 27, no. 3, pp.
  379--423, 1948.

\bibitem{Hu2022}
Hu~Xueqi, Huang Qiusheng, Shi Zhengyi, Li~Siyuan, Gao Changxin, Sun Li, and
  Li~Qingli,
\newblock ``Style transformer for image inversion and editing,''
\newblock {\em arXiv preprint arXiv:2203.07932}, 2022.

\bibitem{Dinh2022}
Dinh Tan~M., Tran Anh~Tuan, Nguyen Rang, and Hua Binh-Son,
\newblock ``Hyperinverter: Improving stylegan inversion via hypernetwork,''
\newblock {\em arXiv preprint arXiv:2112.00719}, 2022.

\bibitem{encodinginstyle}
Elad Richardson, Yuval Alaluf, Or~Patashnik, Yotam Nitzan, Yaniv Azar, Stav
  Shapiro, and Daniel Cohen-Or,
\newblock ``Encoding in style: a stylegan encoder for image-to-image
  translation,''
\newblock in {\em Proceedings of the IEEE/CVF Conference on Computer Vision and
  Pattern Recognition}, 2021, pp. 2287--2296.

\bibitem{wei2021simplebase}
Tianyi Wei, Dongdong Chen, Wenbo Zhou, Jing Liao, Weiming Zhang, Lu~Yuan, Gang
  Hua, and Nenghai Yu,
\newblock ``A simple baseline for stylegan inversion,''
\newblock {\em arXiv preprint arXiv:2104.07661}, 2021.

\bibitem{styleganimporveing}
Tero Karras, Samuli Laine, Miika Aittala, Janne Hellsten, Jaakko Lehtinen, and
  Timo Aila,
\newblock ``Analyzing and improving the image quality of stylegan,''
\newblock in {\em Proceedings of the IEEE/CVF Conference on Computer Vision and
  Pattern Recognition}, 2020, pp. 8110--8119.

\bibitem{sauer2022stylegan}
Axel Sauer, Katja Schwarz, and Andreas Geiger,
\newblock ``Stylegan-xl: Scaling stylegan to large diverse datasets,''
\newblock {\em arXiv preprint arXiv:2202.00273}, 2022.

\bibitem{JPEG}
G.~K. Wallace,
\newblock ``The jpeg still picture compression standard,''
\newblock {\em IEEE Trans. on Consum. Electron.}, vol. 38, no. 1, pp.
  xviii–xxxiv, Feb. 1992.

\bibitem{JPEG2000}
Majid Rabbani,
\newblock ``{JPEG2000: Image Compression Fundamentals, Standards and
  Practice},''
\newblock {\em Journal of Electronic Imaging}, vol. 11, no. 2, 2002.

\bibitem{balle2017end_iclr}
Johannes Ball{\'e}, Valero Laparra, and Eero~P Simoncelli,
\newblock ``End-to-end optimized image compression,''
\newblock in {\em 5th International Conference on Learning Representations,
  ICLR 2017}, 2017.

\bibitem{balle2017end}
Johannes Ball{\'e}, Valero Laparra, and Eero~P Simoncelli,
\newblock ``End-to-end optimization of nonlinear transform codes for perceptual
  quality,''
\newblock in {\em 2016 Picture Coding Symposium (PCS)}. IEEE, 2016, pp. 1--5.

\bibitem{ganextreme}
Eirikur Agustsson, Michael Tschannen, Fabian Mentzer, Radu Timofte, and Luc
  Van~Gool,
\newblock ``Generative adversarial networks for extreme learned image
  compression,''
\newblock in {\em 2019 IEEE/CVF International Conference on Computer Vision
  (ICCV)}, 2019, pp. 221--231.

\bibitem{mentzer2020highgancomp}
Fabian Mentzer, George~D Toderici, Michael Tschannen, and Eirikur Agustsson,
\newblock ``High-fidelity generative image compression,''
\newblock {\em Advances in Neural Information Processing Systems}, vol. 33,
  2020.

\bibitem{minnen2018joint}
David Minnen, Johannes Ball{\'e}, and George Toderici,
\newblock ``Joint autoregressive and hierarchical priors for learned image
  compression,''
\newblock in {\em NeurIPS}, 2018.

\bibitem{cheng2020learned}
Zhengxue Cheng, Heming Sun, Masaru Takeuchi, and Jiro Katto,
\newblock ``Learned image compression with discretized gaussian mixture
  likelihoods and attention modules,''
\newblock in {\em Proceedings of the IEEE/CVF Conference on Computer Vision and
  Pattern Recognition}, 2020, pp. 7939--7948.

\bibitem{balle2018variational}
Johannes Ball{\'e}, David Minnen, Saurabh Singh, Sung~Jin Hwang, and Nick
  Johnston,
\newblock ``Variational image compression with a scale hyperprior,''
\newblock in {\em International Conference on Learning Representations}, 2018.

\bibitem{li2018learning}
Mu~Li, Wangmeng Zuo, Shuhang Gu, Debin Zhao, and David Zhang,
\newblock ``Learning convolutional networks for content-weighted image
  compression,''
\newblock in {\em Proceedings of the IEEE Conference on Computer Vision and
  Pattern Recognition}, 2018, pp. 3214--3223.

\bibitem{cheng2018deep}
Zhengxue Cheng, Heming Sun, Masaru Takeuchi, and Jiro Katto,
\newblock ``Deep convolutional autoencoder-based lossy image compression,''
\newblock in {\em 2018 Picture Coding Symposium (PCS)}. IEEE, 2018, pp.
  253--257.

\bibitem{cheng2019deep}
Zhengxue Cheng, Heming Sun, Masaru Takeuchi, and Jiro Katto,
\newblock ``Deep residual learning for image compression.,''
\newblock in {\em CVPR Workshops}, 2019, p.~0.

\bibitem{toderici2017full}
George Toderici, Damien Vincent, Nick Johnston, Sung Jin~Hwang, David Minnen,
  Joel Shor, and Michele Covell,
\newblock ``Full resolution image compression with recurrent neural networks,''
\newblock in {\em Proceedings of the IEEE Conference on Computer Vision and
  Pattern Recognition}, 2017, pp. 5306--5314.

\bibitem{johnston2018improved}
Nick Johnston, Damien Vincent, David Minnen, Michele Covell, Saurabh Singh,
  Troy Chinen, Sung~Jin Hwang, Joel Shor, and George Toderici,
\newblock ``Improved lossy image compression with priming and spatially
  adaptive bit rates for recurrent networks,''
\newblock in {\em Proceedings of the IEEE Conference on Computer Vision and
  Pattern Recognition}, 2018, pp. 4385--4393.

\bibitem{gan}
Ian Goodfellow, Jean Pouget-Abadie, Mehdi Mirza, Bing Xu, David Warde-Farley,
  Sherjil Ozair, Aaron Courville, and Yoshua Bengio,
\newblock ``Generative adversarial nets,''
\newblock {\em Advances in neural information processing systems}, vol. 27,
  2014.

\bibitem{rippel2017real}
Oren Rippel and Lubomir Bourdev,
\newblock ``Real-time adaptive image compression,''
\newblock in {\em International Conference on Machine Learning}. PMLR, 2017,
  pp. 2922--2930.

\bibitem{generativecompression}
Shibani Santurkar, David Budden, and Nir Shavit,
\newblock ``Generative compression,''
\newblock in {\em 2018 Picture Coding Symposium (PCS)}, 2018, pp. 258--262.

\bibitem{tschannen2018deep}
Michael Tschannen, Eirikur Agustsson, and Mario Lucic,
\newblock ``Deep generative models for distribution-preserving lossy
  compression,''
\newblock in {\em Proceedings of the 32nd International Conference on Neural
  Information Processing Systems}, 2018, pp. 5933--5944.

\bibitem{msssim_met}
Z.~Wang, E.P. Simoncelli, and A.C. Bovik,
\newblock ``Multiscale structural similarity for image quality assessment,''
\newblock in {\em The Thrity-Seventh Asilomar Conference on Signals, Systems
  Computers, 2003}, 2003, vol.~2, pp. 1398--1402 Vol.2.

\bibitem{zhang2018unreasonable}
Richard Zhang, Phillip Isola, Alexei~A Efros, Eli Shechtman, and Oliver Wang,
\newblock ``The unreasonable effectiveness of deep features as a perceptual
  metric,''
\newblock in {\em Proceedings of the IEEE conference on computer vision and
  pattern recognition}, 2018, pp. 586--595.

\bibitem{johnson2016perceptual}
Justin Johnson, Alexandre Alahi, and Li~Fei-Fei,
\newblock ``Perceptual losses for real-time style transfer and
  super-resolution,''
\newblock in {\em European conference on computer vision}. Springer, 2016, pp.
  694--711.

\bibitem{superresolution}
Christian Ledig, Lucas Theis, Ferenc Huszár, Jose Caballero, Andrew
  Cunningham, Alejandro Acosta, Andrew Aitken, Alykhan Tejani, Johannes Totz,
  Zehan Wang, and Wenzhe Shi,
\newblock ``Photo-realistic single image super-resolution using a generative
  adversarial network,''
\newblock in {\em 2017 IEEE Conference on Computer Vision and Pattern
  Recognition (CVPR)}, 2017, pp. 105--114.

\bibitem{dosovitskiy2016generating}
Alexey Dosovitskiy and Thomas Brox,
\newblock ``Generating images with perceptual similarity metrics based on deep
  networks,''
\newblock {\em Advances in neural information processing systems}, vol. 29, pp.
  658--666, 2016.

\bibitem{styletransferpercept}
Leon~A. Gatys, Alexander~S. Ecker, and Matthias Bethge,
\newblock ``Image style transfer using convolutional neural networks,''
\newblock in {\em 2016 IEEE Conference on Computer Vision and Pattern
  Recognition (CVPR)}, 2016, pp. 2414--2423.

\bibitem{chen2020perceptually}
Li-Heng Chen, Christos~G Bampis, Zhi Li, Andrey Norkin, and Alan~C Bovik,
\newblock ``Perceptually optimizing deep image compression,''
\newblock {\em arXiv preprint arXiv:2007.02711}, 2020.

\bibitem{h264}
T.~Wiegand, G.J. Sullivan, G.~Bjontegaard, and A.~Luthra,
\newblock ``Overview of the h.264/avc video coding standard,''
\newblock {\em IEEE Transactions on Circuits and Systems for Video Technology},
  vol. 13, no. 7, pp. 560--576, 2003.

\bibitem{hevc}
Gary~J. Sullivan, Jens-Rainer Ohm, Woo-Jin Han, and Thomas Wiegand,
\newblock ``Overview of the high efficiency video coding (hevc) standard,''
\newblock {\em IEEE Transactions on Circuits and Systems for Video Technology},
  vol. 22, no. 12, pp. 1649--1668, 2012.

\bibitem{lu2019dvc}
Guo Lu, Wanli Ouyang, Dong Xu, Xiaoyun Zhang, Chunlei Cai, and Zhiyong Gao,
\newblock ``Dvc: An end-to-end deep video compression framework,''
\newblock in {\em Proceedings of the IEEE/CVF Conference on Computer Vision and
  Pattern Recognition}, 2019, pp. 11006--11015.

\bibitem{dosovitskiy2015flownet}
Alexey Dosovitskiy, Philipp Fischer, Eddy Ilg, Philip Hausser, Caner Hazirbas,
  Vladimir Golkov, Patrick Van Der~Smagt, Daniel Cremers, and Thomas Brox,
\newblock ``Flownet: Learning optical flow with convolutional networks,''
\newblock in {\em Proceedings of the IEEE international conference on computer
  vision}, 2015, pp. 2758--2766.

\bibitem{lin2020m}
Jianping Lin, Dong Liu, Houqiang Li, and Feng Wu,
\newblock ``M-lvc: multiple frames prediction for learned video compression,''
\newblock in {\em Proceedings of the IEEE/CVF Conference on Computer Vision and
  Pattern Recognition}, 2020, pp. 3546--3554.

\bibitem{hu2020improving}
Zhihao Hu, Zhenghao Chen, Dong Xu, Guo Lu, Wanli Ouyang, and Shuhang Gu,
\newblock ``Improving deep video compression by resolution-adaptive flow
  coding,''
\newblock in {\em European Conference on Computer Vision}. Springer, 2020, pp.
  193--209.

\bibitem{feng2020learned}
Runsen Feng, Yaojun Wu, Zongyu Guo, Zhizheng Zhang, and Zhibo Chen,
\newblock ``Learned video compression with feature-level residuals,''
\newblock in {\em Proceedings of the IEEE/CVF Conference on Computer Vision and
  Pattern Recognition Workshops}, 2020, pp. 120--121.

\bibitem{interframe}
Abdelaziz Djelouah, Joaquim Campos, Simone Schaub-Meyer, and Christopher
  Schroers,
\newblock ``Neural inter-frame compression for video coding,''
\newblock in {\em 2019 IEEE/CVF International Conference on Computer Vision
  (ICCV)}, 2019, pp. 6420--6428.

\bibitem{hu2021fvc}
Zhihao Hu, Guo Lu, and Dong Xu,
\newblock ``Fvc: A new framework towards deep video compression in feature
  space,''
\newblock in {\em Proceedings of the IEEE/CVF Conference on Computer Vision and
  Pattern Recognition}, 2021, pp. 1502--1511.

\bibitem{Liu_2021_CVPR}
Bowen Liu, Yu~Chen, Shiyu Liu, and Hun-Seok Kim,
\newblock ``Deep learning in latent space for video prediction and
  compression,''
\newblock in {\em Proceedings of the IEEE/CVF Conference on Computer Vision and
  Pattern Recognition (CVPR)}, June 2021, pp. 701--710.

\bibitem{wang2021one}
Ting-Chun Wang, Arun Mallya, and Ming-Yu Liu,
\newblock ``One-shot free-view neural talking-head synthesis for video
  conferencing,''
\newblock in {\em Proceedings of the IEEE/CVF Conference on Computer Vision and
  Pattern Recognition}, 2021, pp. 10039--10049.

\bibitem{Zhou_2021_CVPR}
Hang Zhou, Yasheng Sun, Wayne Wu, Chen~Change Loy, Xiaogang Wang, and Ziwei
  Liu,
\newblock ``Pose-controllable talking face generation by implicitly modularized
  audio-visual representation,''
\newblock in {\em Proceedings of the IEEE/CVF Conference on Computer Vision and
  Pattern Recognition (CVPR)}, June 2021, pp. 4176--4186.

\bibitem{Liuvideoappl}
Ming-Yu Liu, Xun Huang, Jiahui Yu, Ting-Chun Wang, and Arun Mallya,
\newblock ``Generative adversarial networks for image and video synthesis:
  Algorithms and applications,''
\newblock {\em Proceedings of the IEEE}, vol. 109, no. 5, pp. 839--862, 2021.

\bibitem{zakharov2019few}
Egor Zakharov, Aliaksandra Shysheya, Egor Burkov, and Victor Lempitsky,
\newblock ``Few-shot adversarial learning of realistic neural talking head
  models,''
\newblock in {\em Proceedings of the IEEE/CVF International Conference on
  Computer Vision}, 2019, pp. 9459--9468.

\bibitem{Zhang2020DAVDNetDA}
Xi~Zhang, Xiaolin Wu, Xinliang Zhai, Xianye Ben, and Chengjie Tu,
\newblock ``Davd-net: Deep audio-aided video decompression of talking heads,''
\newblock {\em 2020 IEEE/CVF Conference on Computer Vision and Pattern
  Recognition (CVPR)}, pp. 12332--12341, 2020.

\bibitem{lowbandcompress}
Dahu Feng, Yan Huang, Yiwei Zhang, Jun Ling, Anni Tang, and Li~Song,
\newblock ``A generative compression framework for low bandwidth video
  conference,''
\newblock in {\em 2021 IEEE International Conference on Multimedia Expo
  Workshops (ICMEW)}, 2021, pp. 1--6.

\bibitem{oquab2021low}
Maxime Oquab, Pierre Stock, Daniel Haziza, Tao Xu, Peizhao Zhang, Onur Celebi,
  Yana Hasson, Patrick Labatut, Bobo Bose-Kolanu, Thibault Peyronel, et~al.,
\newblock ``Low bandwidth video-chat compression using deep generative
  models,''
\newblock in {\em Proceedings of the IEEE/CVF Conference on Computer Vision and
  Pattern Recognition}, 2021, pp. 2388--2397.

\bibitem{papamakarios2019normalizing}
George Papamakarios, Eric Nalisnick, Danilo~Jimenez Rezende, Shakir Mohamed,
  and Balaji Lakshminarayanan,
\newblock ``Normalizing flows for probabilistic modeling and inference,''
\newblock {\em arXiv preprint arXiv:1912.02762}, 2019.

\bibitem{stylegan}
Tero Karras, Samuli Laine, and Timo Aila,
\newblock ``A style-based generator architecture for generative adversarial
  networks,''
\newblock in {\em Proceedings of the IEEE/CVF Conference on Computer Vision and
  Pattern Recognition}, 2019, pp. 4401--4410.

\bibitem{radford2016unsupervised}
Alec {Radford}, Luke {Metz}, and Soumith {Chintala},
\newblock ``Unsupervised representation learning with deep convolutional
  generative adversarial networks,''
\newblock in {\em ICLR 2016 : International Conference on Learning
  Representations 2016}, 2016.

\bibitem{pim}
Sangnie Bhardwaj, Ian Fischer, Johannes Ball\'{e}, and Troy Chinen,
\newblock ``An unsupervised information-theoretic perceptual quality metric,''
\newblock in {\em Advances in Neural Information Processing Systems},
  H.~Larochelle, M.~Ranzato, R.~Hadsell, M.~F. Balcan, and H.~Lin, Eds. 2020,
  vol.~33, pp. 13--24, Curran Associates, Inc.

\bibitem{karras2018progressive}
Tero Karras, Timo Aila, Samuli Laine, and Jaakko Lehtinen,
\newblock ``Progressive growing of gans for improved quality, stability, and
  variation,''
\newblock in {\em International Conference on Learning Representations}, 2018.

\bibitem{filmpac}
Filmpac,
\newblock ``{Filmpac database},'' \url{https://filmpac.com/}, 2022.

\bibitem{mead}
Kaisiyuan Wang, Qianyi Wu, Linsen Song, Zhuoqian Yang, Wayne Wu, Chen Qian, Ran
  He, Yu~Qiao, and Chen~Change Loy,
\newblock ``Mead: A large-scale audio-visual dataset for emotional talking-face
  generation,''
\newblock in {\em European Conference on Computer Vision}. Springer, 2020, pp.
  700--717.

\bibitem{realnvp}
Laurent Dinh, Jascha Sohl-Dickstein, and Samy Bengio,
\newblock ``Density estimation using real nvp,''
\newblock {\em ICLR 2017: International Conference on Learning
  Representations}, 2017.

\bibitem{begaint2020compressai}
Jean B{\'e}gaint, Fabien Racap{\'e}, Simon Feltman, and Akshay Pushparaja,
\newblock ``Compressai: a pytorch library and evaluation platform for
  end-to-end compression research,''
\newblock {\em arXiv preprint arXiv:2011.03029}, 2020.

\bibitem{duda2013asymmetric}
Jarek Duda,
\newblock ``Asymmetric numeral systems: entropy coding combining speed of
  huffman coding with compression rate of arithmetic coding,''
\newblock {\em arXiv preprint arXiv:1311.2540}, 2013.

\bibitem{VTM}
.,
\newblock ``{Versatile Video Coding},'' \url{https://jvet.hhi.fraunhofer.de/},
  2020.

\bibitem{h265}
Gary~J. Sullivan, Jens-Rainer Ohm, Woo-Jin Han, and Thomas Wiegand,
\newblock ``Overview of the high efficiency video coding (hevc) standard,''
\newblock {\em IEEE Transactions on Circuits and Systems for Video Technology},
  vol. 22, no. 12, pp. 1649--1668, 2012.

\bibitem{AV1}
AOM AV1,
\newblock ``{Versatile Video Coding},''
  \url{https://aomedia.googlesource.com/aom/}, 2018.

\bibitem{oneshotgit}
One-ShotFree-ViewNeuralTalkingHeadSynthesis,
\newblock
  ``\url{https://github.com/zhanglonghao1992/One-ShotFree-ViewNeuralTalkingHeadSynthesis},''
  2021.

\bibitem{dvc_git}
DVC,
\newblock ``\url{https://github.com/GuoLusjtu/DVC},'' 2019.

\bibitem{sergei_belousov}
Sergei Belousov,
\newblock ``Mobilestylegan: {A} lightweight convolutional neural network for
  high-fidelity image synthesis,''
\newblock {\em CoRR}, vol. abs/2104.04767, 2021.

\bibitem{BELOUSOV}
Sergei Belousov,
\newblock ``Mobilestylegan.pytorch: Pytorch-based toolkit to compress stylegan2
  model,''
\newblock {\em Software Impacts}, vol. 10, pp. 100115, 2021.

\bibitem{kim_2021}
Hyunsu Kim, Yunjey Choi, Junho Kim, Sungjoo Yoo, and Youngjung Uh,
\newblock ``Exploiting spatial dimensions of latent in gan for real-time image
  editing,''
\newblock in {\em 2021 IEEE/CVF Conference on Computer Vision and Pattern
  Recognition (CVPR)}, 2021, pp. 852--861.

\bibitem{karras2021alias}
Tero Karras, Miika Aittala, Samuli Laine, Erik H{\"a}rk{\"o}nen, Janne
  Hellsten, Jaakko Lehtinen, and Timo Aila,
\newblock ``Alias-free generative adversarial networks,''
\newblock {\em arXiv preprint arXiv:2106.12423}, 2021.

\bibitem{johnson1995continuous}
N.L. Johnson, S.~Kotz, and N.~Balakrishnan,
\newblock {\em Continuous Univariate Distributions, Volume 2},
\newblock Wiley Series in Probability and Statistics. Wiley, 1995.

\end{thebibliography}
